\documentclass[acmtog, nonacm]{acmart}
\acmSubmissionID{208}

\usepackage{booktabs} %

\usepackage{color}
\usepackage[english]{babel}
\usepackage{multirow}
\usepackage{makecell}
\usepackage{graphicx}
\usepackage{url}
\usepackage{subfigure}
\usepackage{dsfont}

\usepackage{amssymb}
\usepackage{hyperref}
\usepackage[toc,page]{appendix}
\usepackage{amsmath}
\usepackage{arydshln}
\usepackage{float}
\usepackage{amsmath}
\usepackage{amssymb}
\usepackage{amsmath,tikz}
\usetikzlibrary{arrows,chains,matrix,positioning,scopes}
\usepackage{booktabs}
\usepackage{natbib}
\usepackage{lipsum}
\usepackage[ruled]{algorithm2e} %

\newcommand{\tabincell}[2]{\begin{tabular}
{@{}#1@{}}#2\end{tabular}}

\SetAlFnt{\small}
\SetAlCapFnt{\small}
\SetAlCapNameFnt{\small}
\SetAlCapHSkip{0pt}

\acmJournal{TOG}

\newcommand{\gl}[1]{{\color{black}#1}}

\newcommand{\wt}[1]{{\color{black}#1}}
\newcommand{\yj}[1]{{\color{black}#1}}
\newcommand{\yyj}[1]{{\color{black}#1}}

\newcommand{\rz}[1]{{\color{black}#1}}
\newcommand{\rzn}[1]{{\color{black}#1}}
\newcommand{\rznn}[1]{{\color{black}#1}}

\newcommand{\togwt}[1]{{\color{black}#1}}

\begin{document}
\title{TM-NET: Deep Generative Networks for Textured Meshes}

\author{Lin Gao}
\affiliation{%
\institution{ICT, CAS and UCAS}
\country{}}
\email{gaolin@ict.ac.cn}

\author{Tong Wu}
\affiliation{%
	\institution{ICT, CAS and UCAS}
	\country{}}
\email{wutong@ict.ac.cn}

\author{Yu-Jie Yuan}
\affiliation{%
	\institution{ICT, CAS and UCAS}
	\country{}}
\email{yuanyujie@ict.ac.cn}

\author{Ming-Xian Lin}
\affiliation{%
	\institution{ICT, CAS and UCAS}
	\country{}}
\email{linmingxian20g@ict.ac.cn}

\author{Yu-Kun Lai}
\affiliation{%
	\institution{Cardiff University}
	\country{}}

\author{Hao Zhang}
\affiliation{%
	\institution{Simon Fraser University}
	\country{}}

\authorsaddresses{
This is the author's version of the work. It is posted here for your personal use. Not for redistribution. }
\begin{abstract}

We introduce TM-NET, a novel deep generative model for synthesizing %
{\em textured meshes\/} in a {\em part-aware\/} manner. Once trained, the network can generate novel textured meshes from scratch or predict textures for a given 3D mesh, \rznn{without image guidance.} Plausible and diverse textures can be generated \rznn{for the same mesh part, while
texture compatibility between parts in the same shape is achieved via conditional generation.}
Specifically, our method produces texture maps for individual shape parts, each as a deformable box, leading to a natural UV map with minimal distortion. The network {\em separately\/} embeds part geometry (via a PartVAE) and part texture (via a TextureVAE) into their respective latent spaces, so as to facilitate learning texture probability distributions conditioned on geometry. We introduce a {\em conditional
autoregressive model\/} for texture generation, which can be conditioned on both part geometry and textures already generated for
other parts to achieve texture \rznn{compatibility}.
To produce high-frequency texture details, our TextureVAE operates in a high-dimensional latent space via dictionary-based vector quantization. We also exploit transparencies in the texture as an effective means to model complex shape structures including topological details.
Extensive experiments demonstrate the plausibility, quality, and diversity of the textures and geometries generated by our network, while avoiding inconsistency issues that are common to novel view synthesis methods.

\end{abstract}


\maketitle

\section{Introduction}
\label{sec:intro}

With rapid advances in deep learning, many recent works propose deep neural networks to learn 3D shape representations~\cite{chen2019-IMNET,park2019-DeepSDF,mescheder2019-Occupancy,pageSAGnet19,AtlasNet2018,li_sig17,Wang2018ocnn,mo2019structurenet,gaosdmnet2019,BSPNet}, generations~\cite{chen2019-IMNET,gaosdmnet2019,3dgan,li_sig17,pageSAGnet19}, and reconstructions from different sources such as images~\cite{Matryoshka,wang2018pixel2mesh,li2021d2im}, point scans~\cite{AtlasNet2018,park2019-DeepSDF}, and sketches~\cite{DeepSketch2Face,lun2017}.
However, state-of-the-art methods still cannot produce 3D shapes with high visual quality, in terms of the realism and richness of the shapes' appearance.
The generation and reconstruction of surface and topological details are especially difficult, due to lack of data and the technical and computational challenges in training deep models for high-resolution geometries.

\begin{figure}[!t]
\centering{\includegraphics[width=\linewidth]{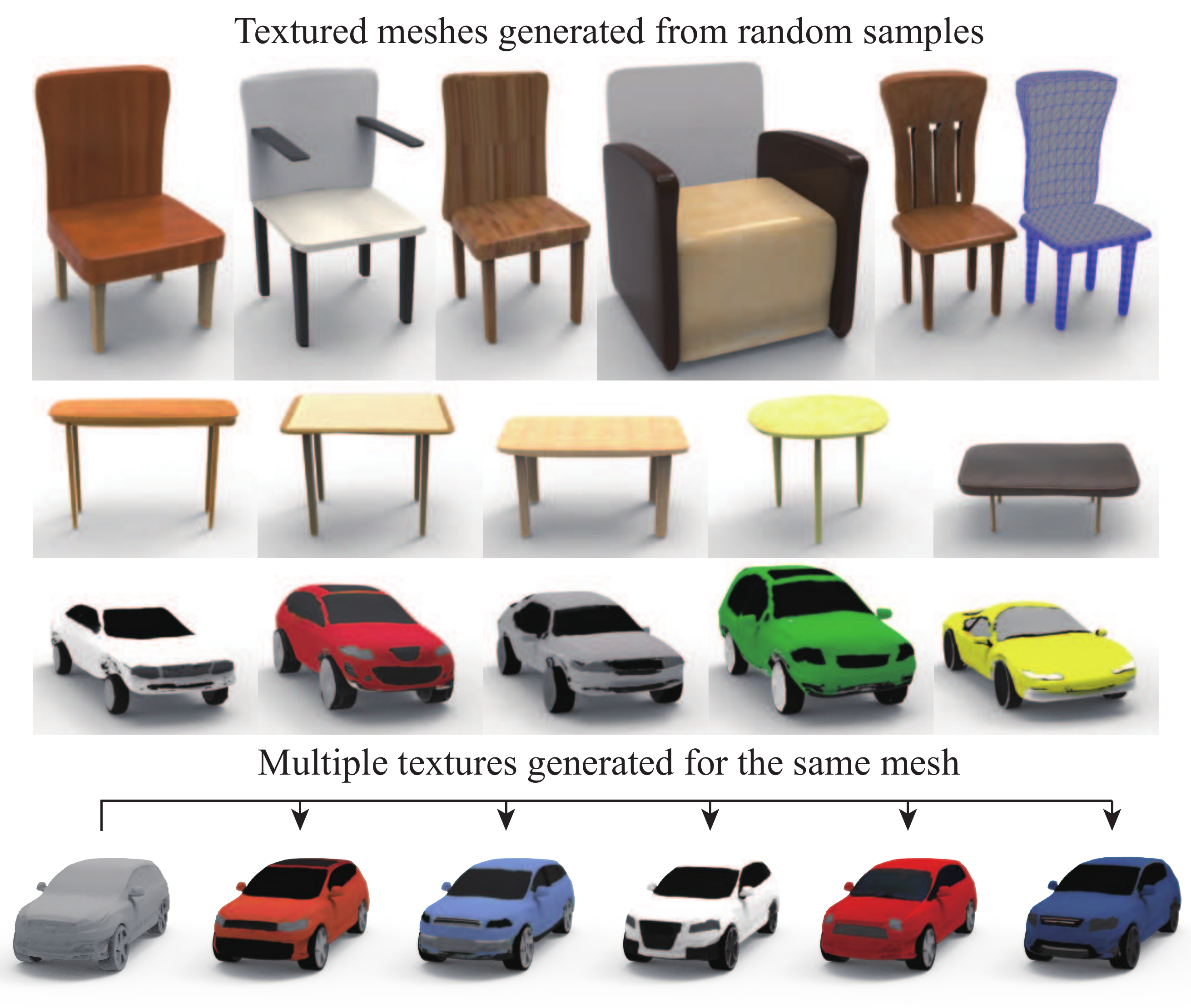}}
\caption{Our deep generative network, TM-NET, \rznn{can generate textured meshes from {\em random samples\/} in a latent space (top) and synthesize {\em multiple\/} textures for the same input shape (bottom).
Our texture generation is part-aware, where different shape parts may be textured differently. Furthermore, even if the underlying mesh is relatively low-resolution (e.g., the top right chair has $<4$K vertices), the generated textures can exhibit the {\em appearance\/} of topological details (e.g., holes between the slats on the chair back).}}
\label{fig:teaser}
\end{figure}

\begin{figure*}[t!]
  \includegraphics[width=0.95\linewidth]{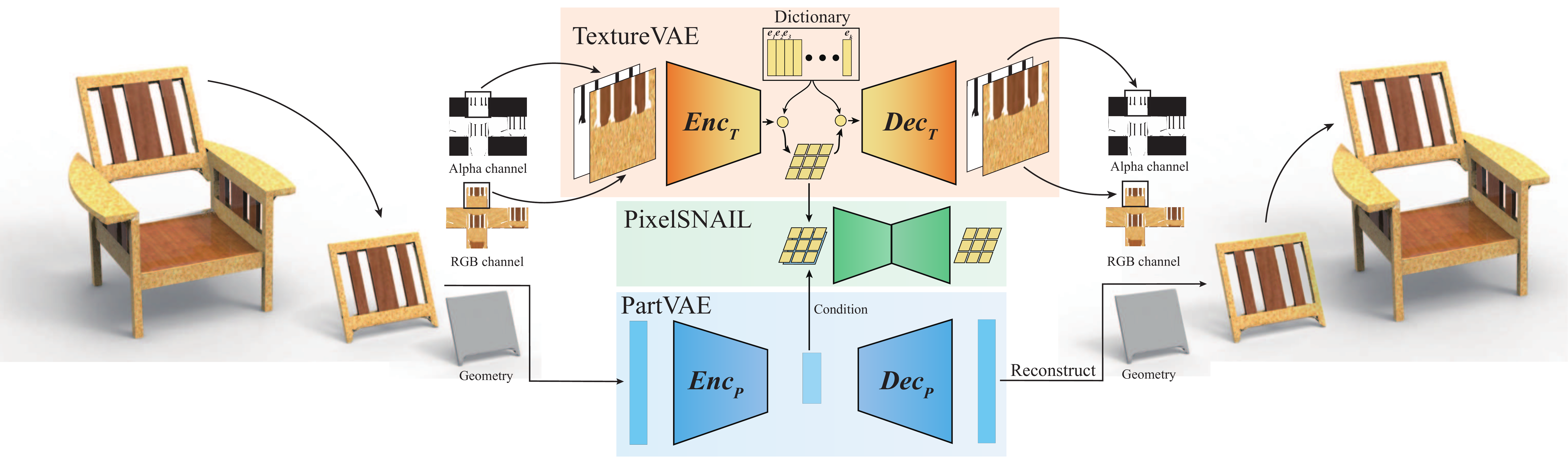}
  \caption{\rzn{An overview of the key components of TM-NET, for textured mesh generation. Each part is encoded using two Variational Autoencoders (VAEs): PartVAE for geometry with $Enc_P$ as the encoder and $Dec_P$ as the decoder, and TextureVAE for texture with $Enc_T$ as the encoder and $Dec_T$ as the decoder.
  For texture generation, TM-NET designs a conditional autoregressive generative model, %
  which takes the latent vector of PartVAE as condition input and outputs discrete feature maps. These feature maps are decoded as texture images for the input mesh geometry.}}
  \label{fig:overview}
\end{figure*}

In the absence of geometric details, a classical graphics ``trick'' is to create the {\em appearance\/} of surface richness
by simulating details in the {\em image space\/}, e.g., via texture or bump mapping. As shown in Figure~\ref{fig:teaser},
textured meshes can exhibit the kind of realism and visual fidelity that geometry alone cannot convey, even when the
underlying mesh resolution is fairly low. However, despite the flourishing of geometric deep learning,
there have been few attempts at developing a deep generative model for {\em textured\/} meshes. To the best of our
knowledge, existing works on textured shape generation have predominantly relied on single- or multi-view image
guidance~\cite{liu2019softras,chen2019dibrender,Raj_2019_CVPR_Workshops}.

In this paper, we introduce TM-NET, a deep generative network for {\em textured meshes},
aiming to fill several gaps in learning-based realistic 3D shape modeling.
First, we seek a {\em generic\/} decoder which can generate textured meshes with or without image guidance.
Next, the network can generate novel textured meshes, \rznn{e.g., from random samples in some latent space,} and
predict textures for a given 3D shape. Last but not least, the generative model should be {\em part-based\/} to allow
different shape parts to be textured differently, which is naturally expected for many real-world objects.

With these goals in mind, our design for TM-NET draws inspiration from SDM-NET~\cite{gaosdmnet2019},
a deep generative network for {\em structured deformable meshes\/}. Specifically, SDM-NET generates meshes formed by parts,
where each part is homeomorphic to a cuboid box and finer-scale geometry of the part can be controlled by deforming the template box.
Architecturally, the network consists of a part-level variational autoencoder (VAE) for learning genus-zero mesh parts and
a structure-part VAE (SP-VAE), which jointly learns part structures and part geometries from a shape collection.

\begin{figure}[t!]
  \includegraphics[width=1\linewidth]{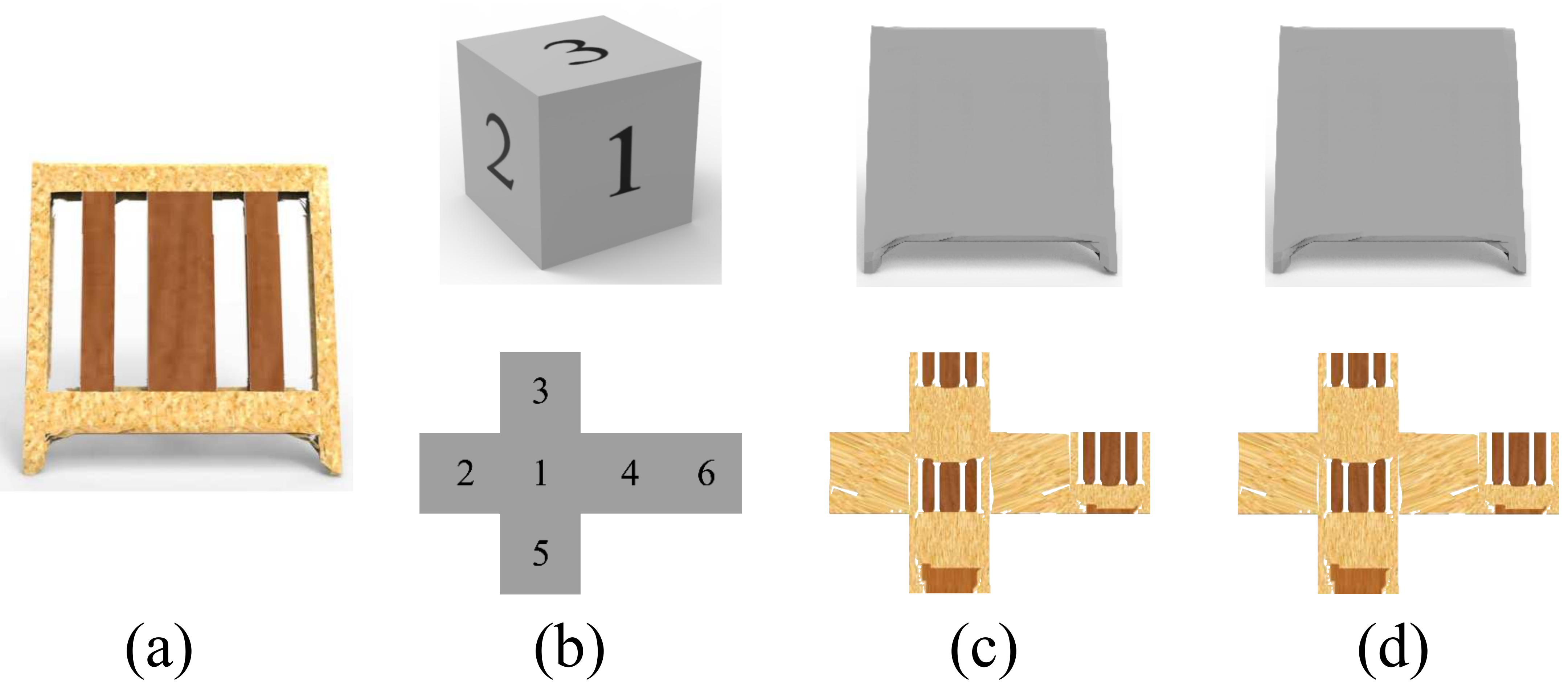}
  \caption{Texture representation with deformed boxes: (a) input model; (b) template box and unfolded UV map where numbers on the box faces help illustrate the unfolding process; (c) geometry and texture representations with the deformable box; (d) decoded geometry and texture.}%
  \label{fig:Representation}
\end{figure}

Figure~\ref{fig:overview} shows the key components of our network TM-NET, with the dataset and training processes detailed in Section~\ref{sec:imp}. Similar to SDM-NET, our model learns a structured mesh representation, where a textured mesh is represented at both the part level and the overall shape level using VAEs.
This allows both geometry and texture to be represented in a well-aligned canonical form: the rich geometry details are approximated and represented by deformations enabled by the template boxes, and these boxes work naturally with the UV-maps for texture mapping. Specifically, we can cut designated edges of a template box and flatten the box by unfolding it onto the UV space that contains texture information; see Figure~\ref{fig:Representation}, for example. The collection of deformable boxes, when mapped to the parameter space, forms a consistent texture atlas even for shapes with substantial structure and/or geometry variations, making effective learning of structured and textured shapes possible.

\rzn{However, the generation of plausible, high-quality, {\em and diverse\/} textured meshes} poses several new challenges:
\begin{itemize}
\item First, the mapping between shape parts and textures is generally far from one-to-one --- the same shape part may possess drastically
different textures and vice versa. \rzn{Thus a joint learning of part geometry and part texture, which resembles the SP-VAE in SDM-NET, would be unsuitable. We need a different and more adaptive encoding approach.}

\item \rzn{Second, a critical requirement to part-based texture representation is to ensure that \rznn{{\em compatible\/}} textures are generated for different parts. Hence, we must add a new layer of dependency, via conditional modeling, into our network design.}

\item Third, to produce textures with high-frequency details, we must address the tendency of VAEs to generate blurry outputs~\cite{VAE-blurry}.

\end{itemize}

\begin{figure}[t!]
  \includegraphics[width=0.99\linewidth]{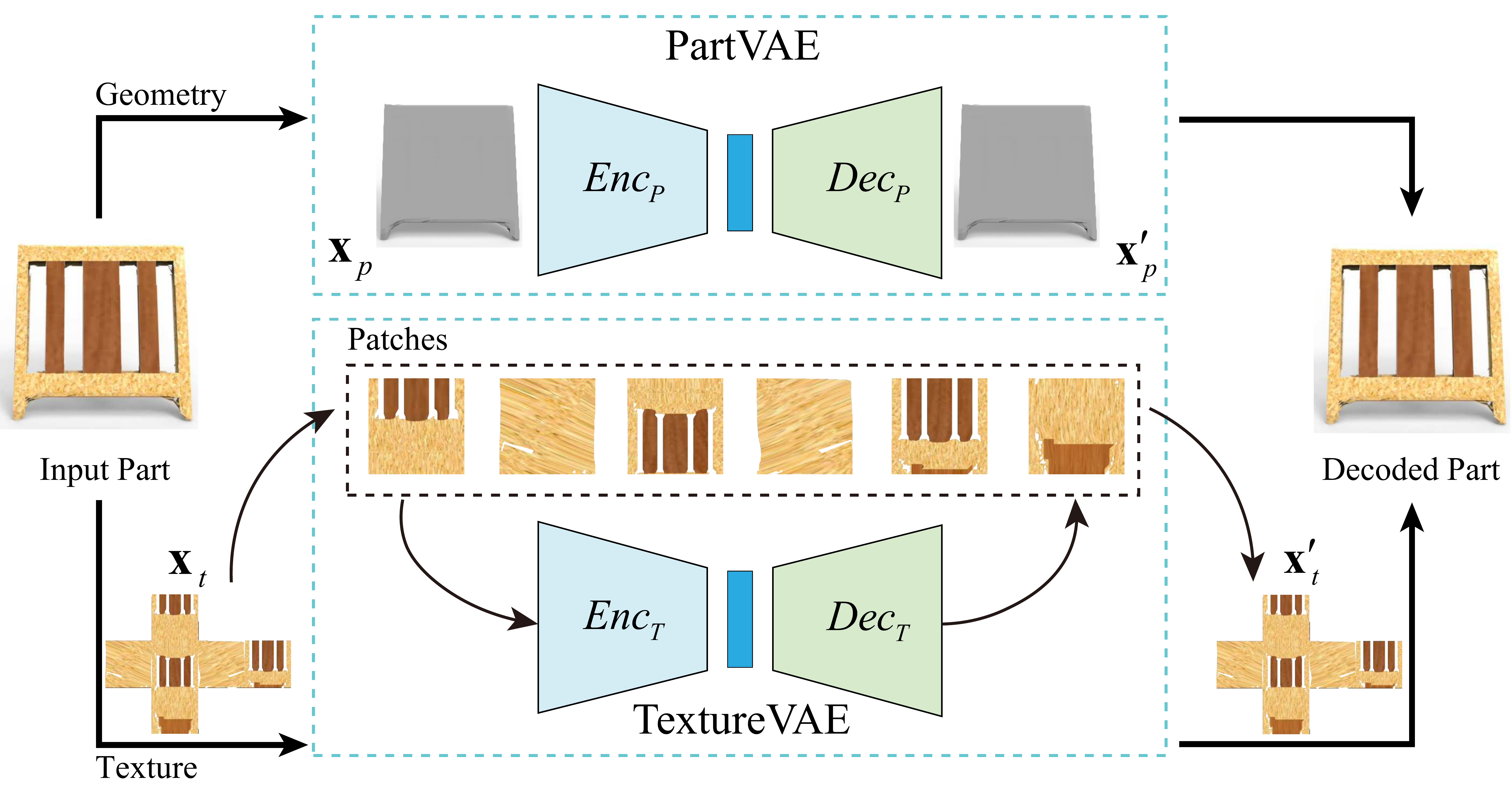}
  \caption{Network architecture for representing a textured part: a PartVAE for
 encoding part geometry and a TextureVAE for texture.}
  \label{fig:TPartVAE}
\end{figure}

\rzn{
To address these challenges, TM-NET {\em separately\/} embeds part geometry (via a PartVAE) and part texture
(via a TextureVAE) into their respective latent spaces (see Figure~\ref{fig:TPartVAE}), as a means to facilitate learning the texture probability distribution conditioned on geometry.
To resolve possible ambiguities due to identical or similar part geometries having different textures, we introduce a {\em conditional autoregressive\/} model based on PixelSNAIL~\cite{chen2018pixelsnail} for texture generation. Within this framework, \rznn{compatibility} between textures generated for different shape parts is \rznn{achieved} via a conditional generative approach. Specifically, we designate a seed part for each
shape category. The network is trained to texture the seed part conditioned on its geometry, while for other parts, the condition input is a concatenation of part geometry and the VGG~\cite{simonyan2014very} image features of the texture of the seed part.
In addition, we add three fully connected layers before feeding the condition into the autoregressive network. Overall, the
learned mapping between geometry and texture is {\em one-to-many\/}.}

\rz{
Last but not least, since part geometry and texture possess different data characteristics, e.g., smooth geometry vs.~high-frequency image details,
the two part-level VAEs have different architectures. In particular, to address the blurry image issue with conventional VAEs,
we adapt VQ-VAE-2~\cite{vqvae,vqvae-2} for learning and generating textured mesh parts. With the
aid of vector quantization (VQ), our approach allows the learning of a high-dimensional texture latent space to
facilitate the generation of quality textures, with high-frequency details and less blurriness.
}

Our main contributions can be summarized as follows:

\begin{itemize}
    \item \rz{To the best of our knowledge, TM-NET is the first deep generative model which learns to synthesize high-quality textured
    meshes in a {\em part-aware\/} manner.}
    \item \rznn{Once trained, our network can generate novel textured meshes from random samples in a latent space or predict textures for a given 3D mesh, with or without image guidance. Plausible and diverse textures can be generated for the same mesh part, with texture
compatibility between parts in the same shape.}

    \item By utilizing texture {\em transparency\/} via the alpha channel, TM-NET is able to generate and reproduce the {\em visual appearance\/} of topological details such as holes in the chair back; see Figure~\ref{fig:teaser}. Visually, the obtained results are superior than those attainable by state-of-the-art generative networks.
\end{itemize}

We study the generative capabilities of TM-NET under several application scenarios, \rznn{including autoencoding and novel generation of textured meshes, shape-conditioned and image-guided shape texturing, as well as textured shape interpolation in the latent space.} The quality of the textured meshes generated by our network is demonstrated both qualitatively and quantitatively. Evaluation is conducted via ablation studies and comparisons to state-of-the-art generative models, including SDM-NET~\cite{gaosdmnet2019}, BSP-Net~\cite{BSPNet}, and Texture Fields~\cite{OechsleICCV2019}.

\section{Related Work}
\label{sec:related}

We cover works most closely related to ours: deep generative models for 3D shapes, texture synthesis, and joint 2D-3D encoding.

\vspace{-5pt}

\paragraph{Deep Generative Models for 3D Shapes.}

With the development of the deep neural networks such as Variational Autoencoders (VAEs)~\cite{kingma2013auto} and Generative Adversarial Networks (GANs)~\cite{NIPS2014_5423}, novel 3D shapes can be effectively generated by learning the distributions of existing data. 
For generating 3D shapes, existing deep generative models produce shapes in different representations, including voxels~\cite{3dgan}, point clouds~\cite{achlioptas18a}, implicit functions~\cite{chen2019-IMNET,park2019-DeepSDF,mescheder2019-Occupancy}, deformable meshes~\cite{meshvae2017}, multi-chart representations e.g. AtlasNet~\cite{AtlasNet2018} and structured representations, e.g. GRASS~\cite{li_sig17}, StructureNet~\cite{mo2019structurenet} and SDM-NET~\cite{gaosdmnet2019}. Recent methods with improved shape representations allow 3D shapes with richer geometric details and more complex structures to be synthesized or reconstructed, e.g.,~\cite{xu2019disn,li2021d2im}. However, none of these methods generate 3D models with textures. Furthermore, textures are crucial for the visual realism and rich appearance of 3D shapes. In this paper, we build our texture generation method on SDM-NET, a recent shape generation method based on structured deformable meshes, where each part is represented by using deformations from a box. The method produces shapes with fine geometric details.

\vspace{-5pt}

\paragraph{Texture synthesis.}

Texture synthesis for 2D images~\cite{Efros1999} and 3D solids~\cite{Kopf2007} have been widely studied in computer graphics. In recent years, neural network based methods have been proposed~\cite{gatys2015texture,Snelgrove:2017:ROD:3145749.3149449,TexSyn18} for the synthesis problem, leading to improved results, although these methods are still restricted to image textures, rather than generating textures over surfaces of 3D shapes.

For surface texture generation, the texture representation strongly relies on the underlying geometry representation. Although point clouds are flexible, and suitable for applications such as segmentation and classification~\cite{Qi2017cvpr,Qi2017nips}, they are too sparse to represent geometry details~\cite{tao2019}, and also suffer from this major issue when used for representing high-resolution textures. For the voxel-based representation, the texture representation based on voxels~\cite{drcTulsiani17} is also limited by the resolution due to the high memory overhead.

Recently, some works propose to generate textures using deep generative models. According to the statistical properties, textures on 3D shapes can be divided into two categories, namely stochastic textures such as wood and marble which can often be compactly modeled using a procedural generation process or some samples, and non-stochastic textures which are generic images on surfaces.
Henzler et al.~\shortcite{henzler2019learning} propose a method to encode stochastic textures from 2D exemplars for synthesizing 3D solid textures. While solid textures are useful for certain applications, they are not associated with shape surfaces. The method is also restricted to stochastic textures, which are insufficient for texturing general 3D objects.

\vspace{-5pt}

\paragraph{Generating textured 3D Shapes.}

To obtain textured 3D shapes, some works use single-view color images as 2D supervision to recover 3D objects with colors. Hu et al.~\shortcite{tao2019} use a two-stage method that first infers the object coordinate map and then uses reprojection onto the input image to recover dense point clouds with colors. The method is able to handle input images of different (and even unseen) object categories; however, it only generates a partial point cloud with visible colored points for a given input image, and multi-view fusion is required to obtain complete shapes. 
\rzn{Other methods, e.g.,~\cite{liu2019softras,chen2019dibrender,pavllo2020convolutional}, rely on differentiable renderers to train networks to predict vertex positions and colors for meshes with a fixed topology.}
These methods rely mainly on the input image to provide constraints, whereas our approach aims to generate textures without image guidance. Moreover, the use of a fixed-topology shape such as a sphere as the starting point also prevents these methods from handling shapes with complicated structures or rich geometric details.

A fundamental challenge for texturing 3D shapes is an effective texture mapping from the 3D surface to some texture (UV) space. UV mapping is usually difficult to be consistently defined across shape collections, and previous UV-mapping based methods mainly deal with models with fixed topology including planar topology (for faces)~\cite{saito2017photorealistic} and spherical topology (for birds)~\cite{cmrKanazawa18}. These representations however cannot be generalized to cope with shapes with complex structures.
Recent work Texture Fields~\cite{OechsleICCV2019} avoids this problem by learning textures using mapping from the 3D texture space. However, the use of 3D solid space is not only expensive, but also harder to learn, as textures of interest are usually only shape surfaces. 
Raj et al.~\shortcite{Raj_2019_CVPR_Workshops} propose a two-stage approach to generating textured meshes, where in the first stage, a network is trained to map 2.5D depth images to texture images, and in the second stage, multiview texture images are fused. However, the method does not give details how to ensure seamless synthesis of texture atlas, and how to ensure sufficient coverage of texture images, especially for shapes with complex structures where self-occlusion is a problem, and shows examples of a single category (cars) which are of spherical topology. Unlike all these existing methods, our approach is able to generate higher-quality textured meshes with more complex structures. Our method can also generate geometry and texture simultaneously, which cannot be achieved with existing methods.

\vspace{-5pt}

\paragraph{Novel view synthesis (NVS)}
Given a single image, humans can easily extrapolate what the object/scene looks like from another view. NVS focuses on generating images from an unknown viewpoint with an input image. Sun et al.~\shortcite{sun2018multiview} takes multiple images from different source viewpoints and predicts a flow field to move the pixels from the source image to the target image. Then it aggregates predictions from different viewpoints to produce the target image. SRN~\cite{sitzmann2019srns} proposes a continuous 3D-structure-aware 3D scene representation that maps the world coordinates into a feature representation which helps preserve the geometry surface from novel viewpoints.
GeLaTO~\cite{martinbrualla2020gelato} moves a step forward to generate rendered images of transparent objects from a new viewpoint and interpolate rendered images from different viewpoints using U-NET~\cite{ronneberger2015u} and AtlasNet~\cite{AtlasNet2018}. However, their intermediate texture is an implicit representation, which cannot be directly mapped onto 3D shapes for typical graphics applications.
Although textured meshes can be rendered from novel viewpoints, our method is fundamentally different from NVS methods, which have been an active research topic in recent years: not only does our method not require a single-view image as input (and can generate textured meshes directly), our method also ensures the rendered images from different views are consistent, which cannot be guaranteed with NVS.

\section{Method}
\label{section::Methodology}

Given a collection of textured 3D meshes, TM-NET encodes the geometry associated with textures for generation and other applications such as conditional shape texturing and textured shape interpolation.
As discussed before, we build on recent SDM-NET~\cite{gaosdmnet2019} which provides structural deformable boxes to represent 3D objects, each corresponding to a semantic part.
Thanks to this representation, it intrinsically provides a way to define consistent texture atlases for objects, even if shapes in the collection may not have identical topology (e.g. a table may have different numbers of legs). Each box can be easily unfolded to the texture domain, with no extra distortion at this stage.
In the following subsections, we will describe our texture representation and how to convert textured 3D models to this representation.
We then describe the details of our novel network architecture, i.e. how the geometry detail determines its texture, first at the part level and finally the object level.

The mesh geometry and structure are learned jointly by one global VAE, to ensure that the generated textured meshes are consistent in both geometry and structure.
\rzn{Since a single part can be textured differently while still being plausible, instead of generating textures conditioned on the input geometry in a one-to-one manner, we learn a probability distribution based on PixelSNAIL~\cite{chen2018pixelsnail}.}
\gl{Once the distribution is learned, we can sample several times on the distribution which will lead to multiple reasonable textures for the input shape. Directly applying this network at the part level could not ensure the compatibility of textures between different parts on the same shape. Hence for each category we choose one part as the seed part (e.g. the table surface for tables).
When modeling texture distributions of other parts, we append the VGG feature of the seed part's texture to the geometry condition.
The texture generation of the seed part is conditional on its geometry, while texture generation of remaining parts depends on both their own geometry and the texture feature of the seed part. This strategy ensures that generated textures for different parts are coherent.}

\begin{figure}[!t]
  \includegraphics[width=1\linewidth]{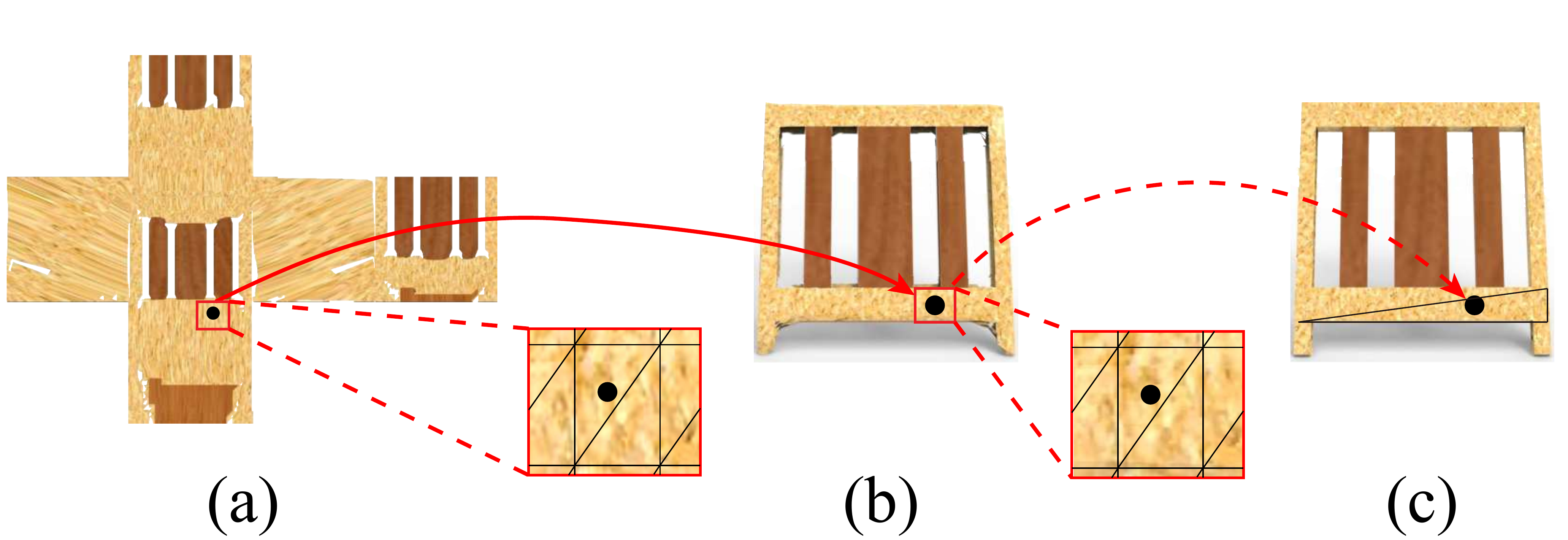}
  \caption{Illustration of the procedure that maps the ground truth texture to  the unfolded box to form a texture image. (a) the texture image to be filled, (b) the deformed box that approximates the input model part, (c) the input model containing ground truth texture.}
  \label{fig:preparation}
\end{figure}

\begin{figure}[!t]
\centering
\begin{tabular}{cc}
	\vspace{3mm}
	\rotatebox{90}{\quad \hspace{4mm} \large{GT}}
	&
    \vspace{-3mm}
	\hspace{-3mm}
    \includegraphics[width=0.30\linewidth]{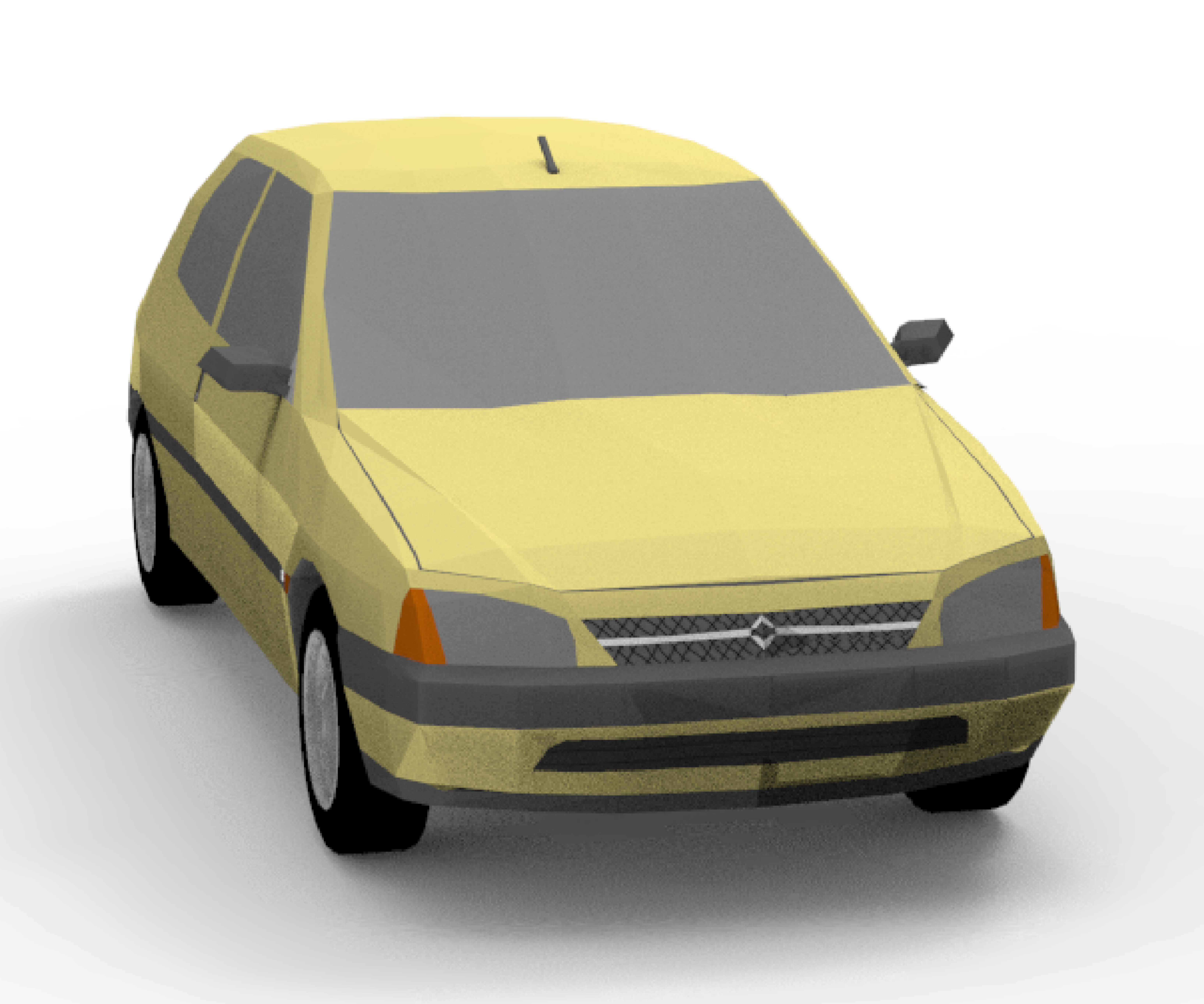}
	\includegraphics[width=0.30\linewidth]{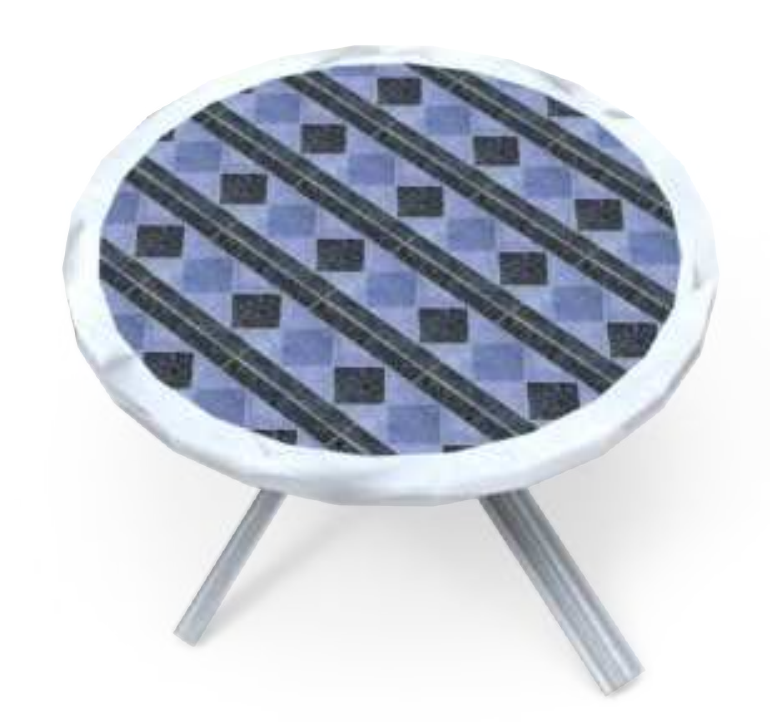}
    \includegraphics[width=0.30\linewidth]{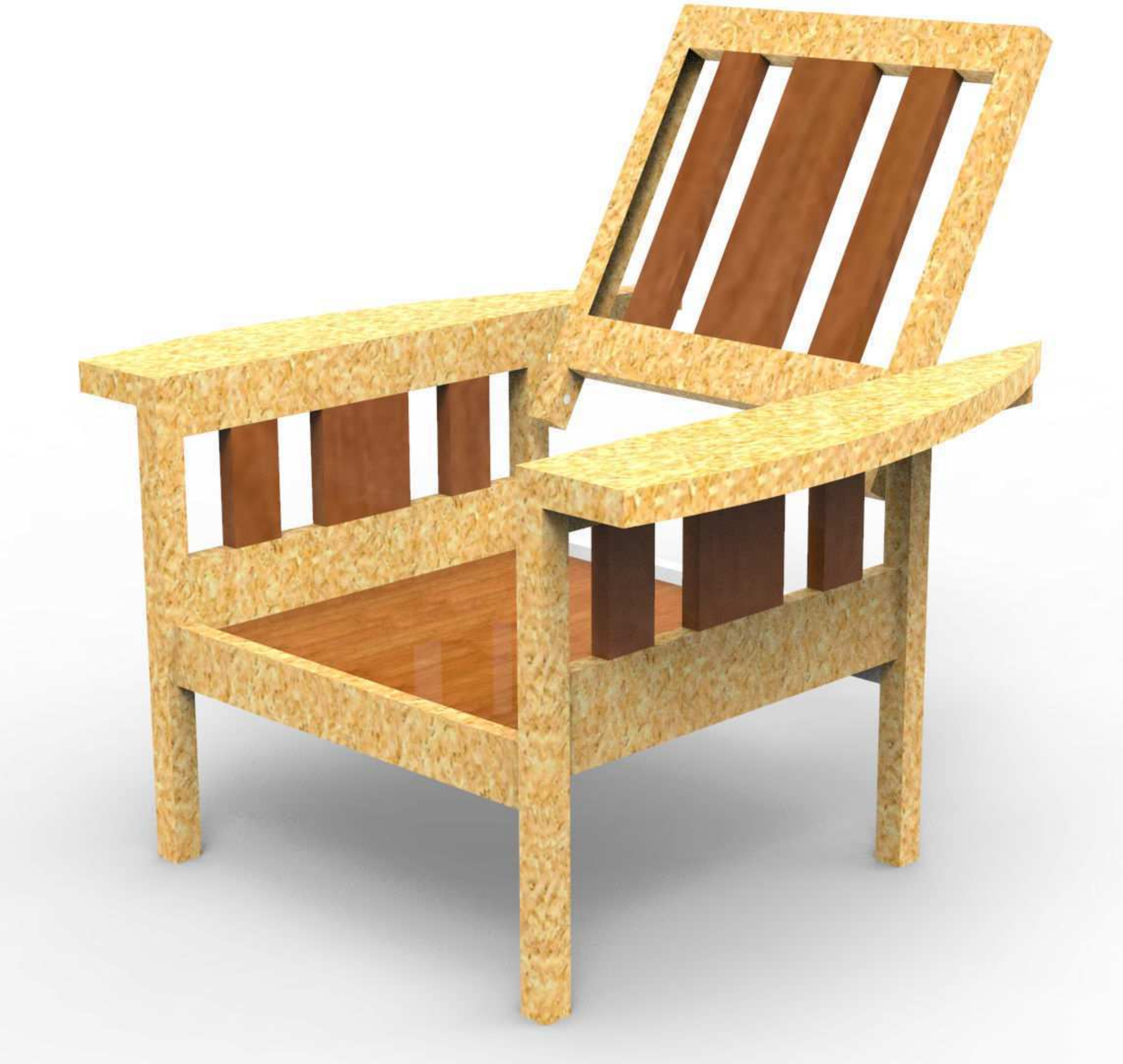}
    \\
    \rotatebox{90}{\quad \hspace{2mm} \large{Our rep.}}
    &
    \vspace{-3mm}
    \hspace{-3mm}
   	\includegraphics[width=0.30\linewidth]{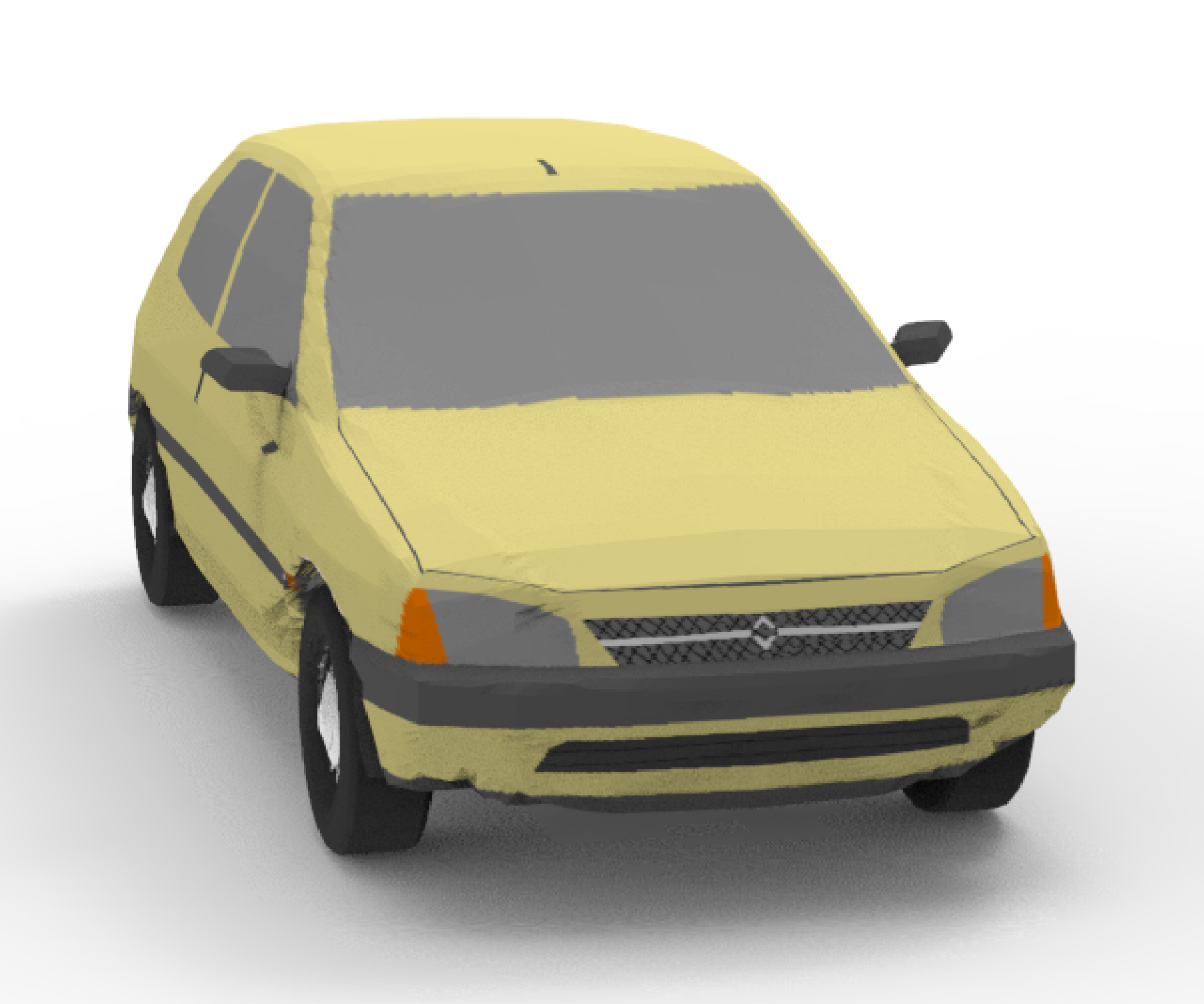}
	\includegraphics[width=0.30\linewidth]{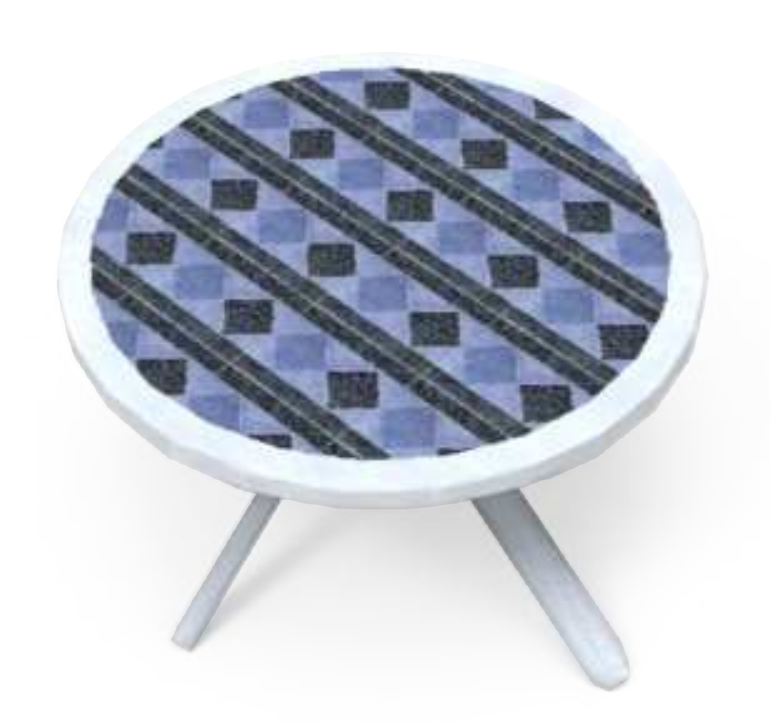}
	\includegraphics[width=0.30\linewidth]{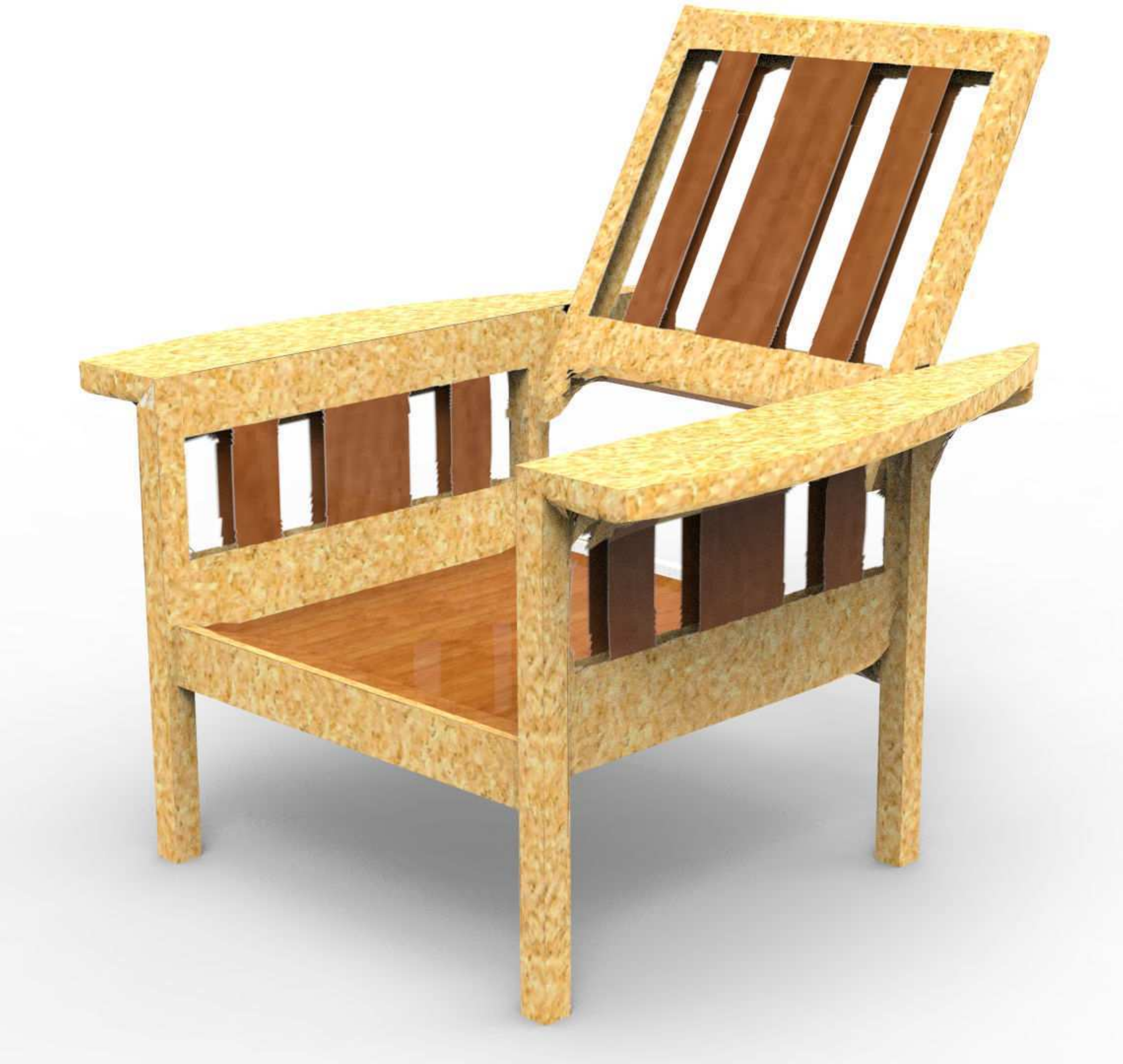}
	 \\
\end{tabular}
  \caption{\rzn{Examples showing our part-based texture representation (bottom row), which is able to retain most fine-level details and visually reproduces the appearance of rich topological structures (e.g., holes in the chair's back) that cannot be represented using the underlying geometry.}}
\label{fig:representation_power}
\end{figure}

\subsection{Part-based Texture Representation}
\label{section::TextCon}

\rzn{To represent textures on a mesh, a key step is to build a texture atlas. Traditionally, such an atlas is generated individually for each mesh, which makes learning difficult.}

Our method is based on SDM-NET~\cite{gaosdmnet2019}, which provides a set of consistently segmented parts across each category. The geometry details of each part are represented by the deformations of a template box. In our setting, template boxes are also served as the domains for surface textures
with predefined UV mapping. To minimize the mapping distortion from the template box to the 2D texture maps, the template box is unfolded in a straightforward manner as illustrated in Figure~\ref{fig:Representation} (b).
Let $I_{i,j}$ be the texture image of the $j^{\rm th}$ part %
on the $i^{\rm th}$ shape. By the unfolding operation, the template box $box_0$ is unfolded and parameterized to the 2D mesh in the UV space to obtain $\widetilde{box}_0$ as shown in Figure~\ref{fig:Representation}, with texture image size $4l \times 3l$ where $l$ is the length of the edge of the template box when mapped to the UV space. 
\rznn{
Throughout the paper, we fix $l$=256, while the texture resolution for a single part is set at $1,024 \times 768$.
Although a larger $l$ would lead to higher-resolution texture images that better represent details, the network training must require more data. Our current setting reflects a quality-cost trade-off.}

\rznn{With the UV mapping between the deformed box $b'_{i,j}$ and the corresponding texture image $I_{i,j}$ as defined above,}
the next step is to fill in each pixel $\tilde{u}$ on the texture image $I_{i,j}$. As shown in Figure~\ref{fig:preparation}, the 2D mesh $\widetilde{box}_0$ gives a partition of the texture image into triangles. Assuming that $\tilde{u}$ belongs to a 2D triangle $\tilde{f}$ on $\widetilde{box}_0$ with barycentric coordinates $bc(\tilde{u})$, we can find the corresponding 3D face $f'$ located on the deformed box $b'_{i,j}$, and the corresponding 3D position $p'$ based on barycentric coordinates $bc(\tilde{u})$ on $f'$. We then project $p'$ onto the input shape to obtain the ground truth color for pixel $\tilde{u}$.

{Note that a genus-zero box cannot geometrically model a high-genus part, e.g., the chair back in Figure~\ref{fig:preparation} with holes, unless a topology-varying deformation is performed. We avoid the need for such a deformation by utilizing the {\em alpha channel\/} of the texture image
to indicate {\em transparency\/}, so as to allow the appearance of a high-genus part to be produced through ray tracing. Specifically, if the projection of $p'$ is not on the surface of the input shape, then the alpha channel for the pixel $\tilde{u}$ is set to 0, indicating total transparency, otherwise it is set to 1 for opaque. When rendering a shape whose parts are all of genus 0 using ray tracing, a ray that hits a point whose alpha value is 0 would continue to travel; see Figures~\ref{fig:representation_power} and~\ref{fig:ReconstructionCompareN} for some examples of such a rendering of decoded meshes by TM-NET. Without such transparency flags, which are encoded into the texture images, previous works such as SDM-NET~\cite{gaosdmnet2019} cannot produce the visual appearance of topological details.

\begin{figure}[!t]
  \includegraphics[width=1\linewidth]{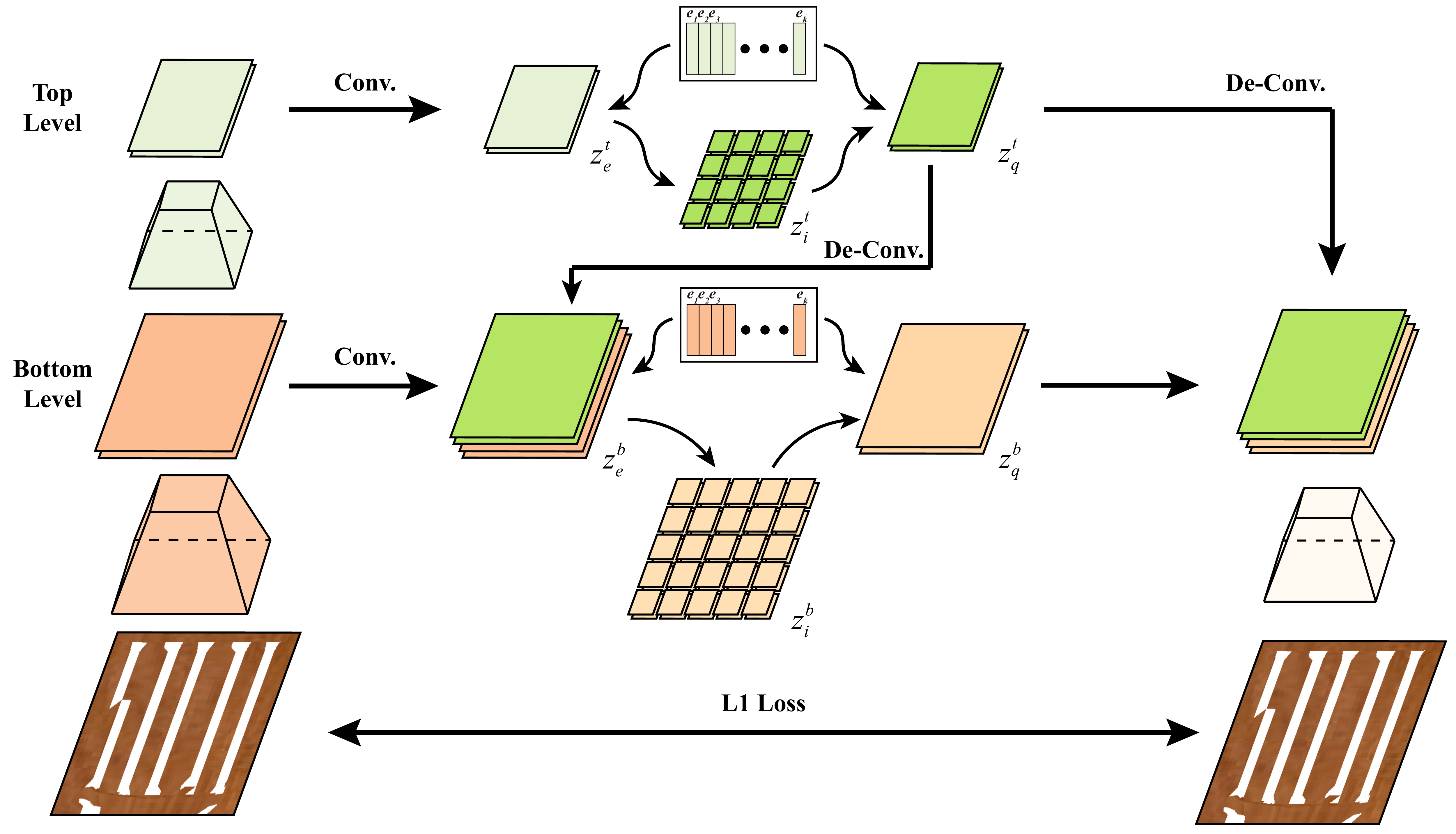}
  \caption{Network architecture of TextureVAE. The encoder maps the input image patch onto two continuous feature maps, top ($t$) and bottom ($b$). Then the dictionary-based vector quantization is performed. The decoder takes the discrete feature maps as input and reconstructs the image.}
  \label{fig:TextureVAE}
\end{figure}

\subsection{Part-level Texture and Geometry Encoding}
\label{section::EncodingofTexturedPart}

\yyj{We first present how the geometry and texture are encoded for a given object part. For the geometry, we use the mesh-based variational autoencoder (PartVAE) in SDM-NET~\cite{gaosdmnet2019} to encode the geometric details. Please refer to~\cite{gaosdmnet2019} for its network structure. As for the texture, we introduce a TextureVAE for encoding texture images. Different networks are needed to address their fundamentally different characteristics.}

Texture often contains high-frequency details but shares similar patterns, which is difficult to effectively encode using traditional VAEs.
\rznn{Our TextureVAE is inspired by VQ (Vector Quantization)-VAE-2~\cite{vqvae,vqvae-2} and its overall architecture is illustrated in Figure~\ref{fig:TextureVAE} and detailed network architecture is shown in the supplementary material.}
The input of TextureVAE is a four-channel texture patch in size 256$\times$256, including three channels of RGB values and one channel of alpha values.
Texture images in the UV space contain large areas of unused pixels (see Figure~\ref{fig:preparation} for an example).
To make texture encoding more efficient, we perform encoding based on $256\times 256$ patches each corresponding to a face of the cube.

\rzn{The encoder in VQ-VAE-2~\cite{vqvae-2} that is capable of encoding 256$\times$256 images is a two-level image pyramid which extracts two feature maps. We refer to them as the top- and bottom-level feature maps $z_e^t$ and $z_e^b$, respectively.
Each feature map $z_e^*$, where $*=b$ or $t$, goes through a vector quantization which maps each feature vector to its closest vector in the dictionary $e^*$ to obtain the discretized feature map $z_q^*$.}
\gl{The dictionary is updated during the training process as in~\cite{vqvae-2}, to capture the data distribution.
The decoder takes the discrete feature maps as input to produce the reconstructed image.
For more information about vector quantization and discrete vectors for quantization, please refer to the appendix of~\cite{vqvae-2}.
This architecture captures multiscale texture characteristics and allows high-frequency details often present in textures to be faithfully represented using a low-dimensional latent vector, thanks to the dictionary-based vector quantization and low-dimensional embedding.}

With PartVAE encoding the geometry and TextureVAE encoding the texture, our method can represent meshes with fine geometry and fine textures.
Moreover, our texture contains transparency information so the texture is also leveraged for its geometric representation ability, enabling our method to represent a single part \rzn{with possibly many holes in it, as shown in Figure~\ref{fig:representation_power}.}

Since the geometry and texture distributions lie in their own manifold spaces, PartVAE and TextureVAE are trained separately. Let $Enc_P(.)$ and $Dec_P(.)$ denote the encoder and decoder of part deformations for representing geometric details,
$Enc_T(.)$ and $Dec_T(.)$ denote the encoder and decoder for the associated texture images, $x_p$ and $x_t$ be the respective input deformation representation and texture images, and $z_p=Enc_P(x_p)$ and $z_t=Enc_T(x_t)$ be the encoded latent vectors of geometry and texture, respectively, where $z_t = (z_q^t, z_q^b)$ is the concatenation of discrete feature maps from the VAEs of the two layers. And finally, let
$x'_p=Dec_P(z_p)$ and $x'_t=Dec_T(z_t)$ be the reconstructed deformation feature vector and texture image, respectively.

The loss function of PartVAE follows \cite{gaosdmnet2019}, so that the TextureVAE would minimize the following loss:
\begin{equation}
    L_{TextureVAE} = L_{recon} + L_{seam} + \alpha_{1}L_{embedding}, %
\end{equation}
where $L_{recon} = \|x_t' - x_t\|_1$ is the reconstruction loss to ensure faithful reconstruction and
$L_{seam}$ penalizes the differences between texture boundary pixels that are in the same template box edge before unfolding to promote synthesizing seamless textures. For more detail about $L_{seam}$, please refer to the supplementary file.

The embedding loss, $L_{embedding}$, is defined as:
\begin{equation}
    L_{embedding} = \sum_{* \in \{t, b\}} \|sg[z_e^*]-e^*\|_2^2+\beta_{1}\|sg[e^*]-z_e^*\|_2^2,
\end{equation}
where $sg$ stands for stop gradient operator which does not influence the forward pass but has zero partial derivatives in the backward pass. The first term aims at updating the embedding space, i.e., making the dictionary $e^*$ closer to the feature maps $z_e^*$ extracted by the encoder. The second term updates the encoder's parameters to make the feature map move towards the embedding space or the dictionary $e^*$. Please refer to \cite{vqvae,vqvae-2} for more detail. $\alpha_{1}$ and $\beta_{1}$ control the relative importance of each loss term.

\subsection{Geometry Guided Texture Generation}
\label{section::GeometryInferTexture}

Mesh geometry and texture could be encoded by PartVAE and TextureVAE respectively.
In many cases, there are clear correlations between geometry and texture, e.g., beams always appear in the front of the car and they are usually white, while tires are black. In other cases however, the same geometry may be textured in different, and equally plausible ways, e.g., color of the car body.

To enable the texture generation network to output different textures for an input shape, we develop an autoregressive model to predict the conditional probability distribution of each pixel in the texture image, and sample different textures from the predicted distribution. Our model design is inspired by PixelSNAIL~\cite{chen2018pixelsnail}, which has shown success for discrete feature map generation by integrating temporal convolution with attention mechanism.

\rzn{PixelSNAIL models the distribution of every single pixel in the discrete representations $z_i^*$.
We adapt the conditional PixelSNAIL with three fully connected layers for feeding the geometry latent vector, as shown in Figure~\ref{fig:pixelsnail}.
Two such different conditional autoregressive models learn the high-dimensional distributions of the two discrete index matrices $z_i^t$ and $z_i^b$. It should be clarified that we learn the data distribution of the entire texture image instead of that of individual texture patches. That is, we concatenate the discrete index matrices of different patches in a texture image together and feed it into the autoregressive model.}

The top level autoregressive model takes the top level index matrix $z_i^t$ as input and the geometry latent vector $z_p$ (for the seed part) or concatenation of geometry latent vector $z_p$ and the seed part texture's VGG feature~\cite{simonyan2014very} (for other parts) as condition, while the bottom level autoregressive model takes the bottom level discrete index matrix $z_i^b$ as input and the top level discrete index matrix $z_i^t$ as condition.

\begin{figure}[!t]
  \includegraphics[width=1\linewidth]{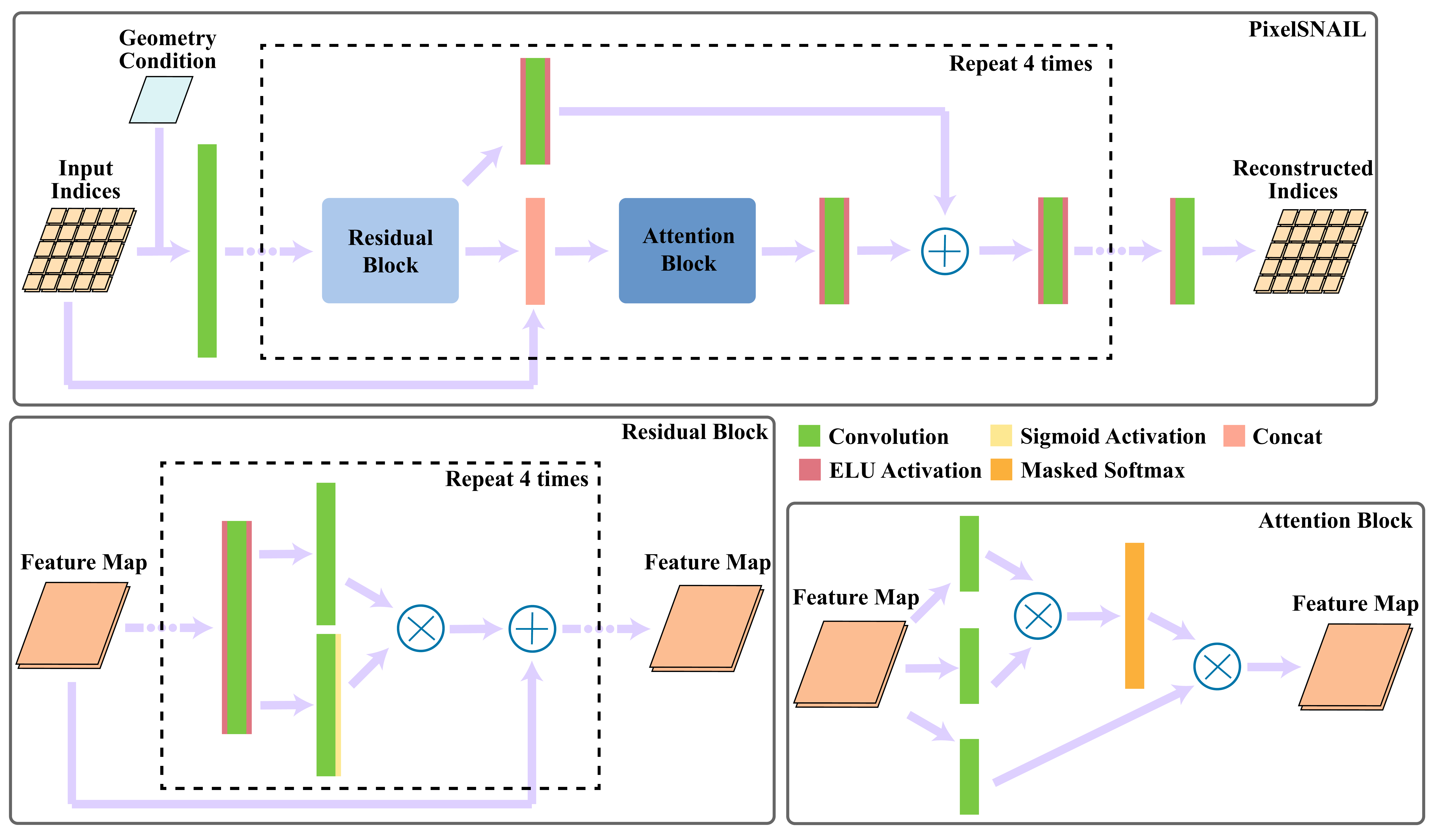}
  \caption{\wt{The architecture of our autoregressive generative model. The network takes an index matrix as input and geometry latent vector as condition and outputs reconstructed index matrix.}}
  \label{fig:pixelsnail}
\end{figure}

To aggregate geometry feature and the seed part texture's VGG feature, three fully connected layers are used to fuse them. For simplicity, let $x$ and $c$ denote the input and the condition of the conditional autoregressive model $f$. In the training process, the conditional autoregressive model minimizes the following loss:
\begin{equation}
    L_{Conditional} = L_{CrossEntropy}(x, f(x, c)).
\end{equation}
Note that we use the SP-VAE from SDM-NET~\cite{gaosdmnet2019} to jointly encode the geometry and structure of different parts to represent the whole 3D shape.

\subsection{Applications}
\label{section::TextureStructureDeformMeshEncoding}

We now present how our method can be applied to various applications with textured meshes.

\vspace{-5pt}

\paragraph{Shape-Conditioned Mesh Texturing.} Our network can be used to generate texture for a given shape automatically. Given a new shape, we first convert it into our representation. Then we can obtain the geometry latent vector of each part in the latent space using PartVAE. \togwt{The autoregressive generative model PixelSNAIL for the seed part models the distribution of the two-level index matrices. Then we sample on the distribution for several times to generate different index matrices $z_i^t$ and $z_i^b$, which are later used as indexes to obtain corresponding code vectors in the codebook to form discrete feature maps $z_q^t$ and $z_q^b$. Finally, the decoder of TextureVAE decodes $z_q^t$ and $z_q^b$ to generate textures. Once the texture of the seed part is generated, we extract its VGG feature  and concatenate it with the geometry of other parts as the condition for their autoregressive models, which will lead to coherent texture for different parts.}

\vspace{-5pt}

\paragraph{Generation and Interpolation.}
Given our deep generative model TM-NET, we can randomly generate textured mesh models and interpolate mesh models in the latent space.
\yyj{
	For textured mesh generation, we first generate the structured meshes with fine geometry by randomly sampling from the SP-VAE latent space.
	For each part we encode the geometry into the latent space of PartVAE.
}
\togwt{
	Then the problem is converted to \rznn{conditional texturing} the generated shape as mentioned above.
	For interpolation, the geometry and structure of two input shapes are interpolated in the latent space of SP-VAE, and the textures are interpolated in the continuous feature maps $z_e^*$ of TextureVAE for generating in-between textured meshes.
}

\vspace{-5pt}

\paragraph{Image-Guided Mesh Generation.}
\label{section::svr}

\rznn{
Our method can be adapted for textured shape generation as guided by a single-view image, where the untextured shape geometry 
is produced by existing methods with a pre-trained geometry prediction network. Specifically,}
we first predict the 3D shape and camera viewpoint from the input image using DISN~\cite{xu2019disn}.
The predicted shape is then segmented and encoded using our PartVAE.
With the segmentation and camera viewpoint, we project all parts onto the image plane to obtain an image mask.
Finally, a pretrained PixelSNAIL network takes the VGG features of the masked input image and the part geometry features as condition to produce the part textures.

\section{Datasets and Network Implementation}
\label{sec:imp}

\gl{In this section, we describe our network architecture and training process in detail. The experiments are performed on a cluster with three deep learning servers, each of which is equipped with an I7 6850K CPU, 64GB RAM and four Nvidia RTX 2080Ti GPUs. Twelve Nvidia RTX 2080Ti GPUs are used for training TM-Net. }

\subsection{Dataset}
\rzn{The dataset used in our experiments consists of segmented meshes with textures from the dataset employed by SDM-Net~\cite{gaosdmnet2019}, which is a subset of ShapeNet Core V2 \cite{shapenet} with part segmentations provided by SDM-Net.
See Table~\ref{tab:Dataset} for the numbers of textured meshes in different categories in our dataset.}

\begin{table}[!t]
\centering
\caption{Numbers of segmented, textured meshes in our experiments.}
    \begin{tabular}{c ||p{0.6cm}<{\centering}  p{0.6cm}<{\centering}  p{0.8cm}<{\centering} p{0.6cm}<{\centering}}
    \hline
    Object category                         & Car  & Chair & Plane & Table \\
    \hline\hline
    \# Segmented, textured meshes & 1,824 & 3,746 & 2,690 & 5,266  \\
    \# Texture patches & {\small 73,482} & {\small 162,984} & {\small 164,766} & {\small 167,628} \\
    \hline
\end{tabular}
\label{tab:Dataset}
\end{table}

\subsection{Network Architecture}
\togwt{Our network is composed of four parts including PartVAE for encoding part geometry, TextureVAE for encoding part texture, conditional PixelSNAIL for generating textures, and SP-VAE for jointly encoding both global structure and part geometries.
The architectures of PartVAE and SP-VAE follow~\cite{gaosdmnet2019}.}

\togwt{As shown in Figure~\ref{fig:TextureVAE}, the encoder of TextureVAE has three convolutional blocks, each followed by one convolution layer to obtain two different levels of continuous feature representations, which then go through vector quantization. The decoder concatenates the discrete latent variables from different levels and outputs the reconstructed texture images through deconvolution blocks. 

The detailed architecture of encoding and decoding networks can be found in the supplementary material. For the hyperparameters in the loss function of TextureVAE, they are the same as those in \cite{vqvae-2}. } \togwt{Further, the conditional texture generation network architecture is the same as the original implementation, except that before putting conditional input into PixelSNAIL, we add three fully connected layers to convert the conditional input into the same size as the input discrete index matrix.}

\togwt{The network architecture of SP-VAE follows the SP-VAE in~\cite{gaosdmnet2019}. Leaky ReLU is set as the activation function. We first train PartVAE and SP-VAE. After that, we train TextureVAE. The conditional autoregressive network is trained at last.}

\subsection{Metrics for Evaluation}
\rznn{To evaluate and compare quality of the generated shapes and textures, we employ} the
Learned Perceptual Image Patch Similarity (LPIPS)~\cite{zhang2018perceptual} and %
Structure Similarity Image Metric (SSIM)~\cite{Zhou2004}, which are \rznn{image similarity measures between input textured shapes and decoded textured meshes.}
Since the texture and geometry are interconnected, we can evaluate their quality together by rendering the textured shapes. Specifically, we render the decoded textured meshes and the input textured meshes in 12 viewpoints and use SSIM to measure the difference of rendered images from the same viewpoint and \yj{under the same lighting}.

\wt{Also, to evaluate the realism of generated textures for a given shape, we use fooling rate which is a human judgment approach proposed in \cite{zhang2016colorful}. This test shows a real image and a synthesized image to a human participant for one second each and asks them to tell which is ``real''. The fooling rate is the possibility of synthesized images being mistaken for real images.}

\subsection{Parameters}

\rzn{Hyper-parameters are fixed for datasets of different object categories. We demonstrate how the results change with the hyper-parameters. 
Each dataset is randomly divided into training and test sets, with a 9:1 split, using the data released from SDM-NET.}
We use Structural Similarity Index (SSIM) (see more in Sec.~\ref{section::ResultsAndEvaluation}) to measure image perceptual similarity between rendered images of input textured meshes and those of decoded textured meshes. 
\wt{$\alpha_1$ and $\beta_1$ are set to 1 and 0.25 following VQ-VAE-2~\cite{vqvae,vqvae-2}. }

\subsection{Training Details}

\wt{Our network is trained stage by stage. On the geometry side, similar to SDM-NET, every semantic part has a corresponding PartVAE. All of these PartVAEs are trained separately. After training the PartVAEs, we then concatenate the latent vector of PartVAE and the structure code as the input to train SP-VAE. The PartVAE is trained with 20,000 iterations and SP-VAE with 120,000 iterations with batch size 128. On the texture side, a single TextureVAE is trained for textures of each object part. The TextureVAE is trained with 400,000 iterations with batch size 
set as 120.} 

\togwt{After the training process of SP-VAE and TextureVAE are finished, we train the top and bottom level PixelSNAILs. The top level PixelSNAIL is trained with 200,000 iterations and bottom level PixelSNAIL is trained with 300,000 iterations with batch size 6. For all the networks, every training batch is randomly selected from the training dataset. The detailed training process is summarized in Algorithm~\ref{algo:TrainingSteps}.
Take the car dataset with seven parts as an example, the whole training process takes about 5 days. Once the networks are trained, generating a textured shape takes about 10 minutes. }

\begin{algorithm}[!t]
\SetAlgoLined{
    1. Train PartVAE for each part\;
    2. Train SP-VAE for each shape\;
    3. Train TextureVAE for each part\;
    4. Load checkpoint and extract PartVAE and TextureVAE latents $z_p$, $z_i^t$ and $z_i^b$ for the seed part\;
    5. Train conditional PixelSNAIL for the seed part in the top and bottom level discrete feature spaces\;
    6. Load checkpoint and extract PartVAE and TextureVAE latents $z_p$, $z_i^t$, $z_i^b$, and VGG feature of the corresponding seed part's texture for other parts\;
    7. Train conditional PixelSNAIL for other parts in the top and bottom level discrete feature spaces.
 }
 \caption{Training steps}
 \label{algo:TrainingSteps}
\end{algorithm}
\section{EXPERIMENTAL RESULTS AND EVALUATIONS}
\label{section::ResultsAndEvaluation}

In this section, we present results of textured mesh \rznn{autoencoding}, novel generation, shape-conditioned and image-guided texturing, as well as latent-space interpolation. We also perform ablation studies to show the benefits of each component in our design.

\subsection{Reconstruction and Generation of Textured Meshes}

\paragraph{Textured Mesh \rznn{Autoencoding}}
First, we qualitatively and quantitatively compare TM-NET with state-of-the-art \rznn{geometry deep learning methods for the {\em shape\/} autoencoding task.} To this end, only the alpha channel is used, while color textures are ignored. In Figure~\ref{fig:ReconstructionCompareN}, we show \rznn{two representative autoencoding results on the test set, where the input shapes possess fine-grained topological details. A comparison with other recent deep models shows that TM-NET and SDM-NET tend to produce the cleanest shape {\em structures\/} due their their structure-aware autoencoding. With the utilization of texture transparencies, TM-NET is the only network that is able to reproduce the appearance of topological details. Table~\ref{tab:GeometryQuantitativeComparison} provides quantitative comparison results, using the multi-view SSIM metric, demonstrating that TM-NET outperforms the other methods over all four major shape categories of ShapeNet.}

For \rznn{autoencoding} {\em textured\/} meshes, \rznn{we are not aware of existing alternatives from the literature.} Hence, we elect to show some qualitative results in Figure~\ref{fig:Reconstruction}, which provides the ground truth meshes with texture, the input textured meshes with our representation, and the decoded textured meshes by TM-NET. \rznn{The example results shown were {\em randomly\/} selected from each shape category to provide a general assessment of how well the TM-NET autoencoding is able to preserve the geometry and texture details and structures. More random results can be found in the supplementary material.} %

\begin{figure}[!t]
  \centering
	{
	    {\includegraphics[width=0.13\linewidth]{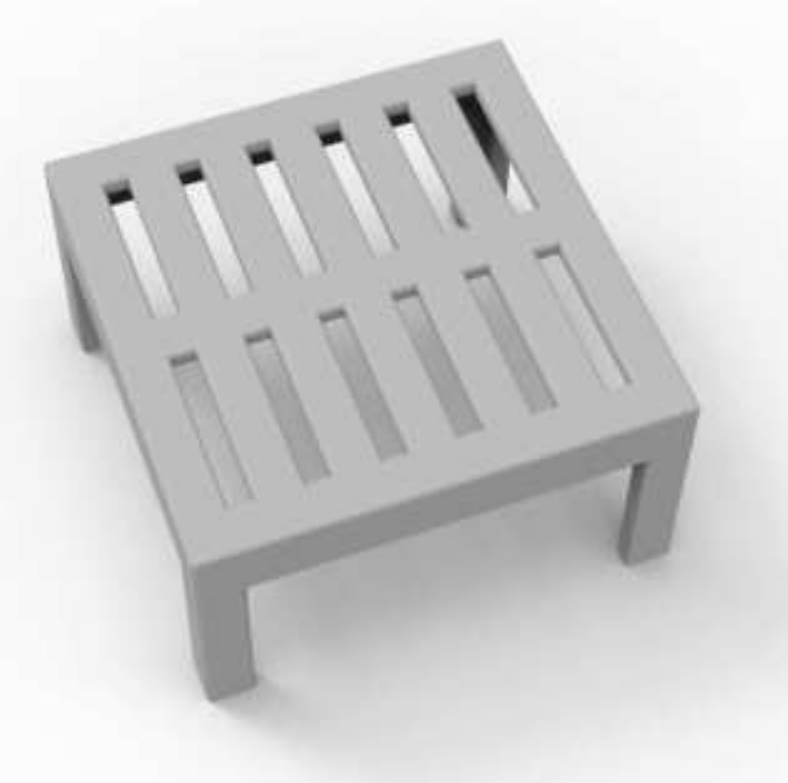}}
		{\includegraphics[width=0.13\linewidth]{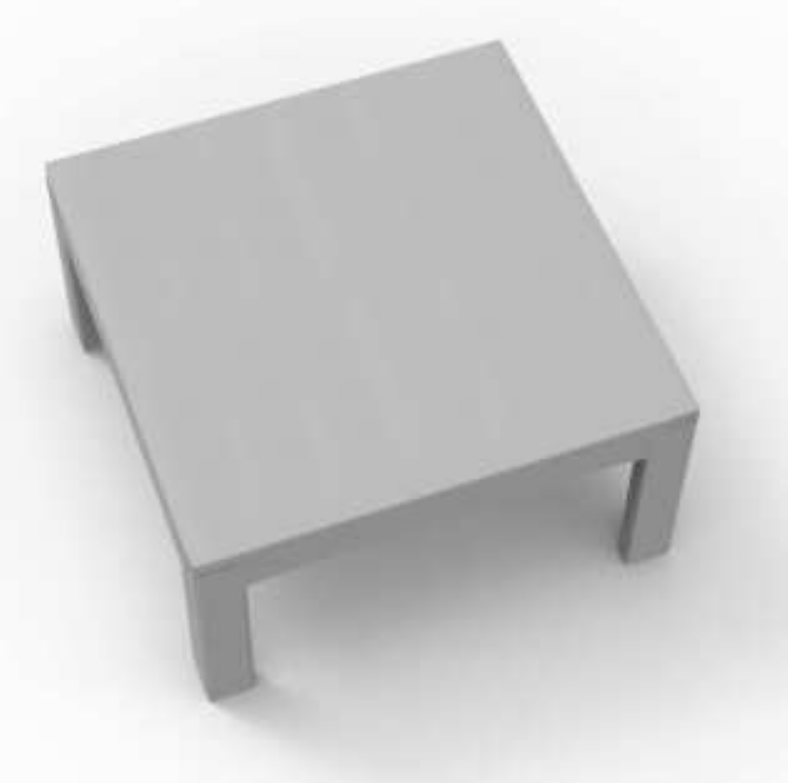}}
		{\includegraphics[width=0.13\linewidth]{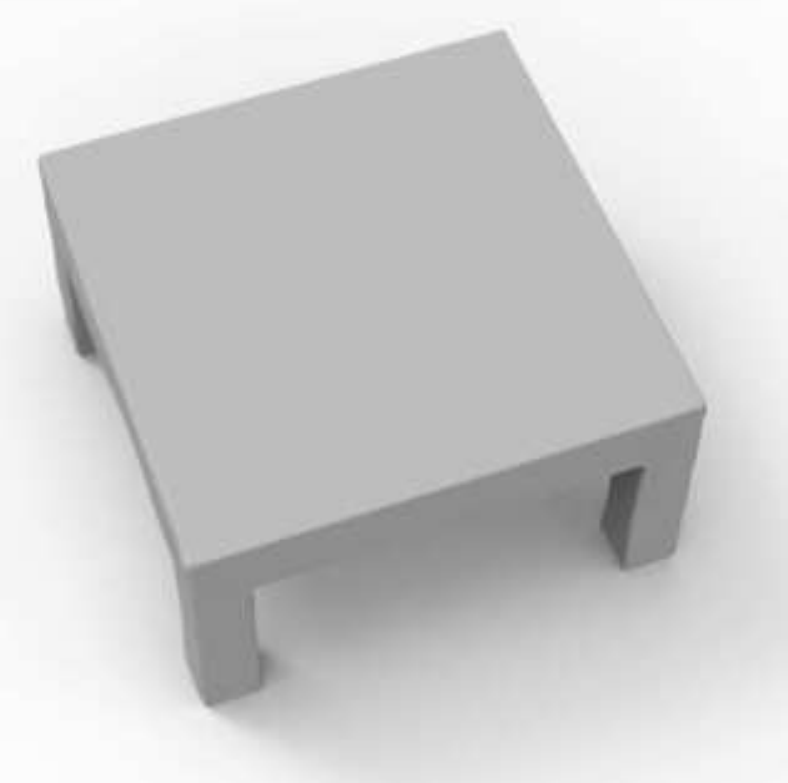}}
		{\includegraphics[width=0.13\linewidth]{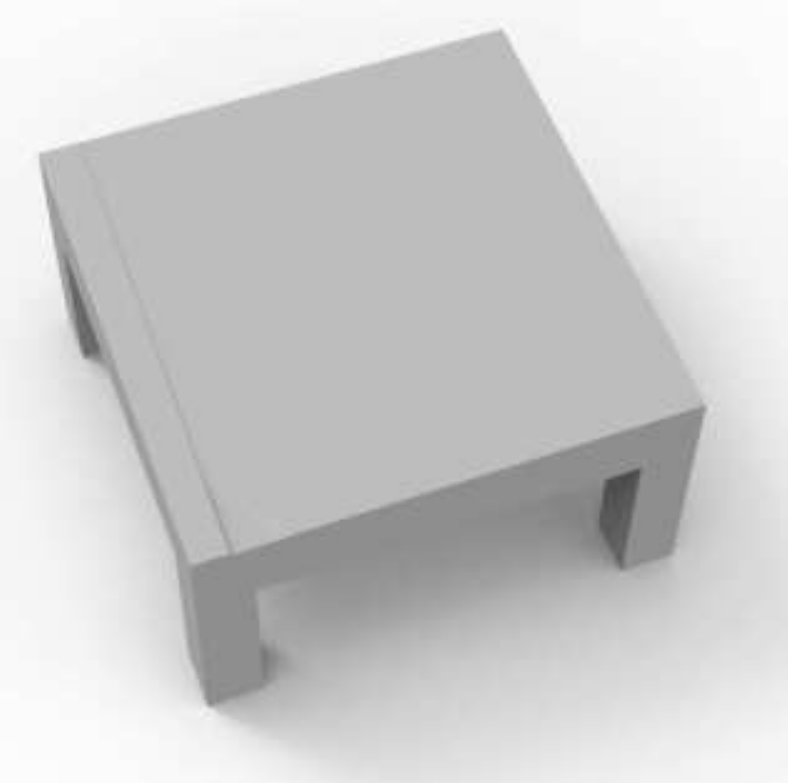}}
		{\includegraphics[width=0.13\linewidth]{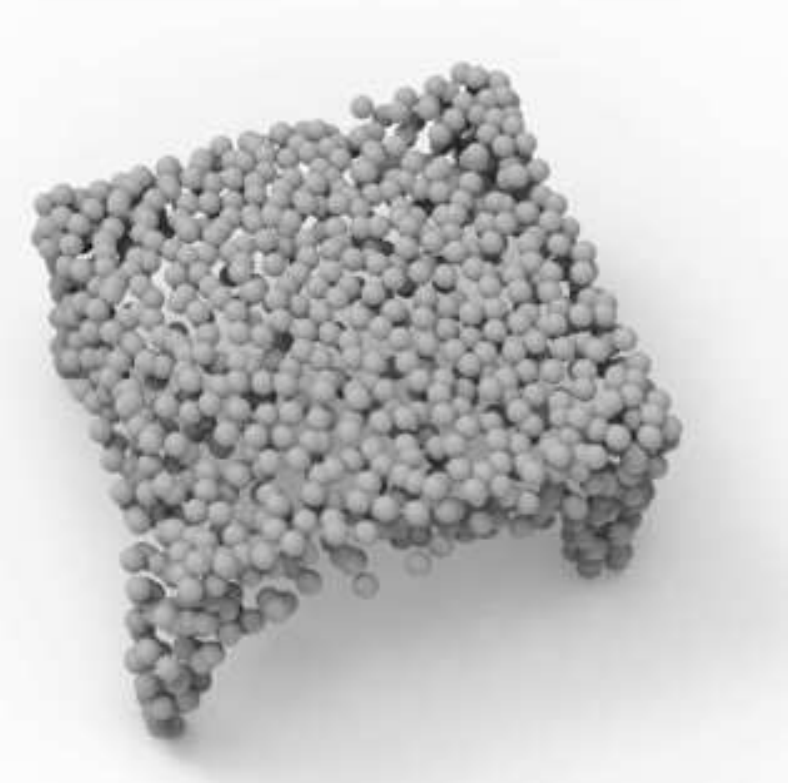}}
		{\includegraphics[width=0.13\linewidth]{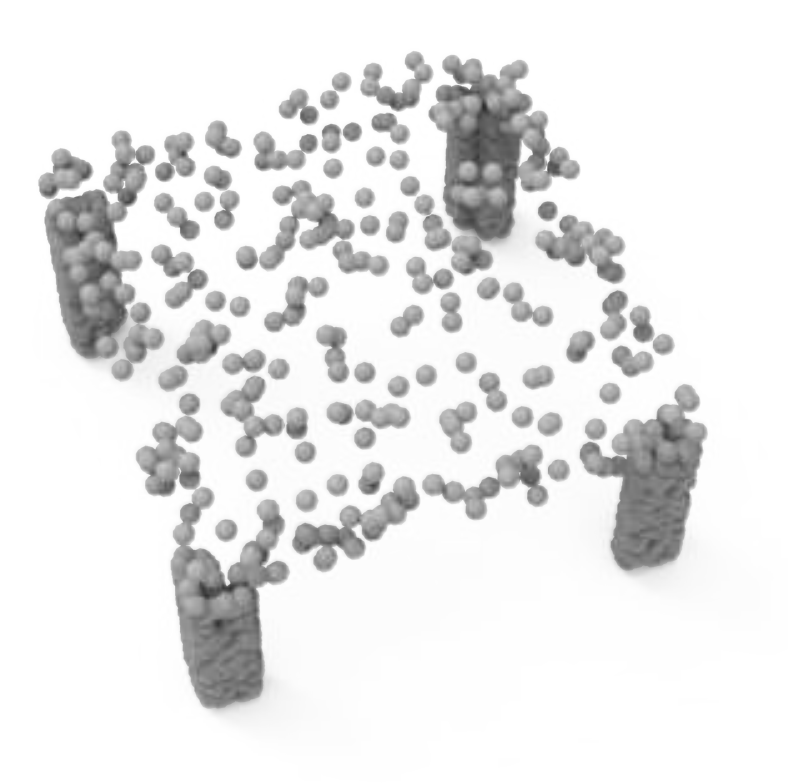}}
		{\includegraphics[width=0.13\linewidth]{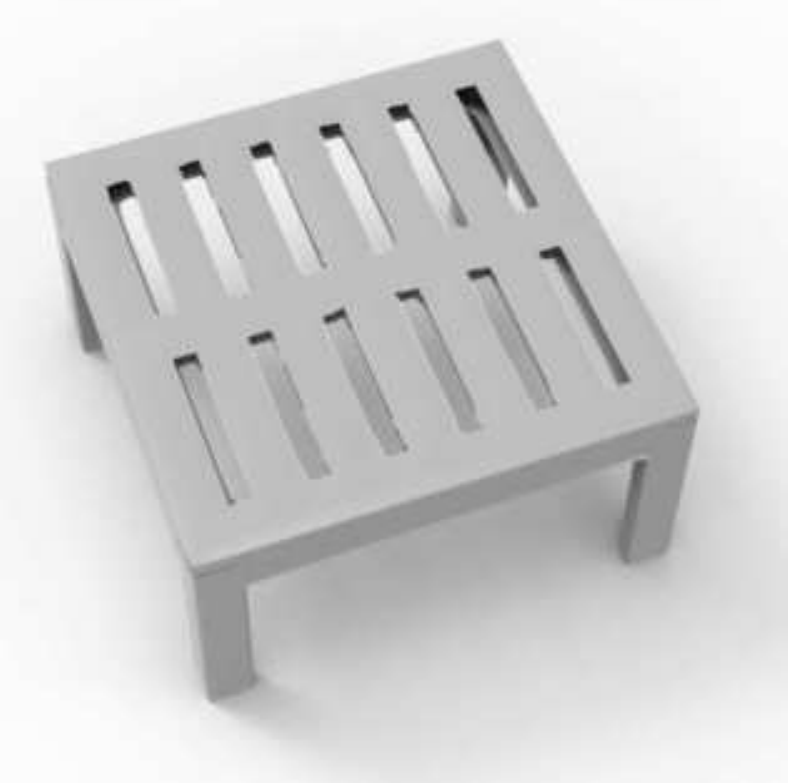}}
\\
		\subfigure[GT.]{\includegraphics[width=0.13\linewidth]{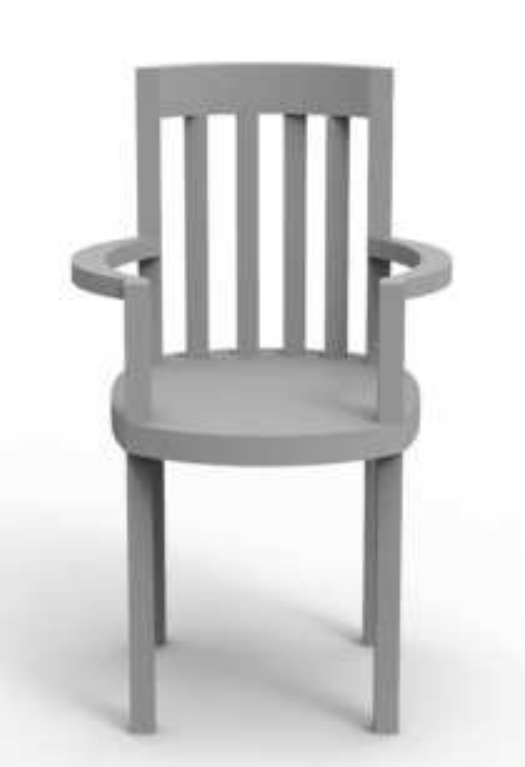}}
		\subfigure[SDM.]{\includegraphics[width=0.13\linewidth]{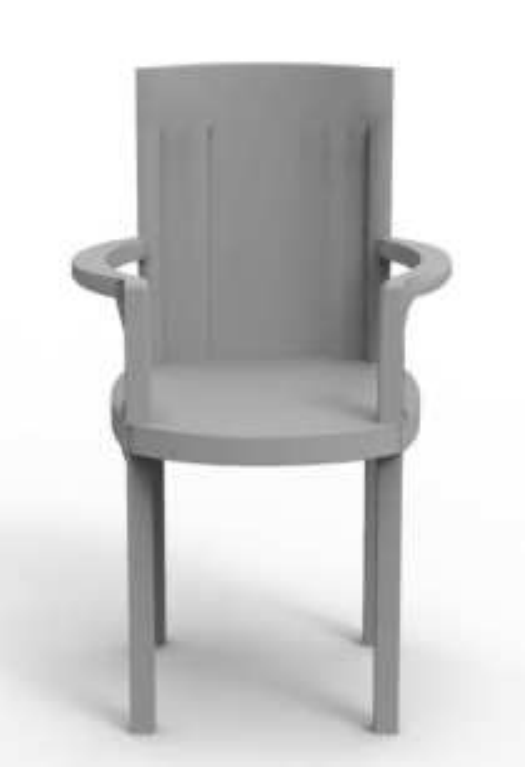}}
		\subfigure[ImN.]{\includegraphics[width=0.13\linewidth]{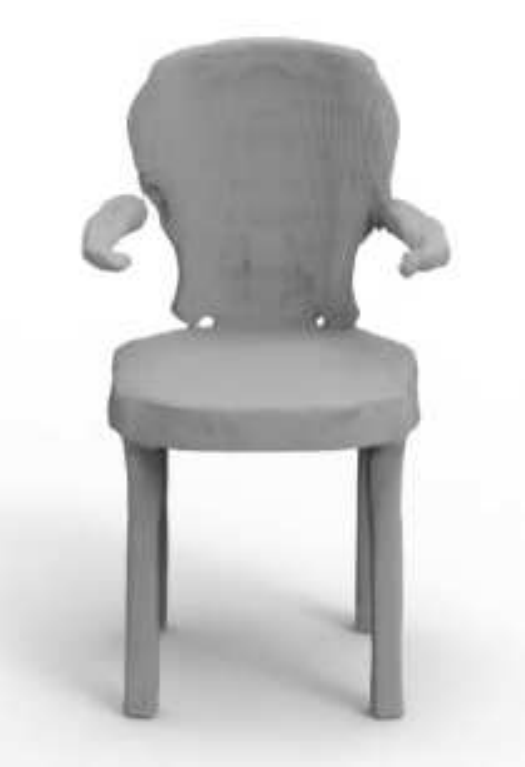}}
		\subfigure[BSP.]{\includegraphics[width=0.13\linewidth]{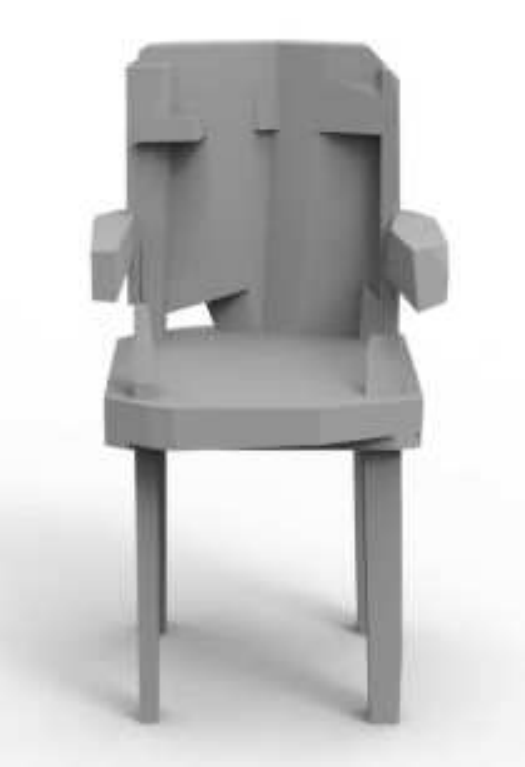}}
		\subfigure[PSG.]{\includegraphics[width=0.15\linewidth]{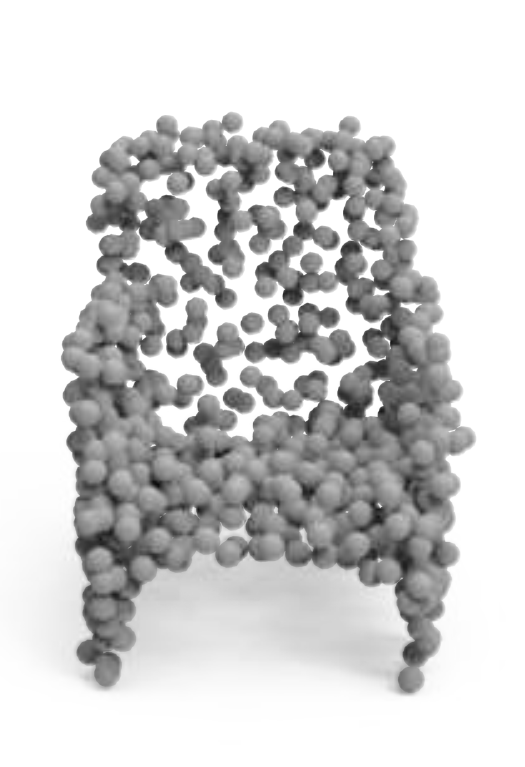}}
		\subfigure[StrN.]{\includegraphics[width=0.13\linewidth]{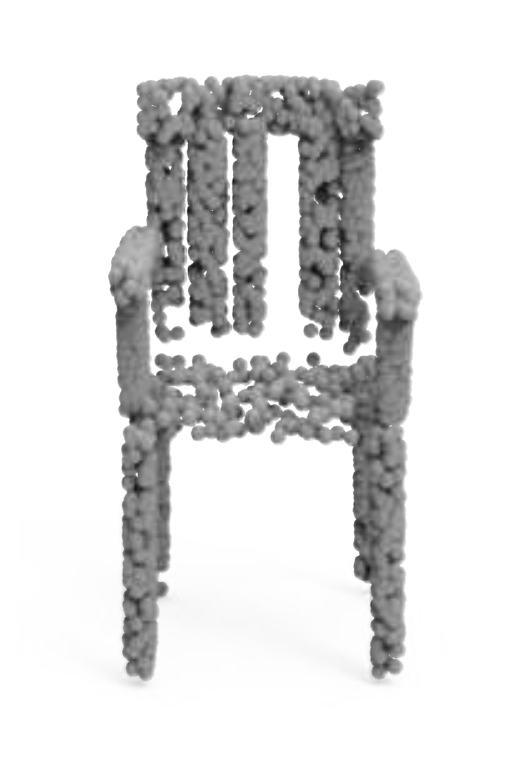}}
		\subfigure[TM-N.]{\includegraphics[width=0.13\linewidth]{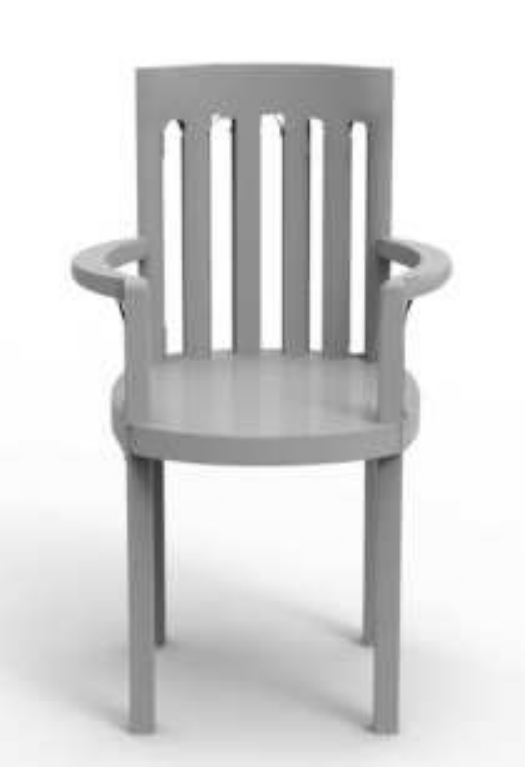}}
	}
  \caption{Comparing shape \rznn{autoencoding} between TM-NET (TM-N) and several state-of-the-art generative models:
SDM=SDM-NET, ImN=IM-NET, BSP = BSP-NET, StrN = StructureNET, with GT = ground truth. Outside of TM-NET, the part-based SDM-NET tends to obtain the best overall shape and structure, but is unable to reproduce topological details such as holes on the
chair back and tabletop. With texturing and transparency, TM-NET is the only method that can produce the appearance of such details.}
\label{fig:ReconstructionCompareN}
\end{figure}

\begin{table}[!t]
  \centering
  \caption{Quantitative comparisons, using the SSIM metric, between TM-NET and other recent deep representations for 3D geometry \rznn{autoencoding}, without color textures. TM-NET outperforms all these other methods over all \rznn{four} shape categories. Note that StructureNET uses PartNet~\cite{mo2018partnet} as input which does not contain the car or plane category.}
    \begin{tabular}{ccccccc}
    \hline
    \multirow{2}{*}{Categories} & &\multicolumn{4}{c}{SSIM metric $\uparrow$} \\
    \multicolumn{1}{c}{} & Car & Chair & Plane & Table \\
    \hline
    \cmidrule{1-7}
     SDM-NET       & 0.901 & 0.915 & 0.919 & 0.925 \\
     \cmidrule{1-7}
     IM-NET        & 0.893 & 0.907 & 0.874 & 0.896 \\
     \cmidrule{1-7}
     BSP-NET       & 0.899 & 0.908 & 0.869 & 0.919 \\
     \cmidrule{1-7}
     PSG          & 0.724 & 0.619  & 0.713  & 0.693 \\
     \cmidrule{1-7}
     StructureNet & -      & 0.803 & -     & 0.798 \\
     \cmidrule{1-7}
     Ours         & \textbf{0.914} & \textbf{0.923} & \textbf{0.920} & \textbf{0.930} \\
    \hline
    \end{tabular}%
  \label{tab:GeometryQuantitativeComparison}%
\end{table}%

\begin{figure}[!t]
  \centering
	{
		{\includegraphics[width=\linewidth]{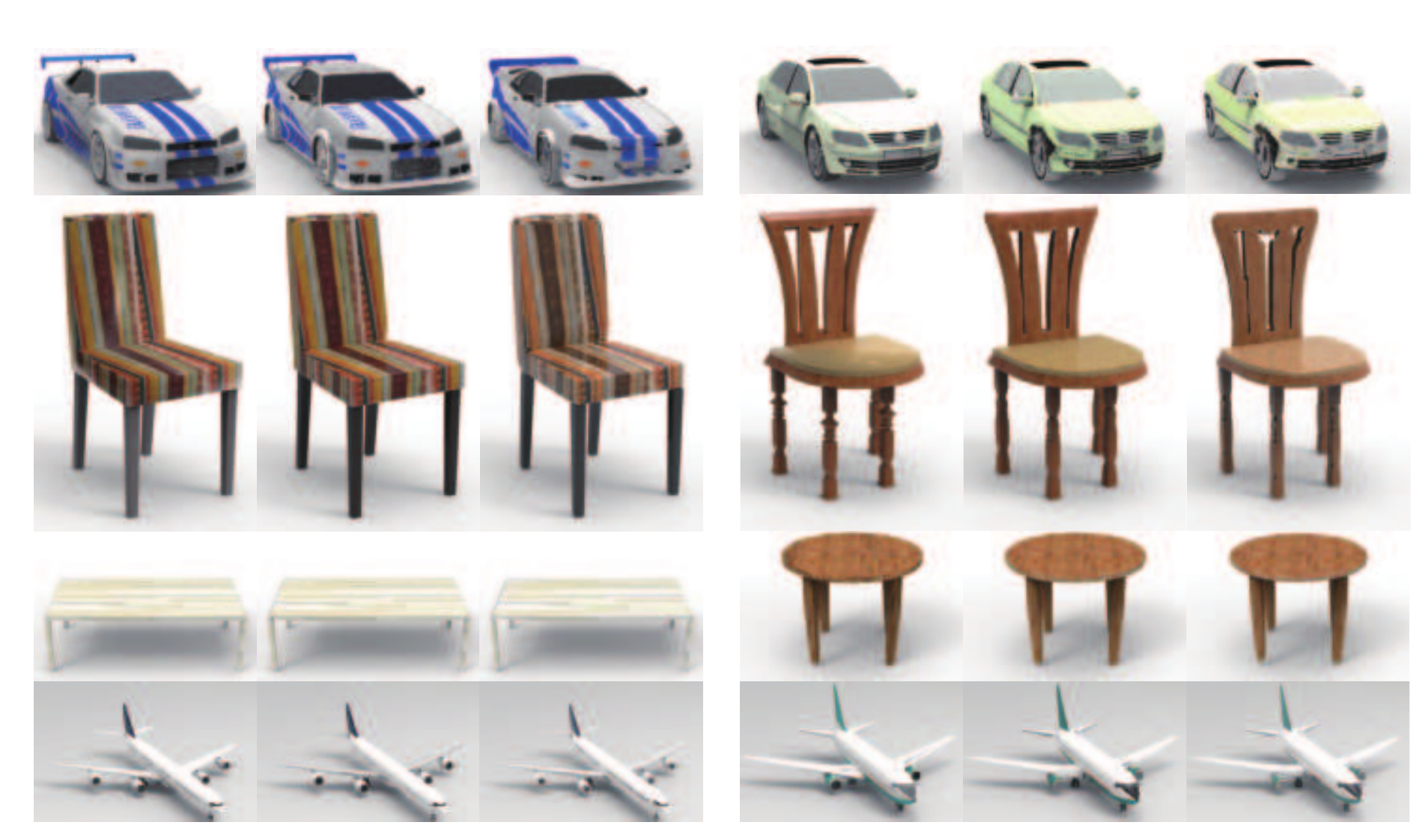}}
{\small GT} \hspace{18pt} {\small Input} \hspace{14pt} {\small Output} \hspace{28pt}
{\small GT} \hspace{18pt} {\small Input} \hspace{14pt} {\small Output}

	}
  \caption{\rznn{Randomly sampled results of} textured mesh \rznn{autoencoding} by TM-Net, on meshes from the test set. These results are better viewed by zooming in. \rznn{In each triplet, the left shows the ground-truth textured shape, the middle is our input representation relaying both geometry and texture features, which are generally well preserved by the decoded mesh (right), demonstrating that TM-Net learns an effective embedding}.
}
  \label{fig:Reconstruction}
\end{figure}

\paragraph{Textured Mesh Generation}

\rznn{
    As a generative model, and compared to SDM-Net~\cite{gaosdmnet2019}, TM-NET can not only output plausible shape structures and geometries but also generate a suitable texture for each part. On the geometry side, our SP-VAE encodes the structural information of different parts together with their geometry. On the texture side, the TextureVAE encodes texture information to ensure texture quality, while the conditional autoregressive model learns the relationship between geometry and texture to ensure compatibility between them.
    To obtain a textured mesh from scratch, we first sample on the latent space of SP-VAE to generate a structured deformable mesh without texture.
    Then similar to the \rznn{shape-conditioned} texturing procedure, we feed the geometry latent vectors into the conditional autoregressive network to predict TextureVAE index matrices.
    Finally, the decoder of TextureVAE takes these predicted index matrices as input to decode plausible textures for the generated shape.
    Figure~\ref{fig:Generation} shows a sampler of output textured meshes using the above procedure for each shape cateogry. Note that
these results were {\em randomly\/} selected, except for the first chair in the first row, which was picked to show an example of using
texture transparency to ``fake'' topological details.

    We qualitatively evaluate the novelty of the generated meshes by showing the closest neighbor shapes from the training set, both in terms of geometry and appearance.
    These results demonstrate that TM-NET is able to generate diverse and novel textured meshes with quality.
    However, the generated car and plane textures tend to be blurrier than those from the other categories. This is because these textures typically contain a mixture of definitive sharp features (e.g., plane windows, car headlights, etc.) and smooth or constant textures (e.g., the color of the remaining parts). Such mixtures are more difficult to encode using the VQ- or compression-based TextureVAE in our framework, as compared to the chair/table textures that have a relatively simpler frequency-domain characterization. More mesh generation results can be found in the supplementary material.

As for a comparison to existing methods, we are not aware of other networks designed for textured {\em 3D\/} shape generation. The work by Zhu et al.~\shortcite{VON} called Visual Object Networks (VON) could be compared in terms of generating different and novel {\em views\/} of object textures. VON does not generate a textured 3D shape, but only 2D images of an object after texture generation from a sampled texture code. As such, it is difficult to maintain texture consistency between multiple views, as shown in Figure~\ref{fig:NVS}, in contrast to results from textured mesh generation by TM-NET.

Table~\ref{tab:FIDComparison} shows a quantitative comparison in Fr\'{e}chet Inception Distance (FID)~\cite{NIPS2017_7240} between VON and TM-NET, where we evaluate the rendered images of the generated textured shapes against those from the training set.
}

\begin{figure}[!t]
\centering
        \includegraphics[width=\linewidth]{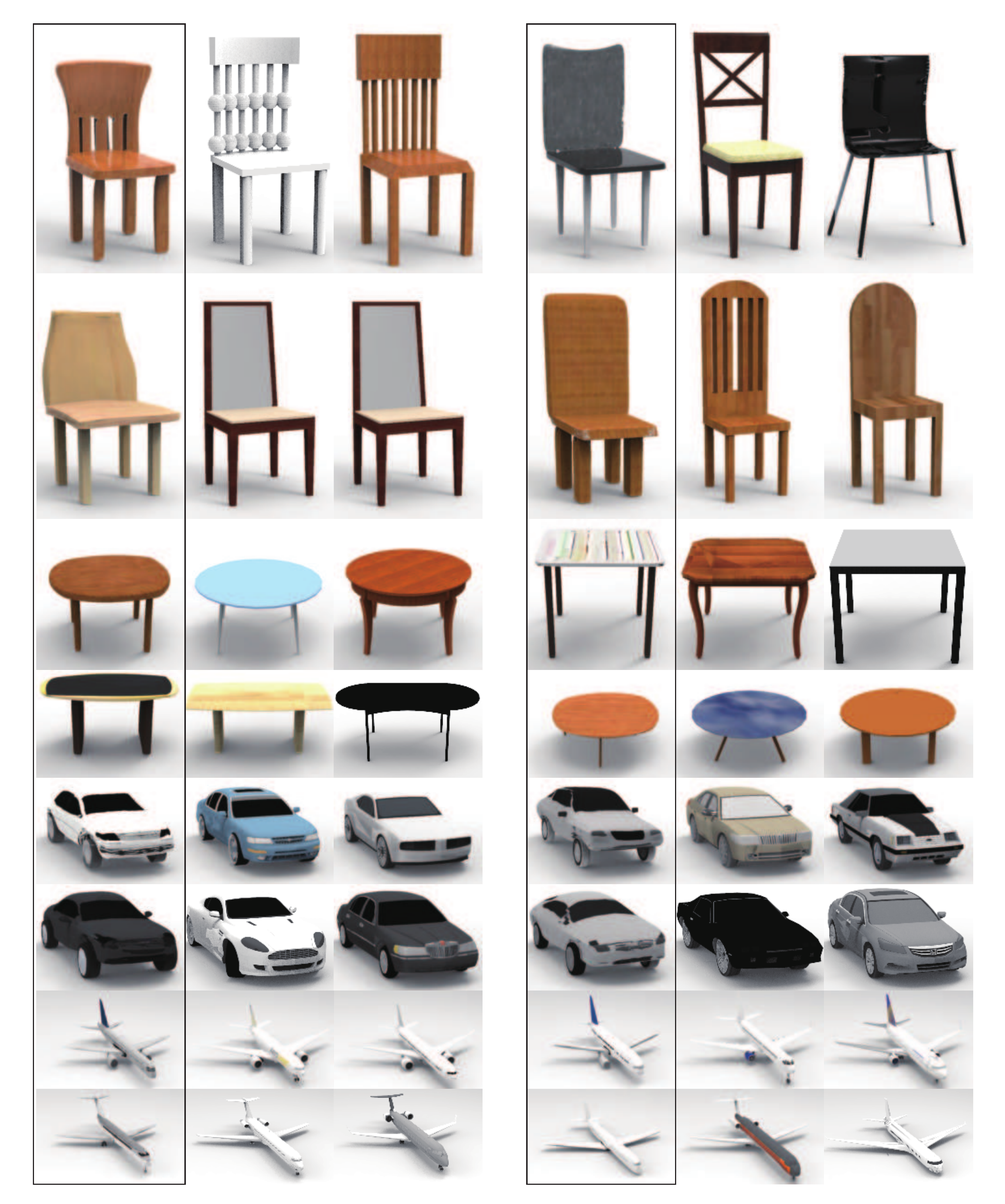}
  \caption{
      \rznn{Randomly sampled results of textured mesh generation by TM-NET (best viewed by zooming in), except for the first chair, which was picked to show texture transparency.
          We show four results per shape category, with more in the supplementary material.
          In each triplet, the left one is the generated textured mesh; the middle shows its closest neighbor from the training set, in terms of shape geometry as measured using Chamfer Distance (CD) on 1,024 uniformly sampled points on each mesh surface; the right shows the closest neighbor from the training set, in terms of appearance as measured by SSIM on multi-view images.
          For each mesh, we normalize it to a unit cube and render it from six fixed viewpoints.
      }
  }
\label{fig:Generation}
\end{figure}

\begin{figure}[!t]
\centering
	{
	\begin{tabular}{cc}
	\vspace{3mm}
	\rotatebox{90}{\quad \hspace{4mm} {\small VON}}
	&
	\vspace{-3mm}
	\hspace{-3mm}
	{\includegraphics[width=0.15\linewidth]{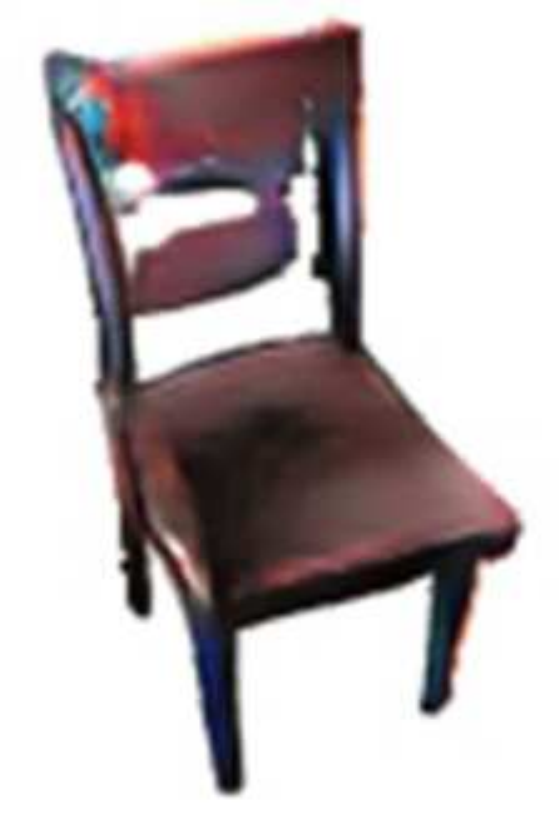}}
	{\includegraphics[width=0.15\linewidth]{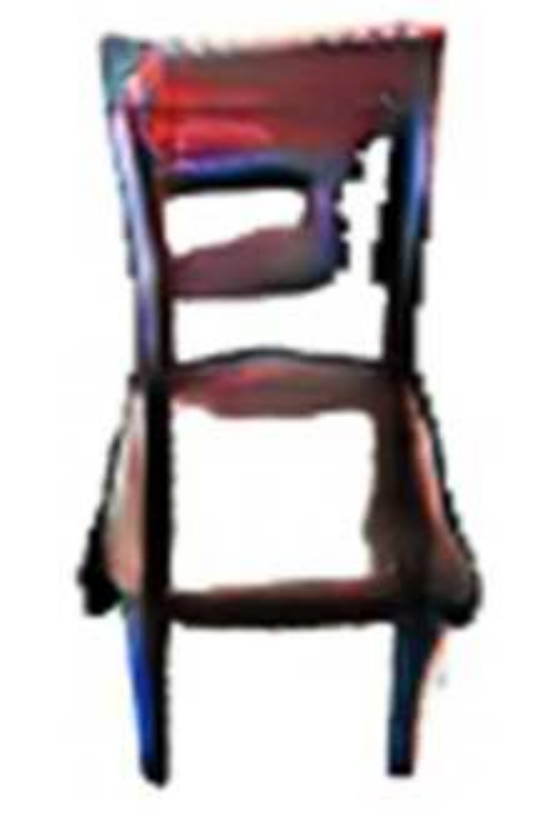}}
	{\includegraphics[width=0.15\linewidth]{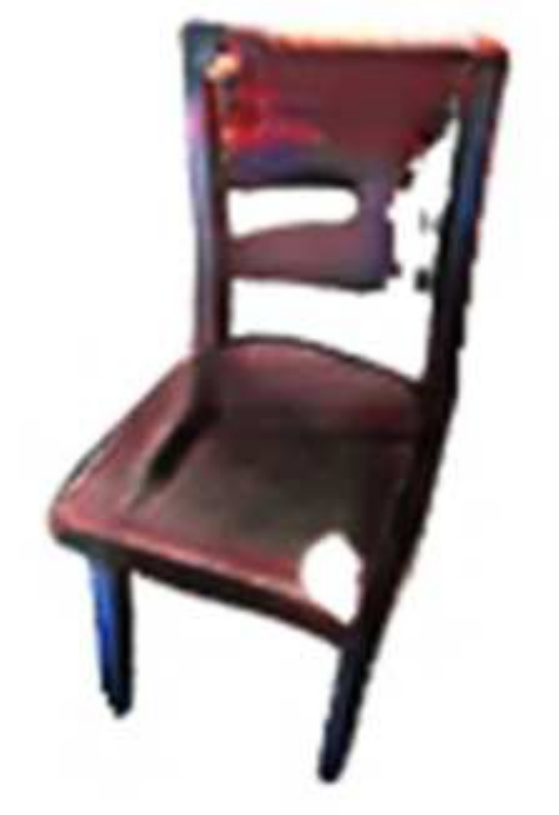}}
  	{\includegraphics[width=0.15\linewidth]{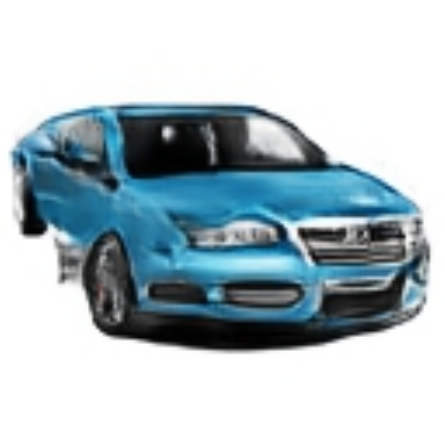}}
    {\includegraphics[width=0.15\linewidth]{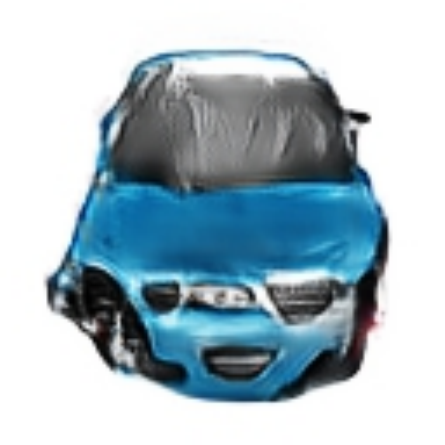}}
    {\includegraphics[width=0.15\linewidth]{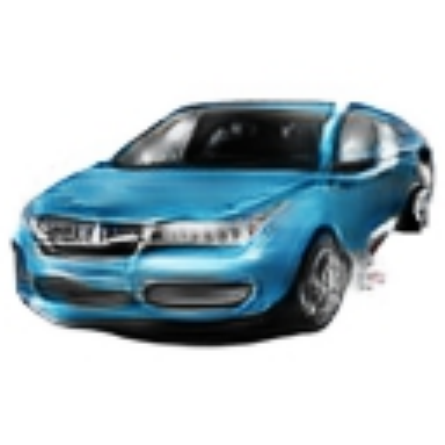}}
	\\
	\rotatebox{90}{\quad \hspace{4mm} {\small TM-NET}}
	&
	\vspace{-3mm}
	\hspace{-3mm}
	{\includegraphics[width=0.15\linewidth]{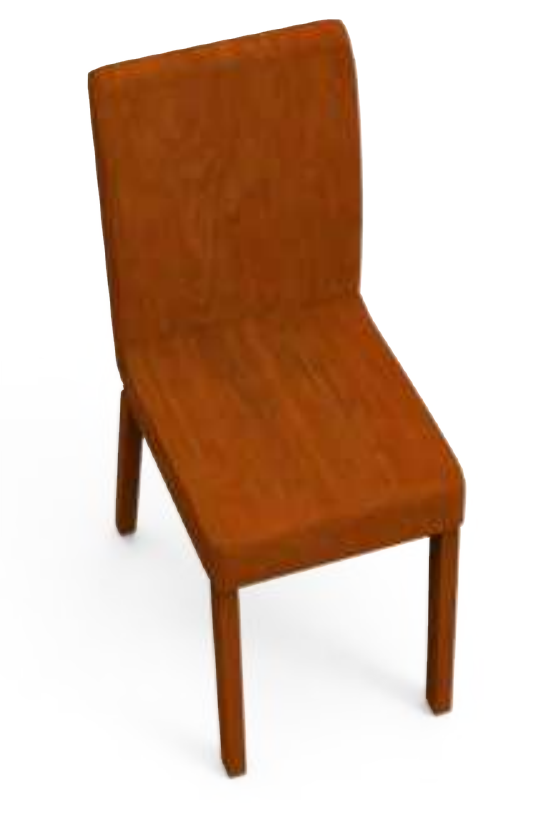}}
	{\includegraphics[width=0.15\linewidth]{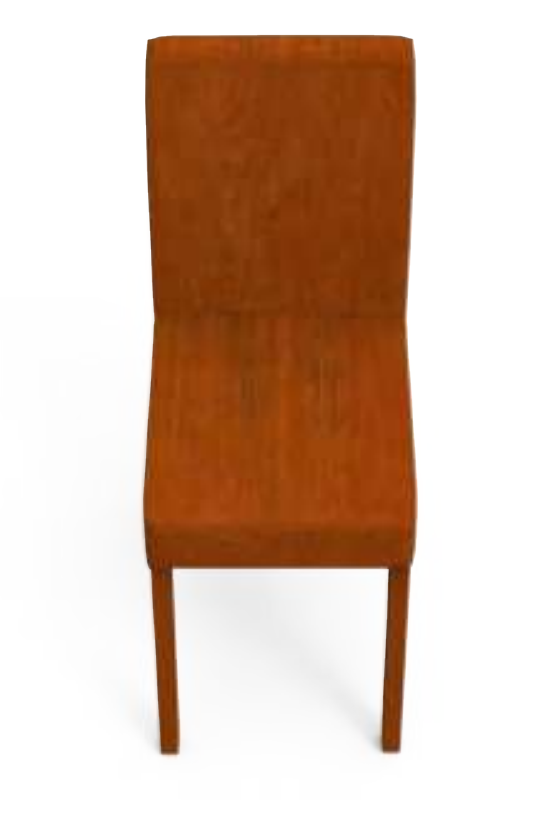}}
	{\includegraphics[width=0.15\linewidth]{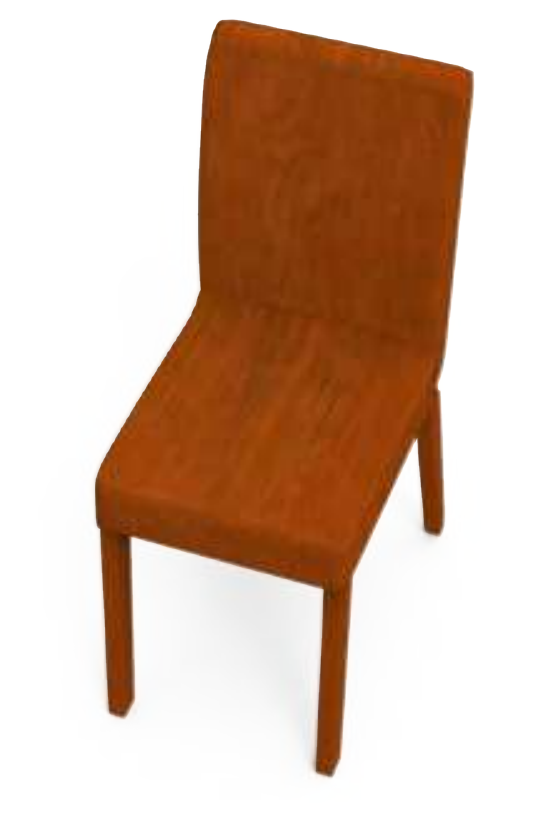}}
	{\includegraphics[width=0.15\linewidth]{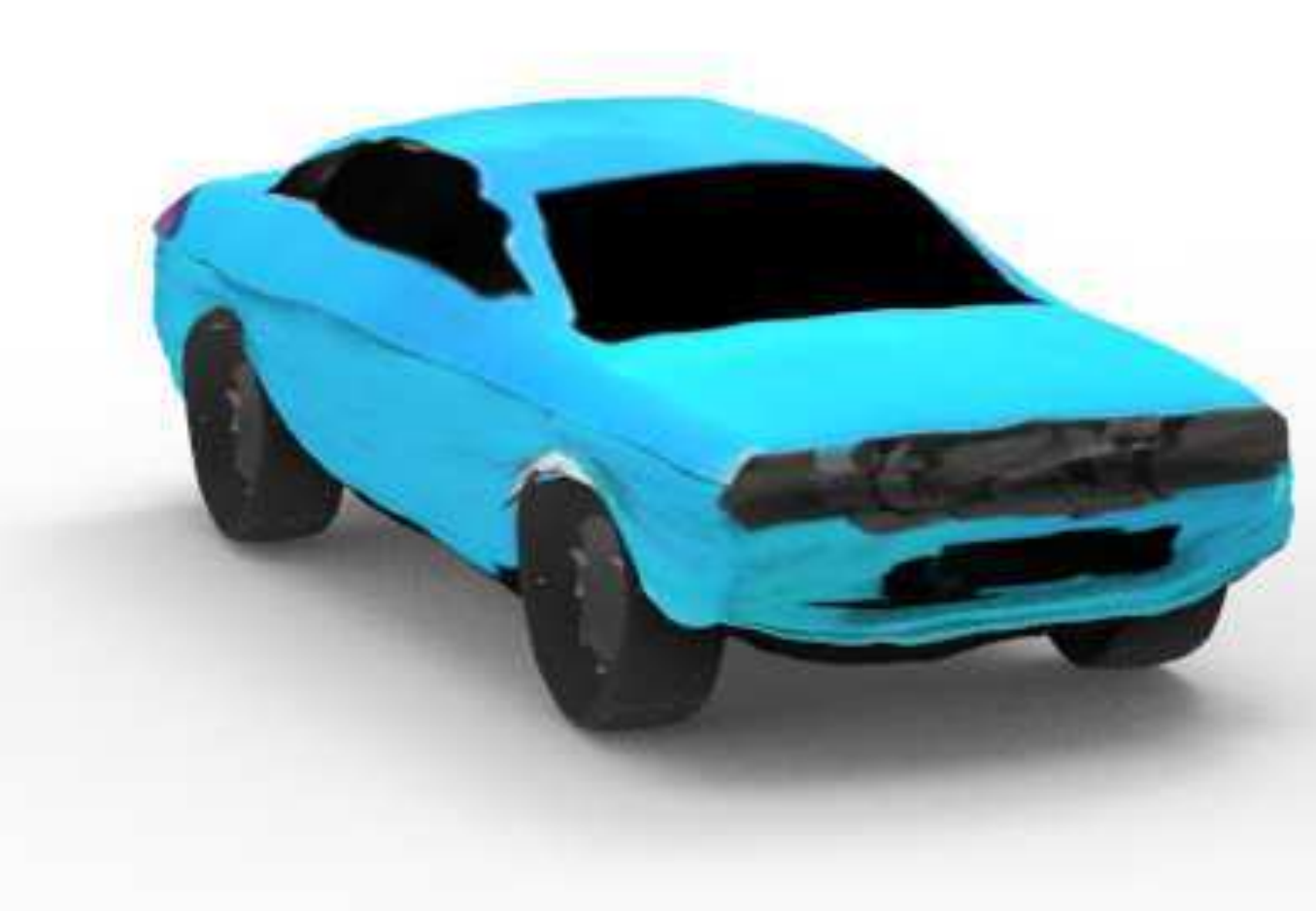}}
    {\includegraphics[width=0.15\linewidth]{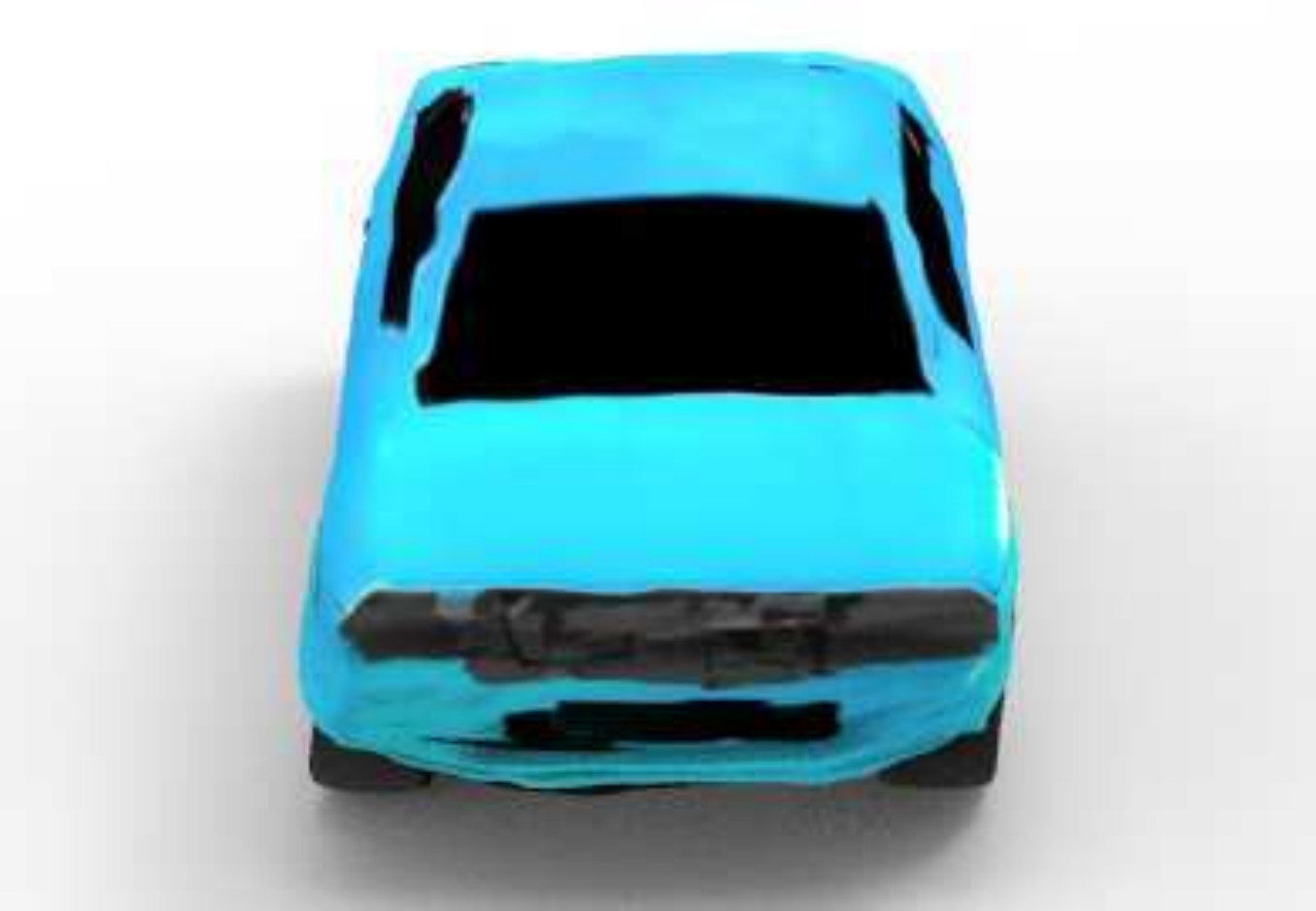}}
    {\includegraphics[width=0.15\linewidth]{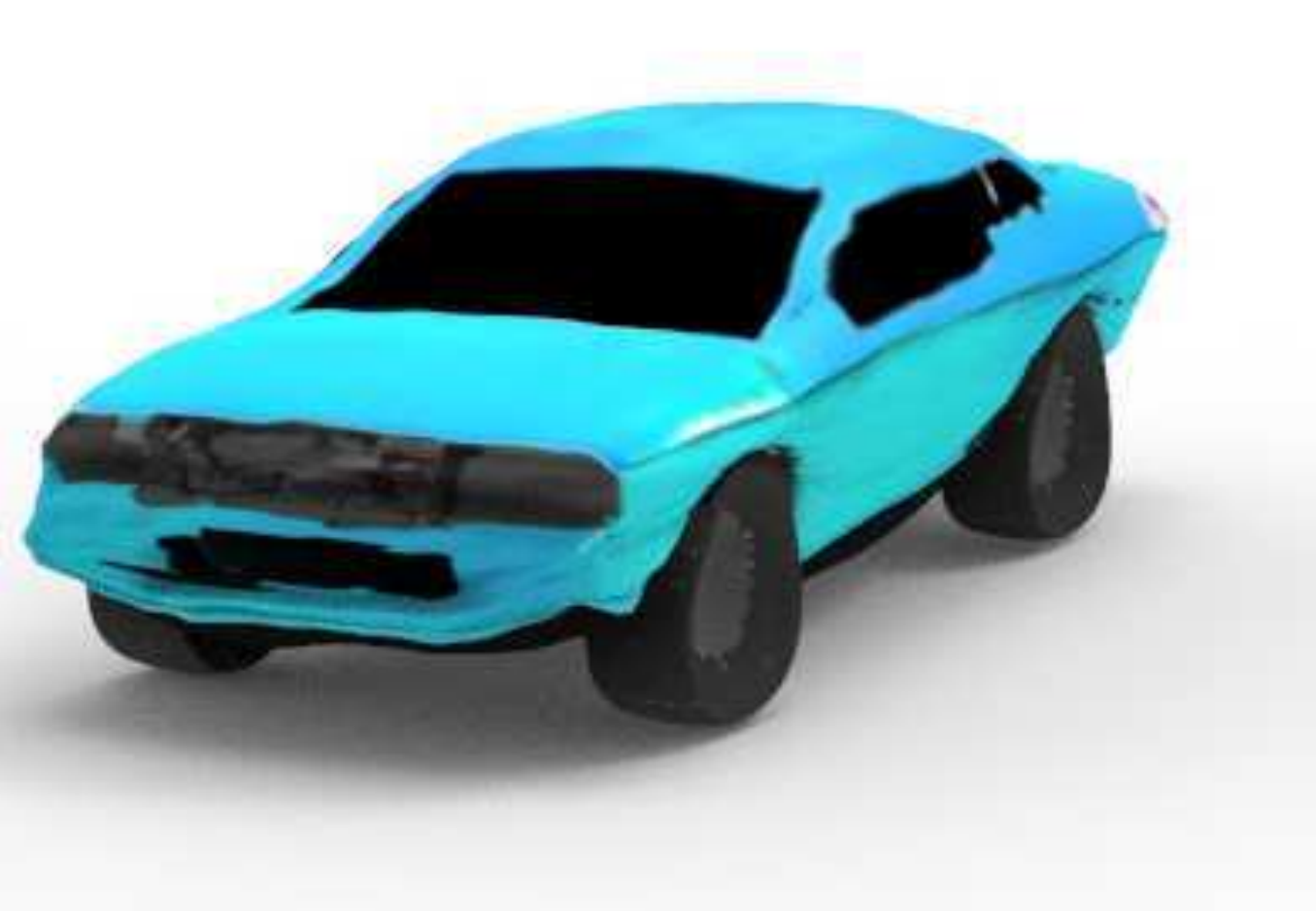}}
	\end{tabular}
    }
	\caption{\rznn{A comparison to VON~\cite{VON} for novel view synthesis of object textures (top row). The two VON results were randomly selected, then for each, we randomly selected a TM-NET result with similar overall color as the VON result. We observe texture inconsistency from different views since each viewpoint was generated separately with no global regularization. In contrast, TM-NET generates an entire textured 3D shape which, when rendered in multiple views, ensures texture consistency.}}
\label{fig:NVS}
\end{figure}

\begin{table}[!t]
    \centering
    \caption{
        \rznn{
            Quantitative comparison between VON and TM-NET in terms of their generative capabilities, measured in {\bf FID $\uparrow$}~\cite{NIPS2017_7240}.
            Note that VON did not contain the plane or table categories.
        }
    }
    \begin{tabular}{ccc cc}
        Method & Car     & Chair   & Plane    & Table   \\
        \hline
        VON    & 83.3    & 51.8    &   -      &   -     \\
        \hline
        TM-NET   & \textbf{56.9}    & \textbf{46.4}  & 34.1     &  42.3   \\
        \hline
    \end{tabular}%
    \label{tab:FIDComparison}
\end{table}%

\paragraph{\rznn{Shape-Conditioned Mesh Texturing}}

\rznn{Given an untextured 3D shape, TM-NET can predict plausible and diverse textures for the shape that are conditioned on the shape's geometry. Figure~\ref{fig:ConditionalAutomaticTexture} shows such shape texturing results, where the same input shape can lead to different
textures, by varying the sampling. The part-by-part texture generation is clearly revealed as different parts of the same shape can also be textured differently (e.g., chair seat vs.~back and tabletop vs.~leg).
One could argue that some part textures generated by TM-NET are not quite compatible, e.g., see the second chair result in the first row. In regards to this, we shall point out that even more drastic texture ``incompatibilities" can be observed from the training set, as shown in Figure~\ref{fig:incomp_chairs}.
In the supplementary material, we show more generated samples for shape texturing, as well as the closest training textures to the generated samples.

We compare our method to Texture Fields (TF)~\cite{OechsleICCV2019}, a state-of-the-art method for texturing an input 3D shape with a texture image sampled from a latent space. The comparisons were done on the same input shapes, from the same viewpoints, and using the same rendering parameters, but the output textures are generally different since both were sampled. We show some visual results in Figure~\ref{fig:texturecompare}, while for a quantitative study, we compare the LPIPS values between different results generated by the same method to demonstrate the diversity, and report fooling rate through a user study to evaluate realism.
As demonstrated by results in Table~\ref{tab:DiversityQuantitativeComparison}, texturing results by TM-Net were deemed to be both more realistic and more diverse than those from Texture Fields.}

\begin{figure}[!t]
    \centering
    \includegraphics[width=0.18\linewidth]{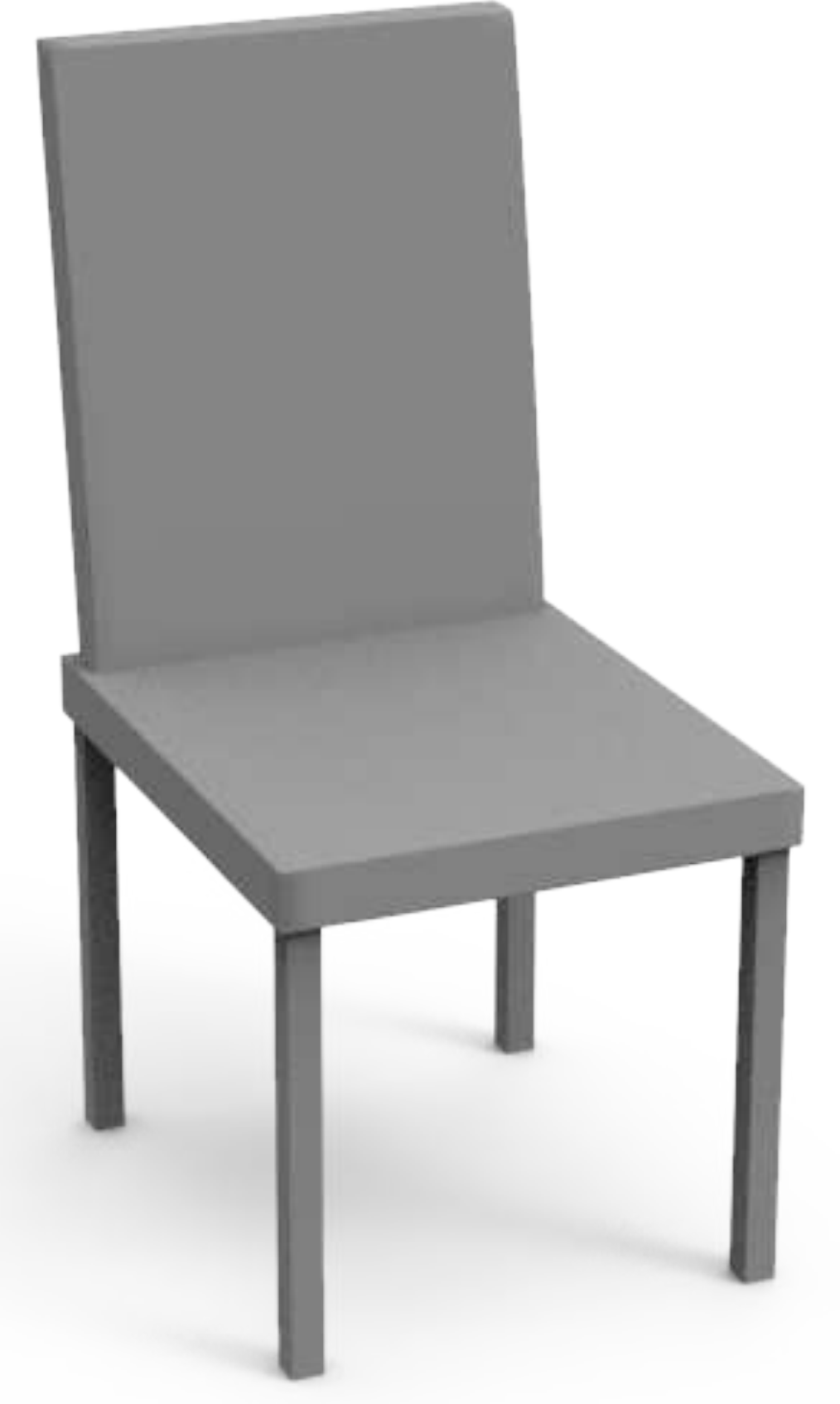}
    \includegraphics[width=0.18\linewidth]{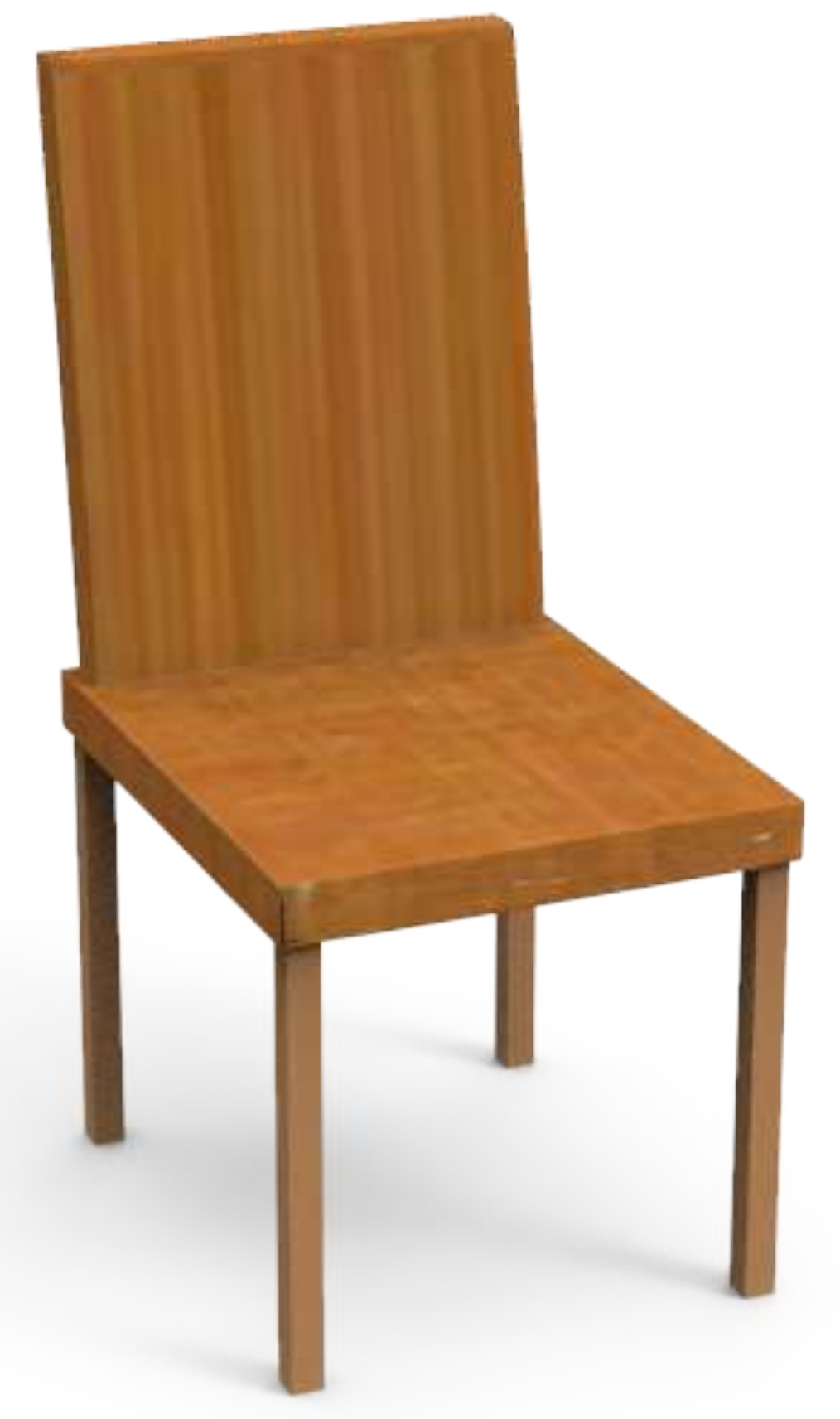}
    \includegraphics[width=0.18\linewidth]{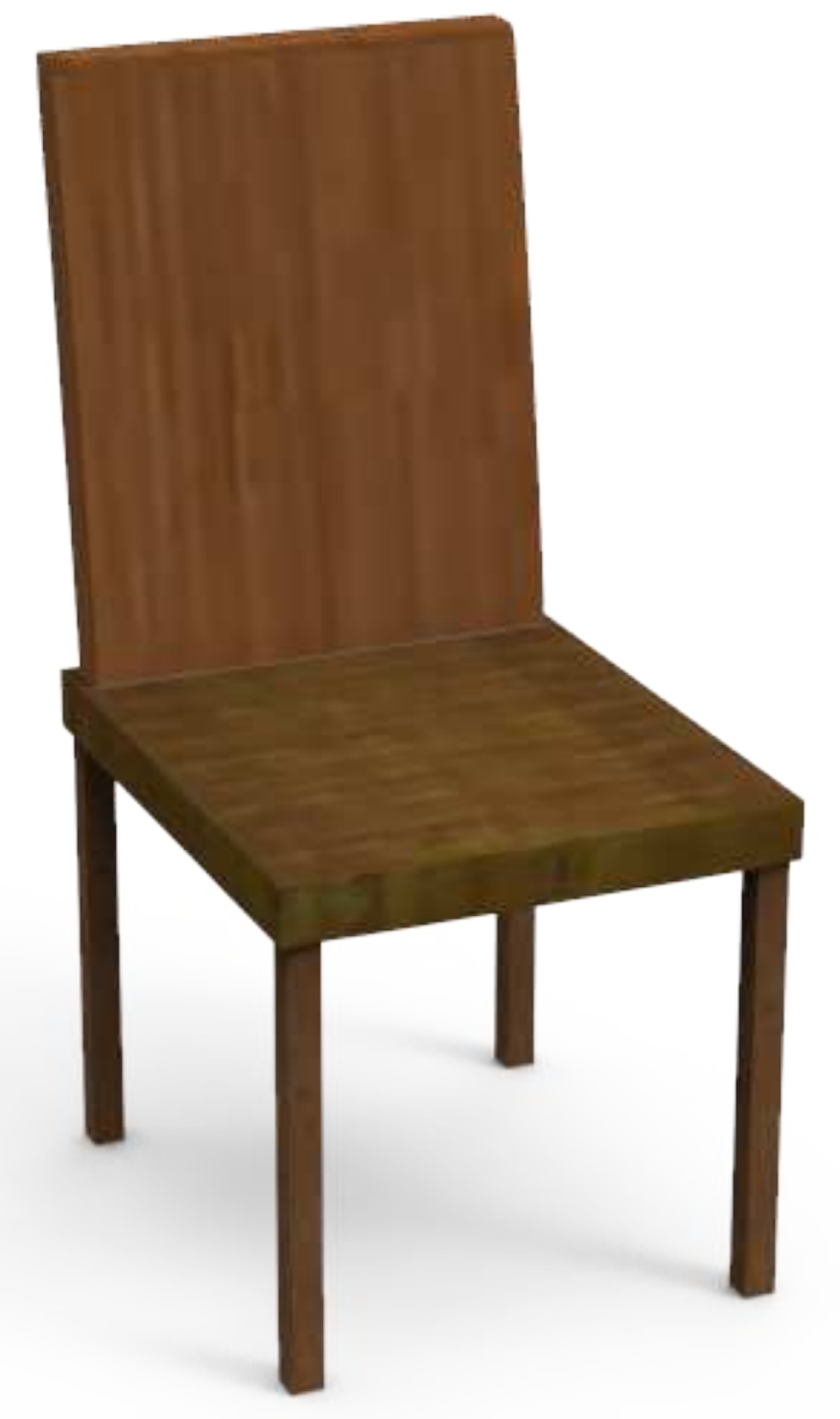}
    \includegraphics[width=0.18\linewidth]{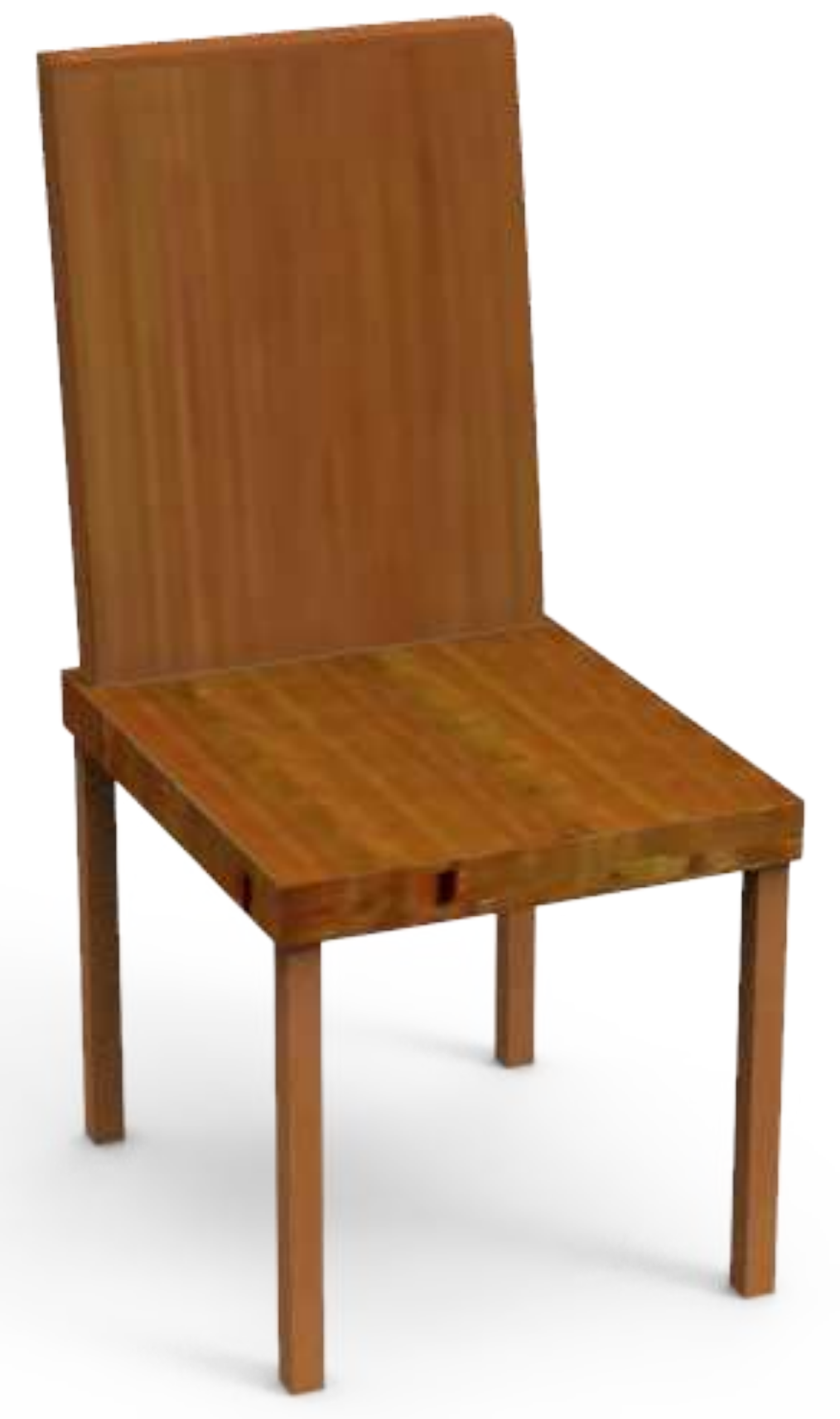}
    \includegraphics[width=0.18\linewidth]{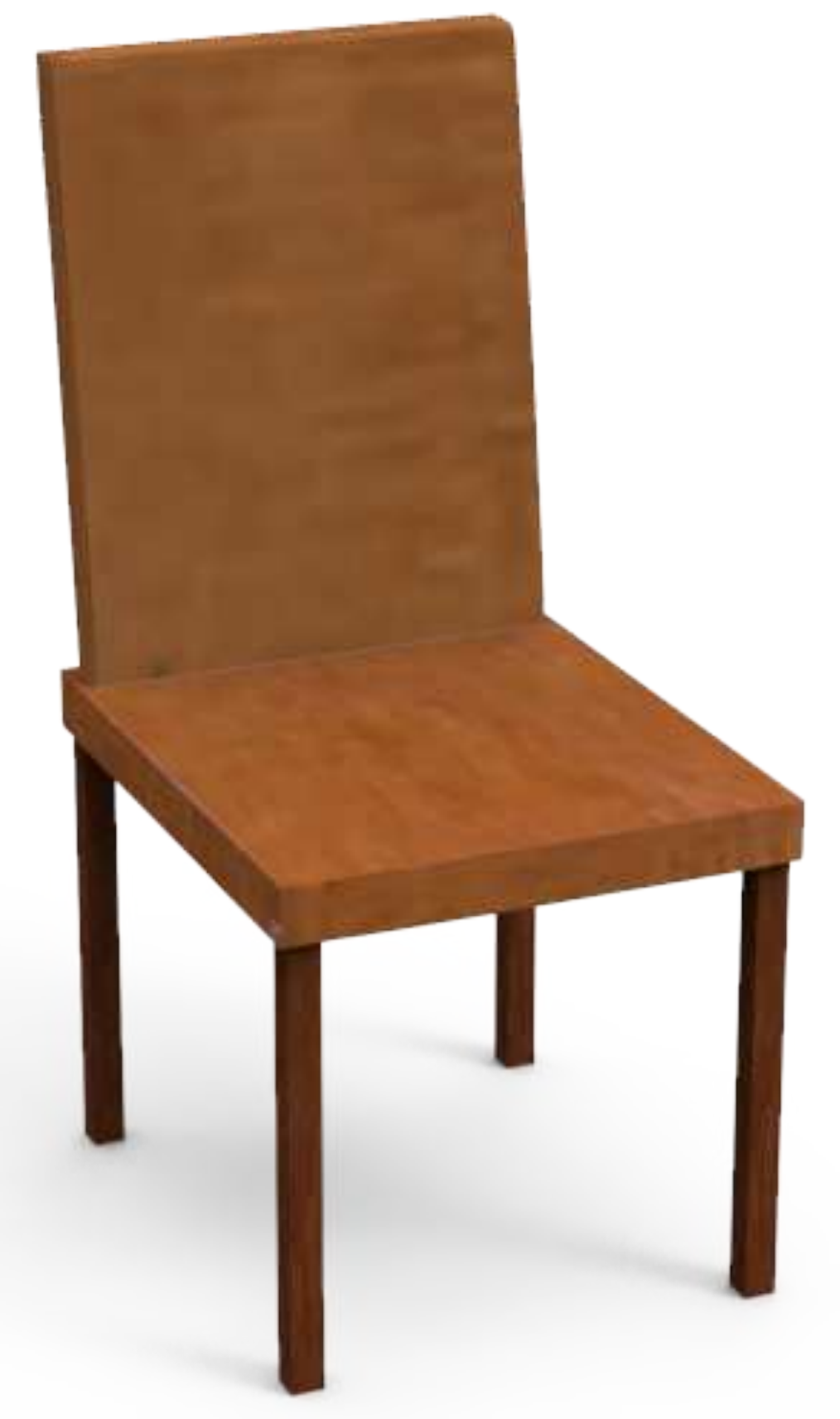}
    \\
    \includegraphics[width=0.18\linewidth]{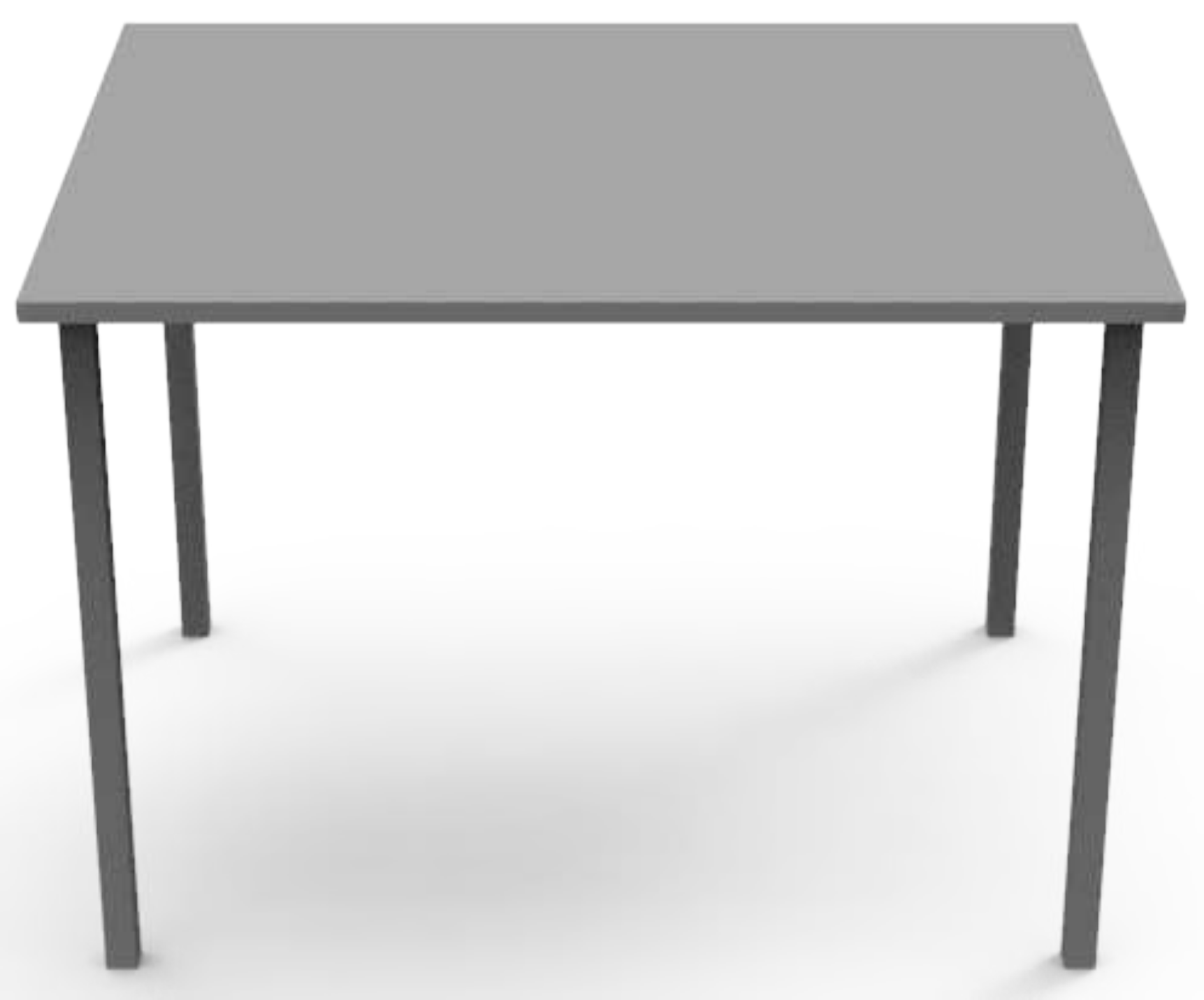}
    \includegraphics[width=0.18\linewidth]{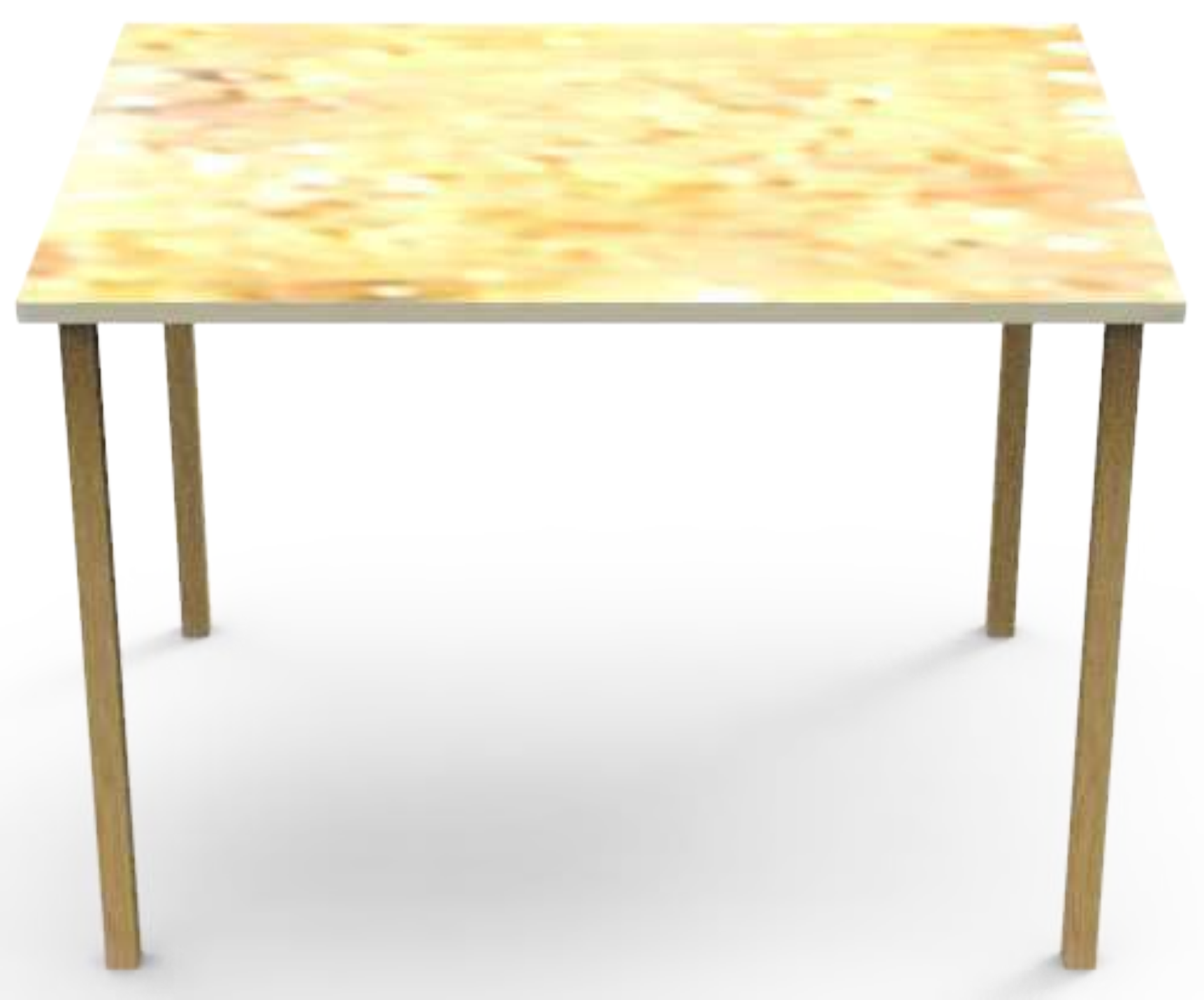}
    \includegraphics[width=0.18\linewidth]{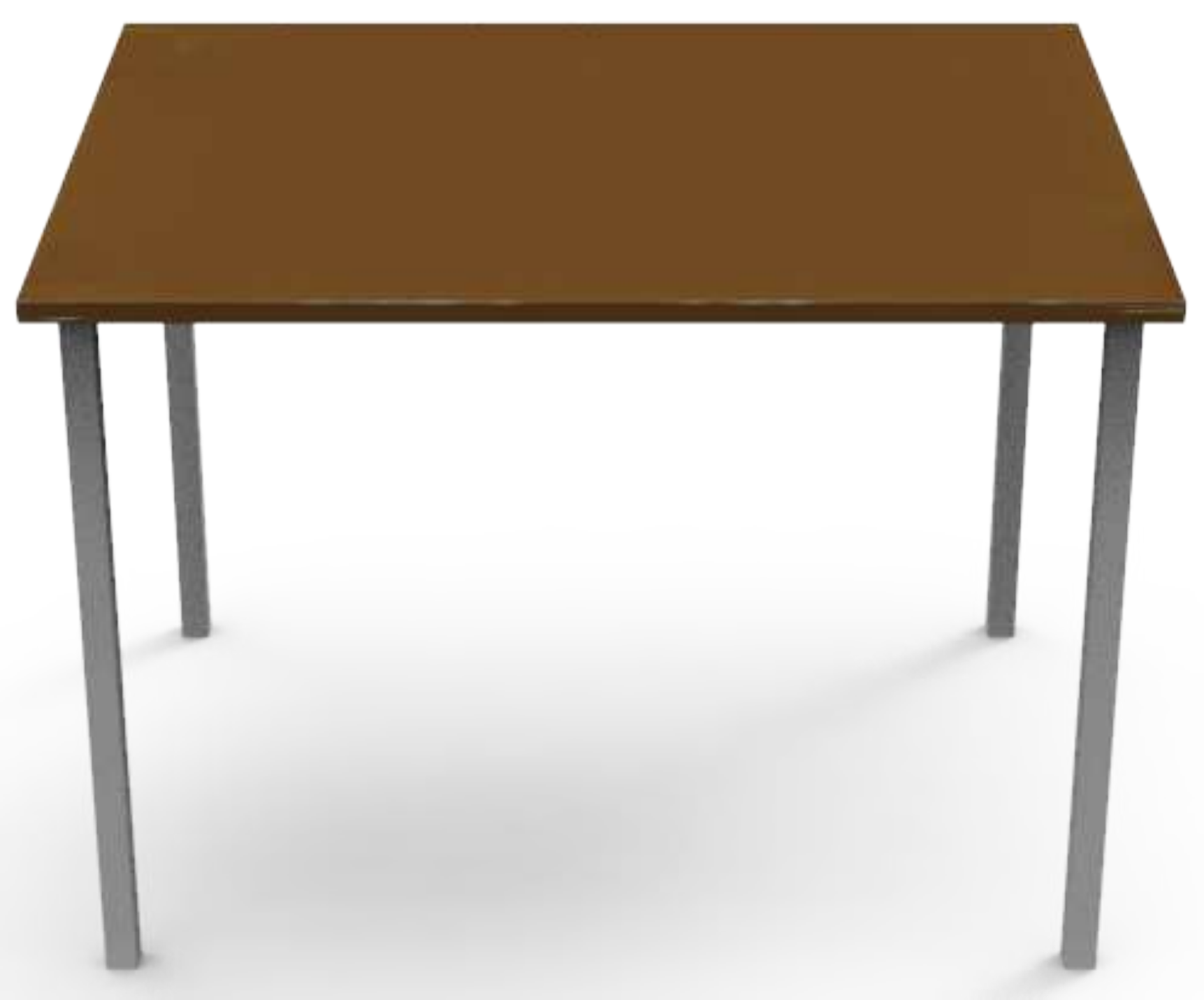}
    \includegraphics[width=0.18\linewidth]{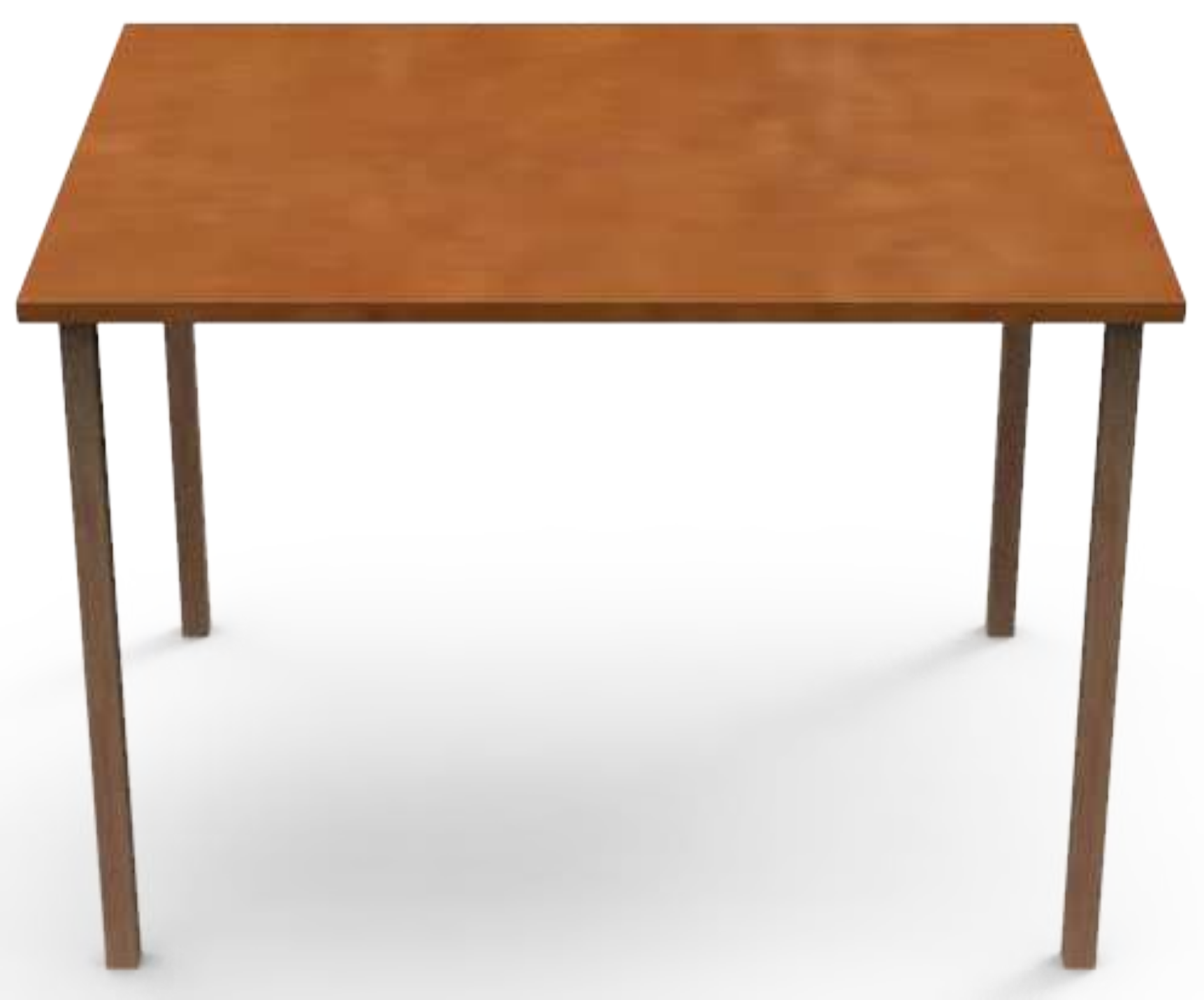}
    \includegraphics[width=0.18\linewidth]{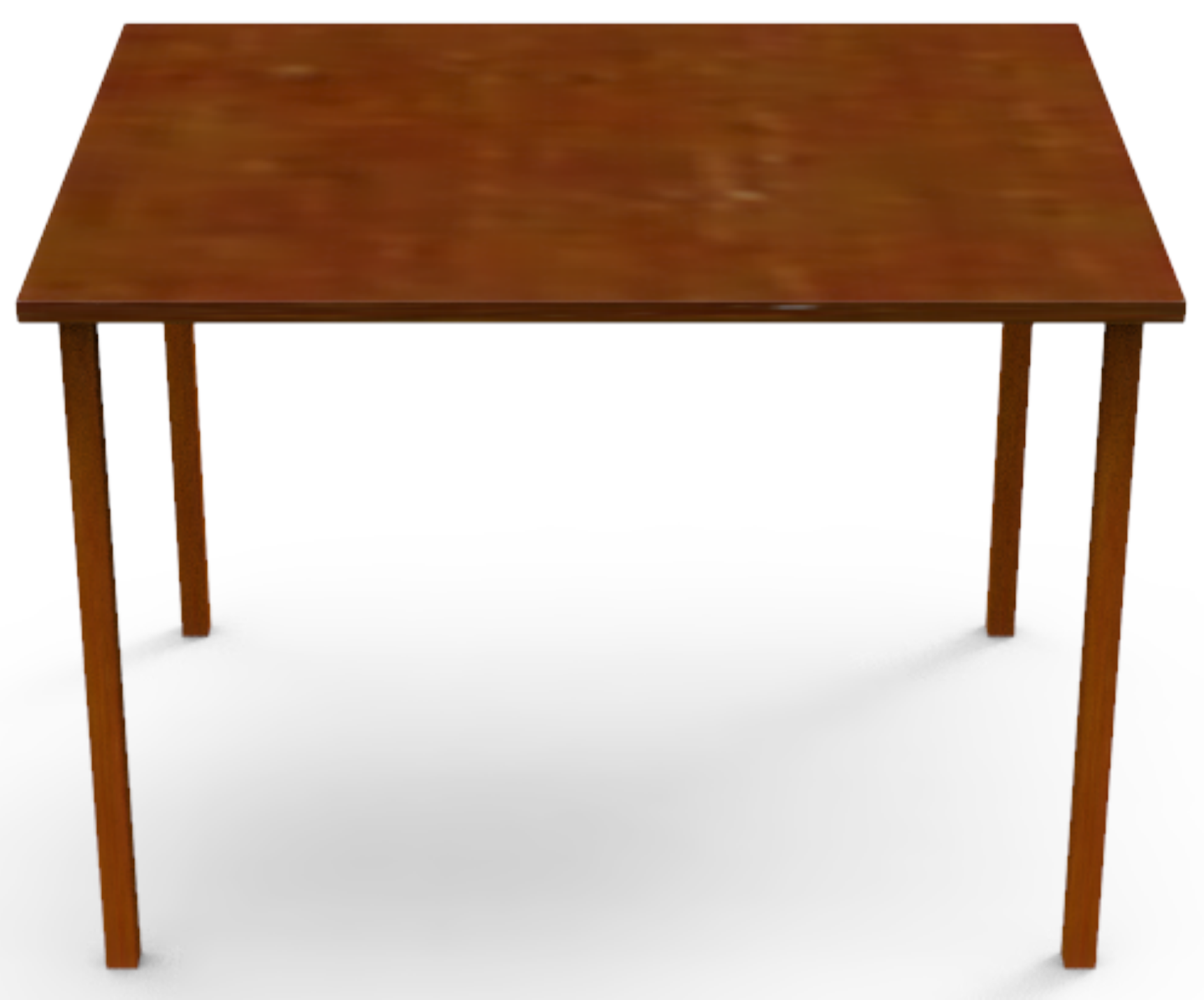}
    \\
    \includegraphics[width=0.18\linewidth]{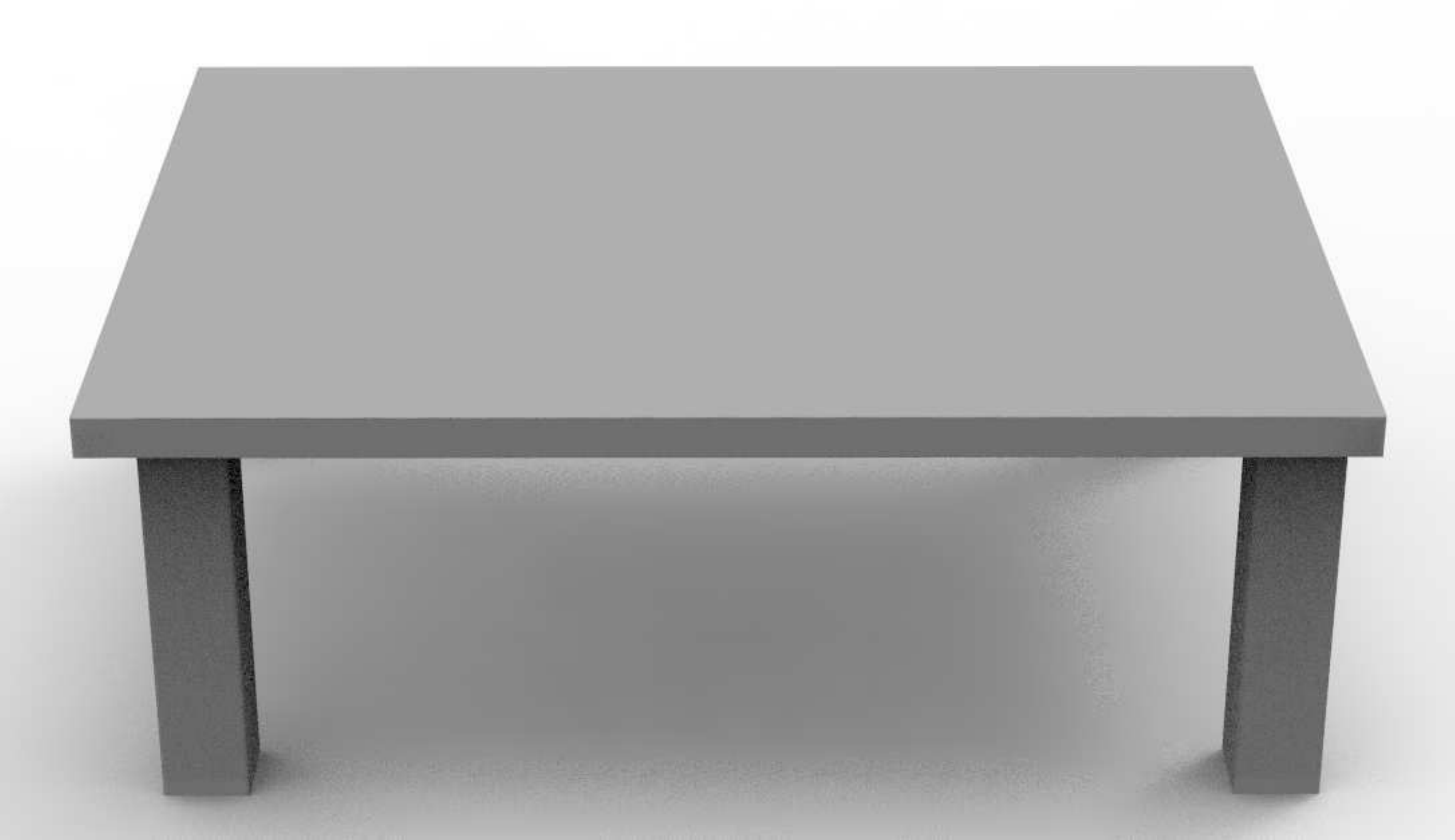}
    \includegraphics[width=0.18\linewidth]{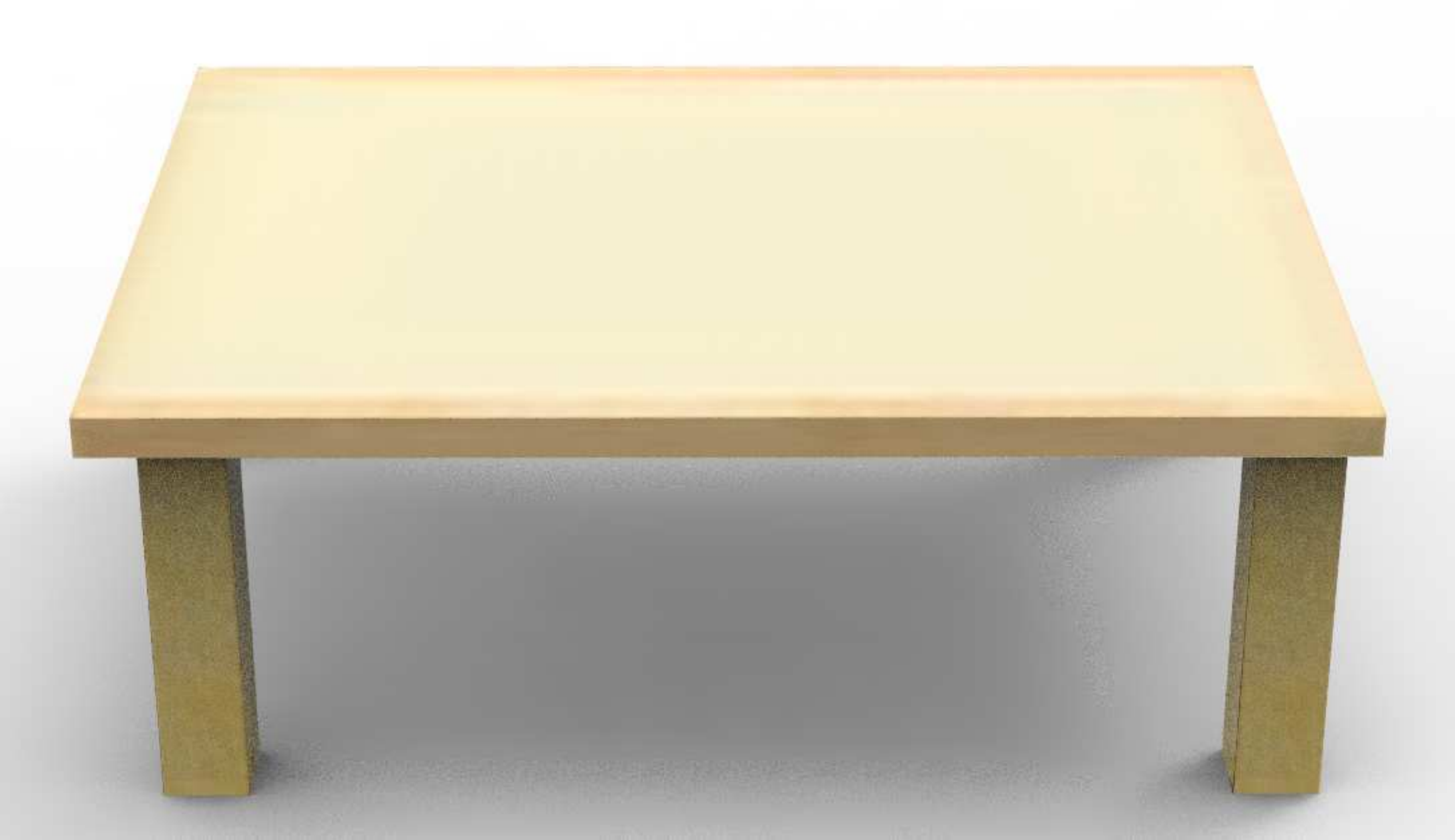}
    \includegraphics[width=0.18\linewidth]{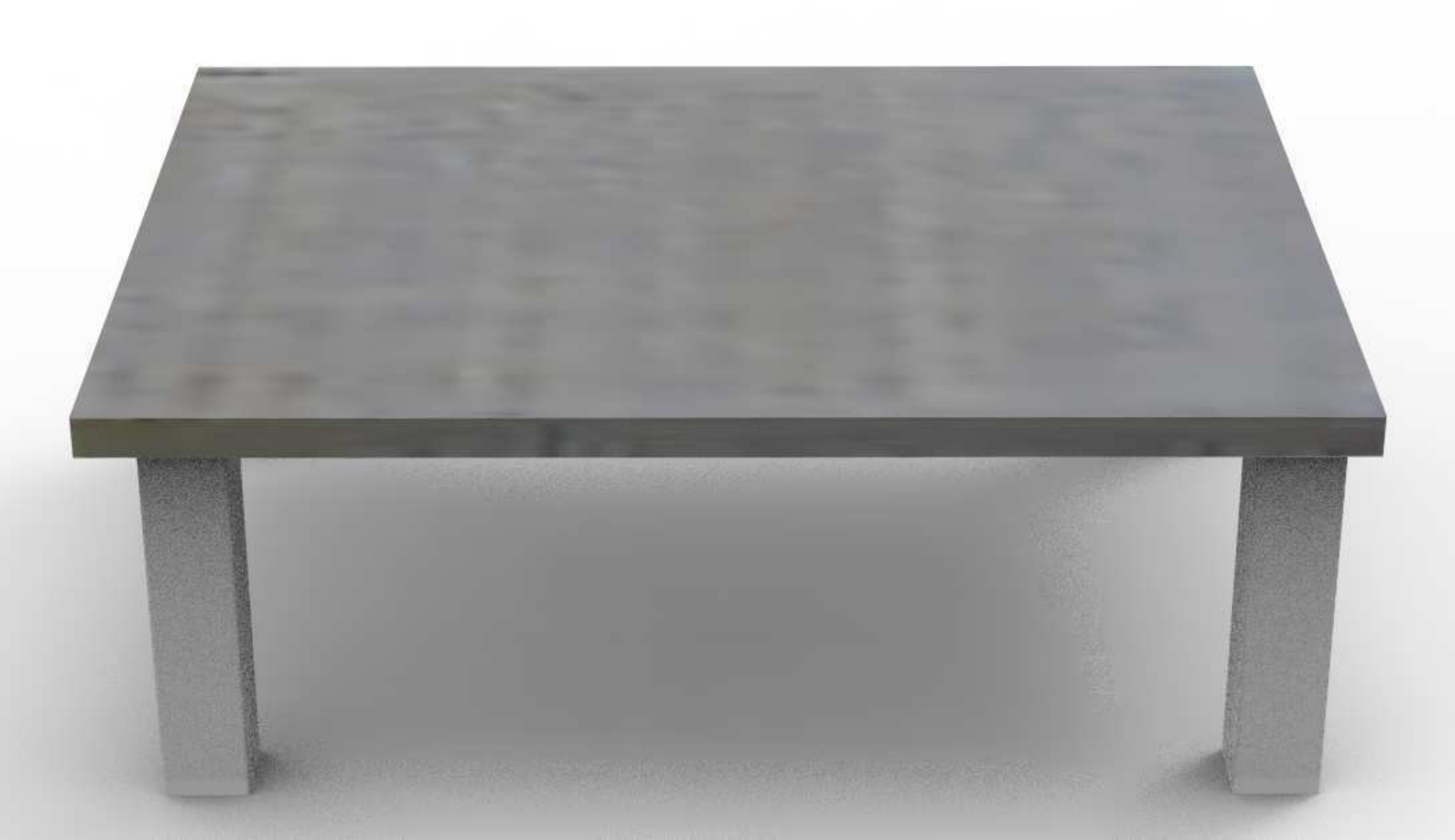}
    \includegraphics[width=0.18\linewidth]{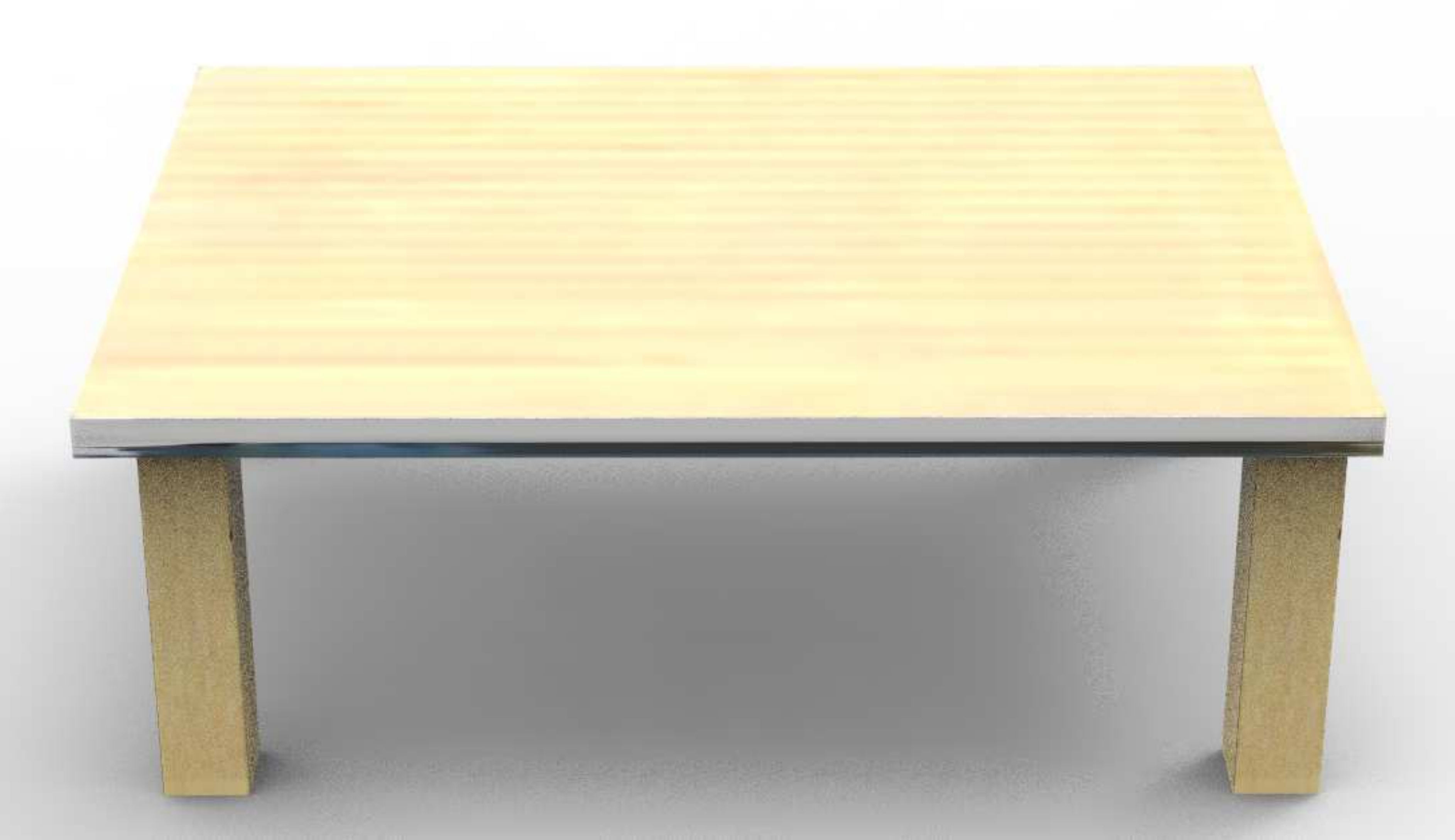}
    \includegraphics[width=0.18\linewidth]{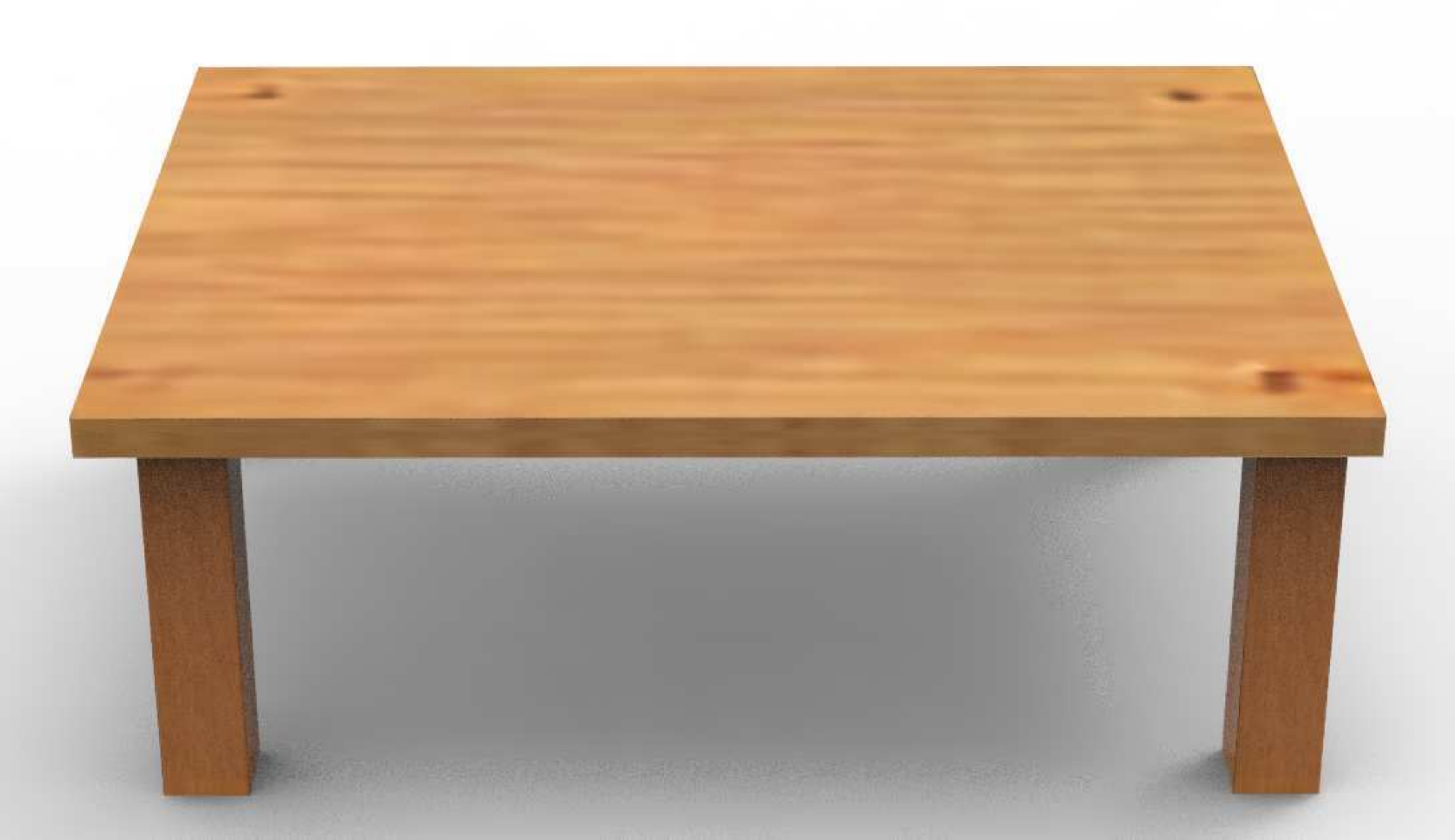}
    \caption{\rznn{Randomly selected shape-conditioned texturing results by four samplings of the learned probability distribution conditioned only on input geometry (left). The seed parts are chair seats and tabletops, respectively. Note the plausible and diverse textures generated by TM-NET, as well as different textures generated for different parts in the same shape.}}
    \label{fig:ConditionalAutomaticTexture}
\end{figure}

\begin{figure}[!t]
    \centering
    \includegraphics[width=0.96\linewidth]{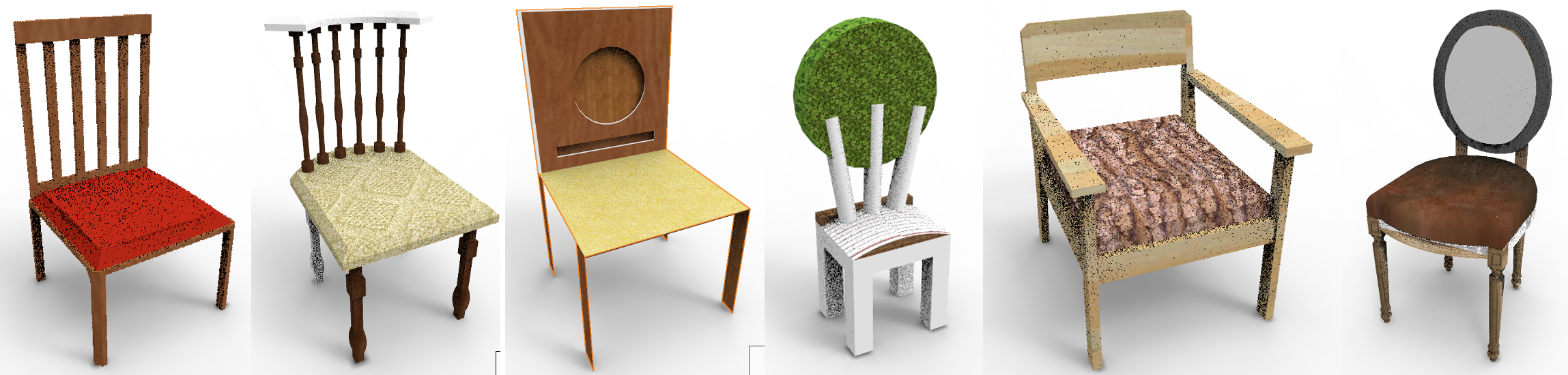}

    \caption{\rznn{Several chair meshes from the training set whose part textures may be perceived as ``incompatible'', yet ``stylish".}}
    \label{fig:incomp_chairs}
\end{figure}

\begin{figure}[!t]
    \centering{
    	\begin{tabular}{cc}
    		\vspace{3mm}
    		\rotatebox{90}{\quad \hspace{2mm} {\small TF}}
    		&
    		\vspace{-3mm}
    		\hspace{-3mm}
    		{\includegraphics[width=0.18\linewidth]{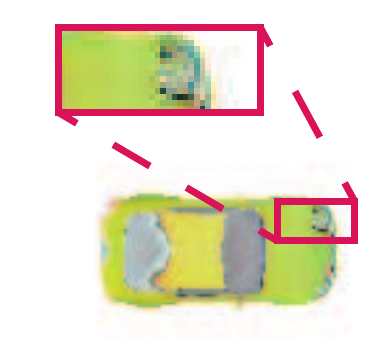}}
    		{\includegraphics[width=0.18\linewidth]{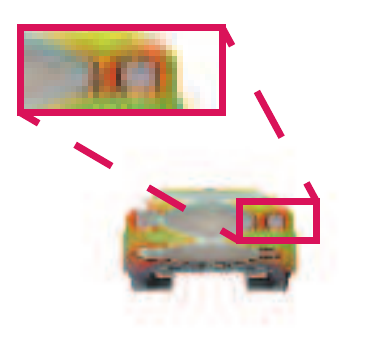}}
    		{\includegraphics[width=0.18\linewidth]{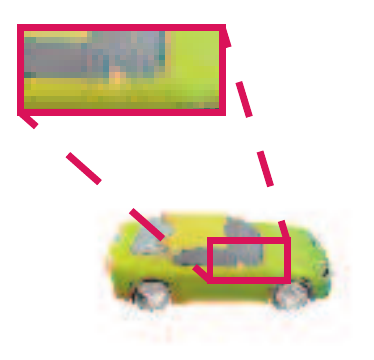}}
    		{\includegraphics[width=0.18\linewidth]{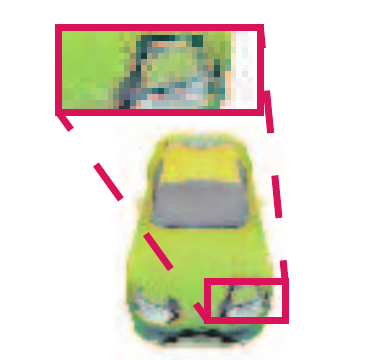}}
    		{\includegraphics[width=0.18\linewidth]{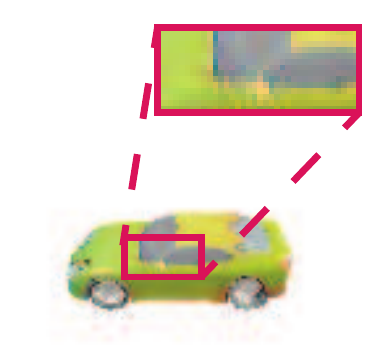}}
    		\\
    		\rotatebox{90}{\quad \hspace{1mm} {\small TM-NET}}
    		&
    		\hspace{-3mm}
    		{\includegraphics[width=0.18\linewidth]{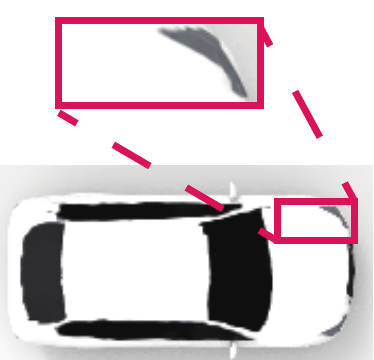}}
    		{\includegraphics[width=0.18\linewidth]{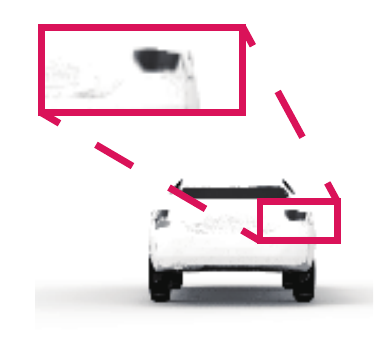}}
    		{\includegraphics[width=0.18\linewidth]{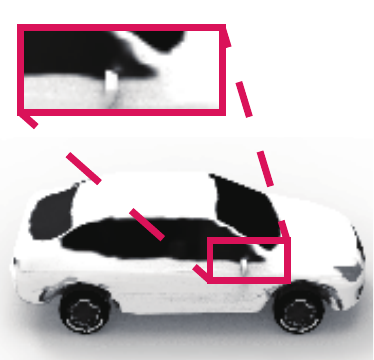}}
    		{\includegraphics[width=0.18\linewidth]{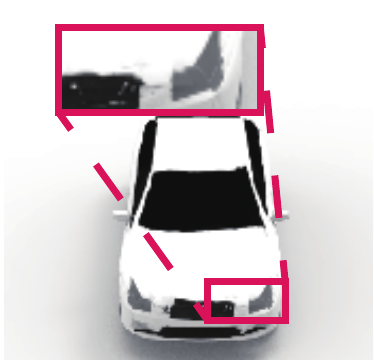}}
    		{\includegraphics[width=0.18\linewidth]{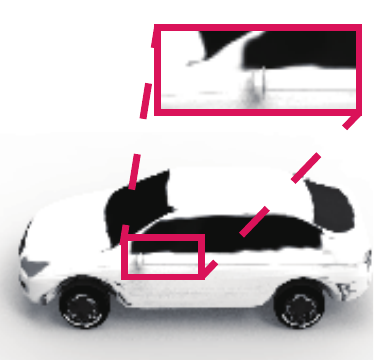}}
    		\\
    		\rotatebox{90}{\quad \hspace{3mm} {\small TF}}
    		&
    		{\includegraphics[width=0.18\linewidth]{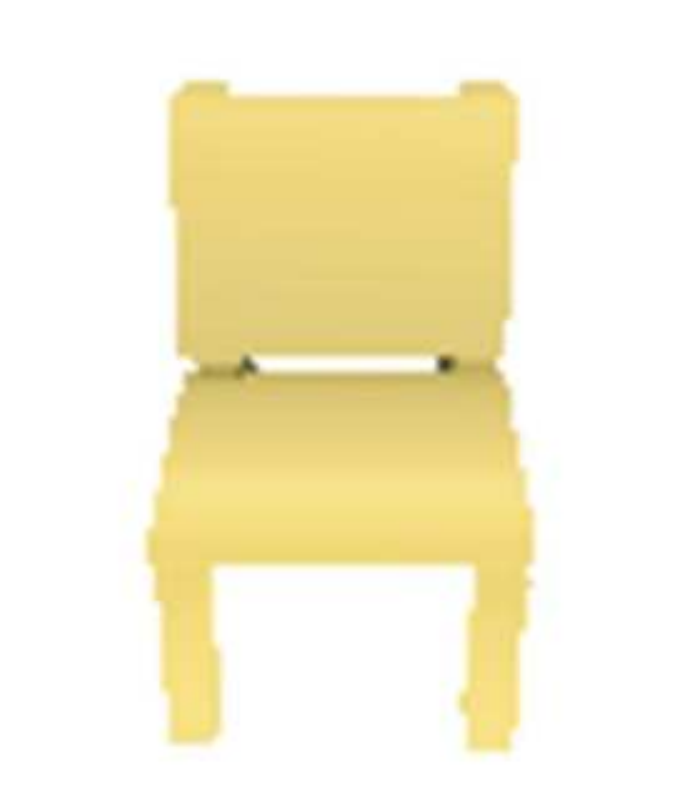}}
    		{\includegraphics[width=0.18\linewidth]{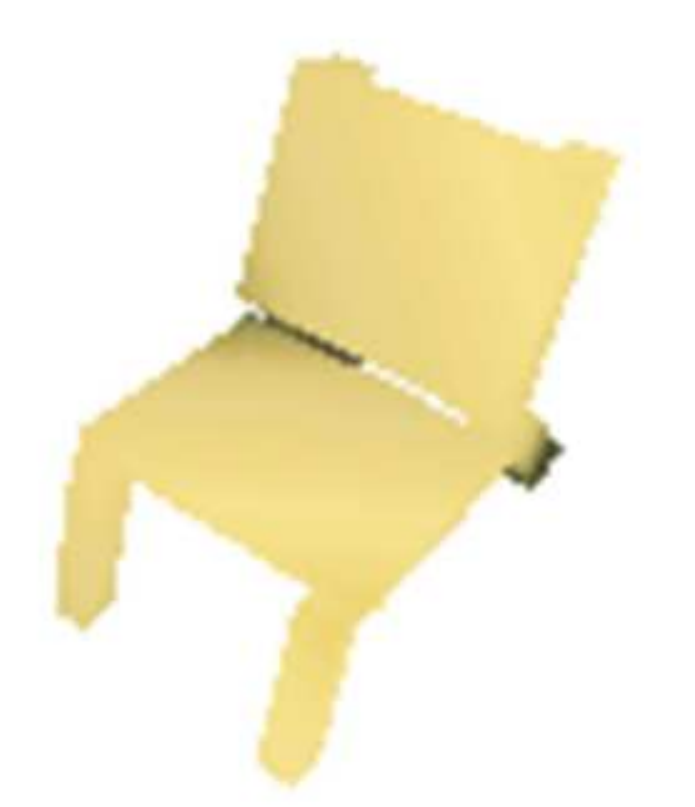}}
    		{\includegraphics[width=0.18\linewidth]{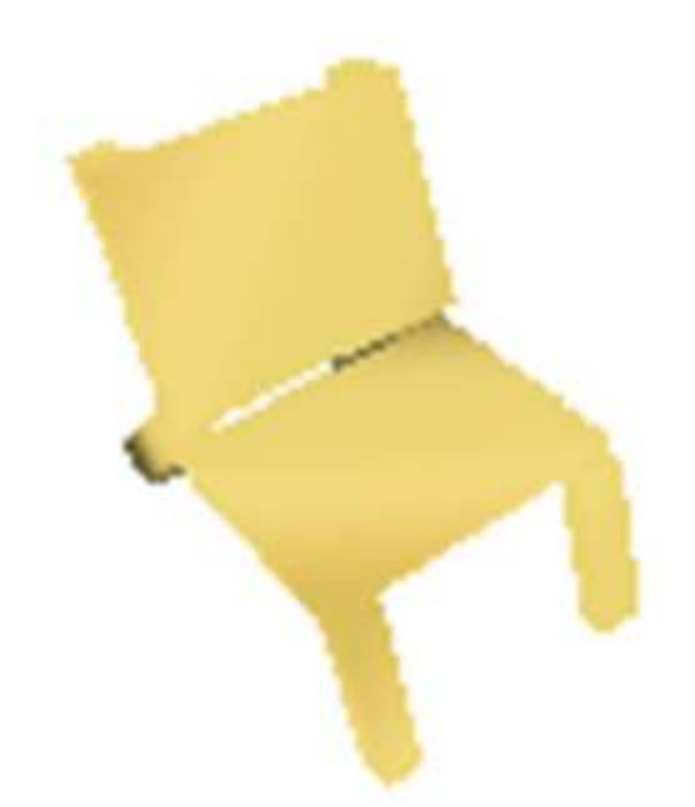}}
    		{\includegraphics[width=0.18\linewidth]{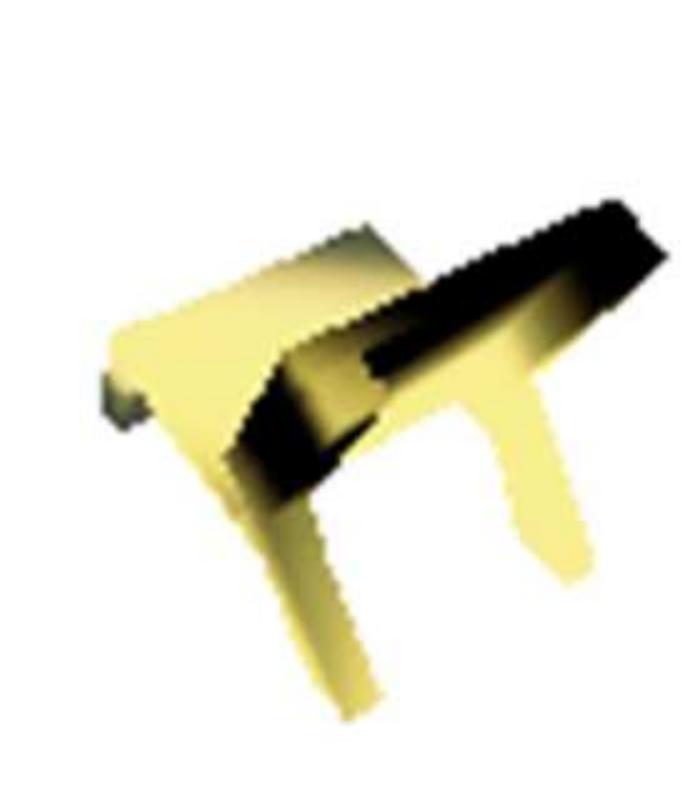}}
    		{\includegraphics[width=0.18\linewidth]{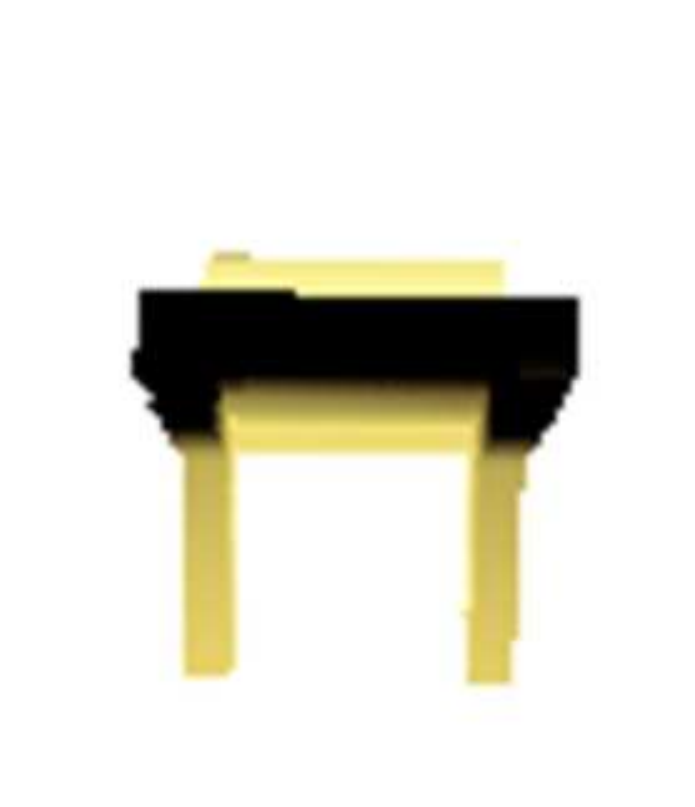}}
    		\\
    		\rotatebox{90}{\quad \hspace{2mm} {\small TM-NET}}
    		&
    		\vspace{-3mm}
    		\hspace{-3mm}
    		{\includegraphics[width=0.18\linewidth]{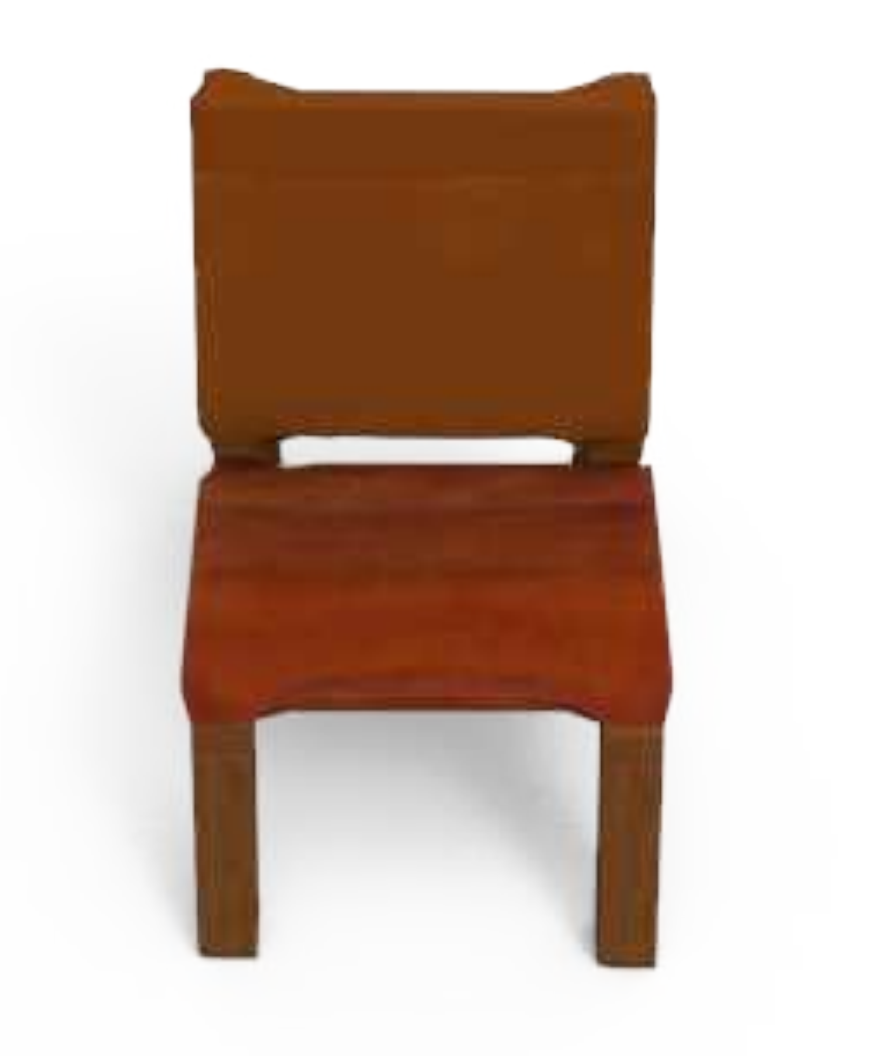}}
    		{\includegraphics[width=0.18\linewidth]{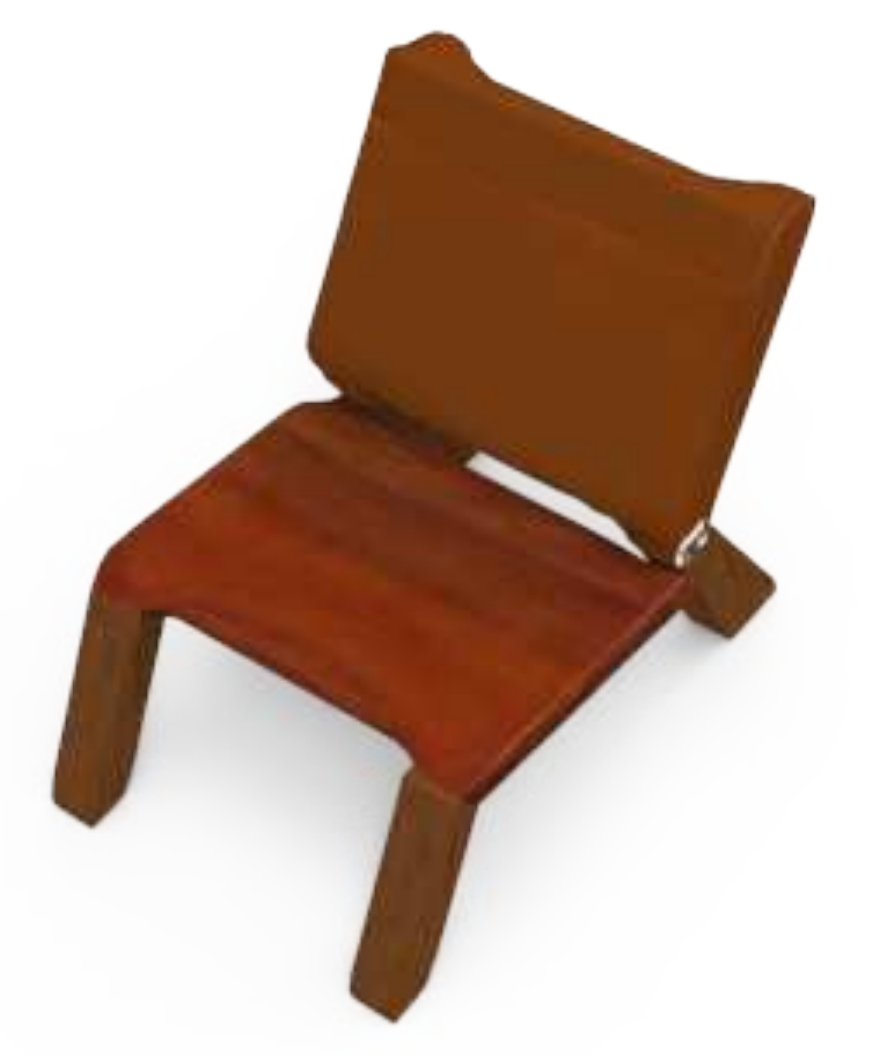}}
    		{\includegraphics[width=0.18\linewidth]{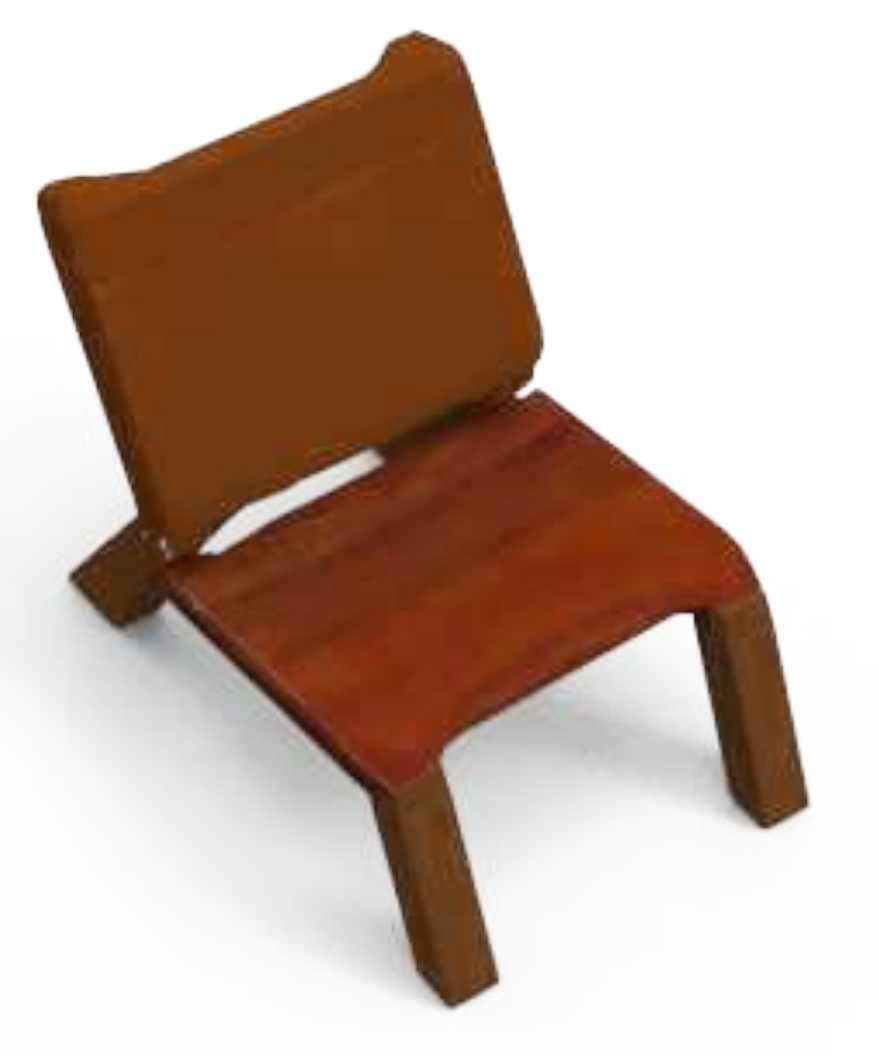}}
    		{\includegraphics[width=0.18\linewidth]{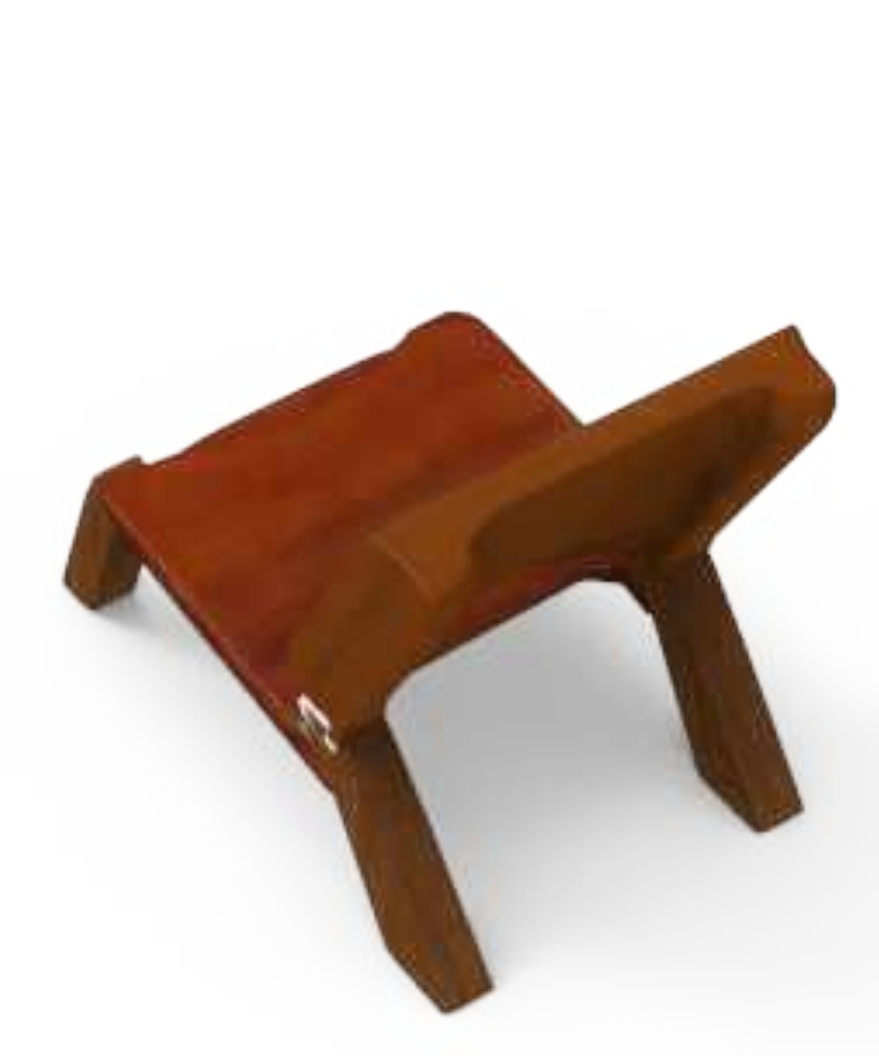}}
    		{\includegraphics[width=0.18\linewidth]{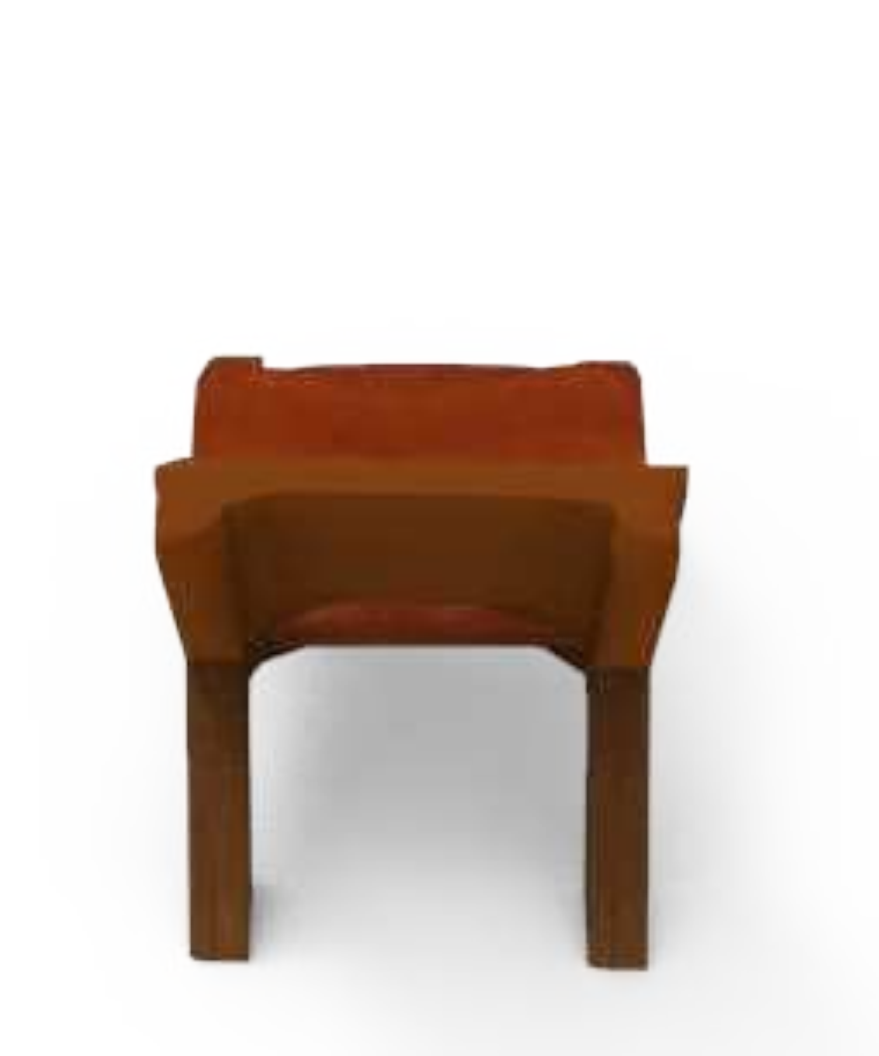}}
    	\end{tabular}
	}
	\caption{Comparison of our \rznn{mesh texturing} with Texture Fields (TF)~\cite{OechsleICCV2019}, on several \rznn{randomly picked} examples.}
\label{fig:texturecompare}
\end{figure}

\begin{table}[!t]
  \centering
  \caption{\rzn{Comparison to Texture Fields~\cite{OechsleICCV2019} in terms of diversity (measured by LPIPS) and realism (measured by fooling rate). As a diversity measure, larger values of LPIPS is better, i.e., more diverse.}} %
    \begin{tabular}{ccc}
     & Realism  & Diversity\\
    \hline
    Method & Fooling rate $\uparrow$  & LPIPS $\uparrow$ \\
    \hline
    Texture Fields  &         23.8\%  & 0.148 \\
    \hline
    Ours            & \textbf{41.5\%} & \textbf{0.263} \\
    \hline
    \end{tabular}%

	\label{tab:DiversityQuantitativeComparison}%
\end{table}%

\paragraph{Image-Guided Mesh Generation}
\rznn{
As described in Section~\ref{section::TextureStructureDeformMeshEncoding}, TM-NET can be adapted for image-guided generation
of textured meshes.
In Figure~\ref{fig:SingleViewReconstruction}, we show several randomly sampled results and compare to DVR~\cite{DVR}. Note that due to the sampling strategy of PixelSNAIL, our method can generate multiple textured shapes from which it would be quite easy to find one that is the closest in appearance to the input image. In general, this result is more faithful to the input than the DVR outputs.}

\begin{figure}[!t]
\centering
	{
		{\includegraphics[width=0.24\linewidth]{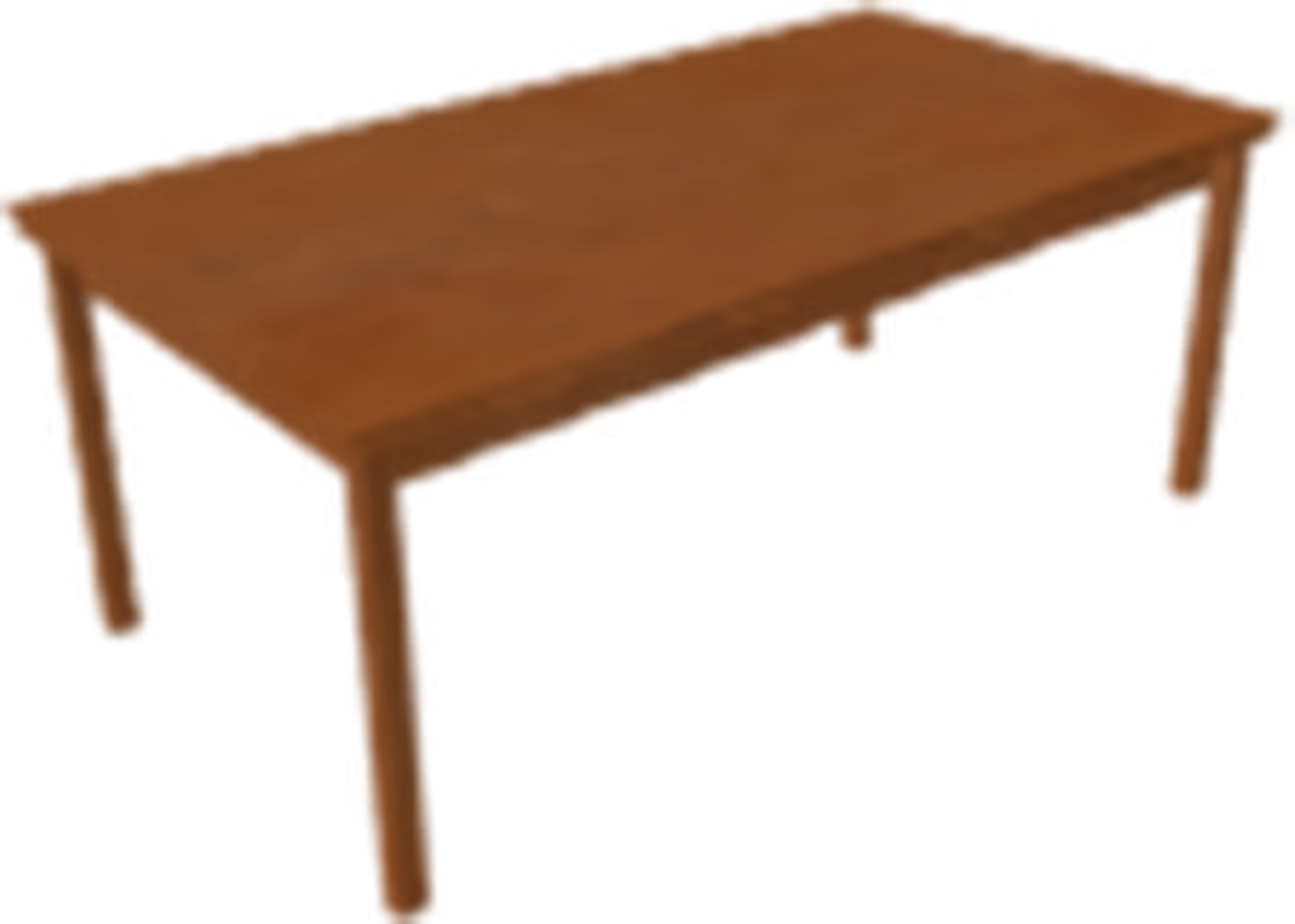}}
		{\includegraphics[width=0.24\linewidth]{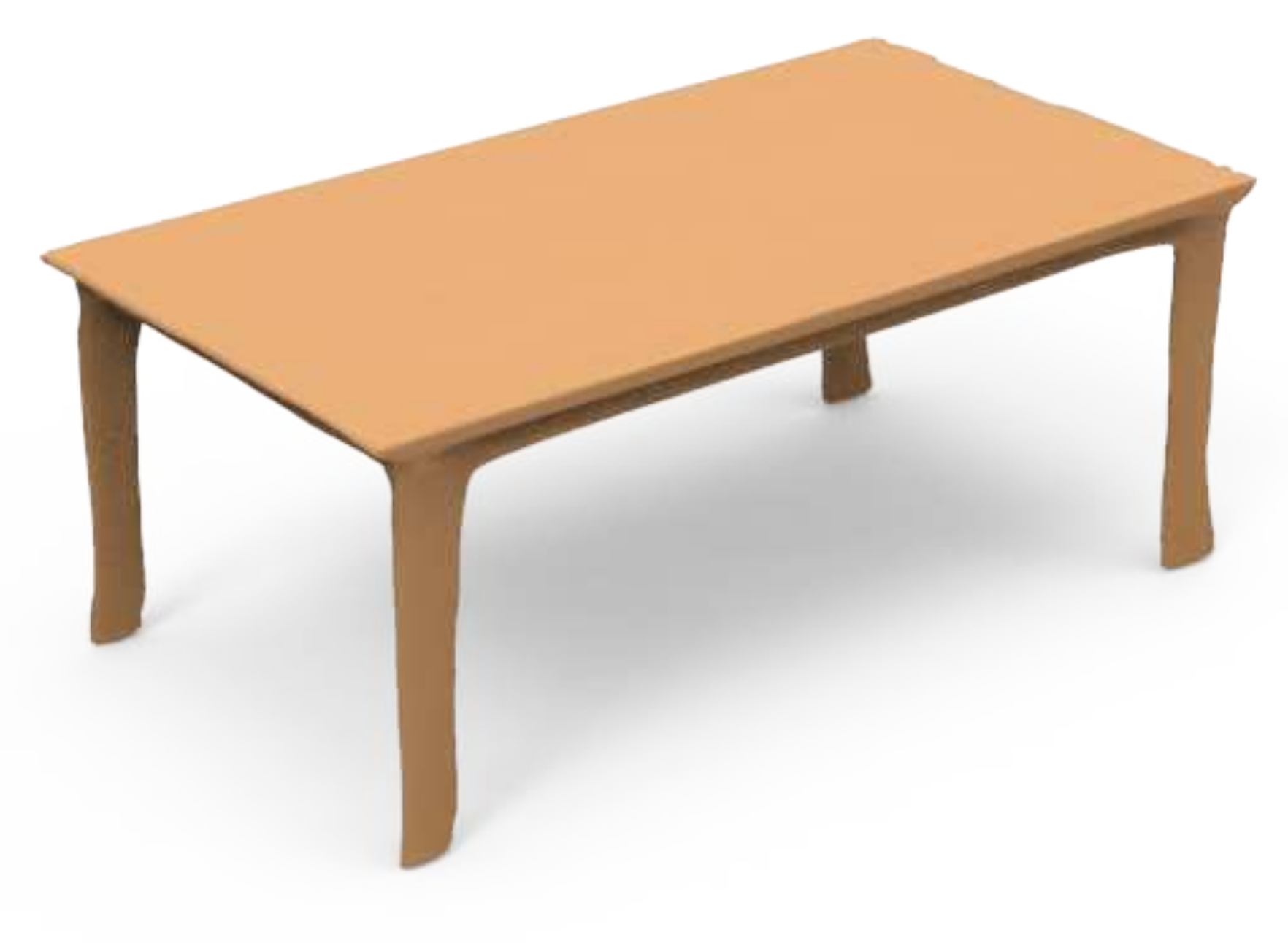}}
		{\includegraphics[width=0.24\linewidth]{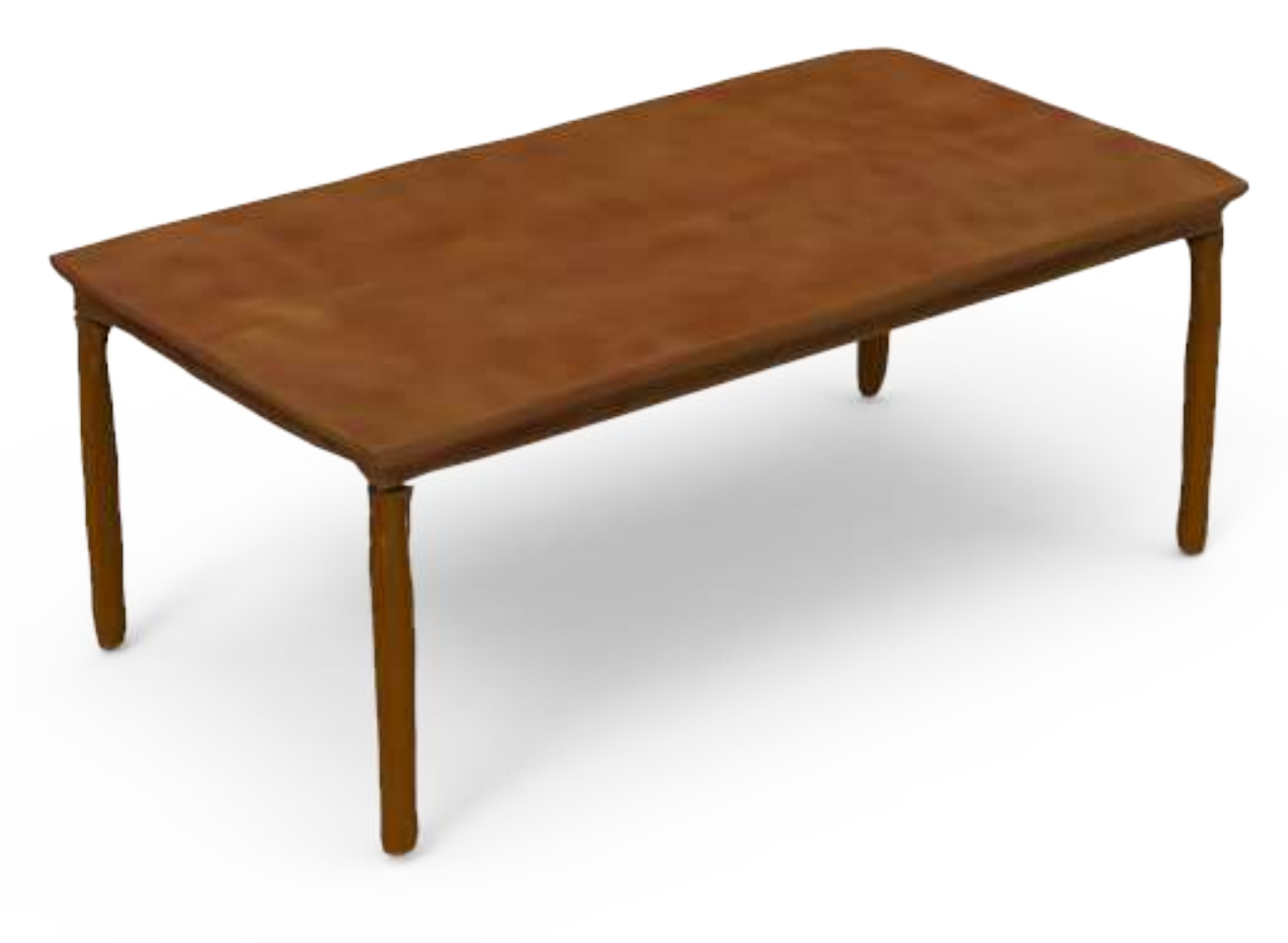}}
		{\includegraphics[width=0.24\linewidth]{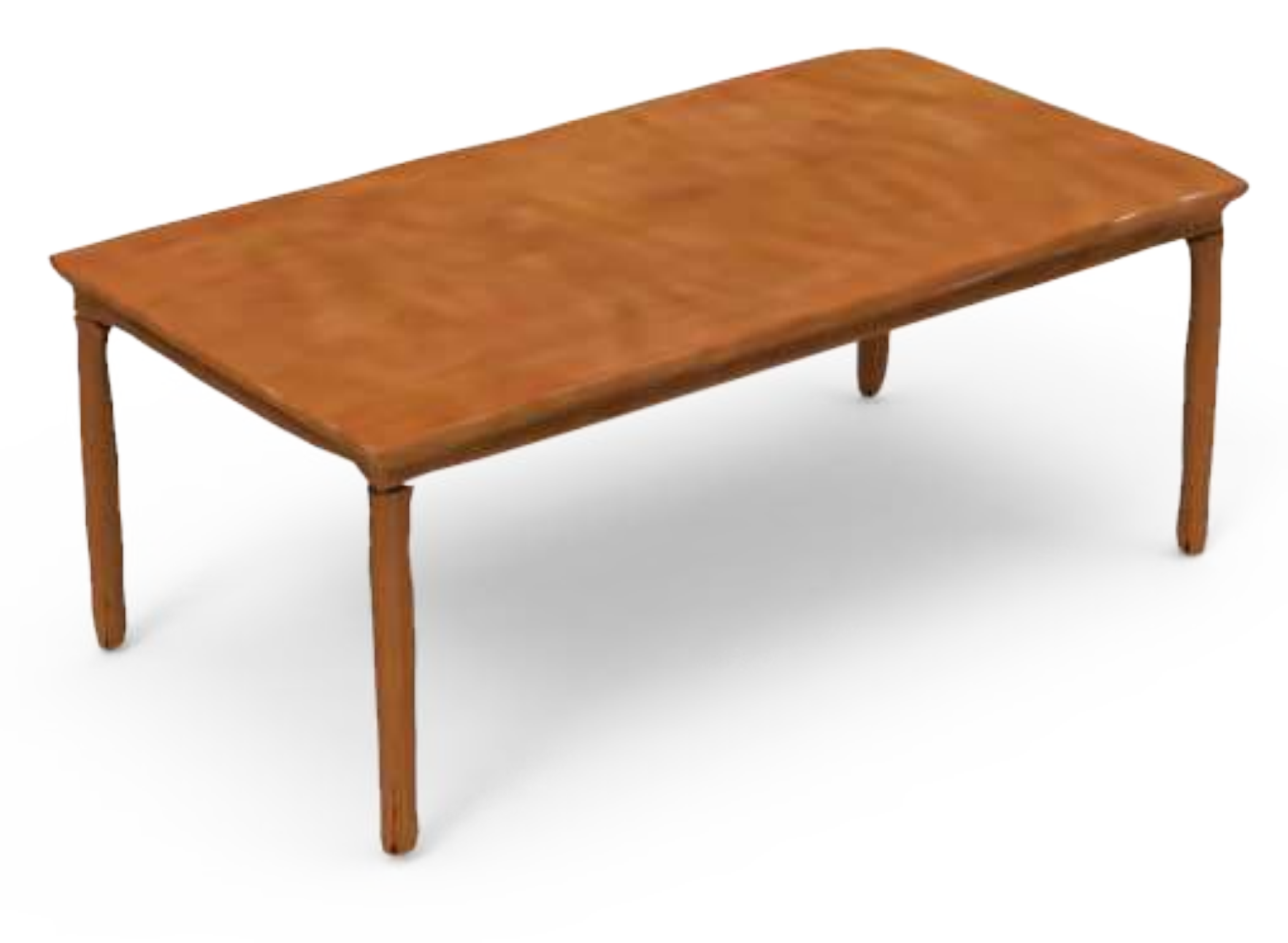}}
		\\
		\subfigure[Image guidence.]{\includegraphics[width=0.24\linewidth]{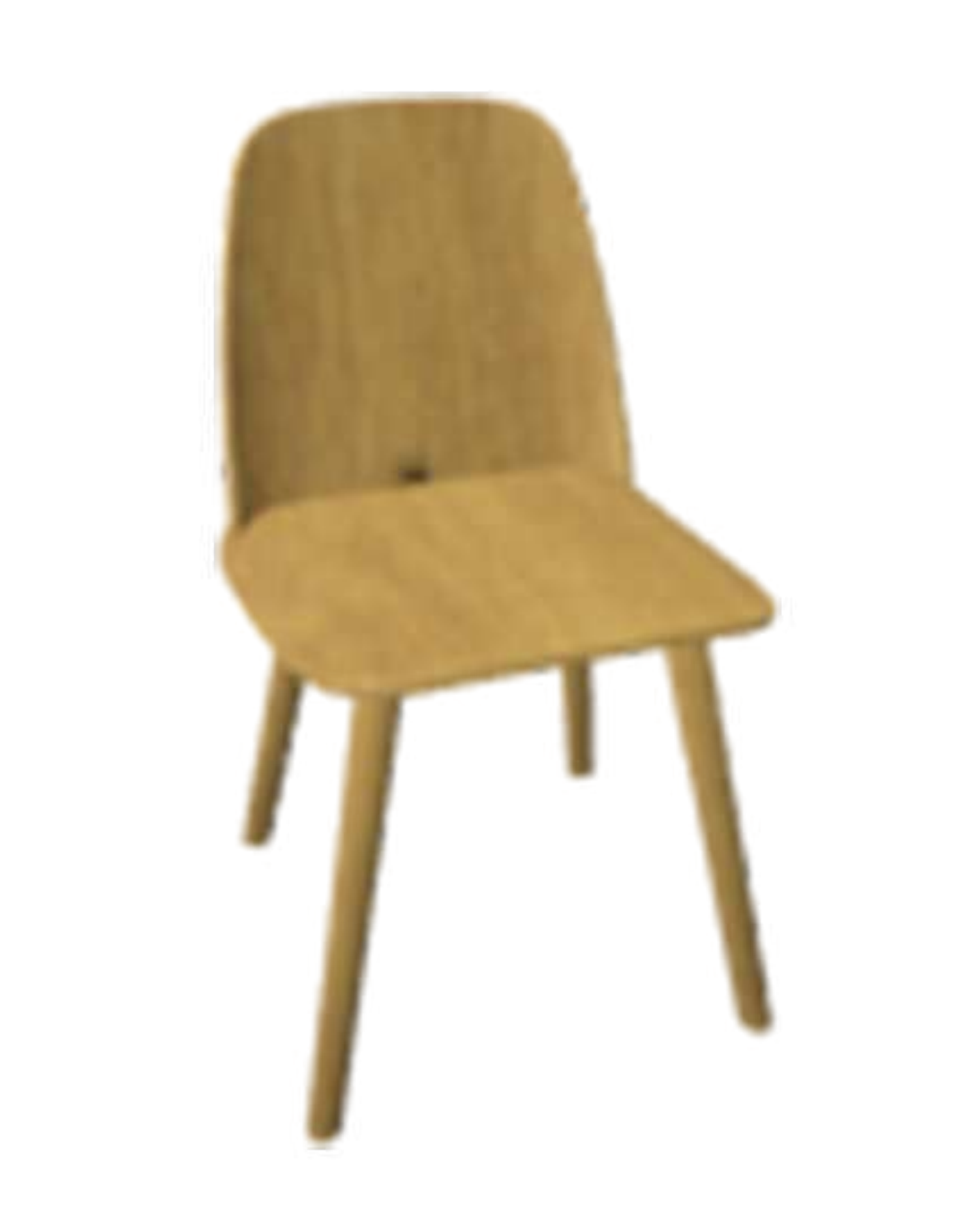}}
 		\subfigure[DVR result.]{\includegraphics[width=0.24\linewidth]{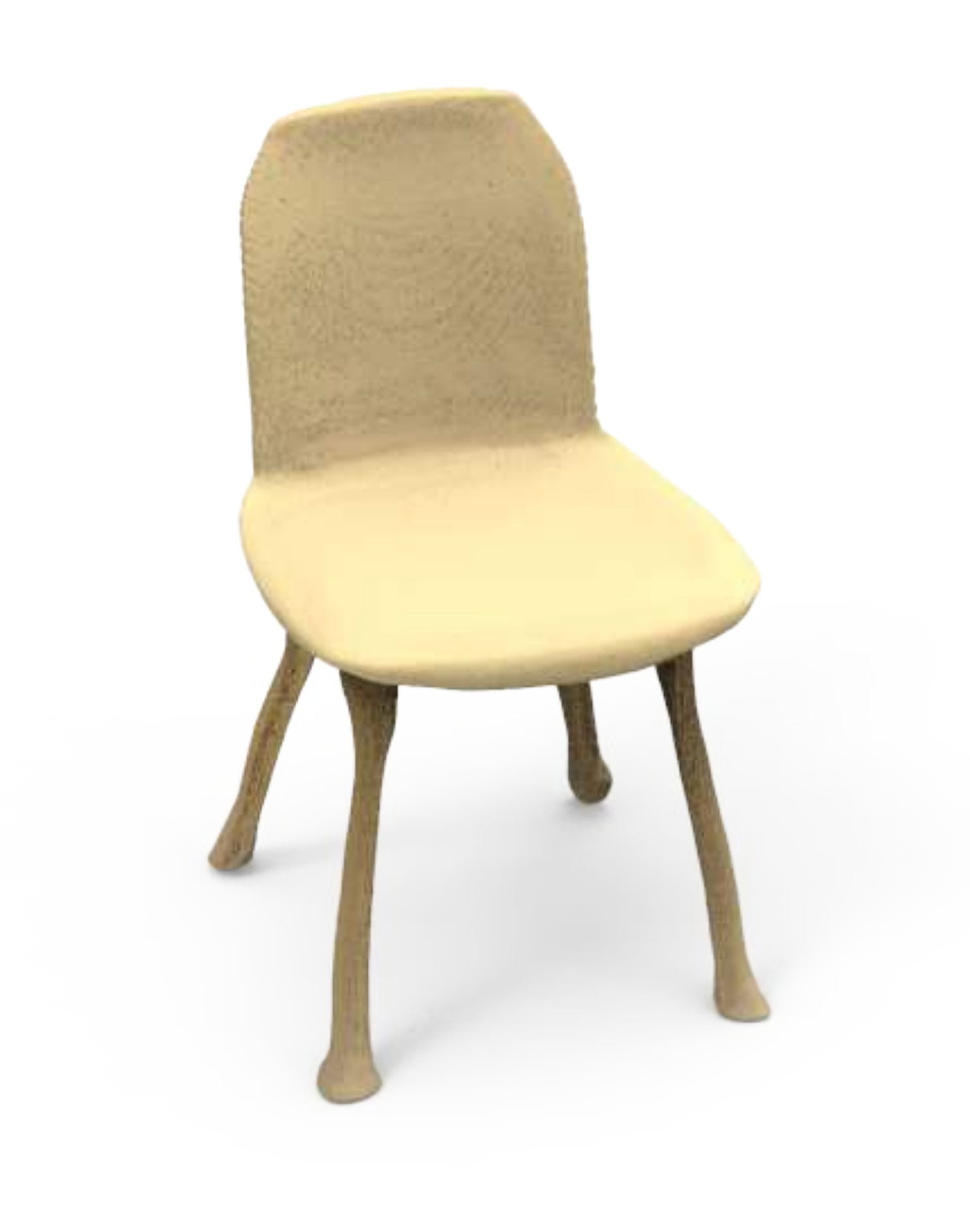}}
 		\subfigure[TM-NET result 1.]{\includegraphics[width=0.24\linewidth]{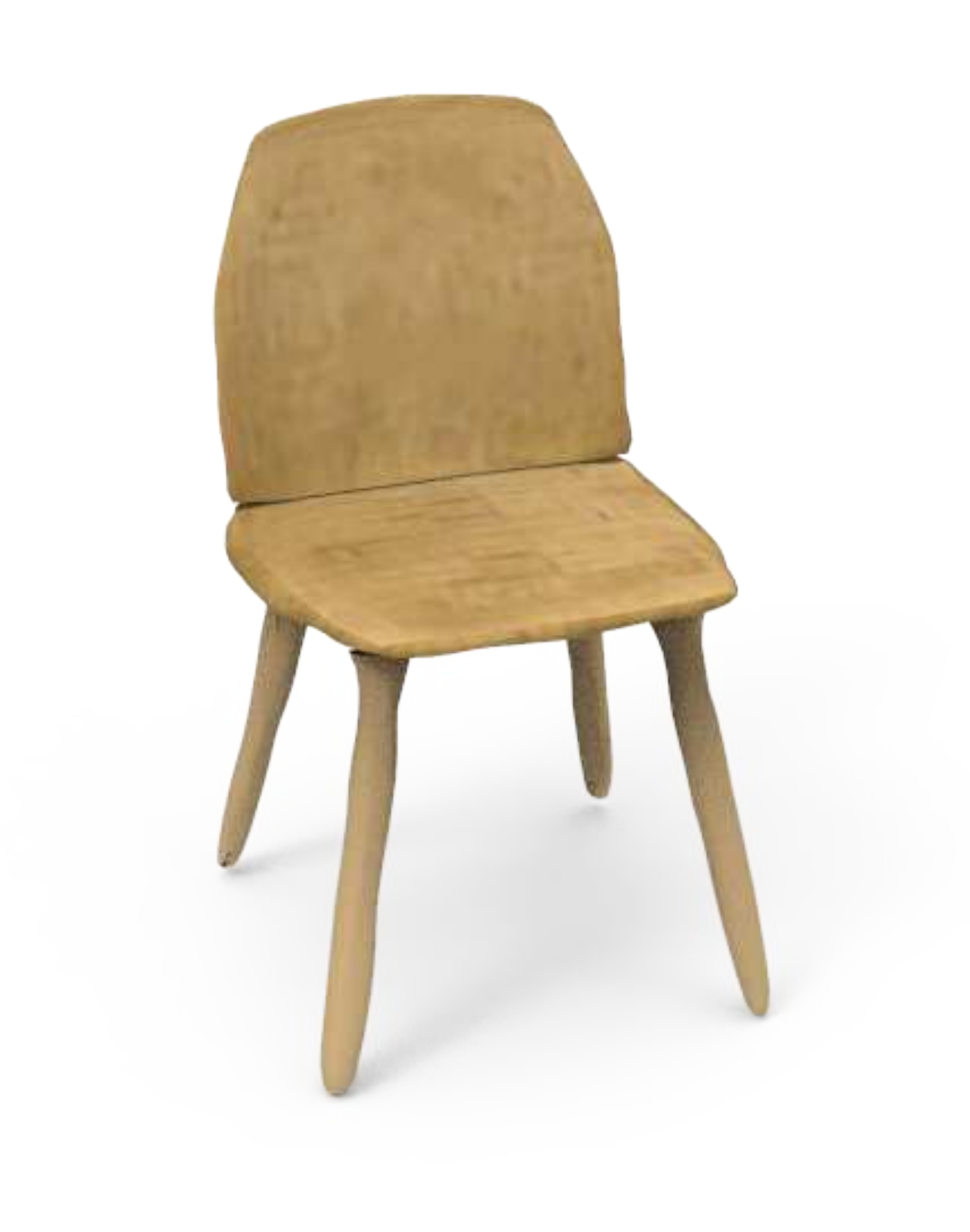}}
 		\subfigure[TM-NET result 2.]{\includegraphics[width=0.24\linewidth]{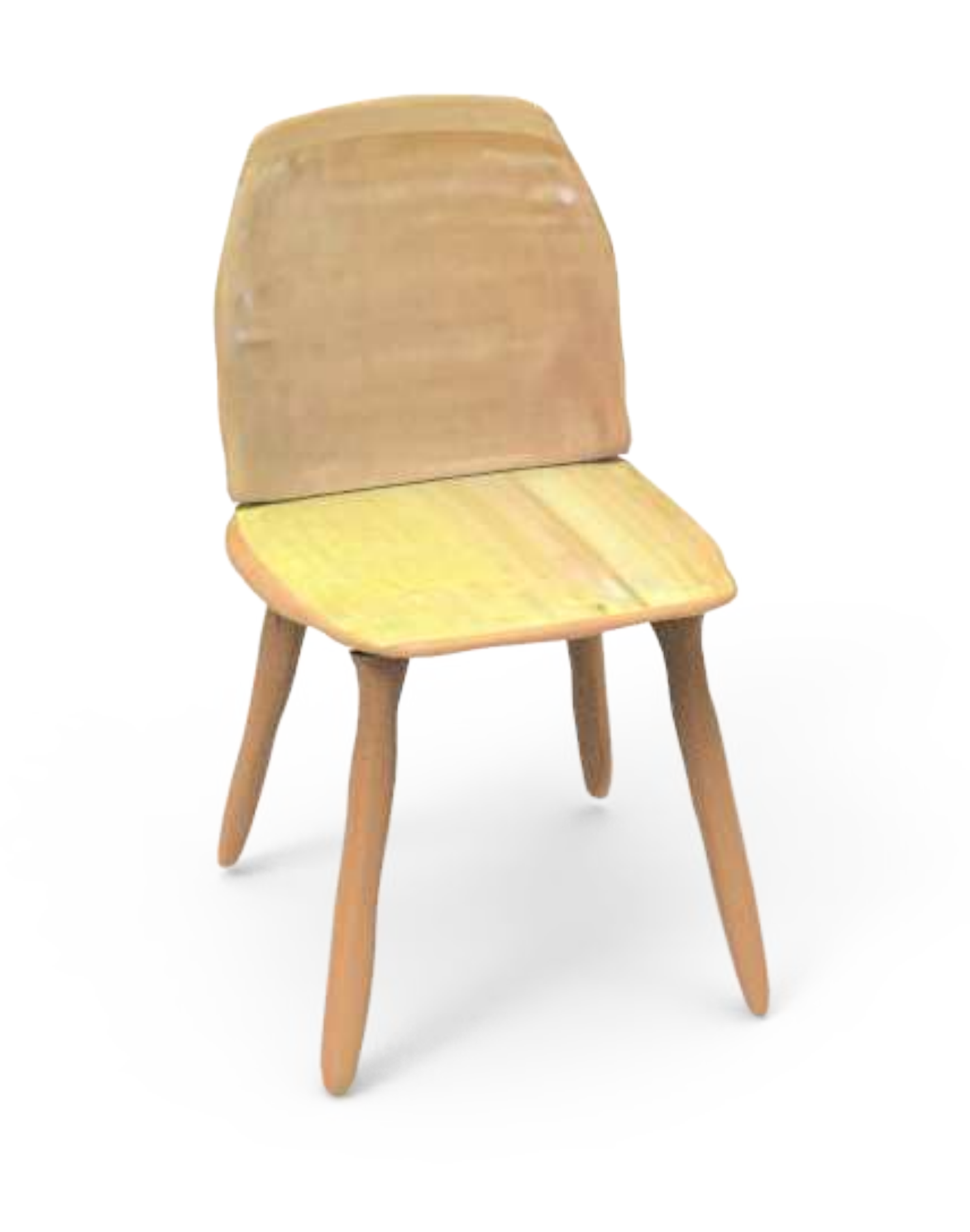}}
	}
    \caption{\rznn{Image-guided generation of textured shapes and a comparison to DVR~\cite{DVR}, on a few randomly sampled results. Our network can generate multiple textured shapes, containing results that are generally closer to and more faithful to the guidance image (c-d).}}    \label{fig:SingleViewReconstruction}
\end{figure}

\paragraph{Textured Shape Interpolation}
\rznn{
In our method, shape geometry and texture are separately encoded into the latent spaces of SP-VAE and TextureVAE, respectively. Hence, we can perform linear interpolation in both latent spaces and then recover the textured meshes by the decoders of SP-VAE and TextureVAE.
In Figure~\ref{fig:Interpolation}, we show a sampler of such interpolation results on chairs, and compare to VON~\cite{VON}, TF~\cite{OechsleICCV2019}, and simple alpha blending between textures on a per-part basis. Note that the original TF implementation only supported texture interpolation. Since their model outputs rendered images using a depth map, a shape code, and a texture code, we obtained the TF interpolation results by interpolating on all these three components.
On the other hand, VON only generates in-between views of textured objects and it is not part-aware.
Since it is a GAN-based model, it is difficult to compare VON to other methods using exactly the same source and target shapes. As a remedy, for each set of comparisons, we first run VON interpolation in both shape and texture latent spaces to obtain a {\em random\/} pair of source and target shapes. We then run TM-NET, TF, and part-based alpha blending using source and target shapes from our {\em test set\/} that are respectively the {\em closest\/}, as measured by SSIM, to the source and target shapes generated by VON.

We can see that compared to TF and VON, the interpolation results by TM-NET are more plausible and faithful in both geometry and texture, demonstrating the advantage of our structured approach and the learned latent spaces. Alpha blending also works well, with results quite close to those from TM-NET. We believe that this can be attributed to the part-by-part texture blending, where each part is simply a box, which facilitates part correspondence.
}

\begin{figure}[!t]
  \centering
  {
    \begin{tabular}{cc}
        \rotatebox{90}{\quad \hspace{2mm} {\small VON}}
        &
        \vspace{-2mm}
        \hspace{-5mm}
        {\includegraphics[width=0.15\linewidth]{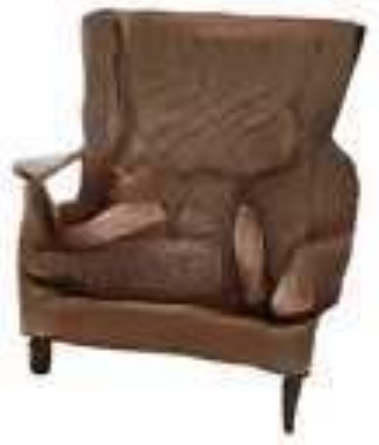}}
        {\includegraphics[width=0.15\linewidth]{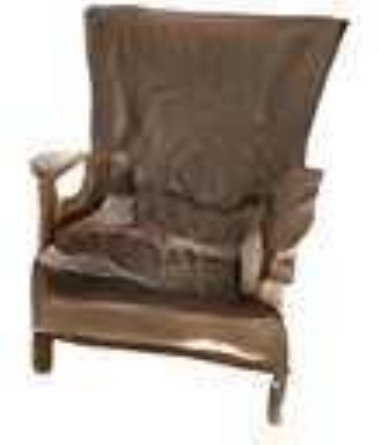}}
        {\includegraphics[width=0.15\linewidth]{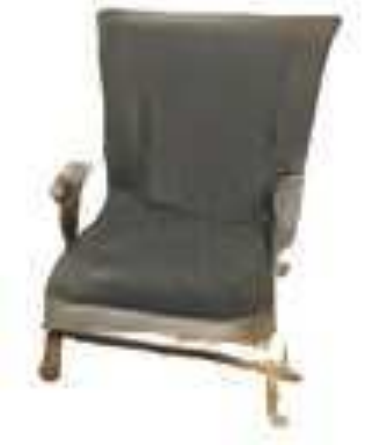}}
        {\includegraphics[width=0.15\linewidth]{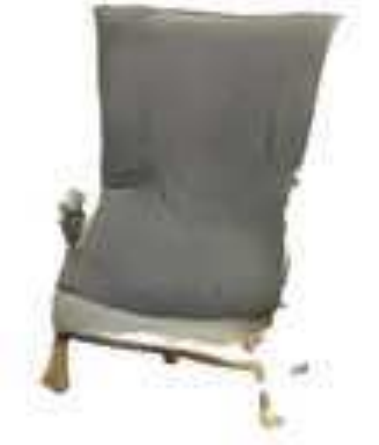}}
        {\includegraphics[width=0.15\linewidth]{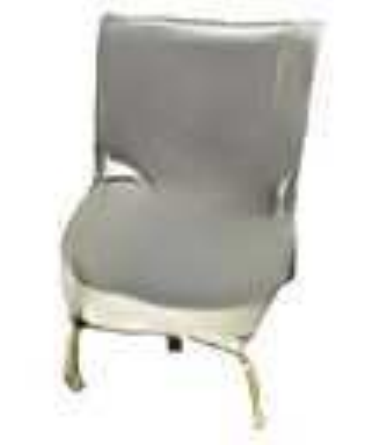}}
        {\includegraphics[width=0.15\linewidth]{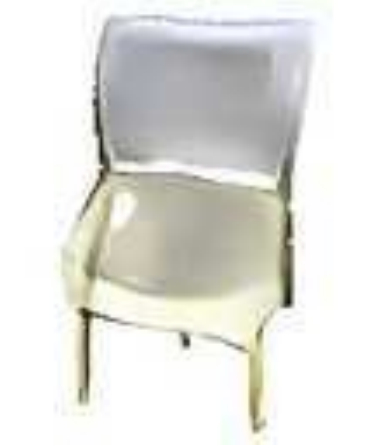}}
        \\
        \rotatebox{90}{\quad \hspace{2mm} {\small TF}}
        &
        \vspace{-1mm}
        \hspace{-5mm}
        {\includegraphics[width=0.15\linewidth]{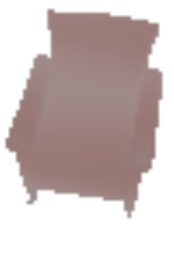}}
        {\includegraphics[width=0.15\linewidth]{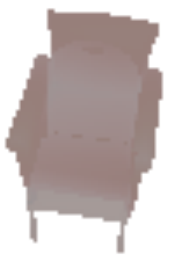}}
        {\includegraphics[width=0.15\linewidth]{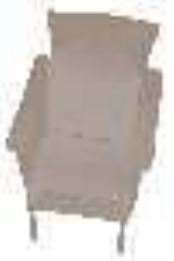}}
        {\includegraphics[width=0.15\linewidth]{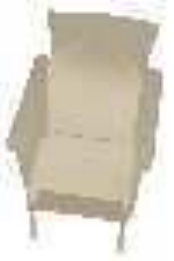}}
        {\includegraphics[width=0.15\linewidth]{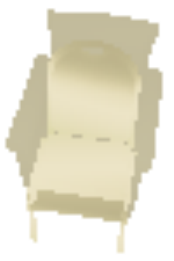}}
        {\includegraphics[width=0.15\linewidth]{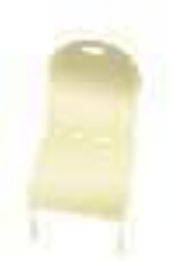}}
        \\
        \rotatebox{90}{\quad \hspace{2mm} {\small Alpha}}
        &
        \vspace{-2mm}
        \hspace{-5mm}
        {\includegraphics[width=0.15\linewidth]{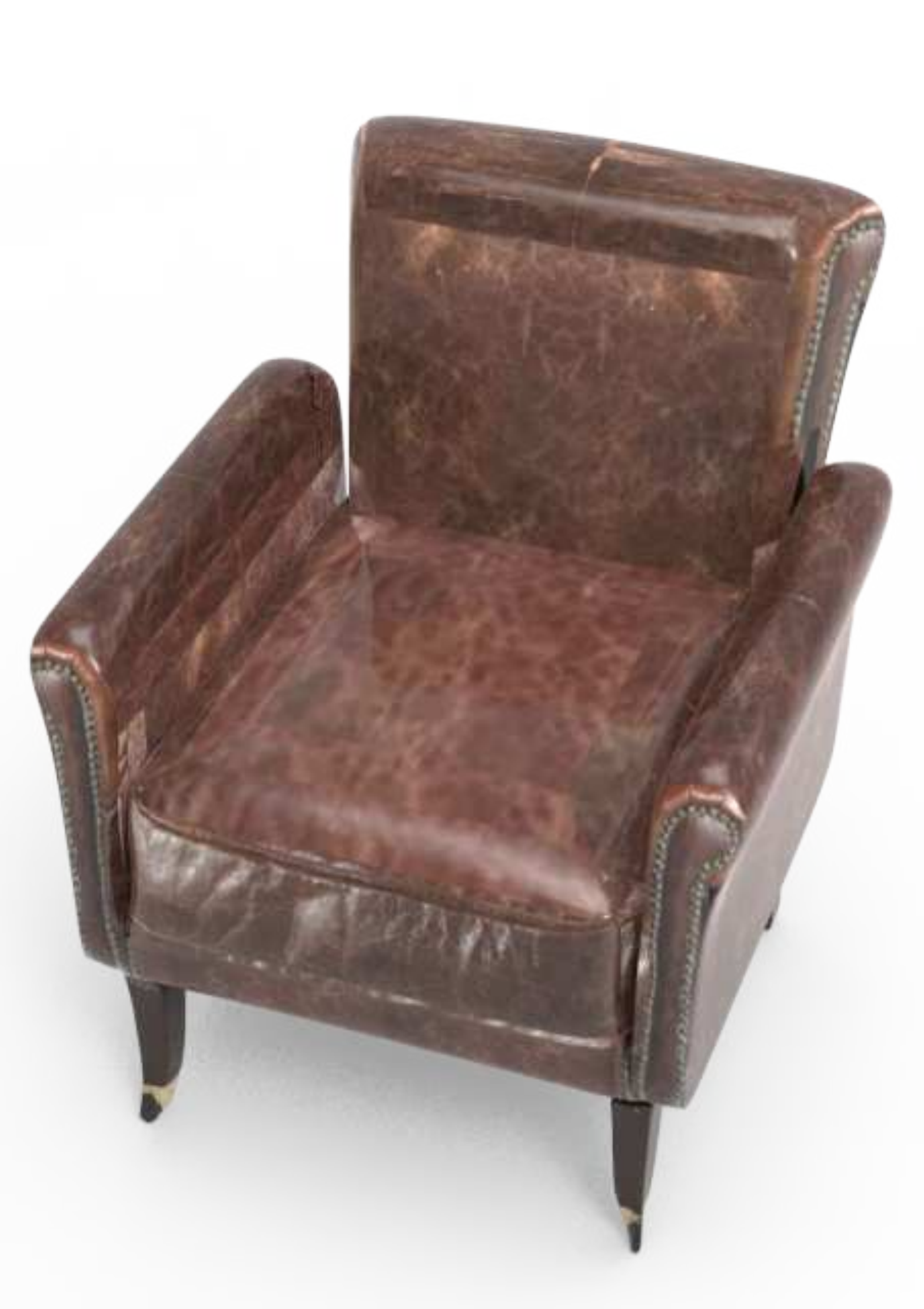}}
        {\includegraphics[width=0.15\linewidth]{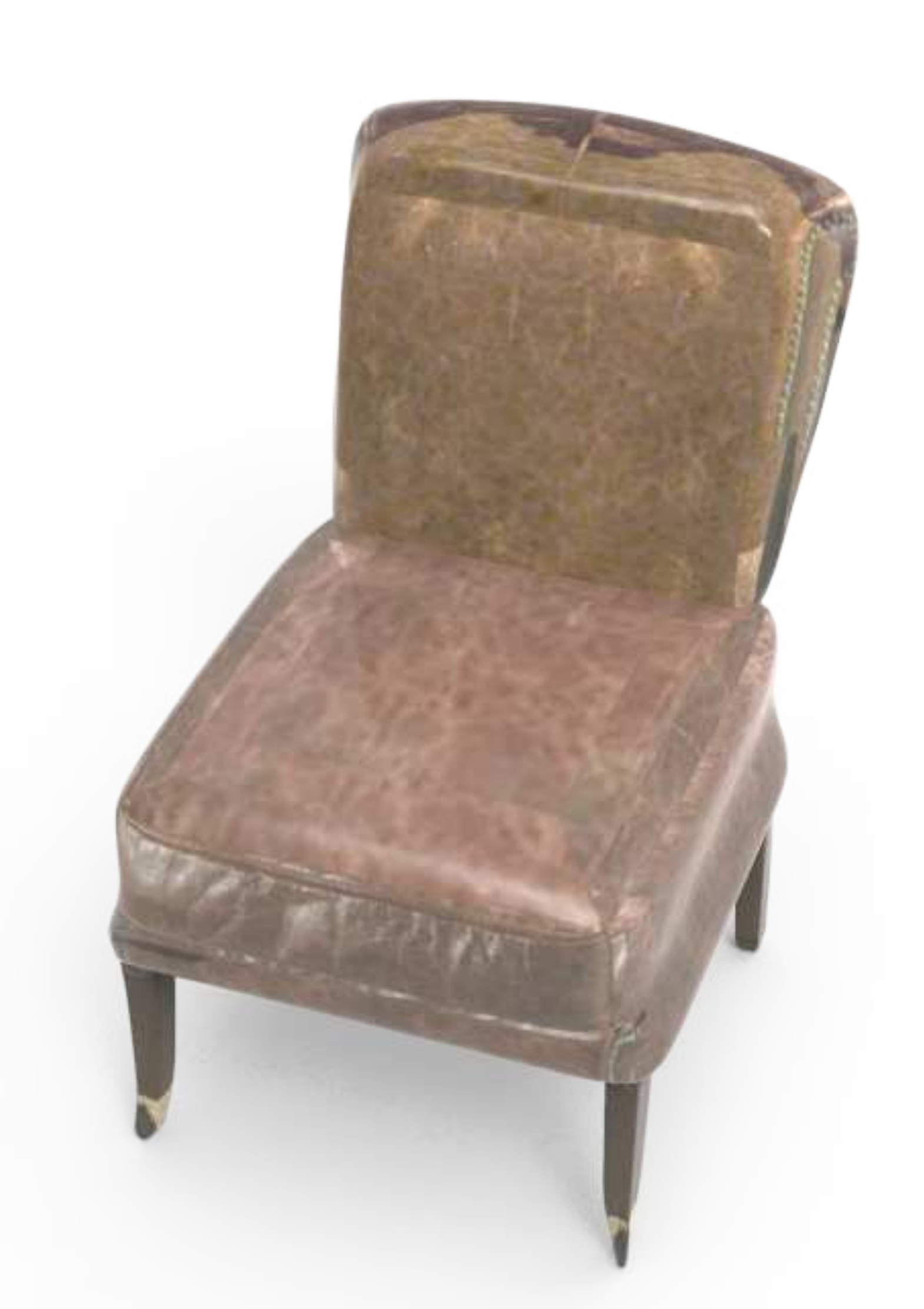}}
        {\includegraphics[width=0.15\linewidth]{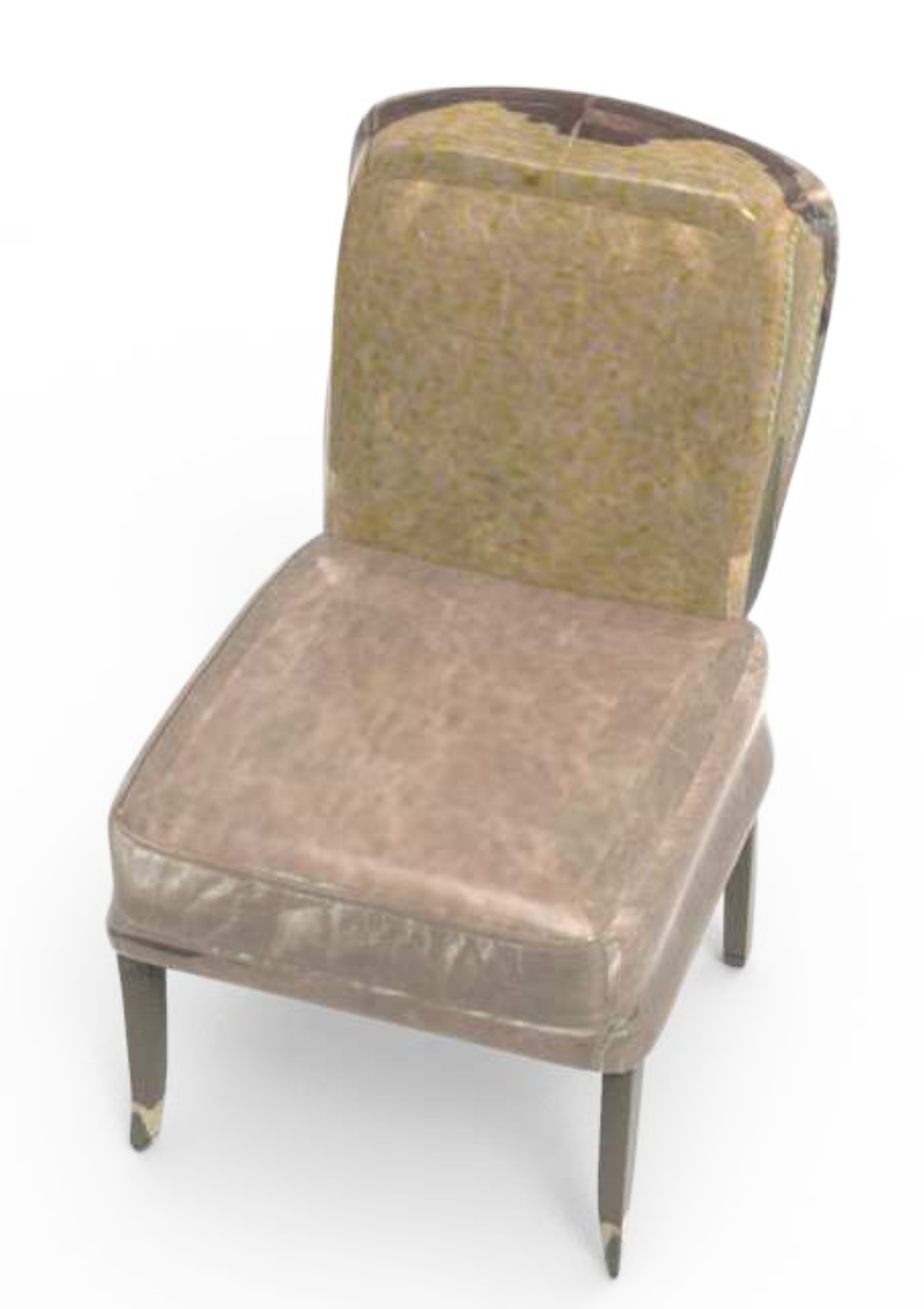}}
        {\includegraphics[width=0.15\linewidth]{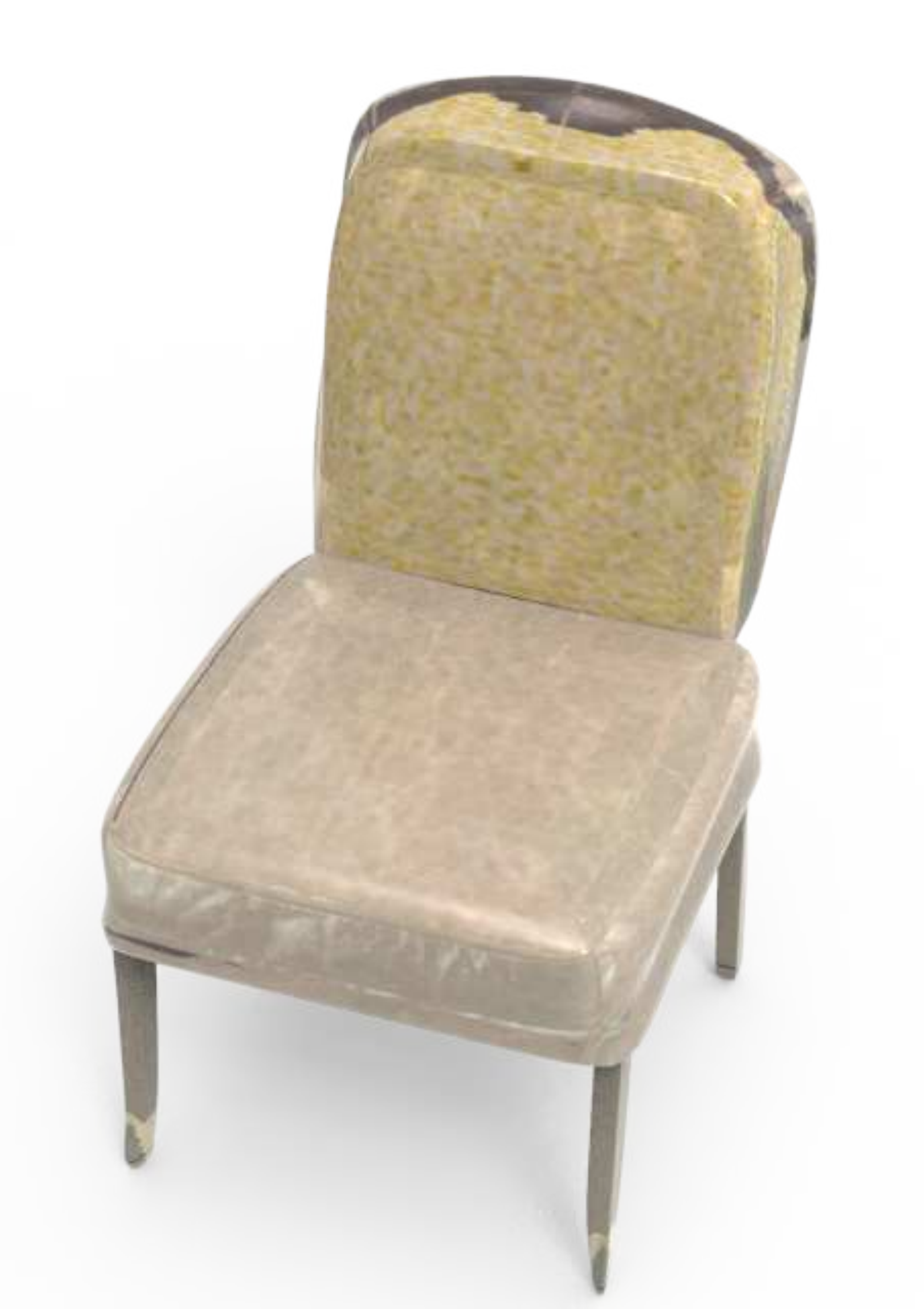}}
        {\includegraphics[width=0.15\linewidth]{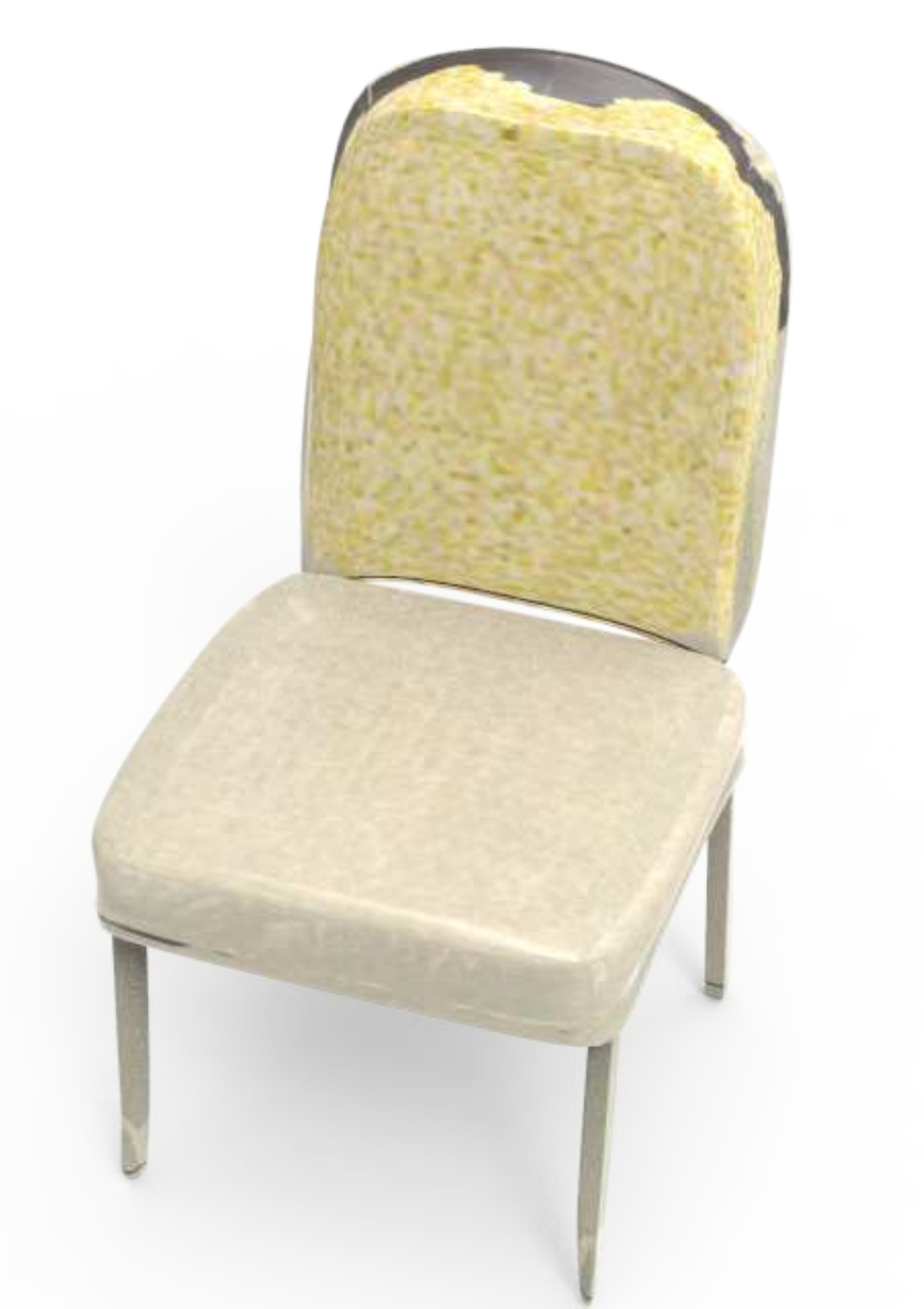}}
        {\includegraphics[width=0.15\linewidth]{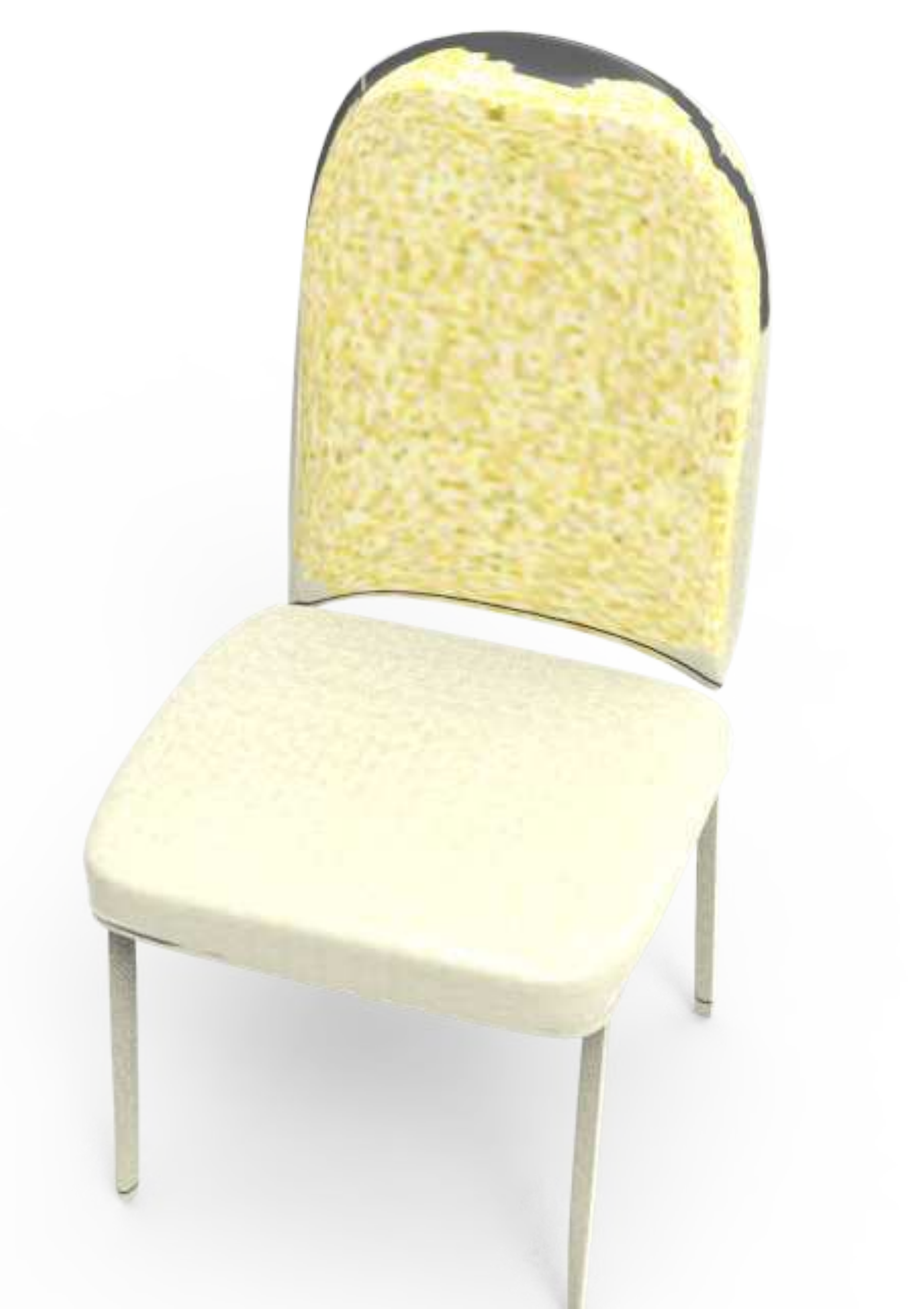}}
        \\
        \rotatebox{90}{\quad \hspace{2mm} {\small TM-NET}}
        &
        \vspace{-2mm}
        \hspace{-5mm}
        {\includegraphics[width=0.15\linewidth]{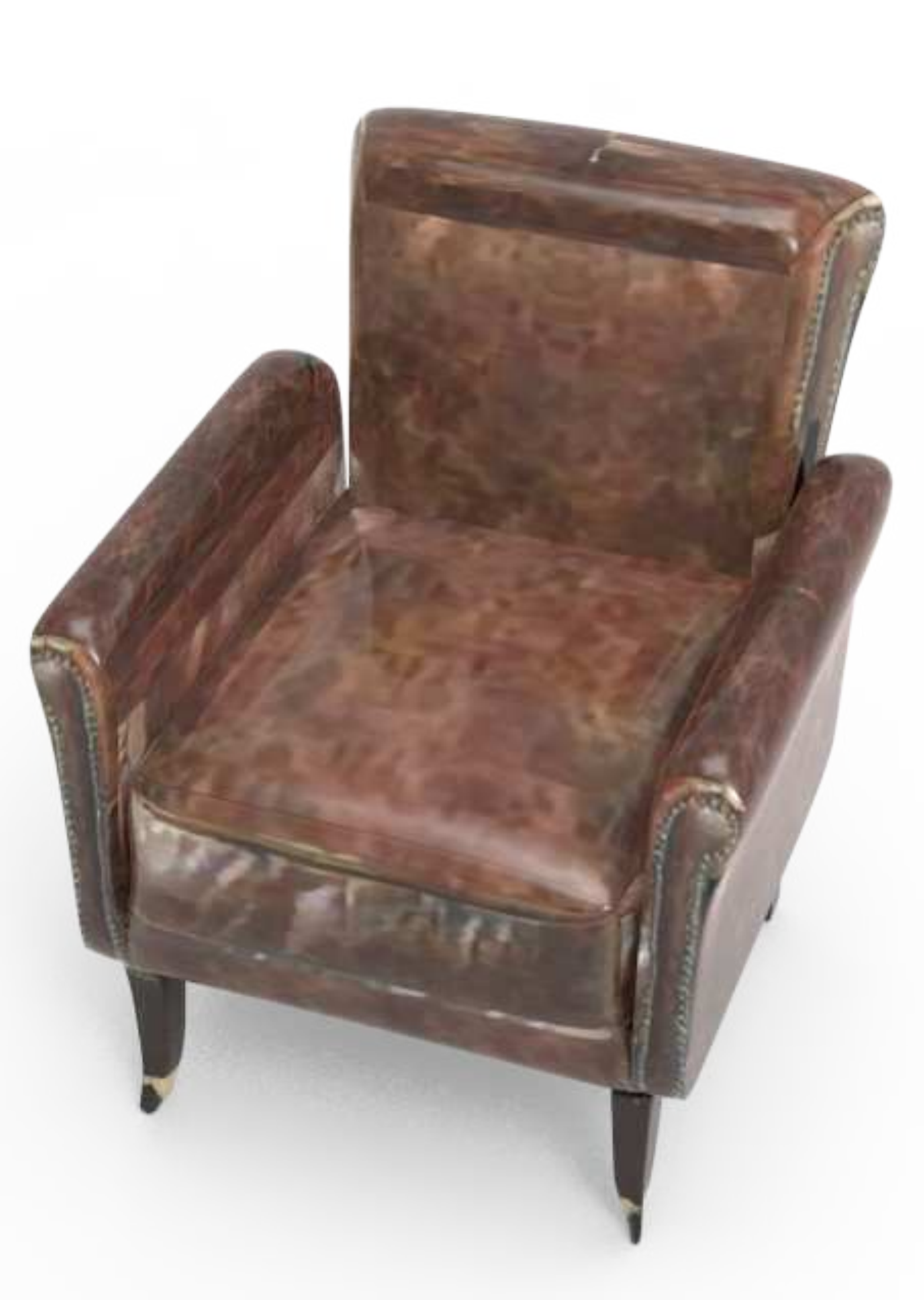}}
        {\includegraphics[width=0.15\linewidth]{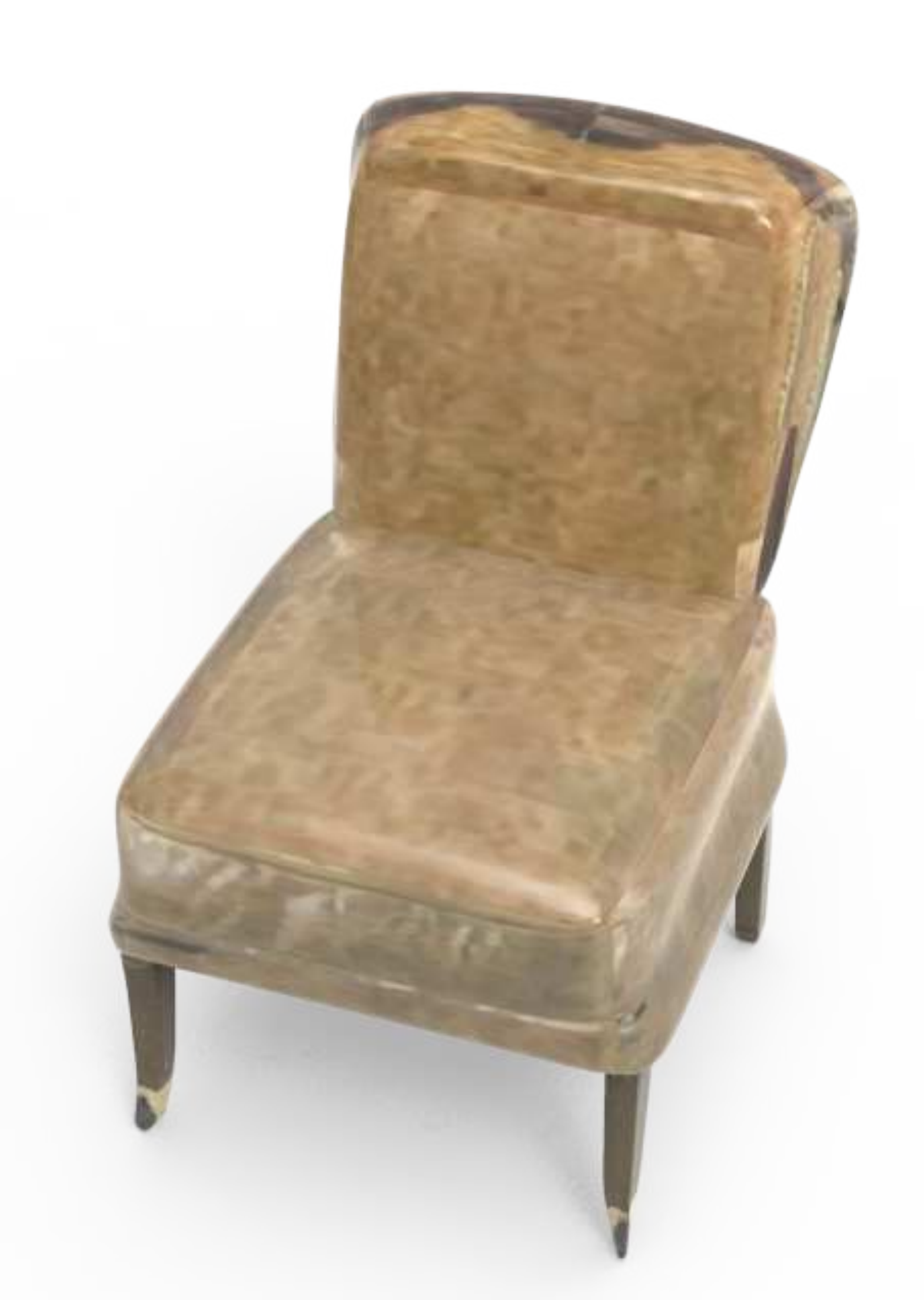}}
        {\includegraphics[width=0.15\linewidth]{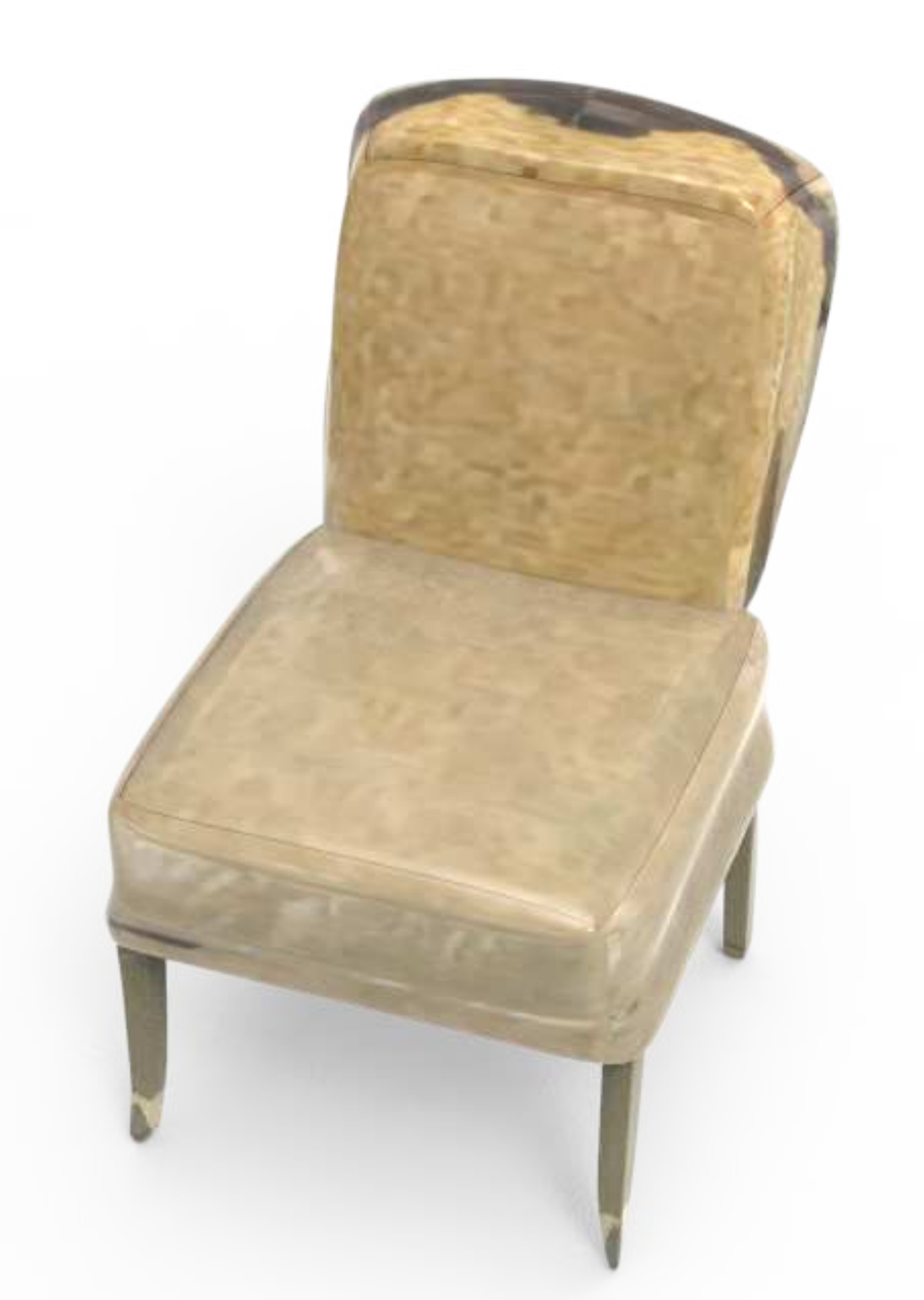}}
        {\includegraphics[width=0.15\linewidth]{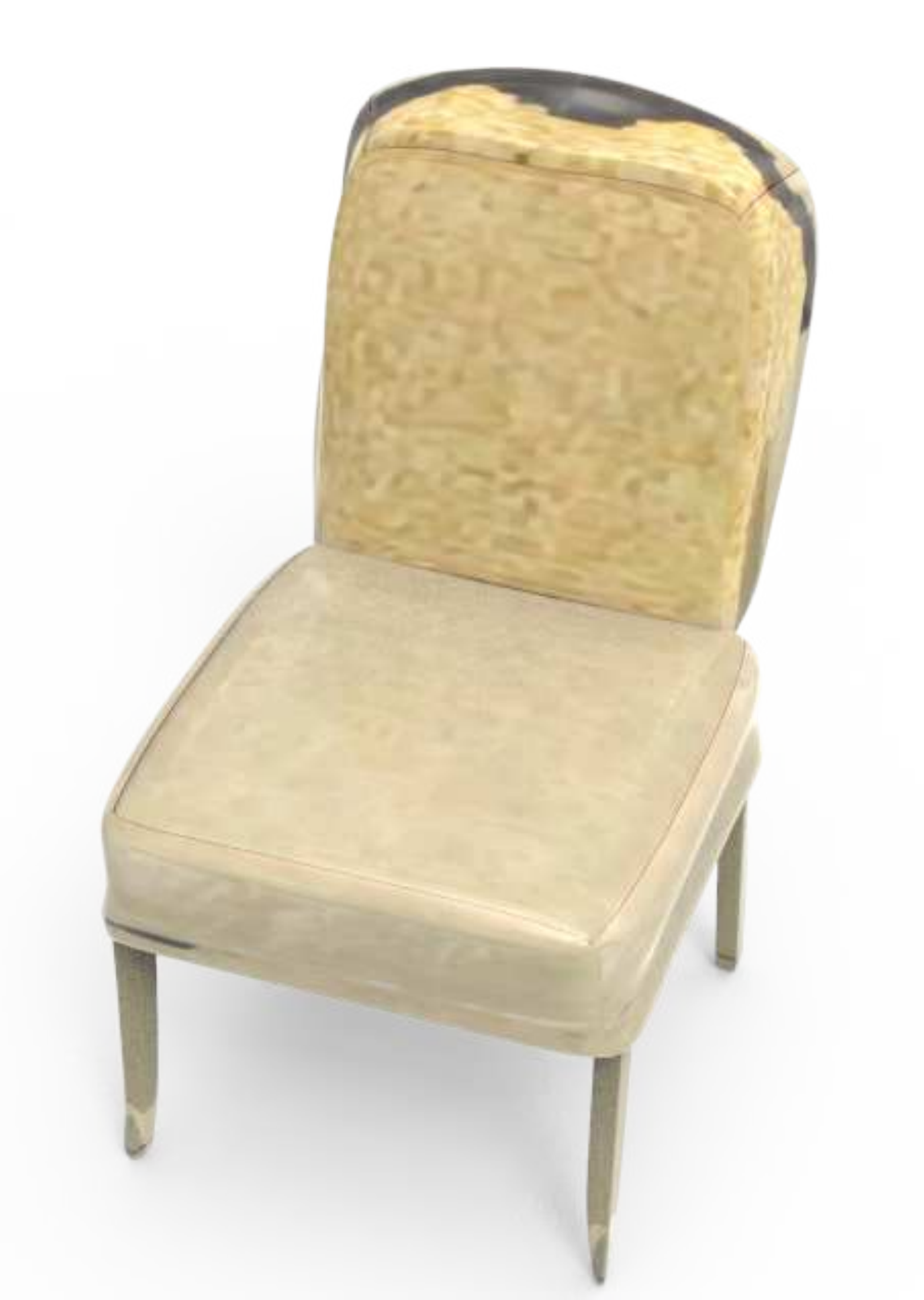}}
        {\includegraphics[width=0.15\linewidth]{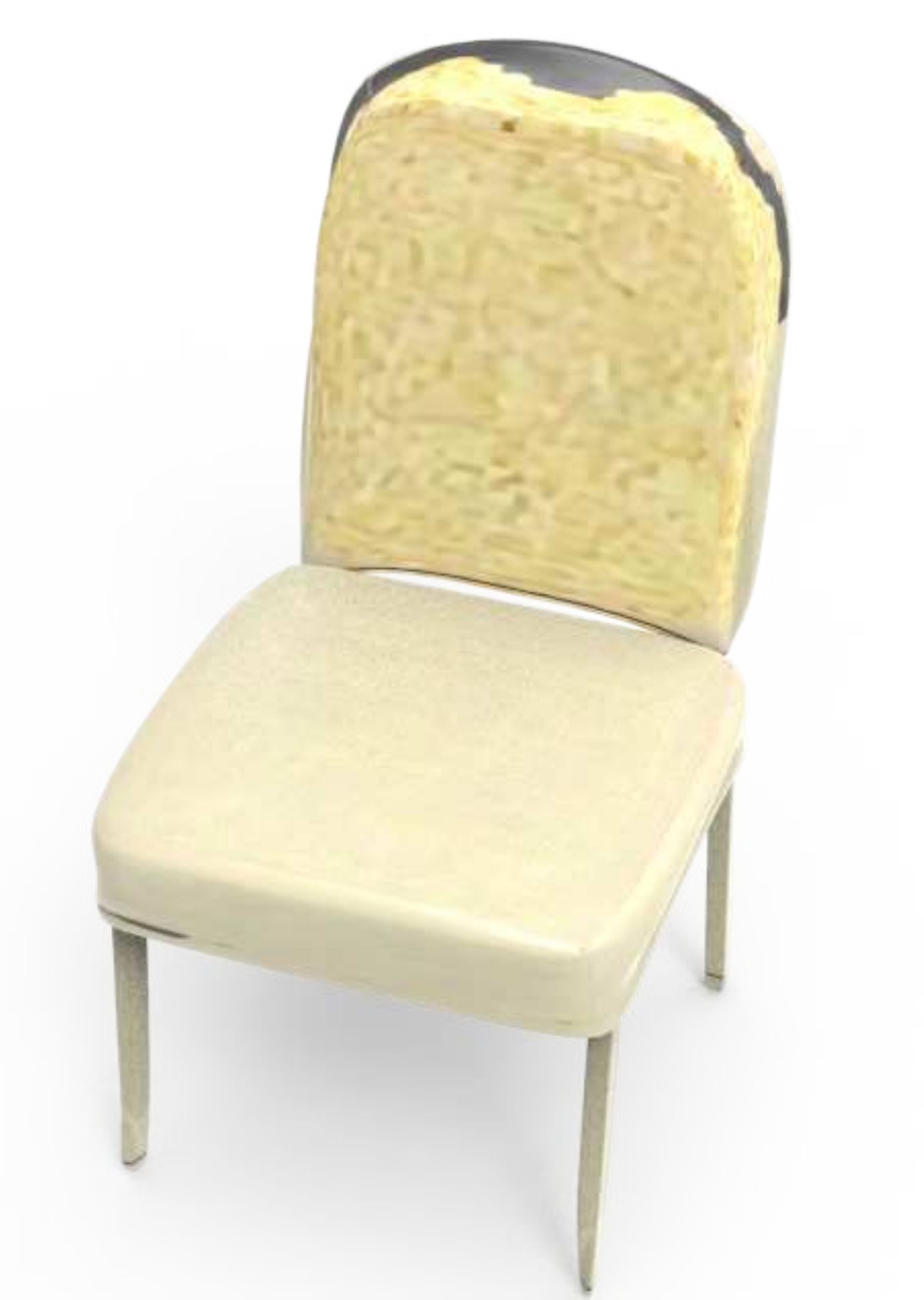}}
        {\includegraphics[width=0.15\linewidth]{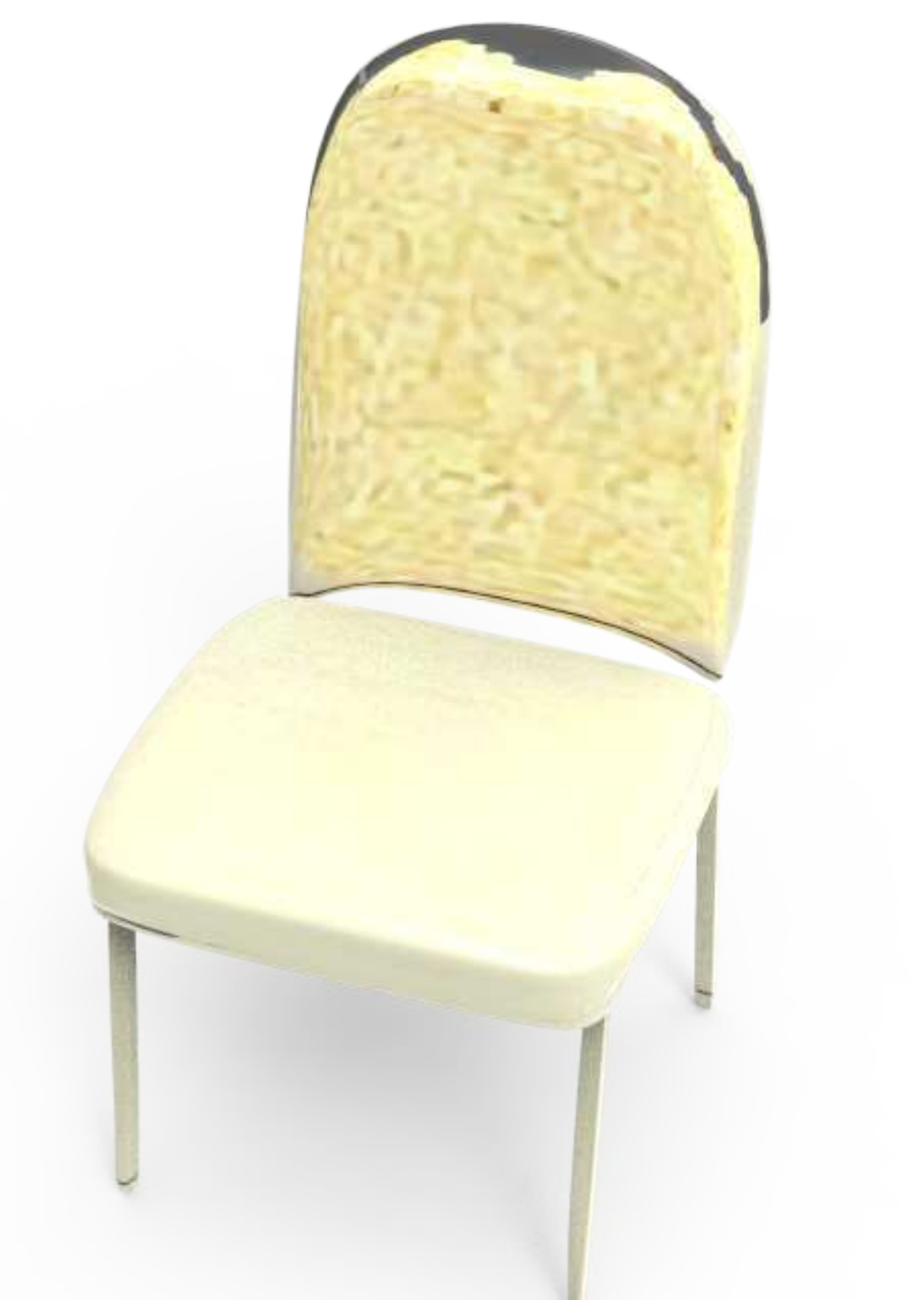}}
        \\
        \rotatebox{90}{\quad \hspace{2mm} {\small VON}}
        &
        \vspace{-2mm}
        \hspace{-5mm}
        {\includegraphics[width=0.15\linewidth]{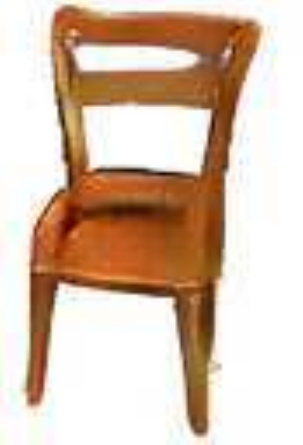}}
        {\includegraphics[width=0.15\linewidth]{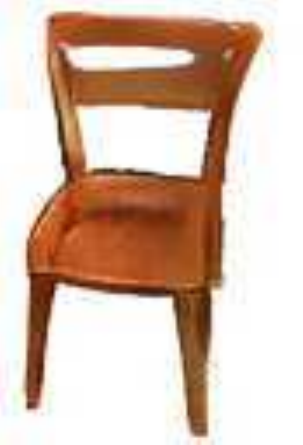}}
        {\includegraphics[width=0.15\linewidth]{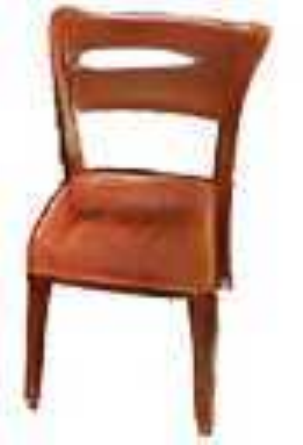}}
        {\includegraphics[width=0.15\linewidth]{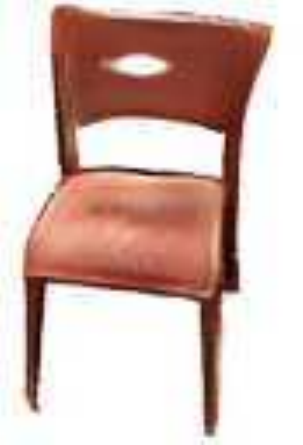}}
        {\includegraphics[width=0.15\linewidth]{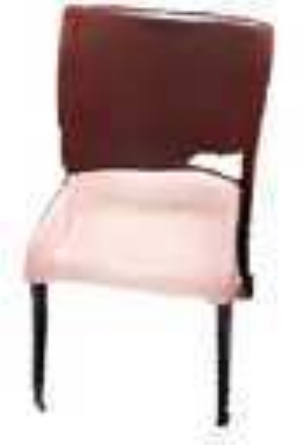}}
        {\includegraphics[width=0.15\linewidth]{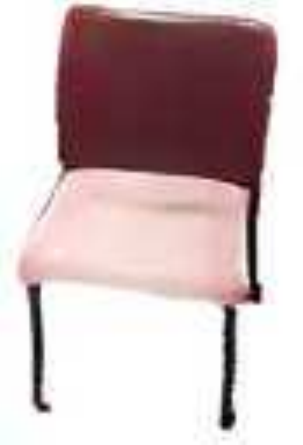}}
        \\
        \rotatebox{90}{\quad \hspace{2mm} {\small TF}}
        &
        \vspace{-1mm}
        \hspace{-5mm}
        {\includegraphics[width=0.15\linewidth]{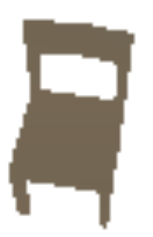}}
        {\includegraphics[width=0.15\linewidth]{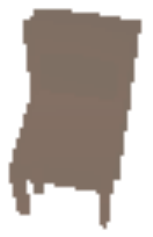}}
        {\includegraphics[width=0.15\linewidth]{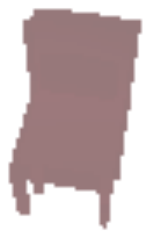}}
        {\includegraphics[width=0.15\linewidth]{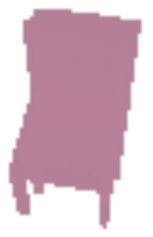}}
        {\includegraphics[width=0.15\linewidth]{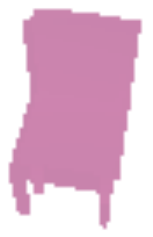}}
        {\includegraphics[width=0.15\linewidth]{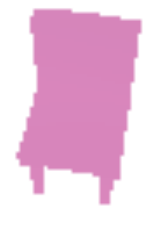}}
        \\
        \rotatebox{90}{\quad \hspace{2mm} {\small Alpha}}
        &
        \vspace{-2mm}
        \hspace{-5mm}
        {\includegraphics[width=0.15\linewidth]{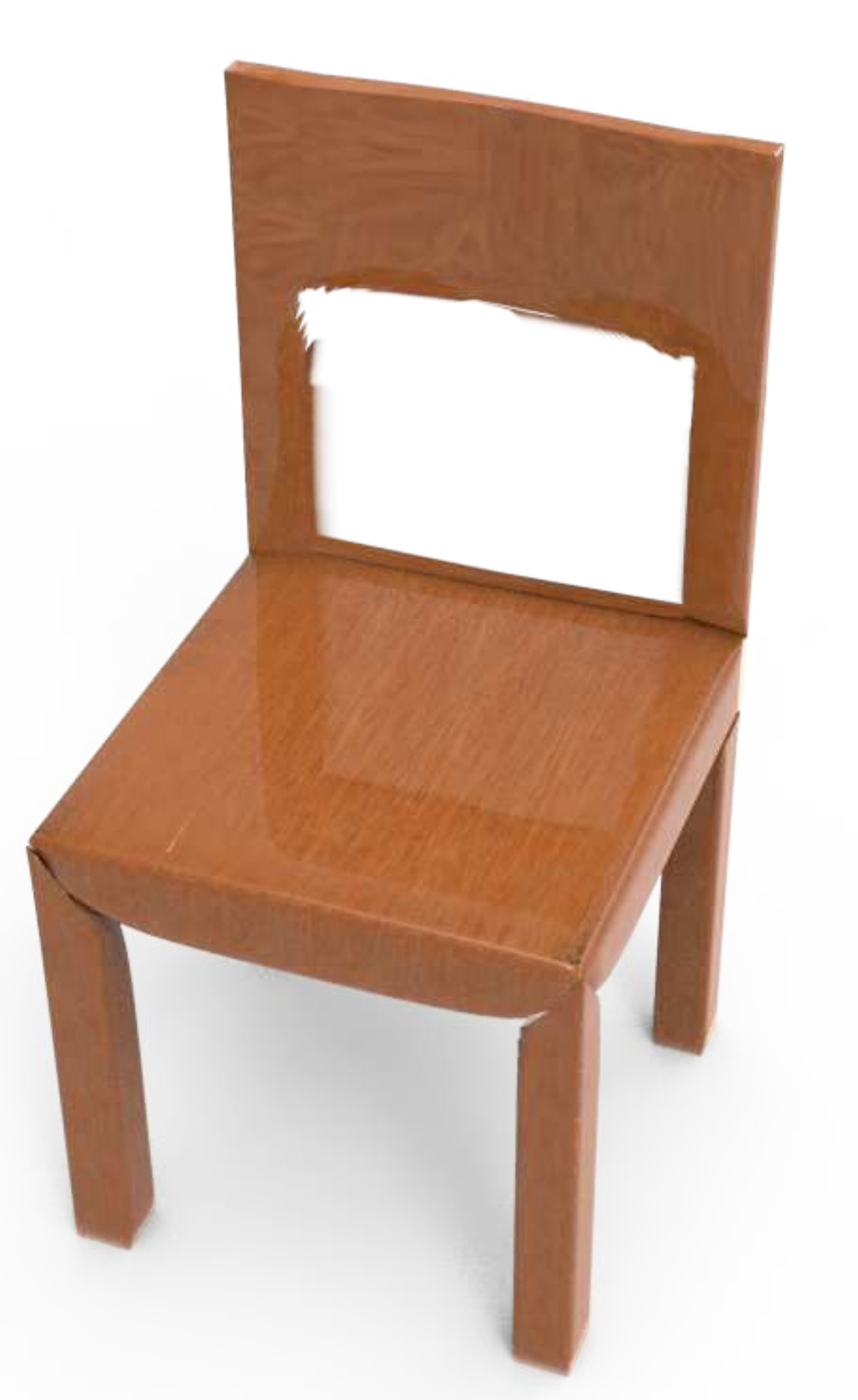}}
        {\includegraphics[width=0.15\linewidth]{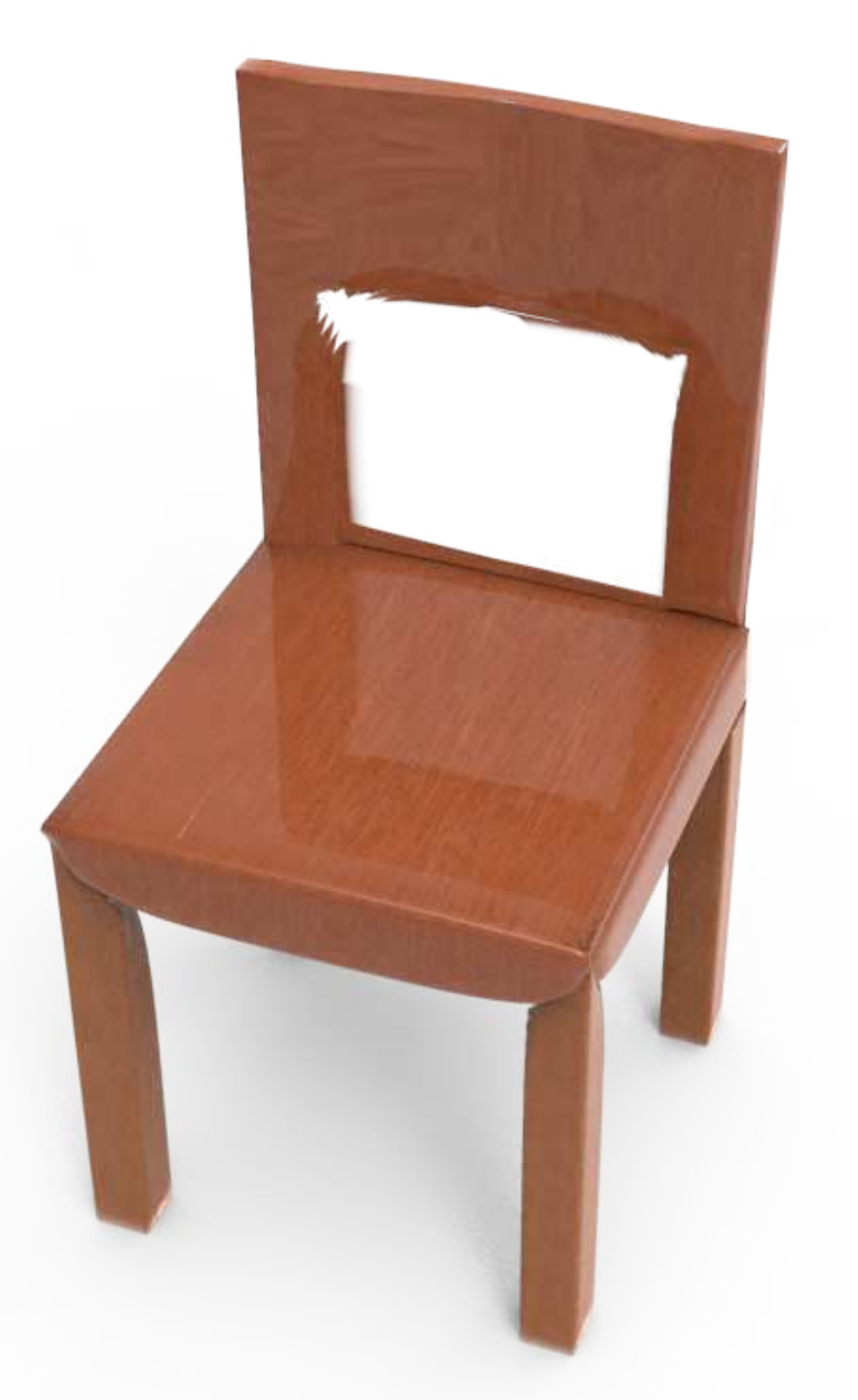}}
        {\includegraphics[width=0.15\linewidth]{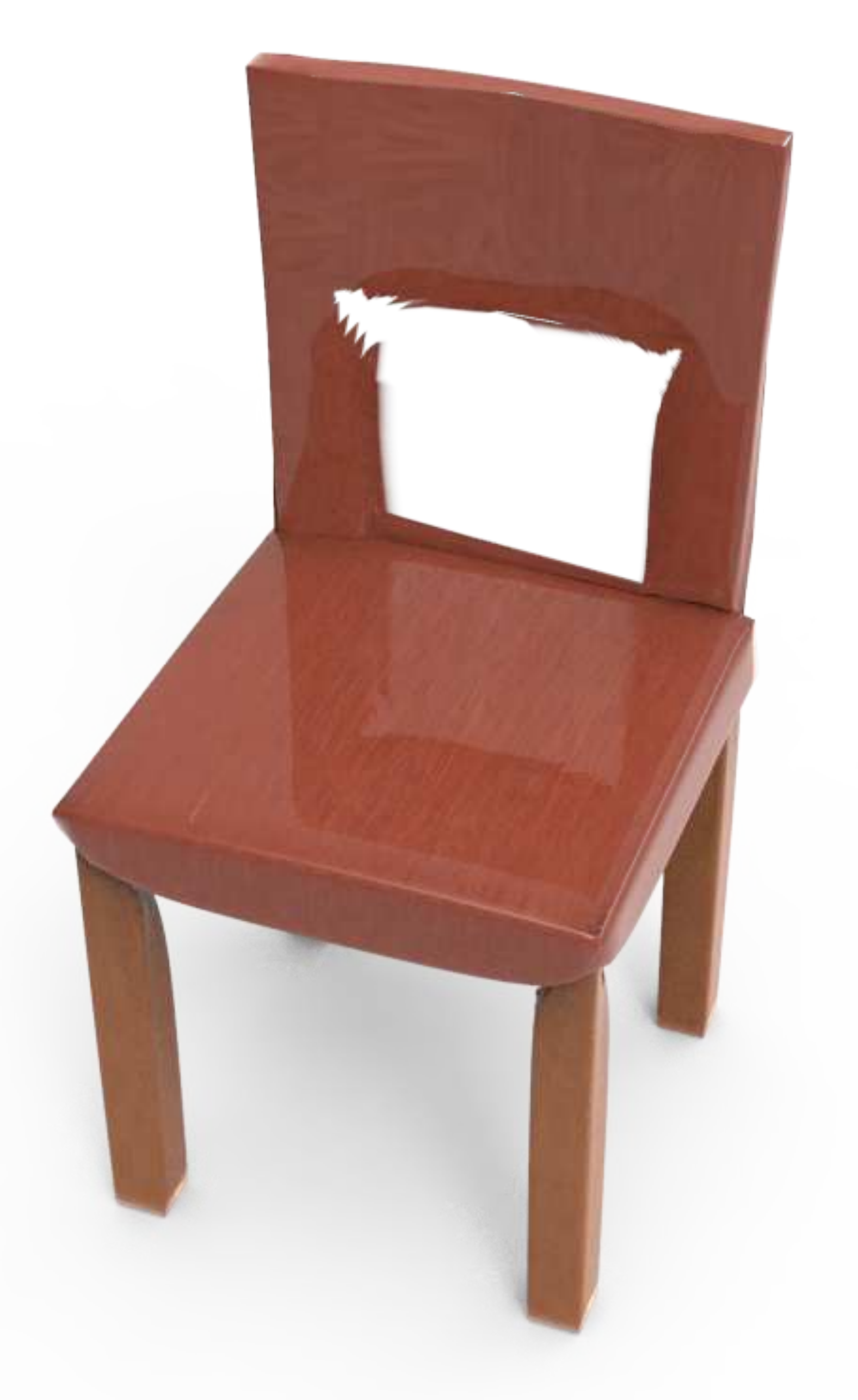}}
        {\includegraphics[width=0.15\linewidth]{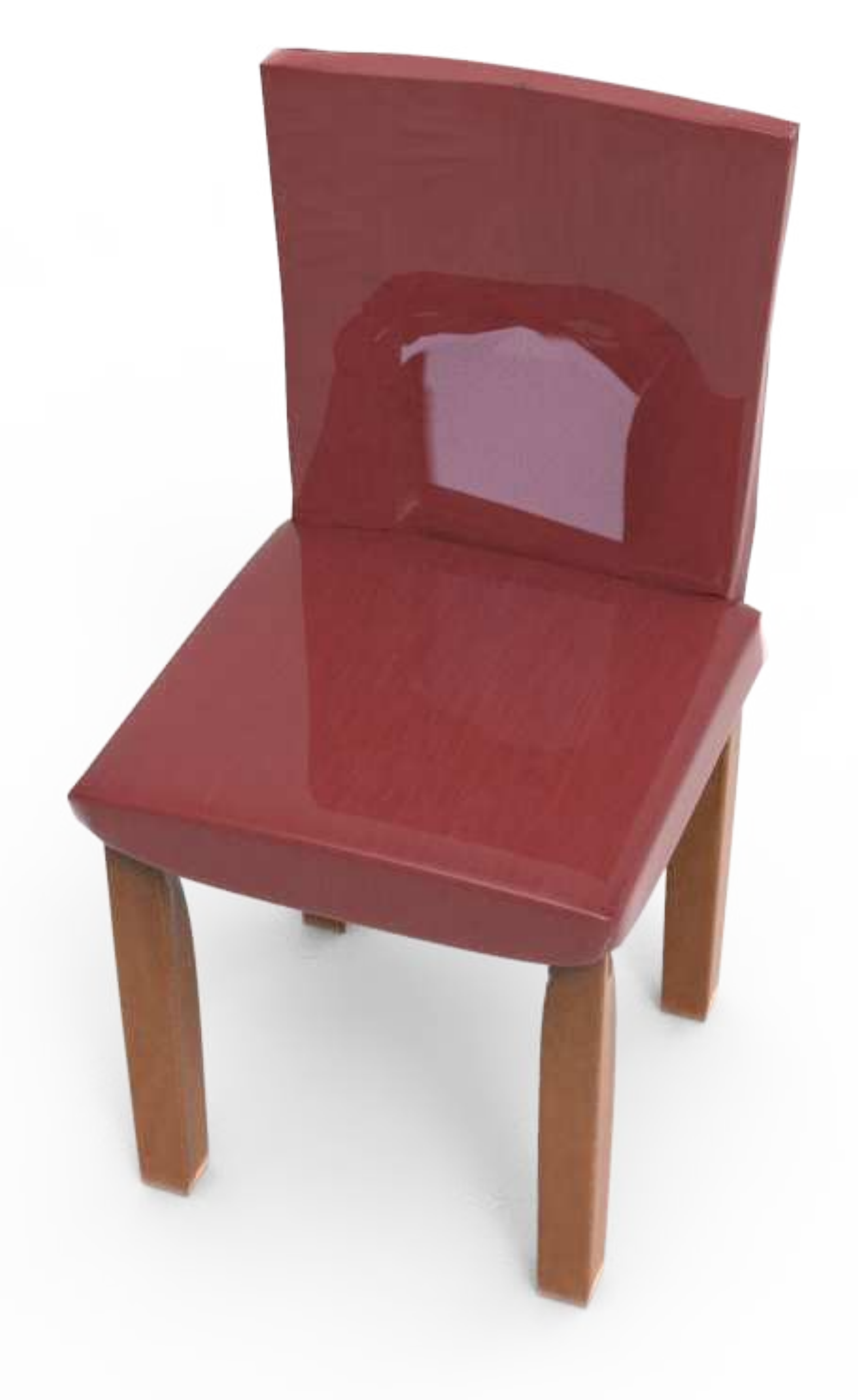}}
        {\includegraphics[width=0.15\linewidth]{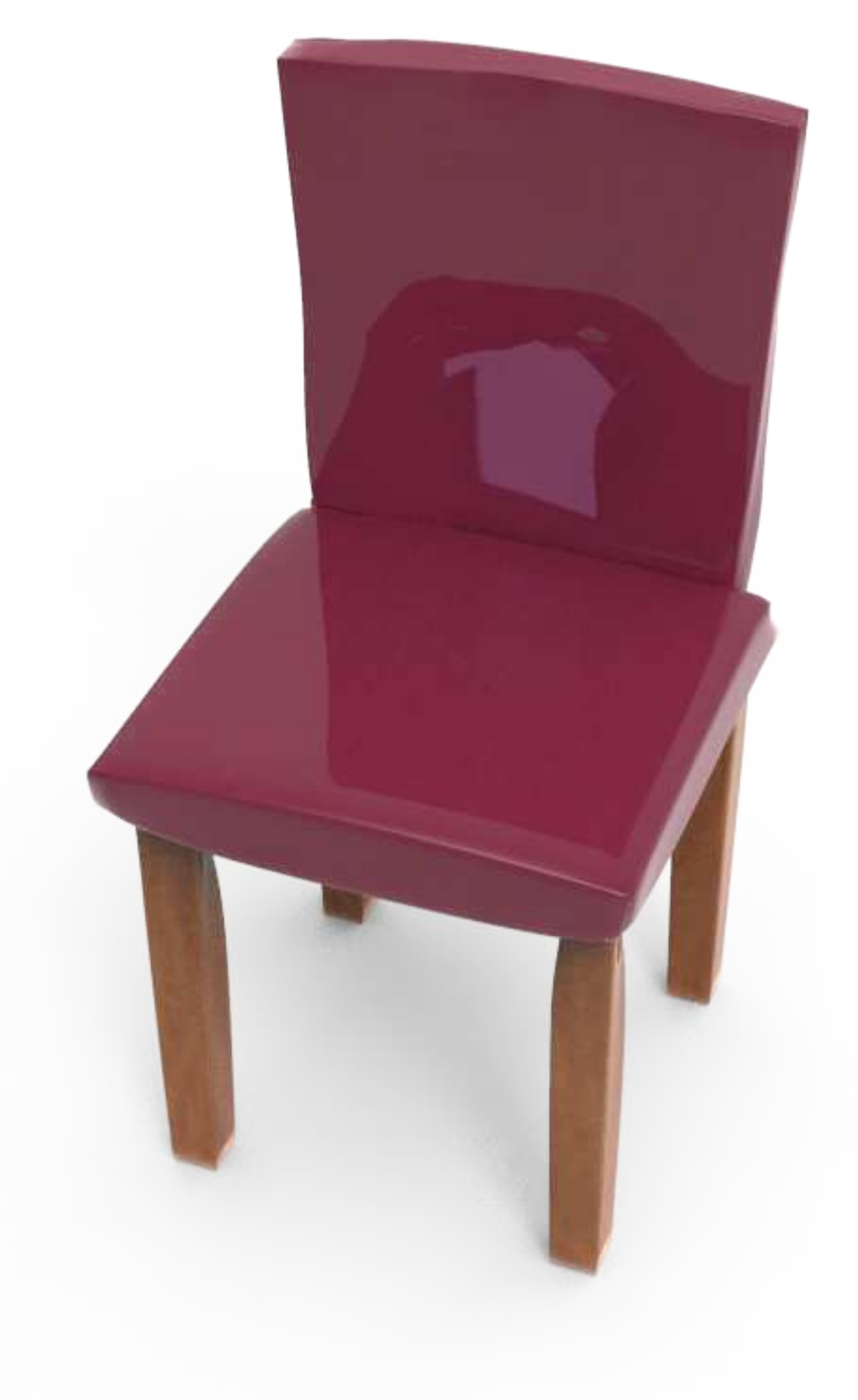}}
        {\includegraphics[width=0.15\linewidth]{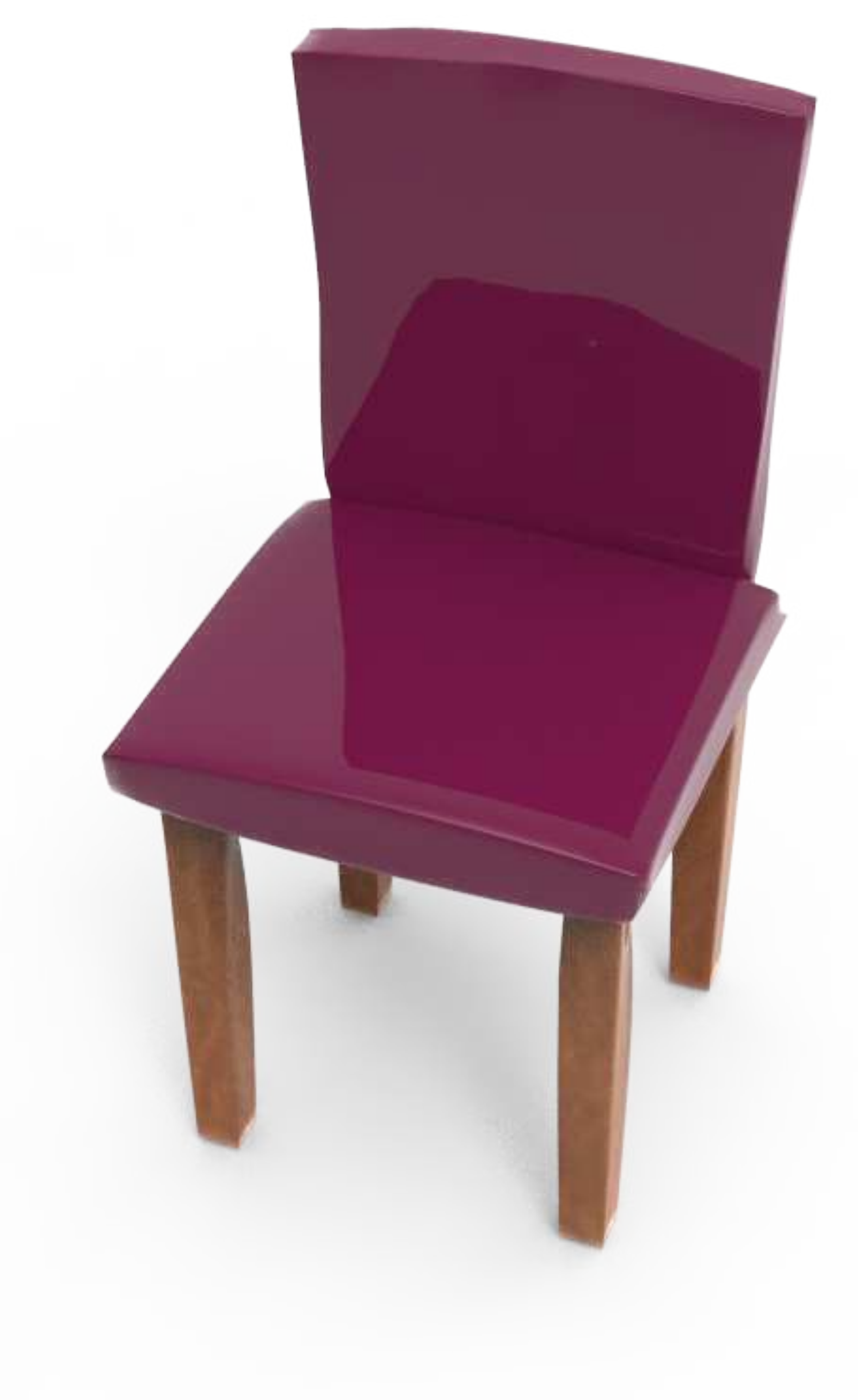}}
        \\
        \rotatebox{90}{\quad \hspace{2mm} {\small TM-NET}}
        &
        \vspace{-2mm}
        \hspace{-5mm}
        {\includegraphics[width=0.15\linewidth]{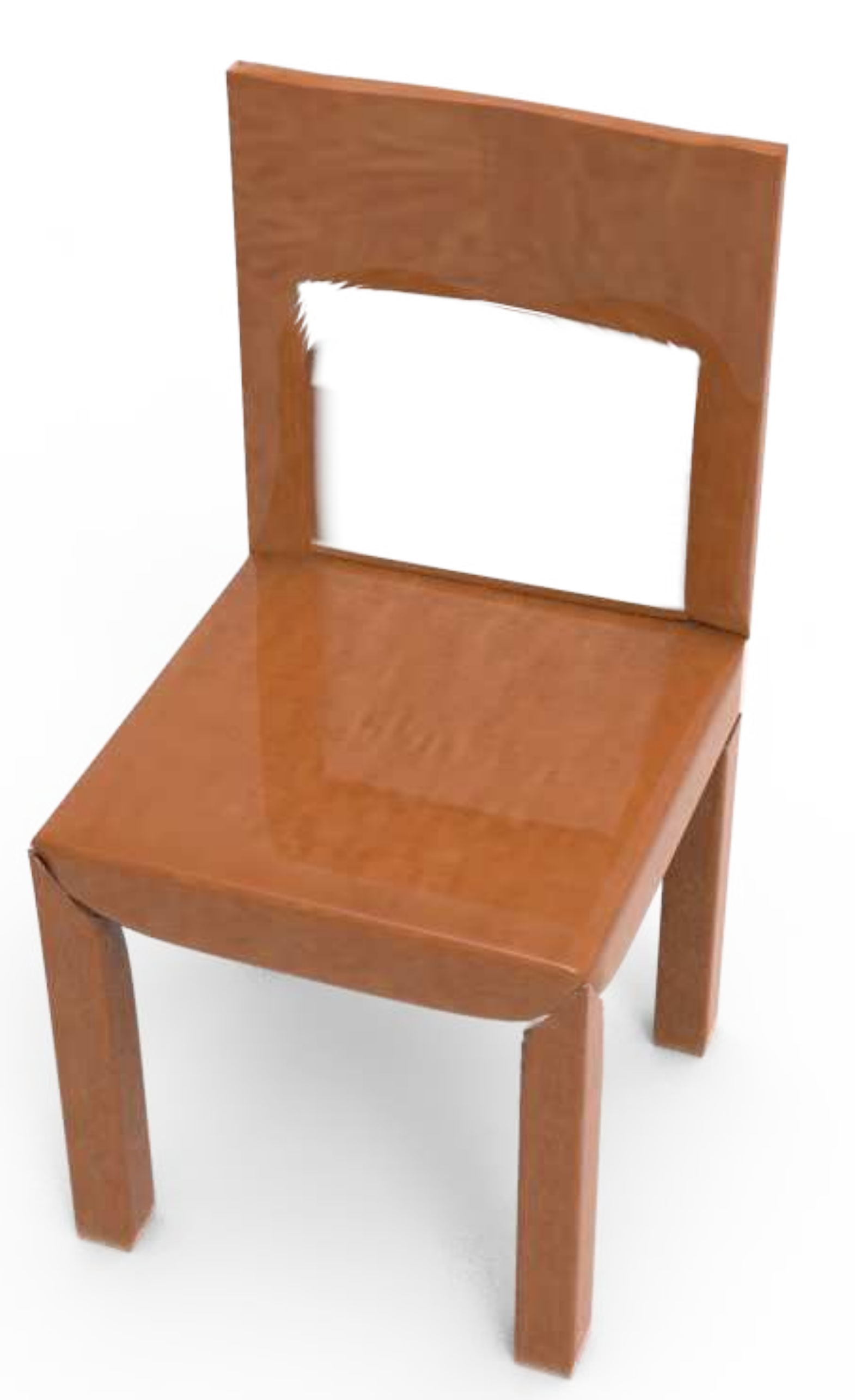}}
        {\includegraphics[width=0.15\linewidth]{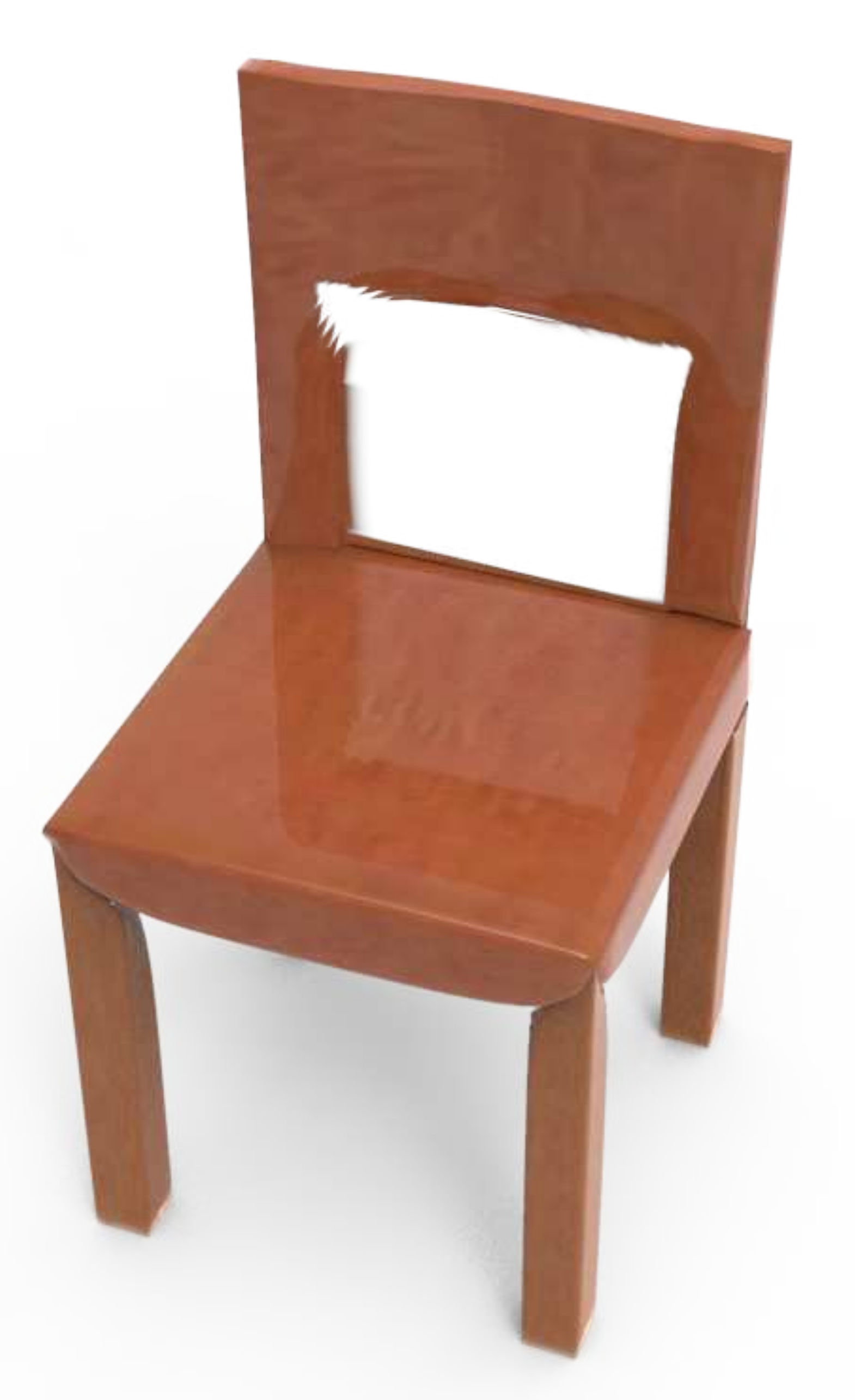}}
        {\includegraphics[width=0.15\linewidth]{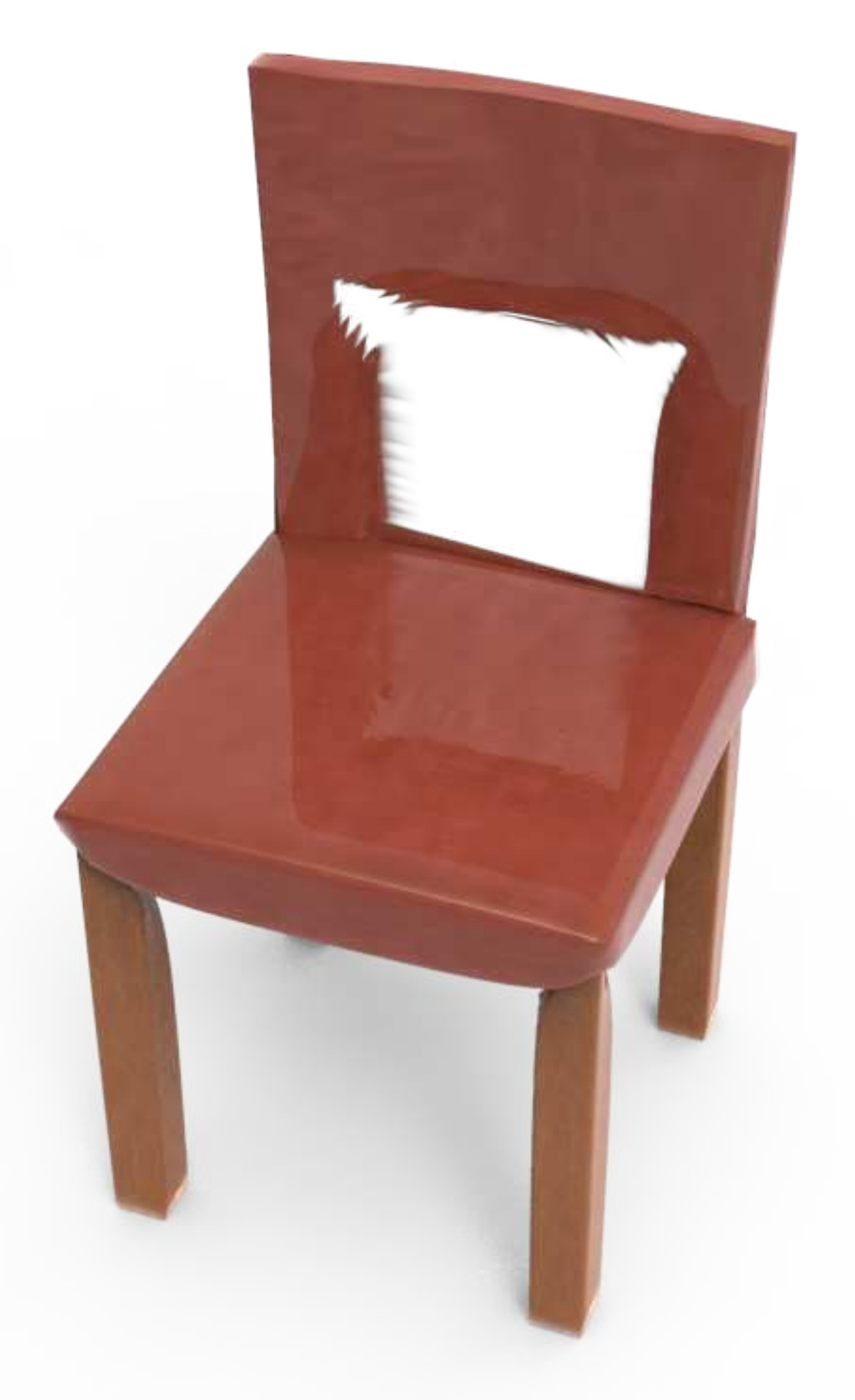}}
        {\includegraphics[width=0.15\linewidth]{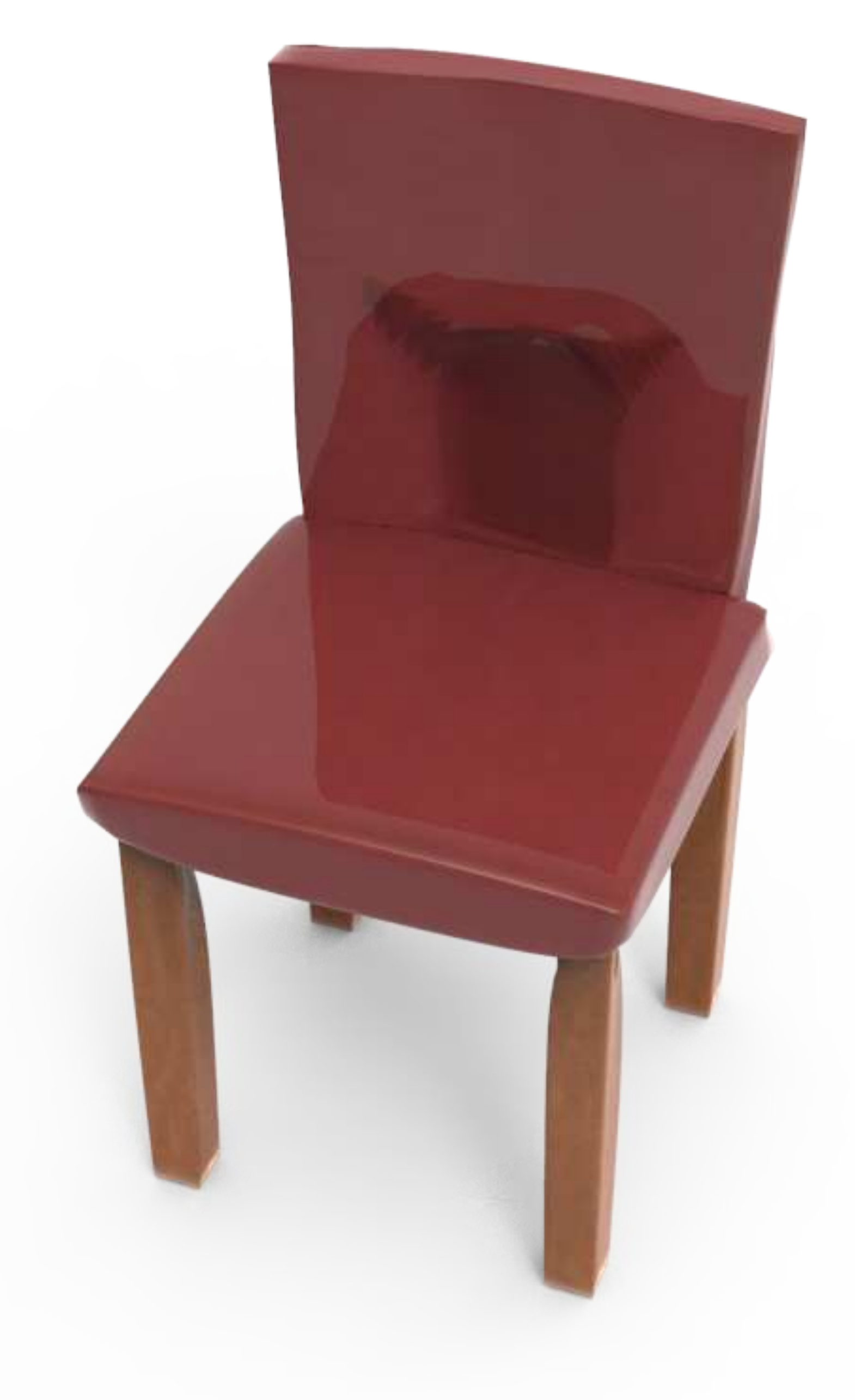}}
        {\includegraphics[width=0.15\linewidth]{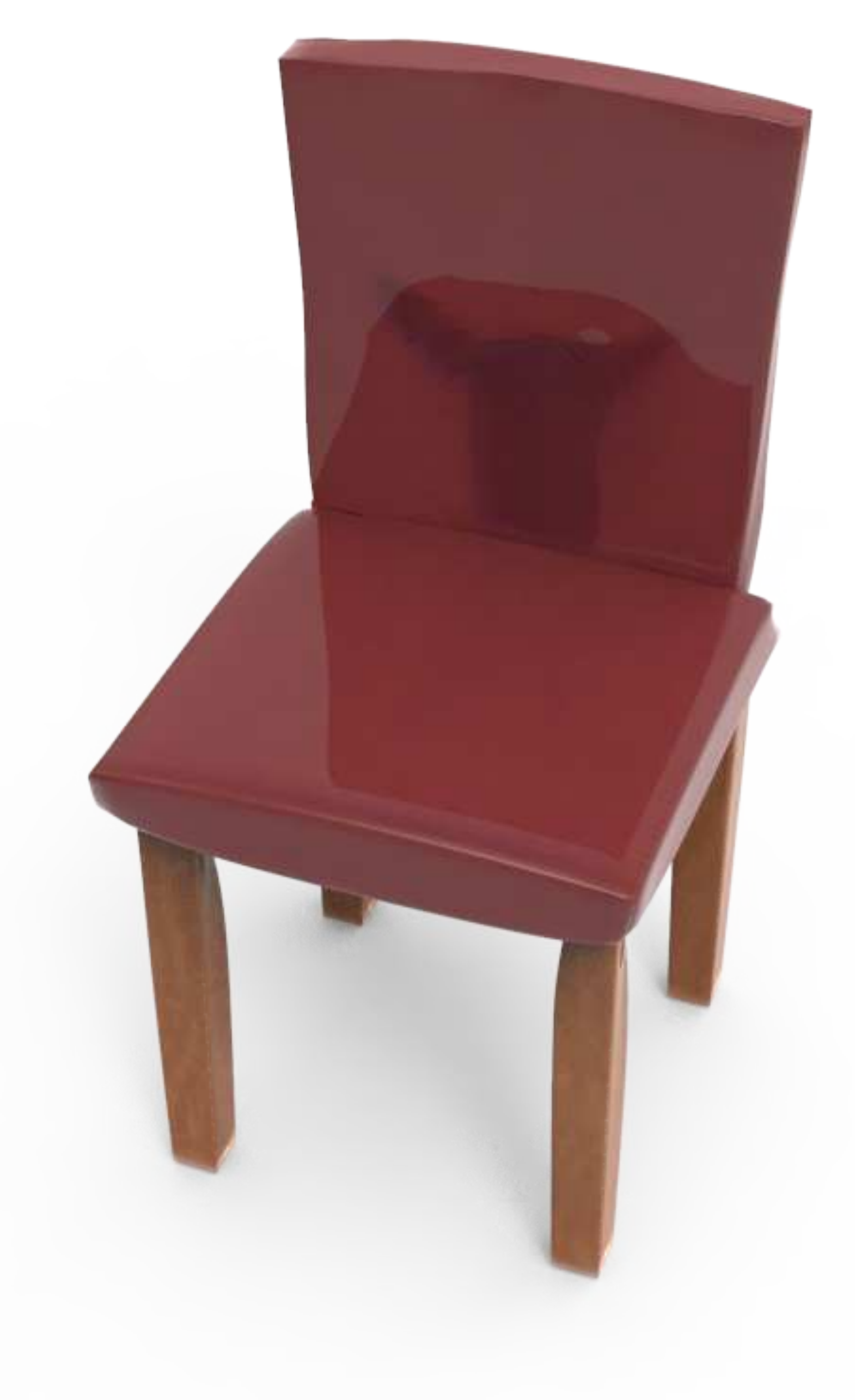}}
        {\includegraphics[width=0.15\linewidth]{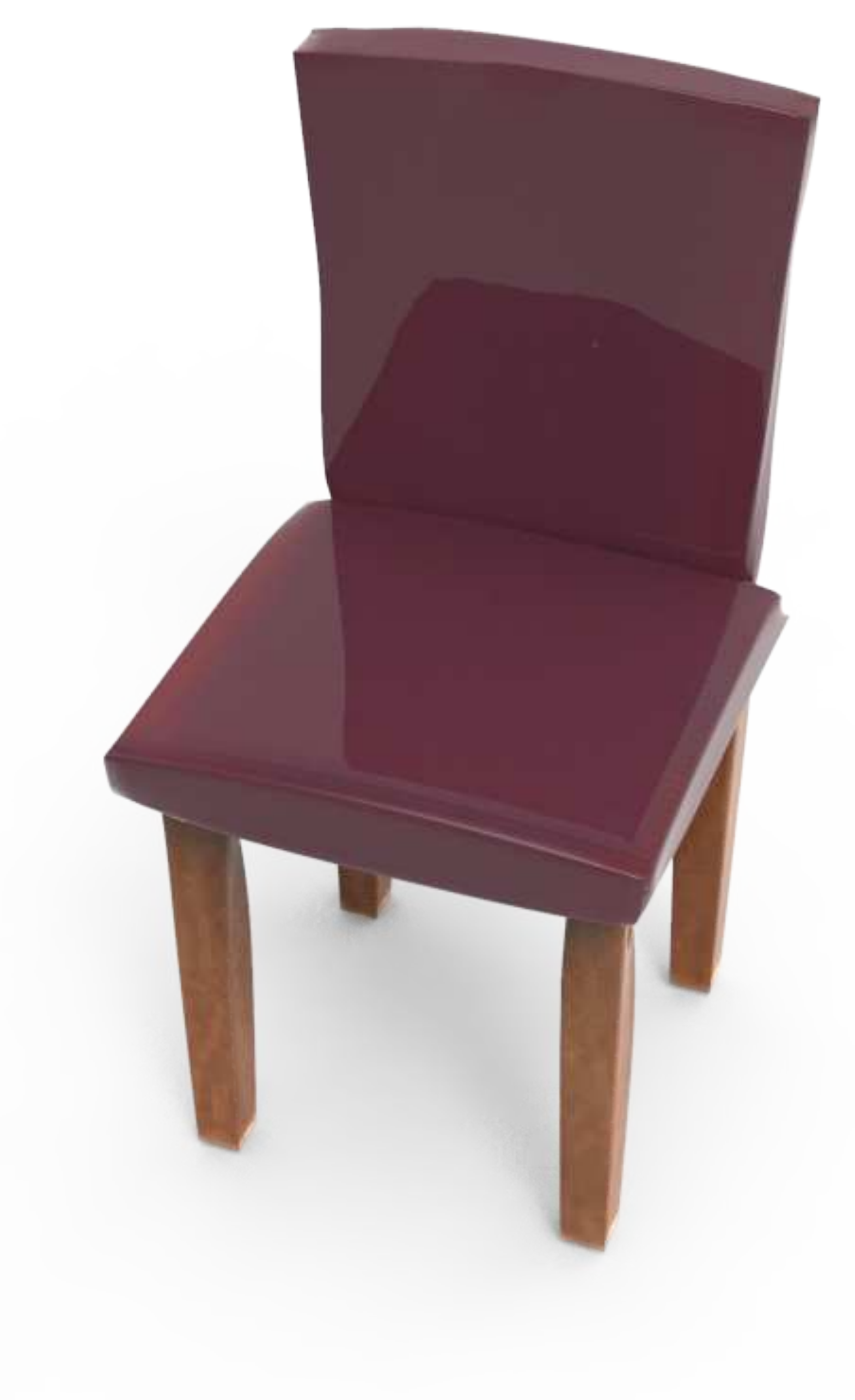}}
    \end{tabular}
  }
\caption{\rznn{Randomly sampled results of latent-space linear interpolation between chairs (first and last columns), comparing TM-NET to VON~\cite{VON}, TF~\cite{OechsleICCV2019}, and part-by-part alpha blending.}}
\label{fig:Interpolation}
\end{figure}

\subsection{Self evaluations}

\paragraph{Dividing texture image into patches}
As mentioned in Sections~\ref{section::EncodingofTexturedPart} and~\ref{section::GeometryInferTexture}, we divide the input texture image into six patches, each corresponding to a face of the template box.
Compared with our divide and conquer approach, a straight-forward strategy is to feed the whole texture image of size $1024 \times 768$ into the TextureVAE. If we feed the whole texture image of size $1024\times768$ into the TextureVAE, we have to adopt a three-level hierarchy TextureVAE and augoregressive model as VQ-VAE-2~\cite{vqvae-2} does and use three index matrices of size $32\times32$, $64\times64$ and $128\times128$, to cope with the higher input resolution.
Here, we compare shape texturing results using these two strategies in Figure~\ref{fig:patch_whole_comparison}. The result shows that by dividing the whole image into patches, the autoregressive model can generate textures with higher quality.

\vspace{-5pt}

\paragraph{Seed Part}
\rznn{To ensure the coherence between textures generated for different shape parts, we introduced the use of seed parts, where the
generation of textures for the remaining parts is conditioned on the seed part's generated texture features and the remaining parts' geometry features. We evaluate how the choice of seed parts influences the generated textures in Figure~\ref{fig:seed_part_choice}, which demonstrates experimentally that while the final texturing results do differ as the seed parts vary, the generated textures remain plausible and exhibit compatability across the whole shapes.}

\begin{figure}[!t]
    \centering
    {\includegraphics[width=0.18\linewidth]{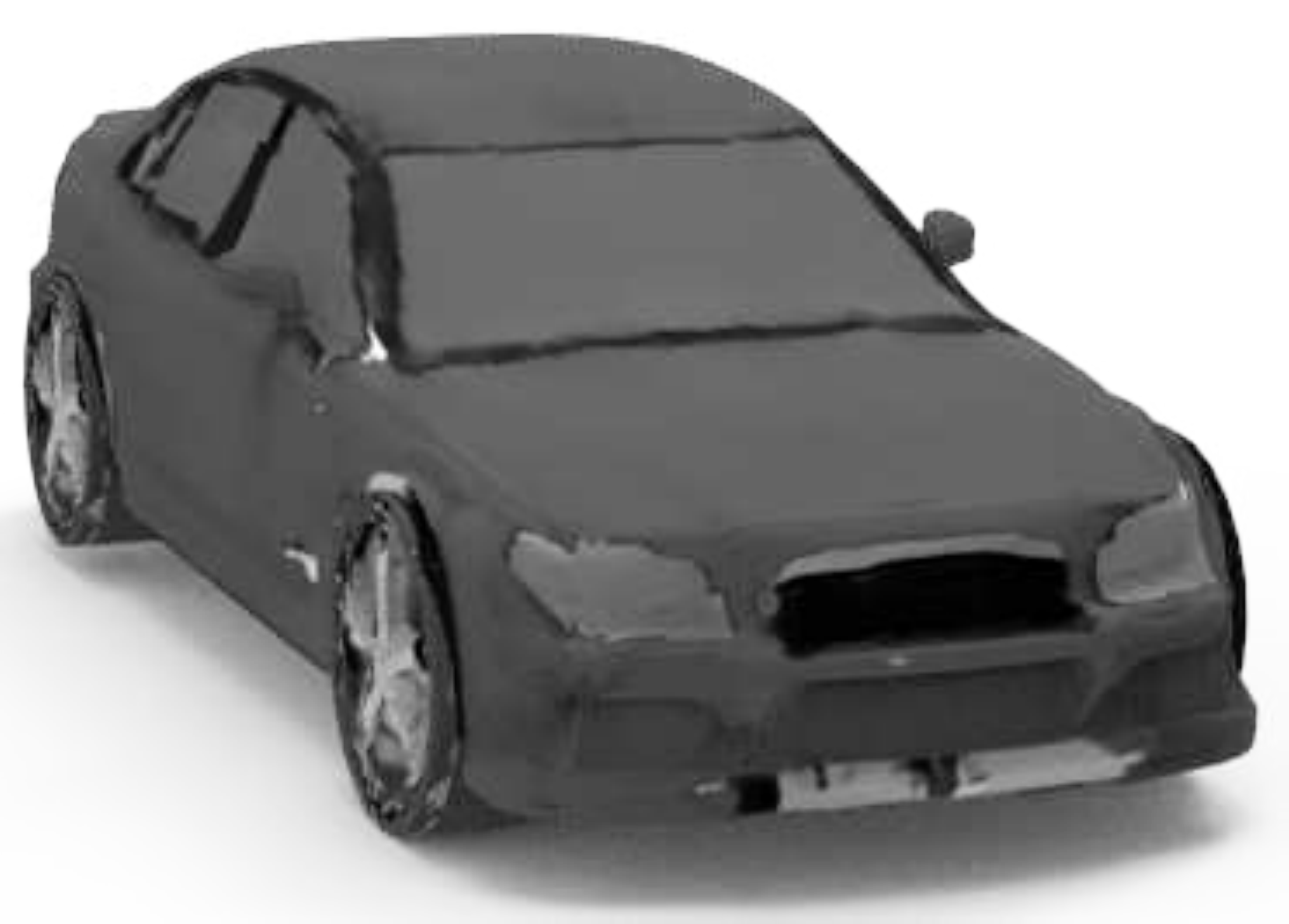}}
    {\includegraphics[width=0.18\linewidth]{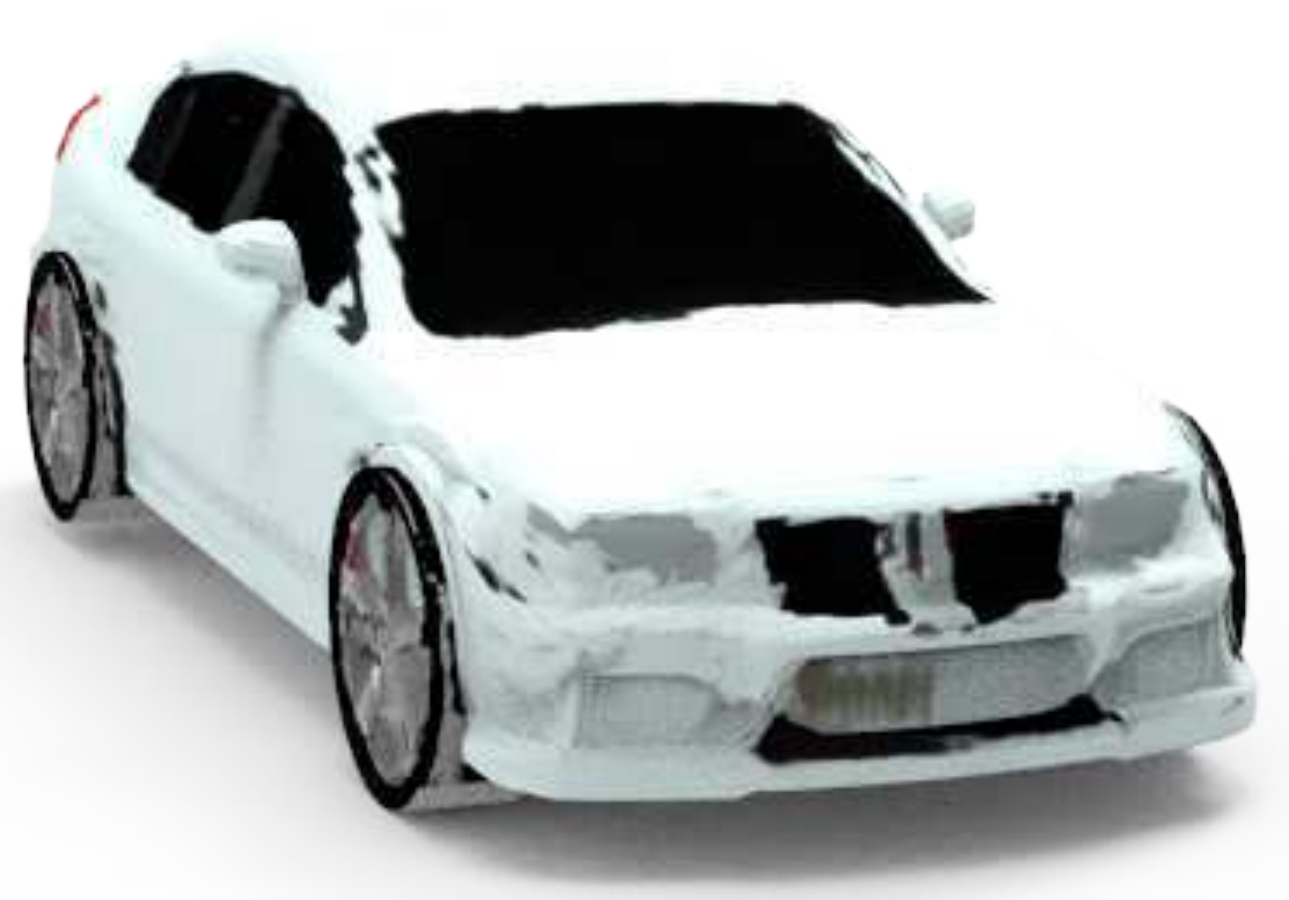}}
    {\includegraphics[width=0.18\linewidth]{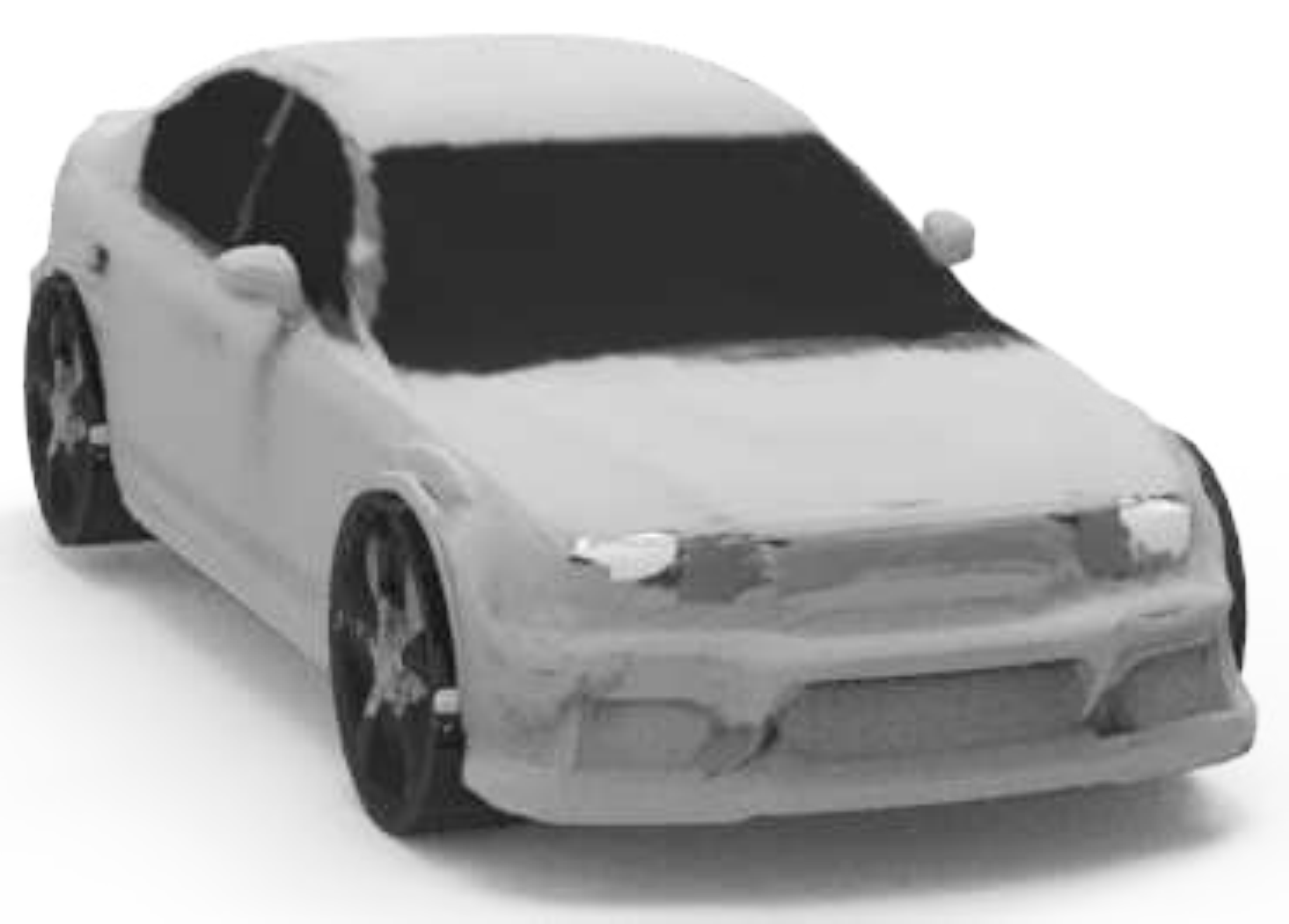}}
    {\includegraphics[width=0.18\linewidth]{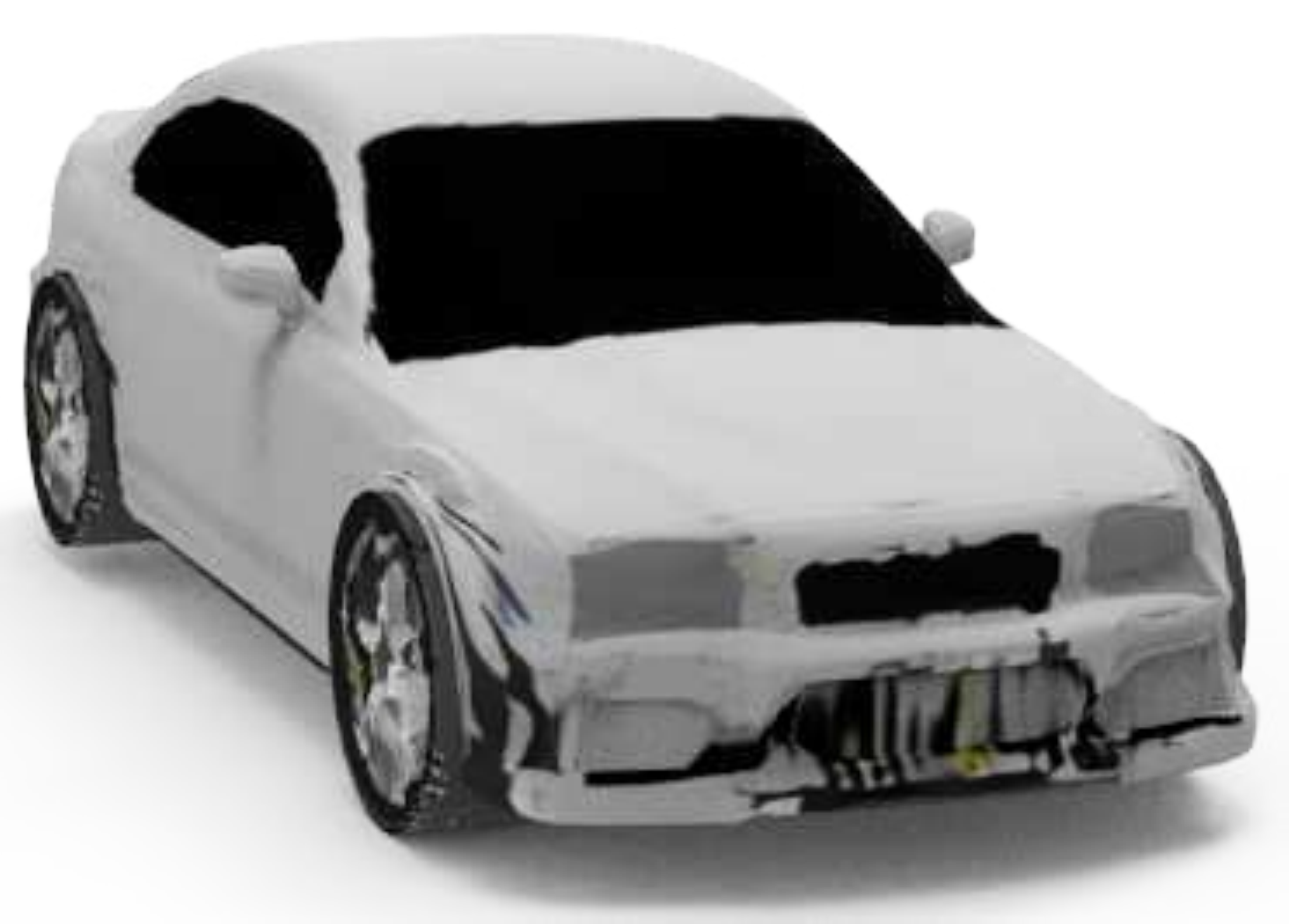}}
    {\includegraphics[width=0.18\linewidth]{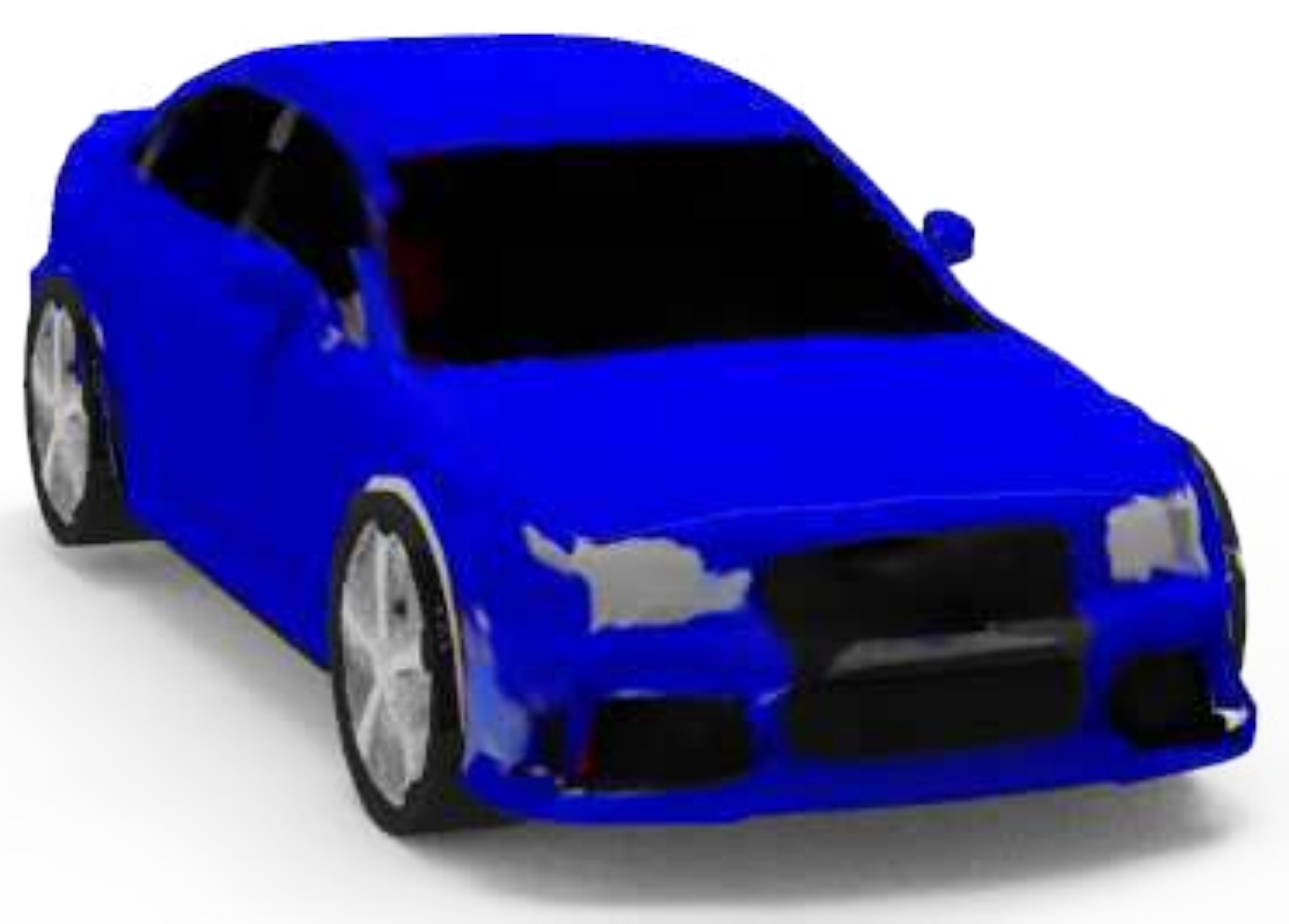}}
    \\
	{\includegraphics[width=0.18\linewidth]{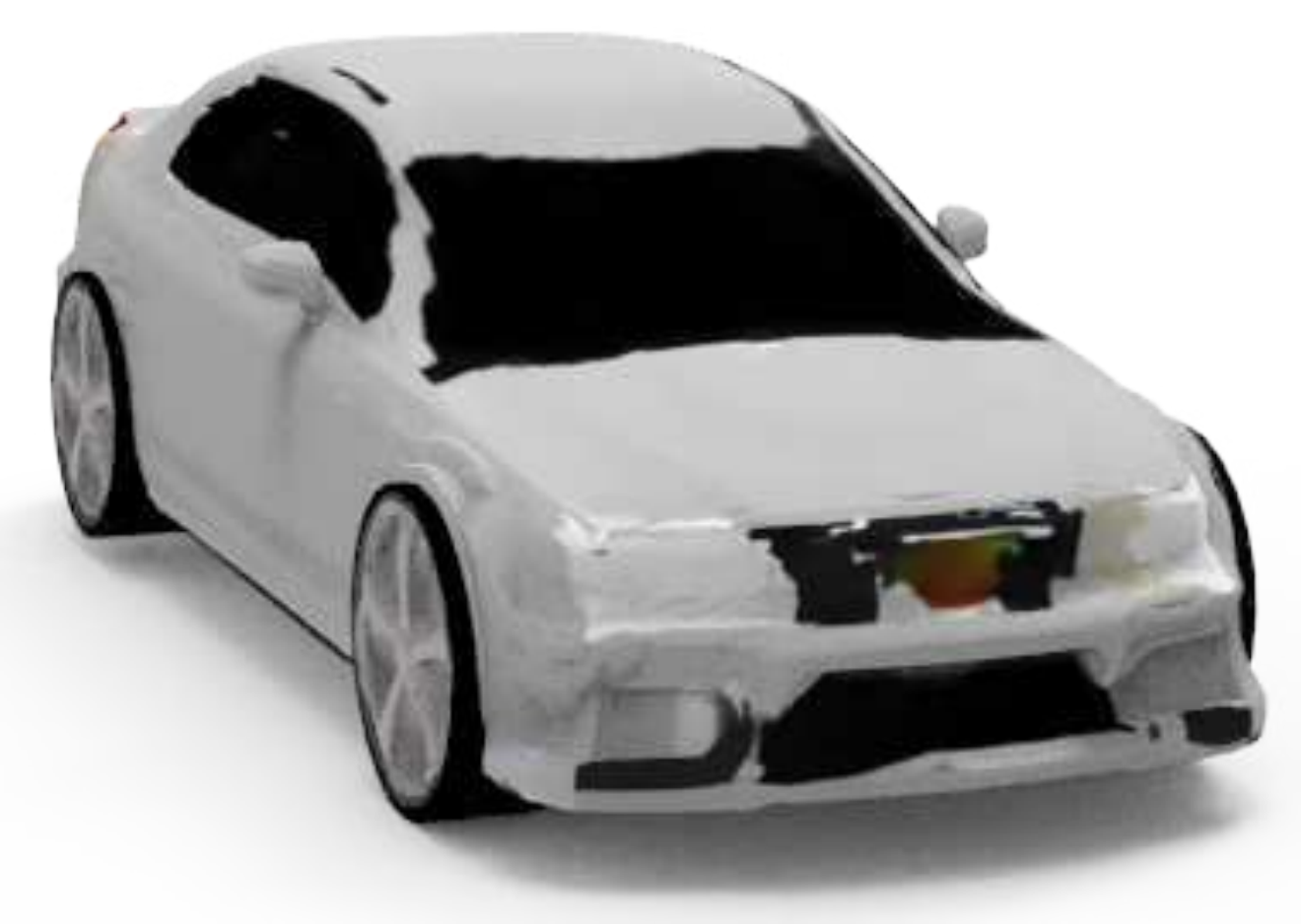}}
	{\includegraphics[width=0.18\linewidth]{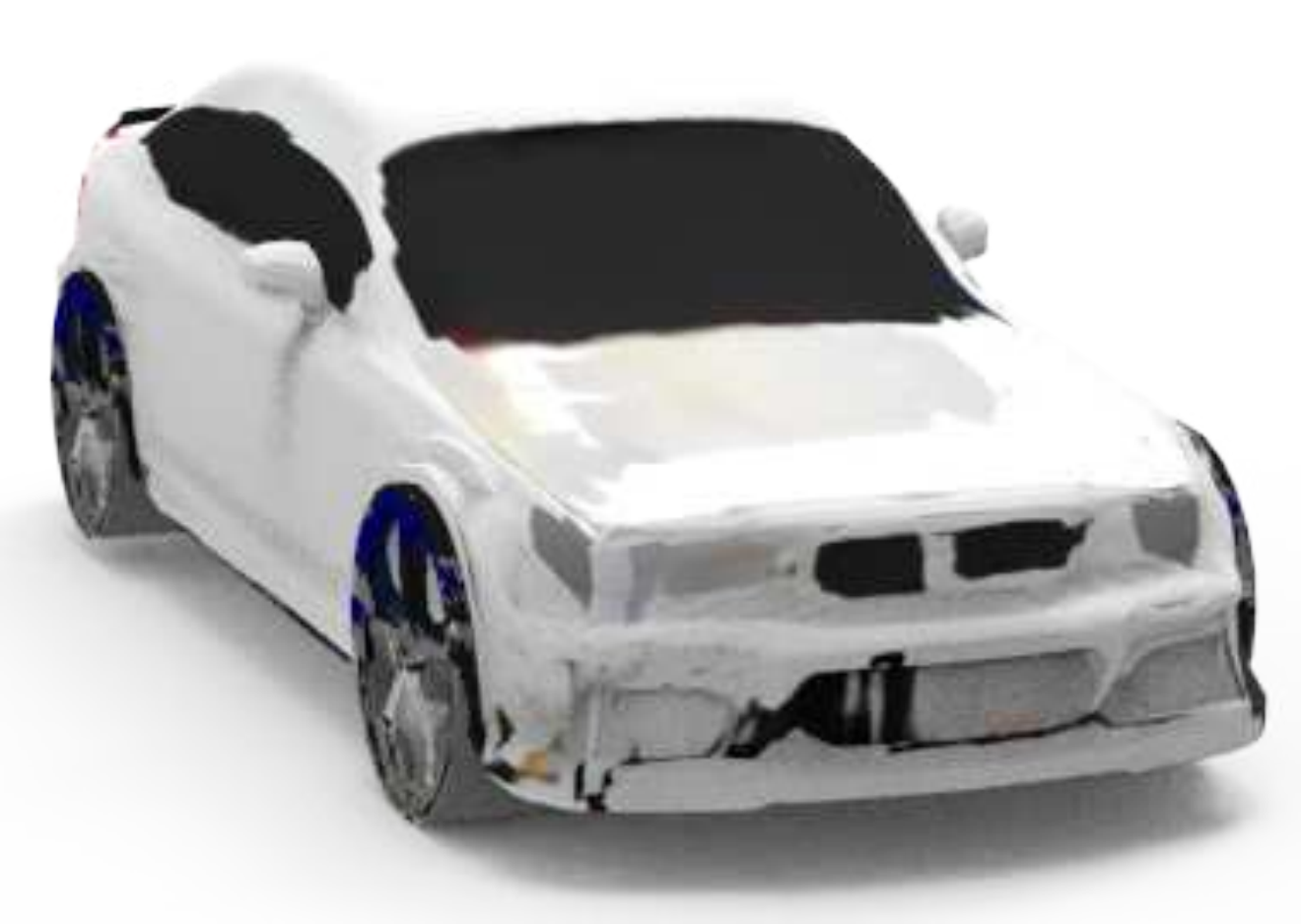}}
    {\includegraphics[width=0.18\linewidth]{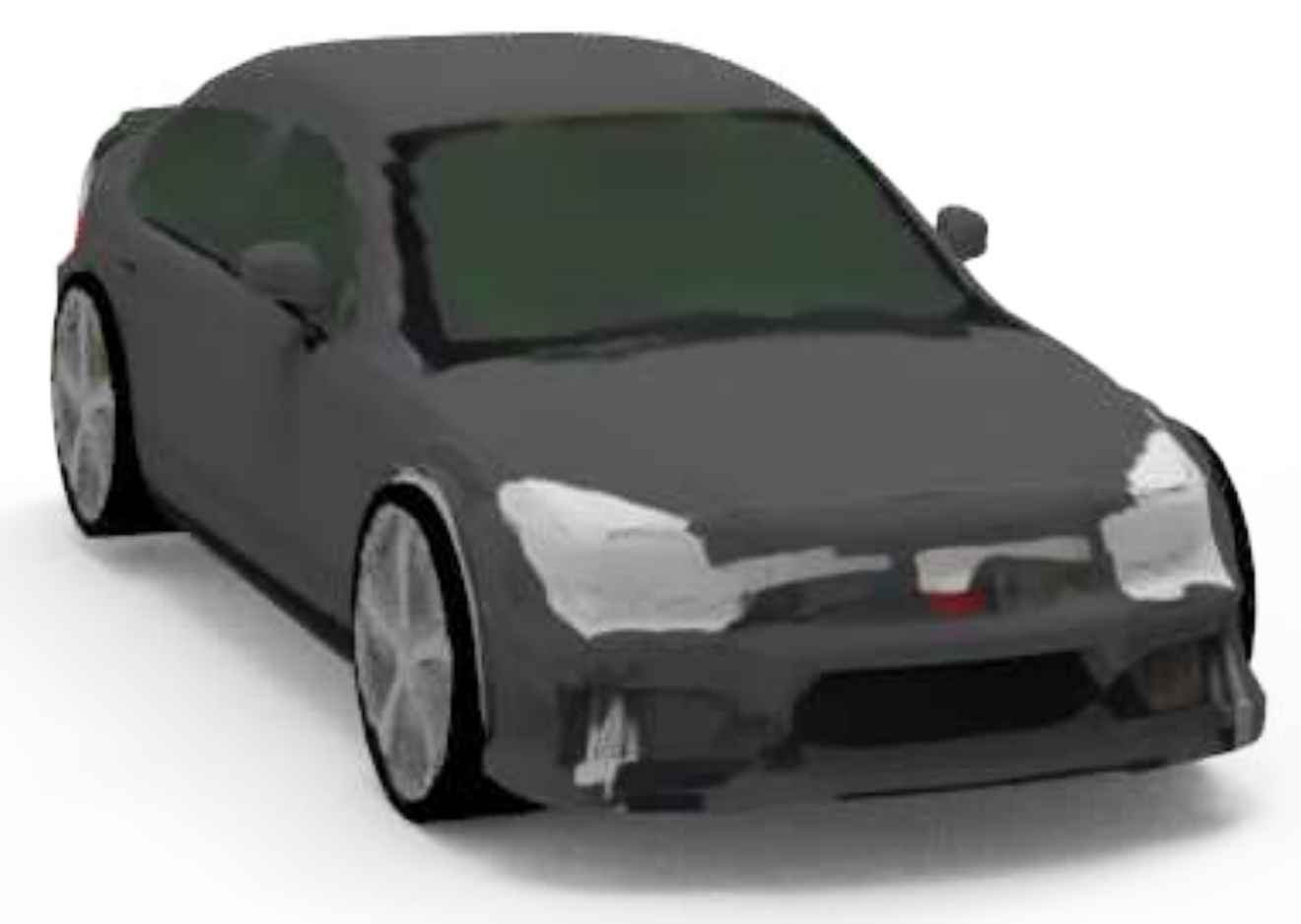}}
    {\includegraphics[width=0.18\linewidth]{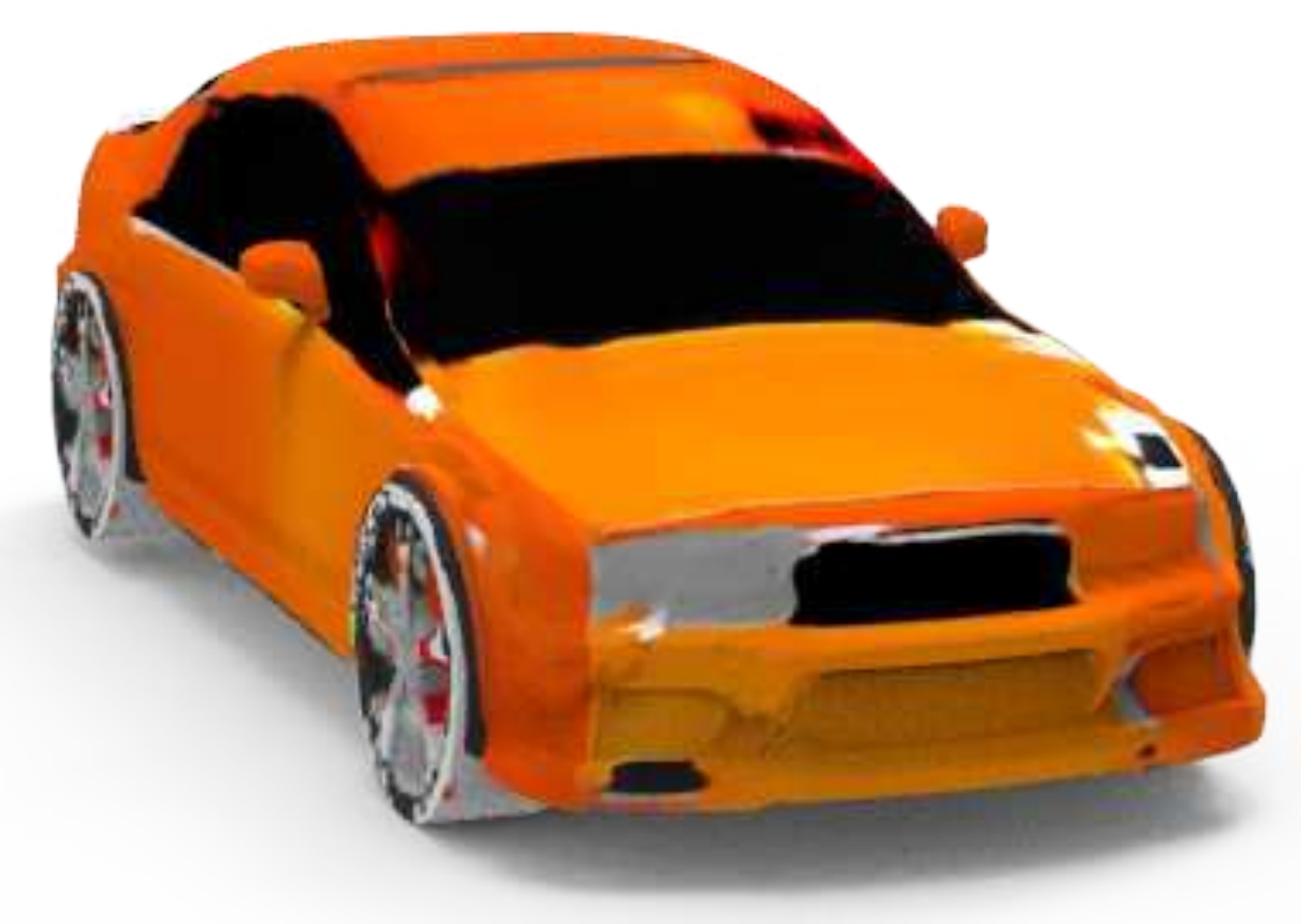}}
    {\includegraphics[width=0.18\linewidth]{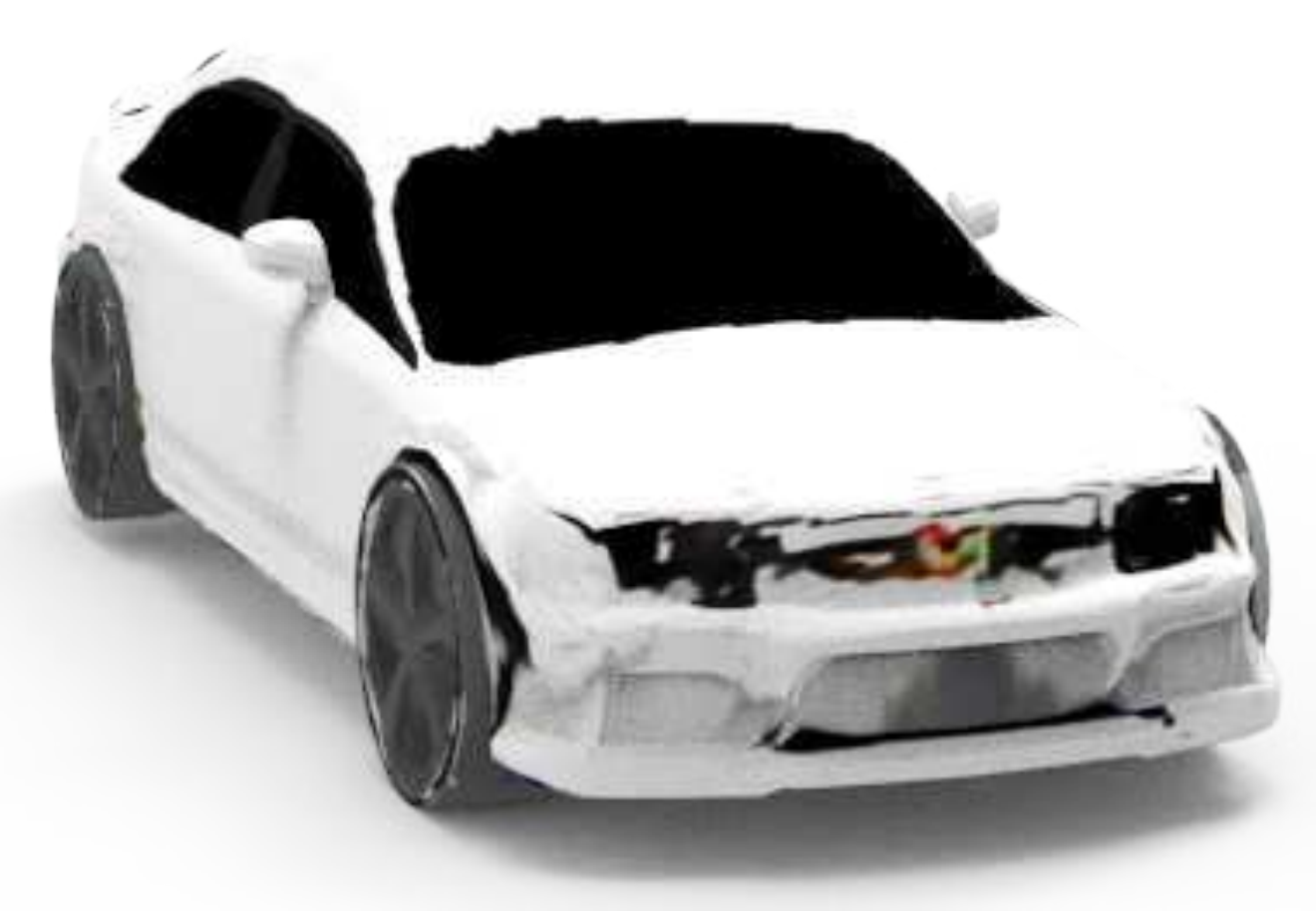}}
    \caption{
    \rzn{Comparison of mesh texturing results. Top row: dividing a texture image into patches. Bottom row: regarding the same texture image as a whole. Results show that with the patch-based, divide-and-conquer approach, %
    TM-NET can generate textures with higher quality.}}
    \label{fig:patch_whole_comparison}
\end{figure}

\begin{figure}[!t]
    \centering
    {\includegraphics[width=0.24\linewidth]{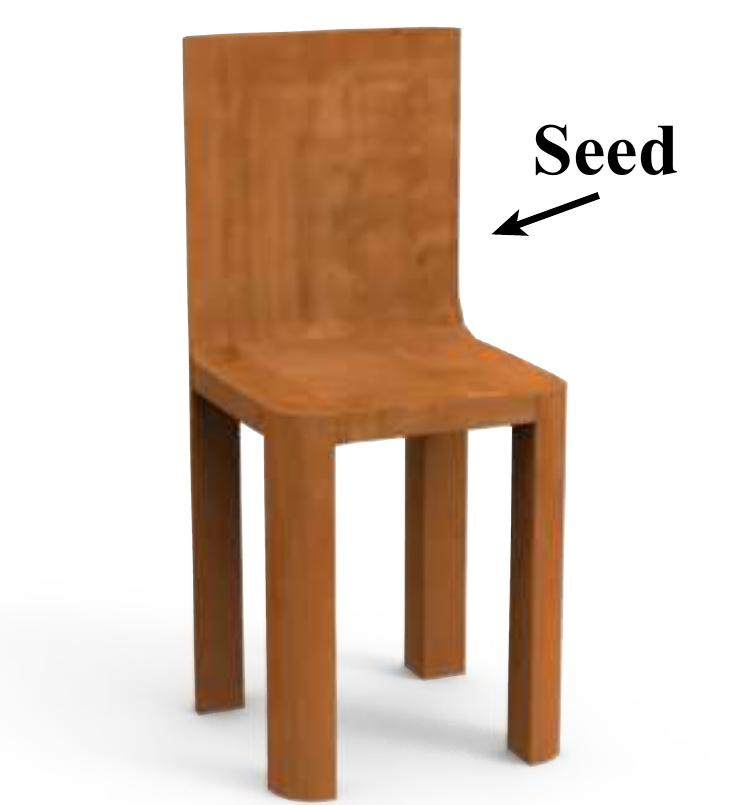}}
    {\includegraphics[width=0.24\linewidth]{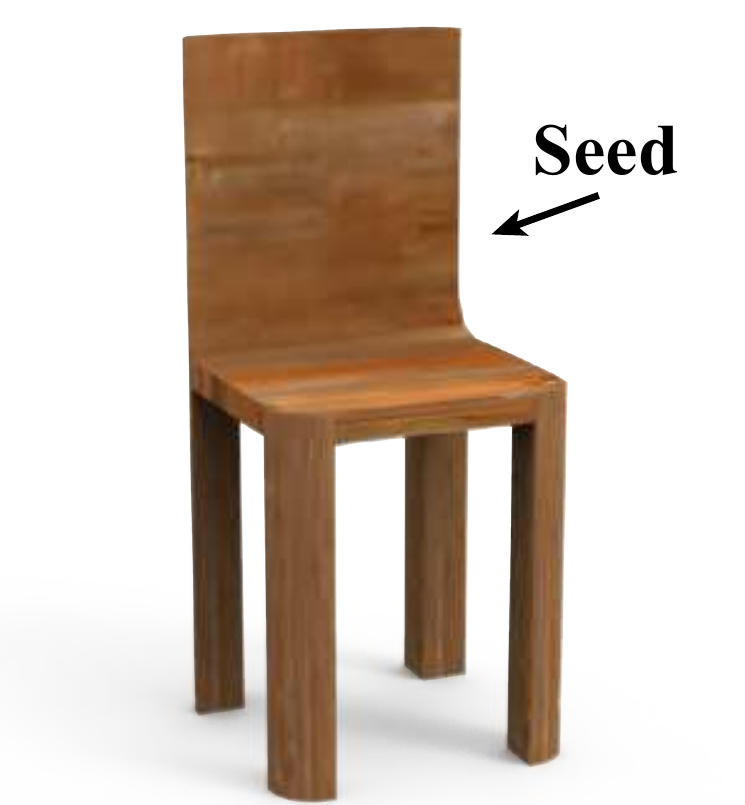}}
    {\includegraphics[width=0.24\linewidth]{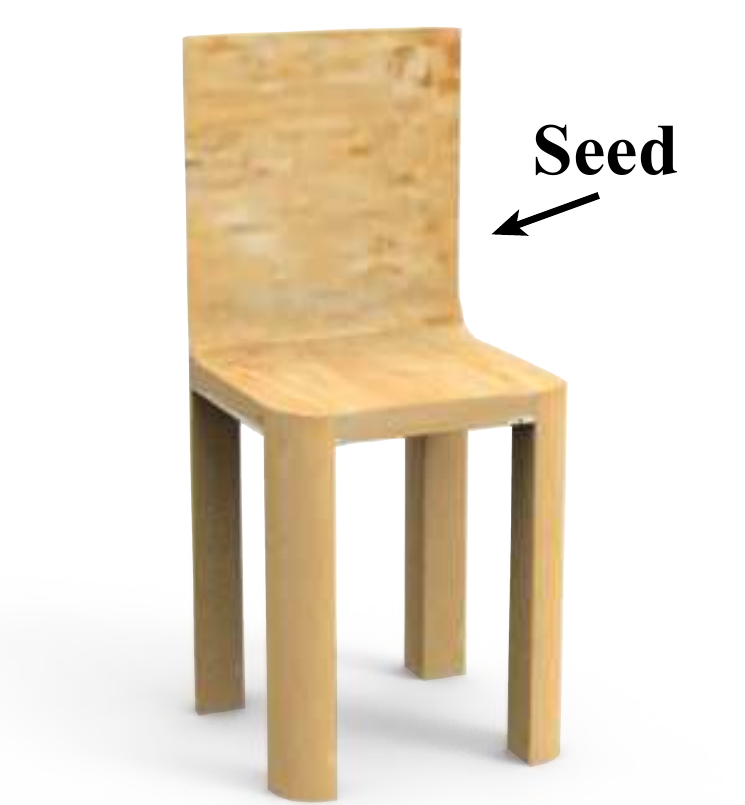}}
    {\includegraphics[width=0.24\linewidth]{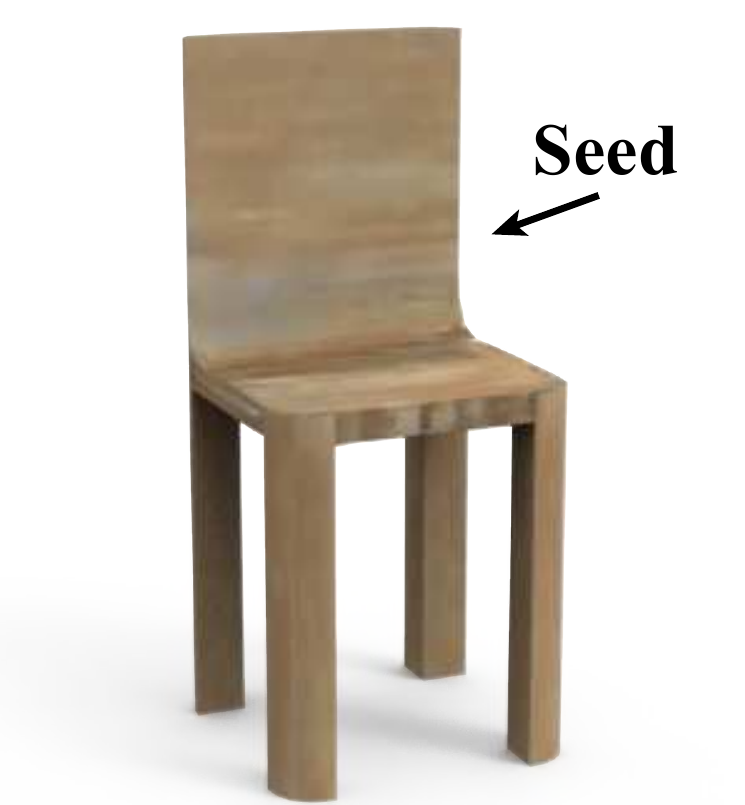}}
    \\
	{\includegraphics[width=0.24\linewidth]{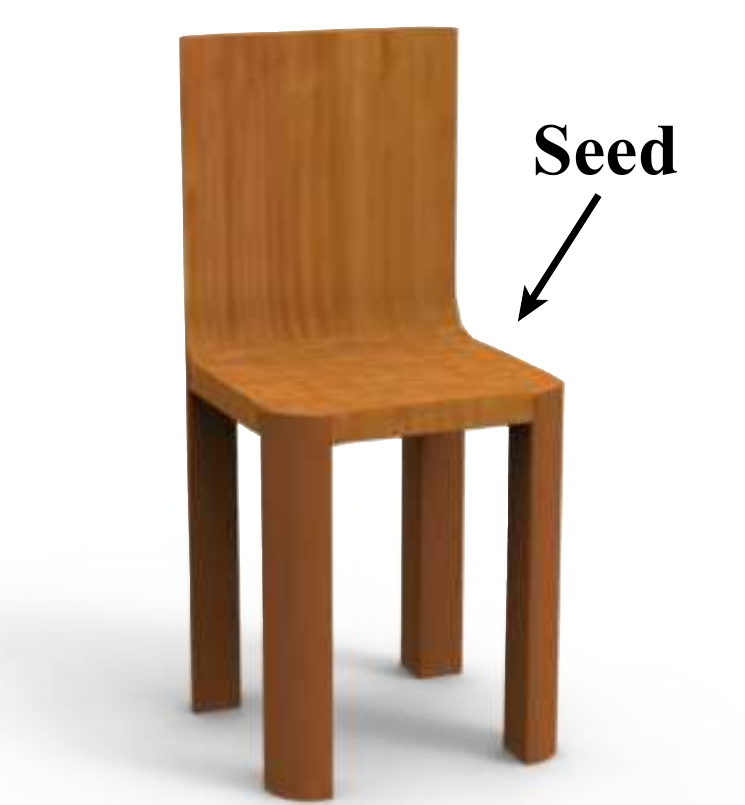}}
	{\includegraphics[width=0.24\linewidth]{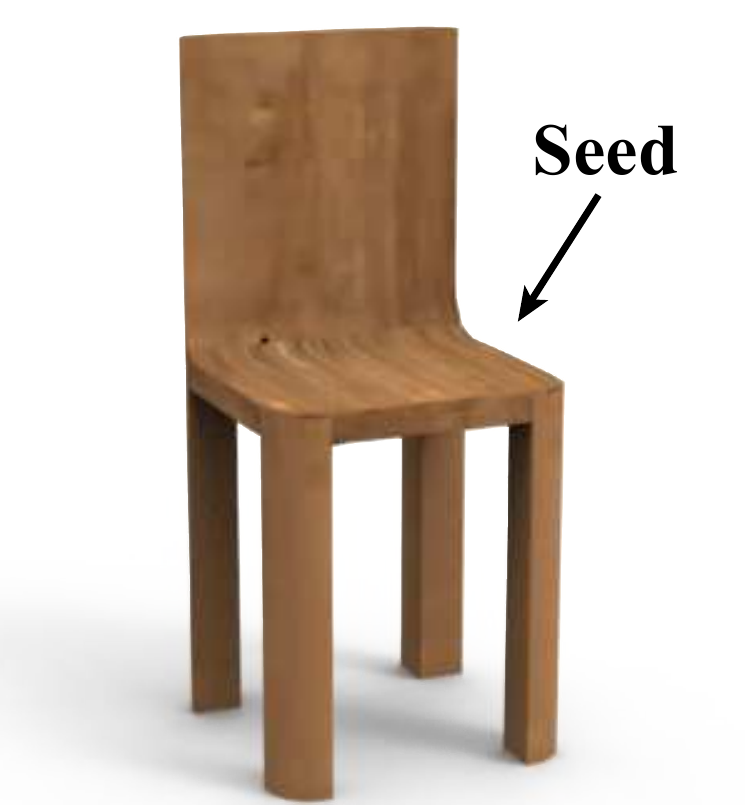}}
    {\includegraphics[width=0.24\linewidth]{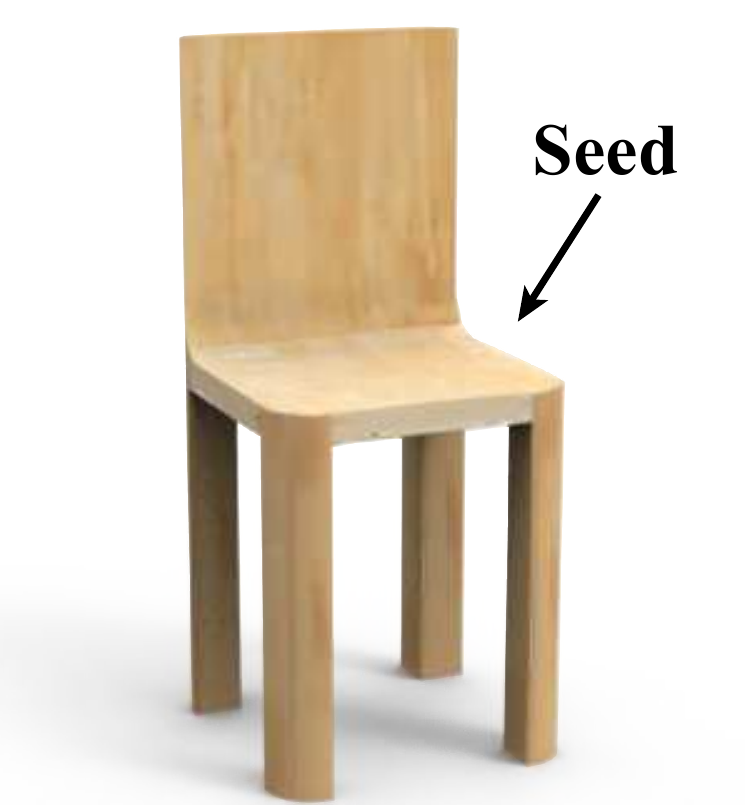}}
    {\includegraphics[width=0.24\linewidth]{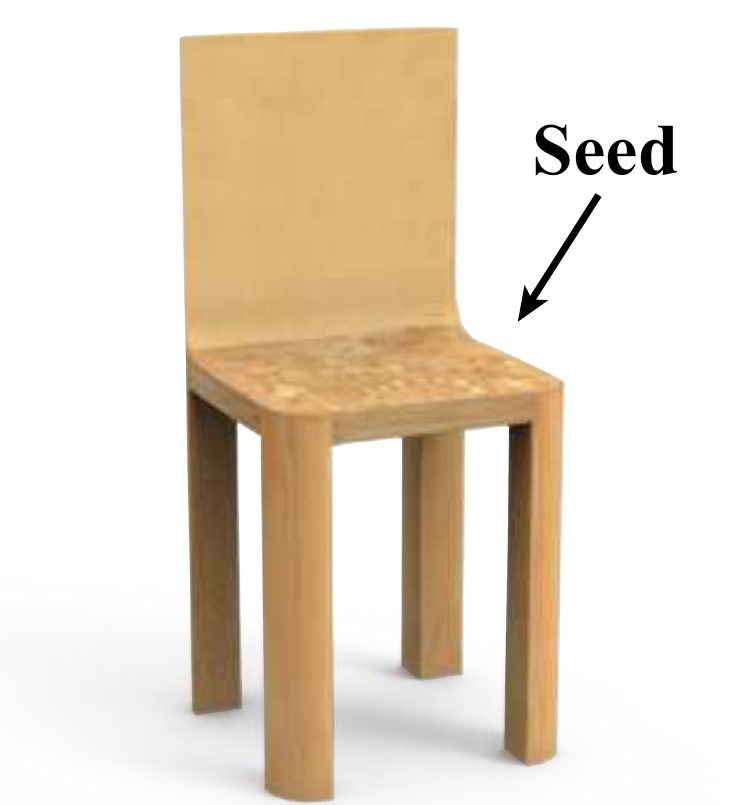}}
    \\
    {\includegraphics[width=0.24\linewidth]{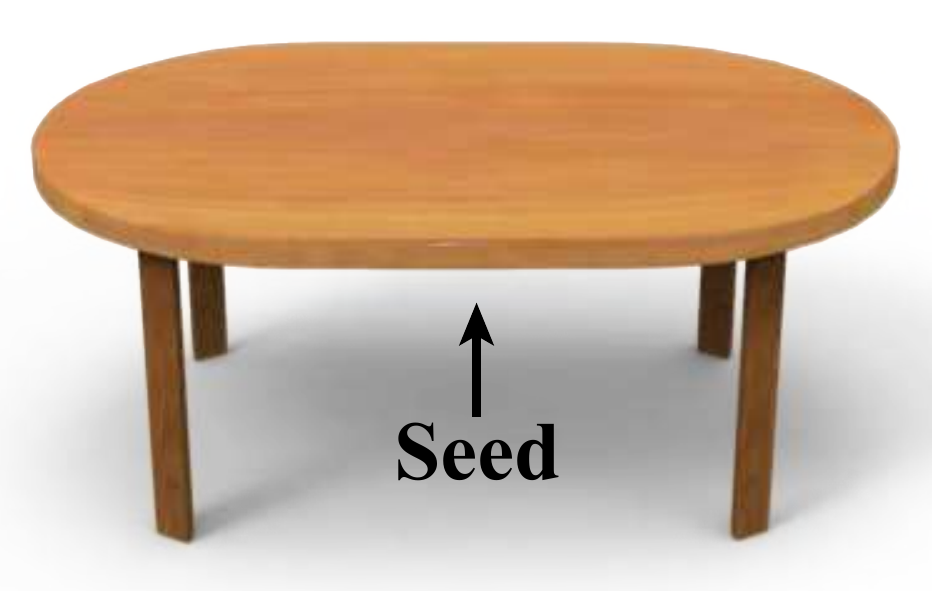}}
    {\includegraphics[width=0.24\linewidth]{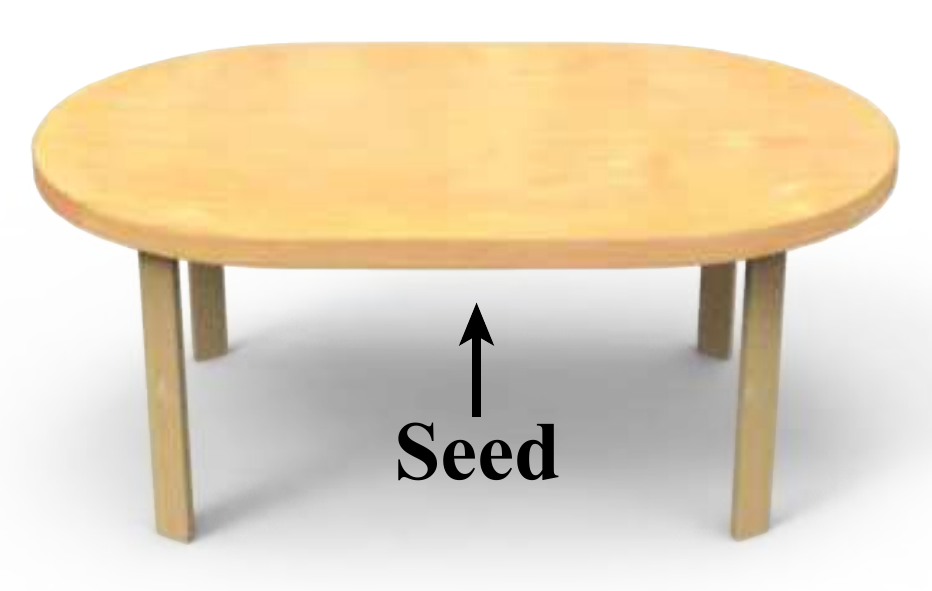}}
    {\includegraphics[width=0.24\linewidth]{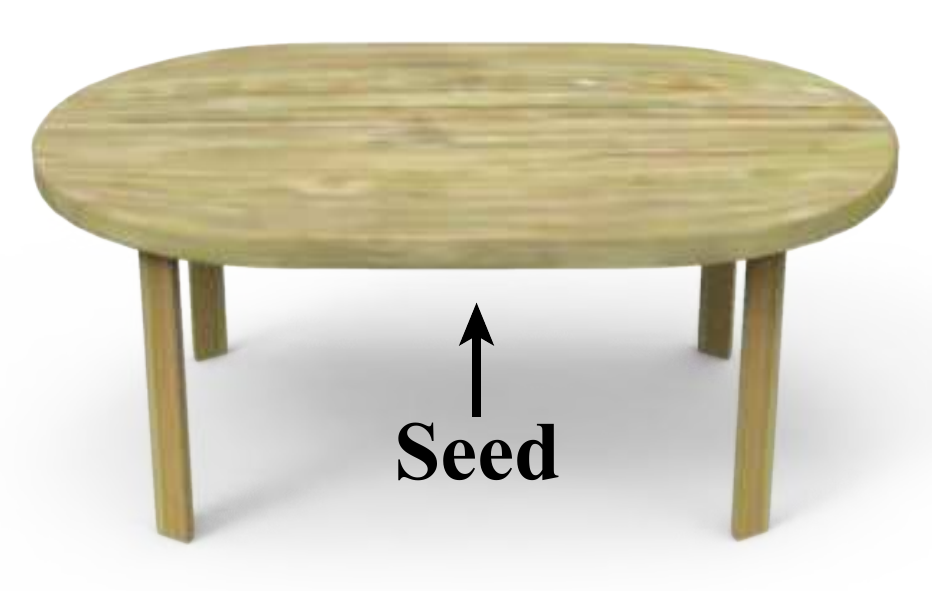}}
    {\includegraphics[width=0.24\linewidth]{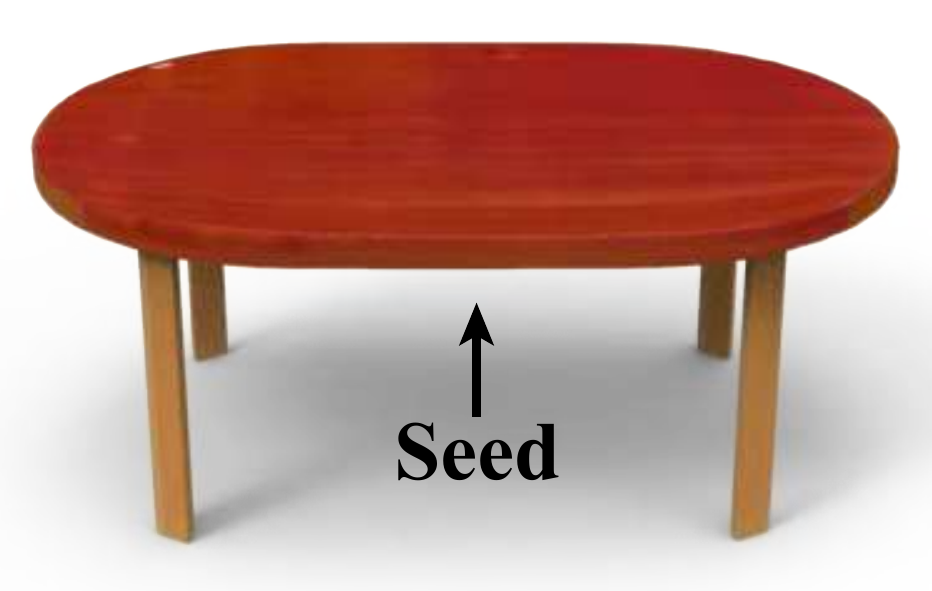}}
    \\
    {\includegraphics[width=0.24\linewidth]{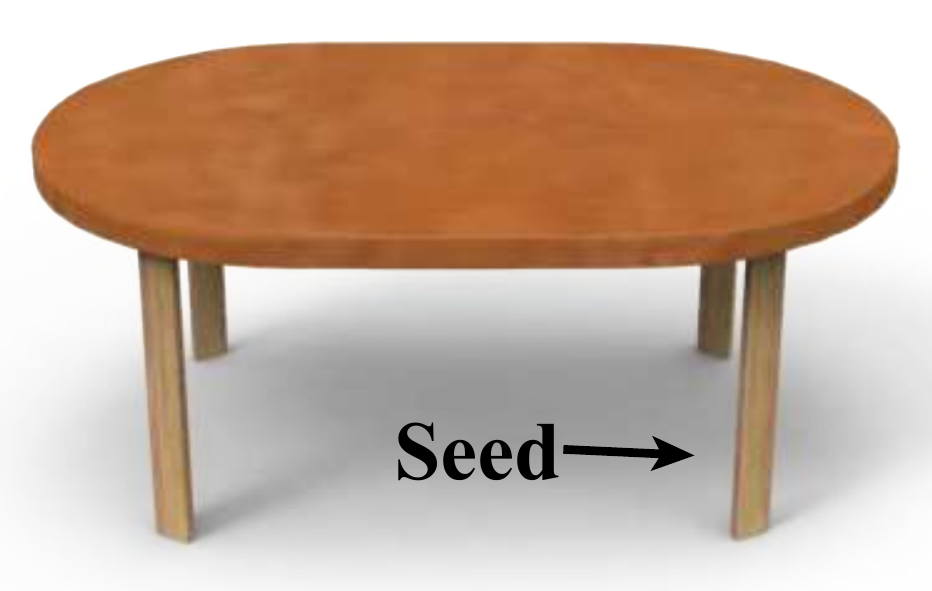}}
    {\includegraphics[width=0.24\linewidth]{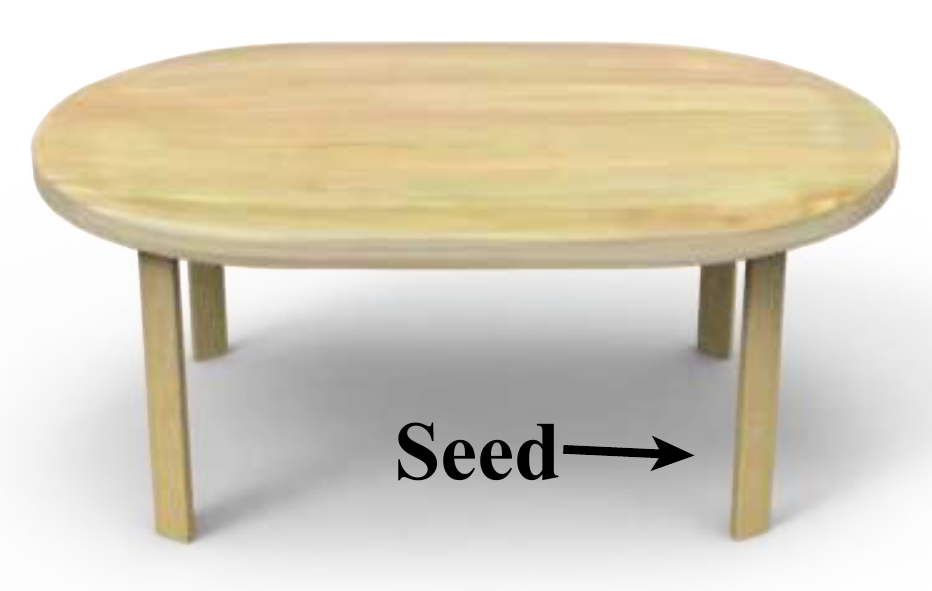}}
    {\includegraphics[width=0.24\linewidth]{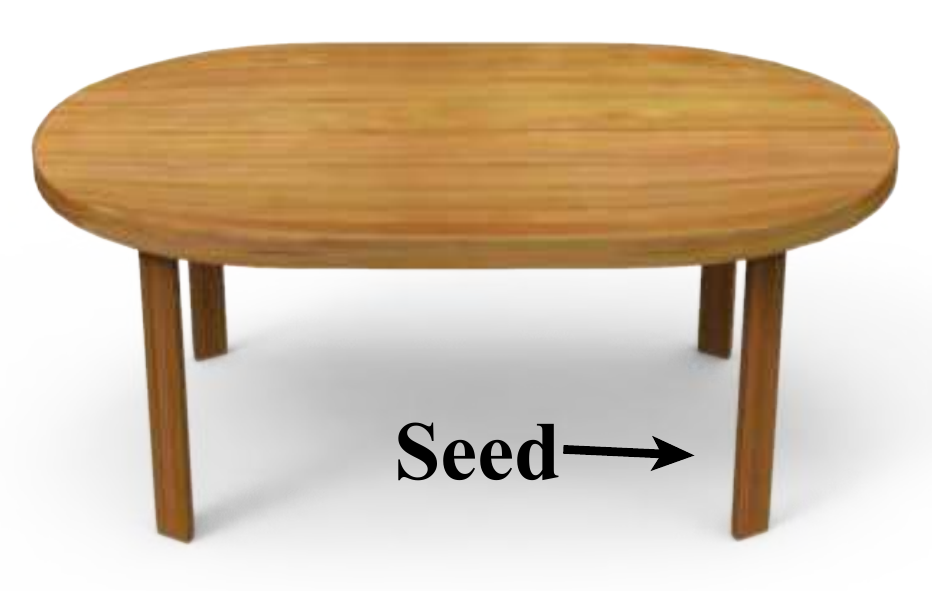}}
    {\includegraphics[width=0.24\linewidth]{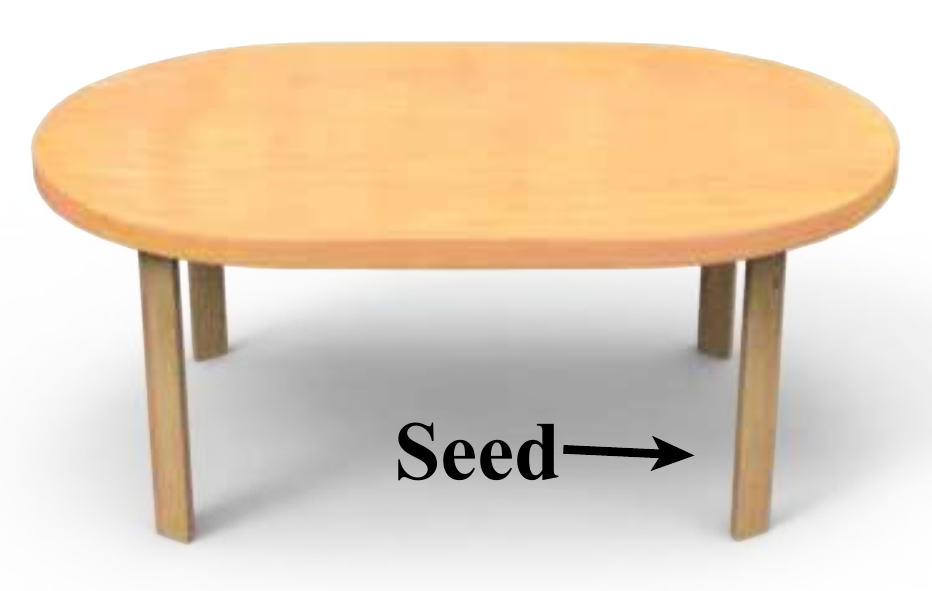}}
    \caption{
    Mesh texturing results on the same shape, \rznn{randomly selected}, by specifying different seed parts. Row 1: chair back as seed; row 2: chair seat as seed; row 3: tabletop as seed; row 4: left front leg as seed.}
    \label{fig:seed_part_choice}
\end{figure}

\vspace{-5pt}

\paragraph{Texture Image Resolution}
\rznn{Finally, we test how TM-NET performs, specifically on the textured mesh autoencoding task, as we vary the resolution of the texture images. As shown in Table~\ref{tab:EvaluationonTextureImageResolution}, when the image resolution is low at $64\times 64$, the SSIM value is low which means that the reconstruction error on the test dataset is high. However, the reconstruction quality does not always improve as the resolution increases, since when the texture resolution is too high, the generalization capability of the network can be compromised. Experimentally, we find that the best quality is attained by setting $l=256$ and the texture resolution for a single part to $1,024 \times 768$; this is the setting we adopt throughout our experiments.}
\begin{table}[!t]
  \centering
    \caption{SSIM measured on rendered images from \rznn{autoencoding the chair test set}, as we vary the texture resolution. Recall that $l$ is the length of the edge of the template box when mapping to UV space; see Section~\ref{section::TextCon}.}
    \begin{tabular}{c||cccc}
		\hline
		\textit{l}               & 64    &  128  & 256            & 512 \\
		\hline\hline
		{SSIM on \rznn{autoencoding} textures} & 0.837 & 0.859 & \textbf{0.928} & 0.905  \\
		\hline
	\end{tabular}
	\label{tab:EvaluationonTextureImageResolution}
\end{table}%

\subsection{Ablation Studies}

We perform several ablation studies to demonstrate the necessity of key components of our network architecture.

\vspace{-5pt}

\paragraph{With vs.~without seam loss}
Since our texture representation relies on unfolding boxes (see Figure~\ref{fig:Representation}), pixels on the boundary during unfolding are mapped to two or more places in the texture space. This may lead to artifacts when pixels on the boundaries are not consistent.
So we introduced a seam loss to prevent this in the TextureVAE network. This loss aims to ensure the unfolded boundaries in the texture images are as consistent as possible. Figure~\ref{fig:AblationSeamLoss} shows the decoded textured results of a chair without and with the seam loss $L_{seam}$. The results show black line artifacts (on legs and seat and highlighted with red rectangles) exist near the unfolded boundaries without the seam loss, and our result with the seam loss does not have such artifacts.

\begin{figure}[!t]
    \centering
    \subfigure[Without seam loss.]{\includegraphics[width=0.48\linewidth]{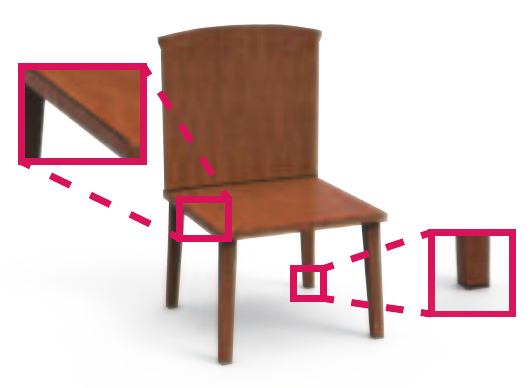}}
	\subfigure[With seam loss.]{\includegraphics[width=0.48\linewidth]{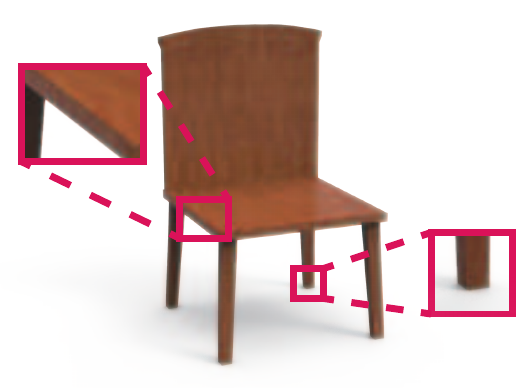}}
    \caption{Comparison of decoded results without and with seam loss $L_{seam}$. \yj{Due to unfolded boundaries of the texture image in our texture representation, the generated texture images may show some artifacts (inconsistency in the unfolded boundaries, as highlighted with red rectangles in (a)). The loss can prevent the artifacts (b). The results demonstrate the seam loss is effective for removing artifacts related to unfolded boundaries.}}
    \label{fig:AblationSeamLoss}
\end{figure}

\vspace{-5pt}

\paragraph{With vs.~without texture part compatibility}
\gl{
As demonstrated in the sections above, the autoregressive model is able to generate reasonable textures for each part. However, generating textures for each part individually can lead to incompatible between different parts. To solve this issue, a greedy strategy is applied to ensure the coherence between different parts. Firstly, one selected part's texture (usually the selected part's center is closest to the shape center) is generated, and then the VGG feature of its texture is extracted and concatenated with other parts' geometry latent code to be the conditional input of other parts' conditional autoregressive network. The comparison of mesh texturing results without and with seed part's VGG feature conditional input is shown Figure~\ref{fig:AblationSimilarityLoss}. We can find that the generated textures are incompatible among parts without the seed part's VGG feature condition input.}

\begin{figure}[!t]
    \centering
	\subfigure[Untextured mesh.]{\includegraphics[width=0.32\linewidth]{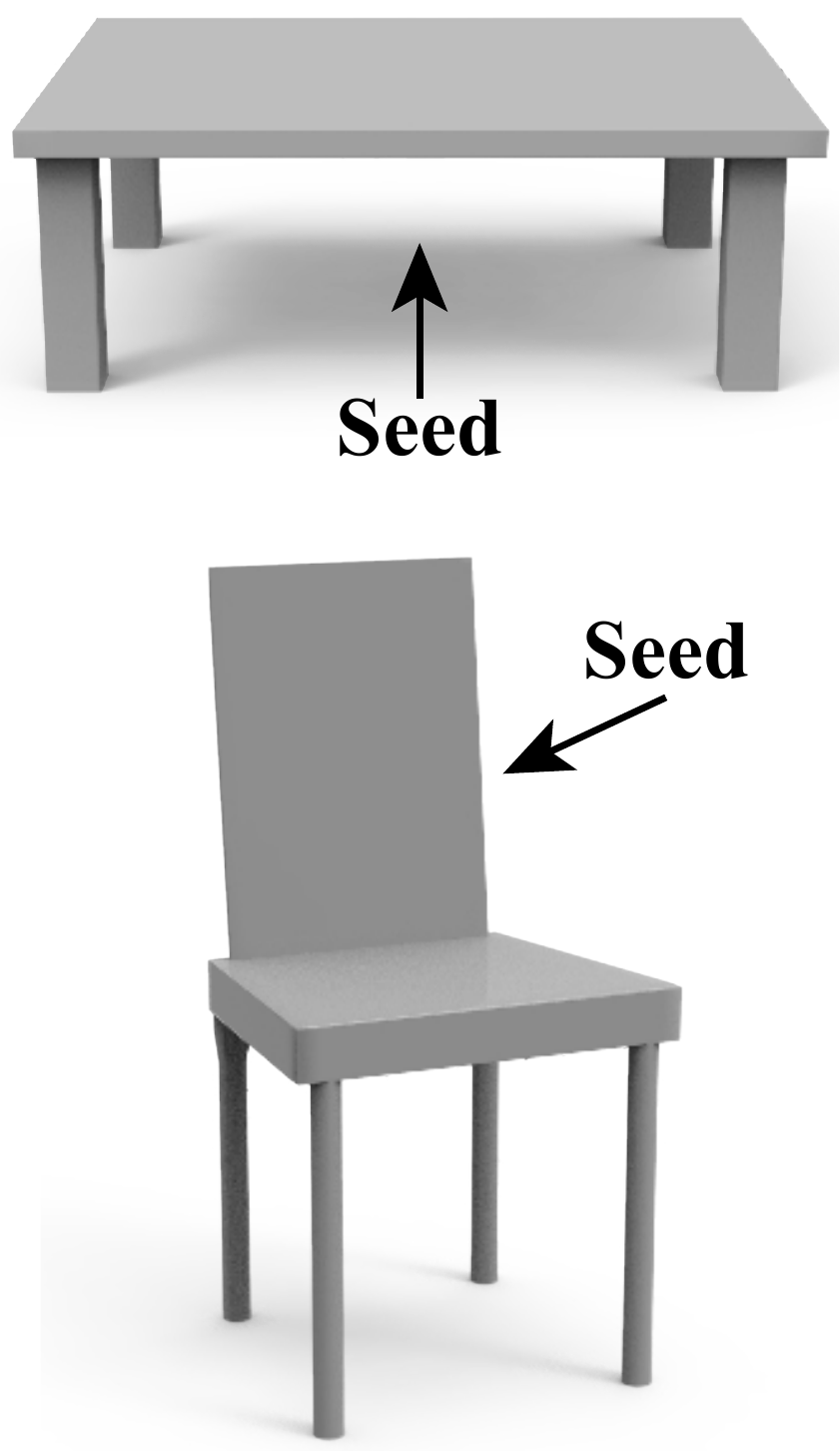}}
	\subfigure[No part compatibility.]
	{\includegraphics[width=0.32\linewidth]{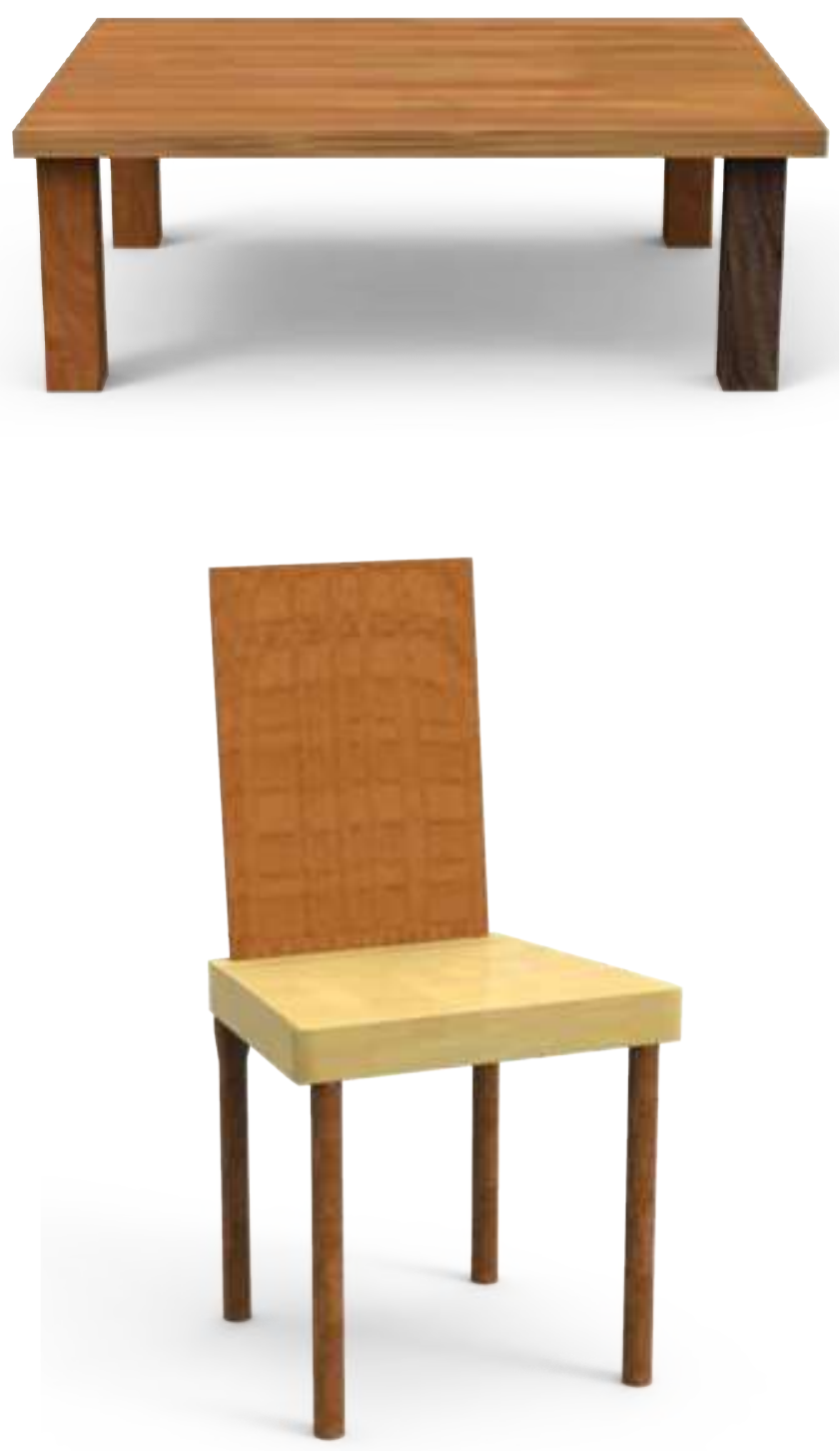}}
	\subfigure[With part compatibility.]{\includegraphics[width=0.32\linewidth]{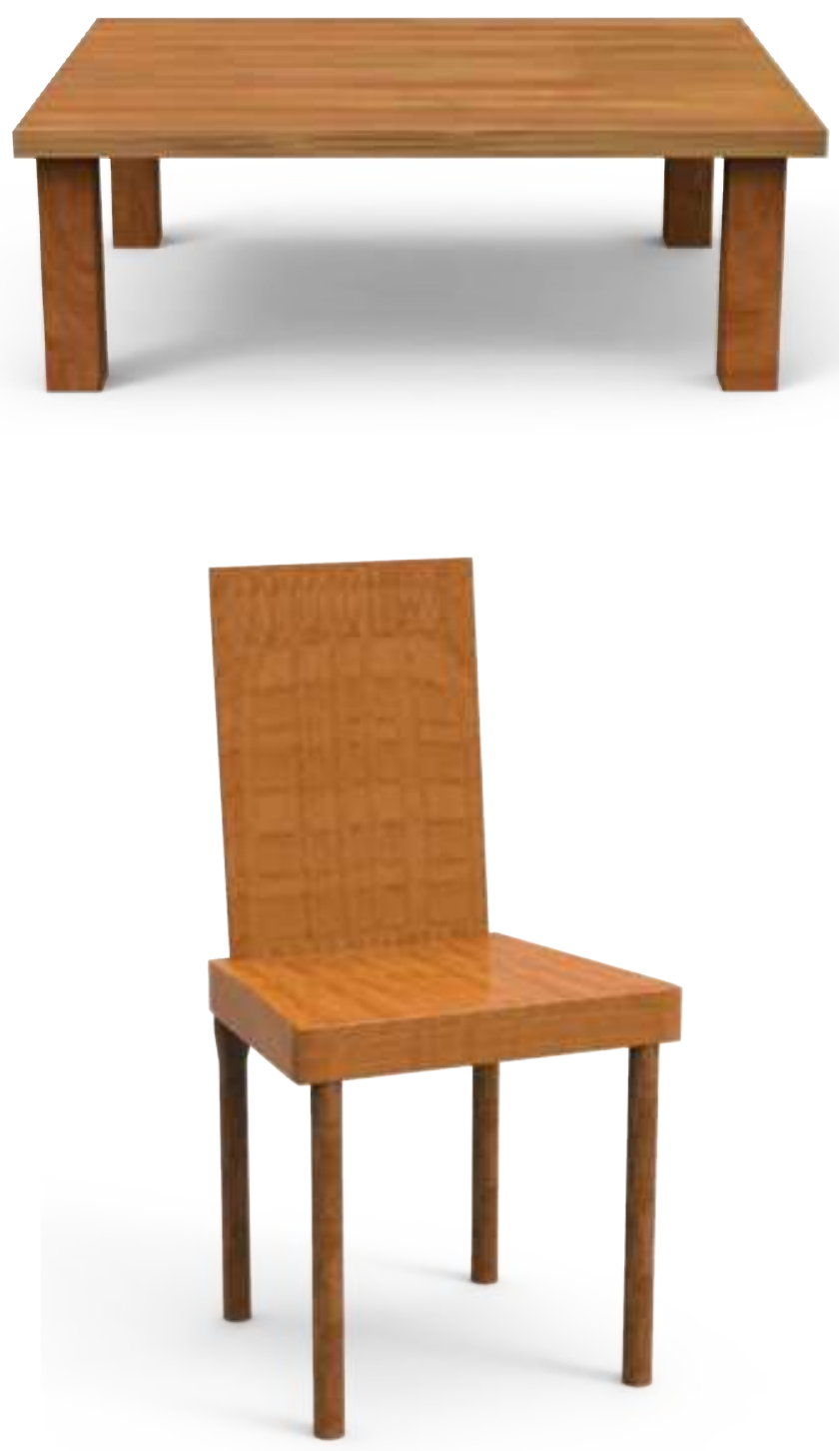}}

    \caption{\rzn{Comparison of mesh texturing without vs.~with part compatibility. Row 1: without using VGG features of the tabletop (the seed part) as conditional input, the texture generated for one of the table legs is inconsistent with the rest. Row 2: with chair back as the seed, the chair seat has an incoherent texture generated if no part compatibility is enforced.}}
    \label{fig:AblationSimilarityLoss}
\end{figure}

\section{Conclusion, limitation, and future work}
\label{sec:conclusion}

We present a \rznn{part-aware} deep generative model, TM-NET, for textured meshes embedded in 3D.
Our network utilizes a simple and effective consistent texture atlas obtained from
unfolding deformable boxes. It separately encodes box deformation and textures
using VAEs at the part level and learns one-to-many relationships between geometry
and texture. Textures are encoded using dictionary-based vector quantization
to allow high-frequency details to be faithfully synthesized.
Our generative model is quite generic and supports several typical applications in the
realm of realistic 3D shape modeling, e.g., synthesizing textures for a given 3D shape,
generation of novel textured meshes, latent-space interpolation, and image-guided
generation of textured shapes, all under a unified framework.

\rznn{As a first attempt at the challenging problem of neural and generative modeling of
both the geometry and texture of 3D shapes, our work is still quite preliminary and limited on
several fronts. To start, we have only explored the dependency of part
texture on shape geometry and structure, without taking into account the correlation
between texture and {\em material\/} properties, which are often tied to {\em object functionality\/}.
For example, to be sat on comfortably, sofa textures are often those of fabrics
and leather. A joint learning of texture, material, and functionality is left for future work.}

\rznn{In general, textures are highly complex signals that can exhibit a wide variety of
frequency-domain characteristics. For example, the wood or leather grain textures on
furniture objects are quite different from image patterns on an airplane. As shown in
Figure~\ref{fig:Generation}, our texture encoding using
VQ-VAEs is not as effective in learning the car and plane textures due to their more
complex and varied frequency-domain characteristics, compared to the furniture
textures.
The resulting artifacts include blurrier decoded textures and cross-fading during
latent-space interpolation (see Figure~\ref{fig:FailureCaseInterpolation}).
In addtition, the VQ-VAE in our TextureVAE does not impose constraints on the intermediate
feature space as the original VAE or normalizing flows~\cite{berthelot2018understanding}, thus it cannot ensure that all
the entries in the codebook are mapped to reasonable textures images.
}

In addition, while TM-NET can represent 3D shapes with high genus and rely on the texturing trick to
create the visual illusion of topological details, using the alpha
channel in texture space, it is not always able to decode the fine-grained
shape details. As shown in Figure~\ref{fig:FailureCase}(a-b), the tabletop contains many small slits.
Our current method cannot reconstruct such geometric details accurately, e.g., some of the
gaps are closed in the decoded shape.
The details around the slits appear over small regions on the texture image, which are difficult to encode and decode precisely.
Still, we believe that the generated textures can provide sufficient constraints to help refine the geometry in a postprocess.
To train a network to accomplish this in an end-to-end manner left for future work.

Another possible failure case occurs for our image-guided generation of textured 3D shapes, as shown in Figure~\ref{fig:FailureCase}(c-d).
The shape generation uses the guidance image as a condition. However, due to the sampling strategy employed by the conditional generative network, the input texture is not guaranteed to be reproduced or even closely approximated. It is worth considering in future work how to reconstruct both shapes and textures more deterministically by making better use of the input image via neural rendering.

\begin{figure}
	\centering
	{
        {\includegraphics[width=0.15\linewidth]{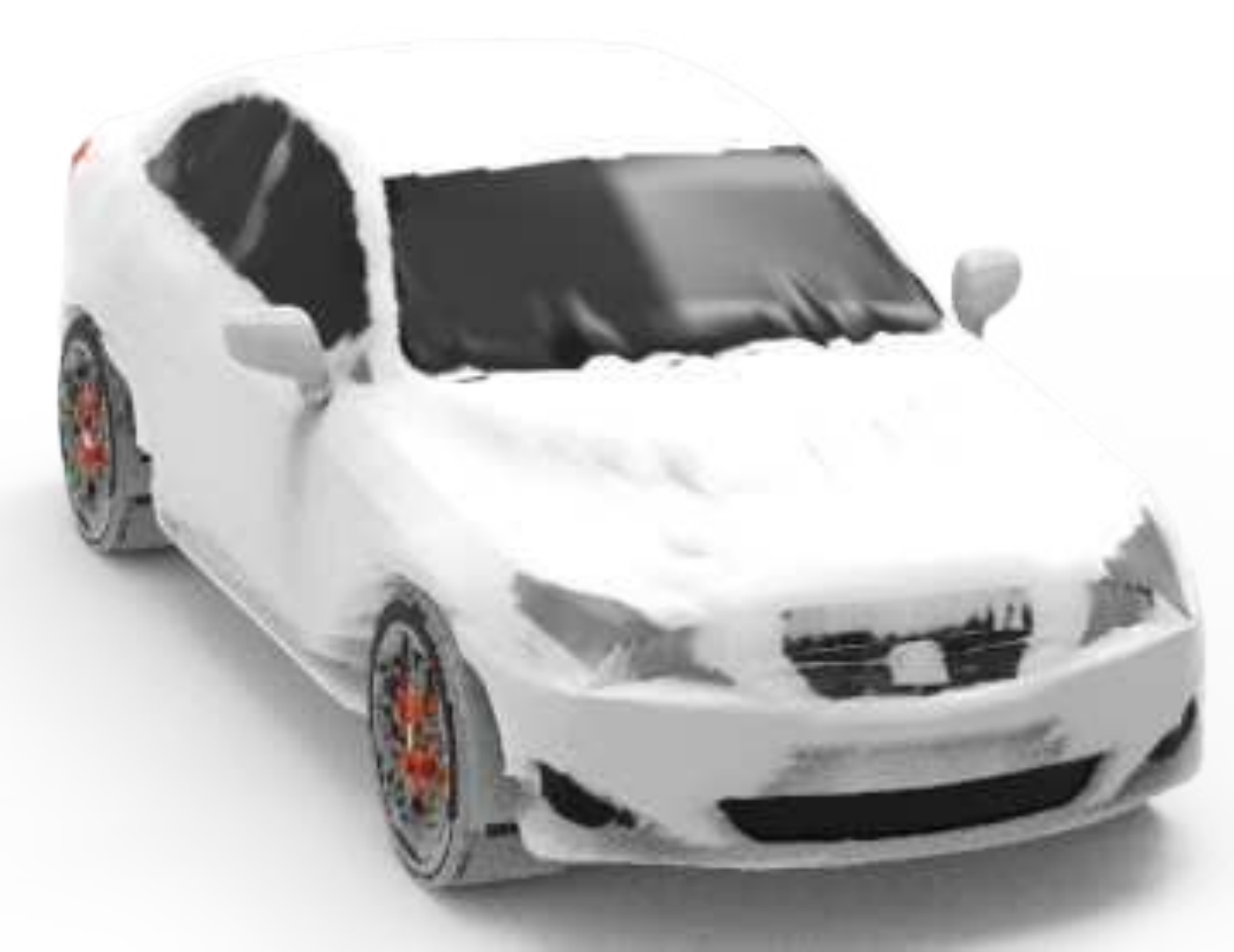}}
        {\includegraphics[width=0.15\linewidth]{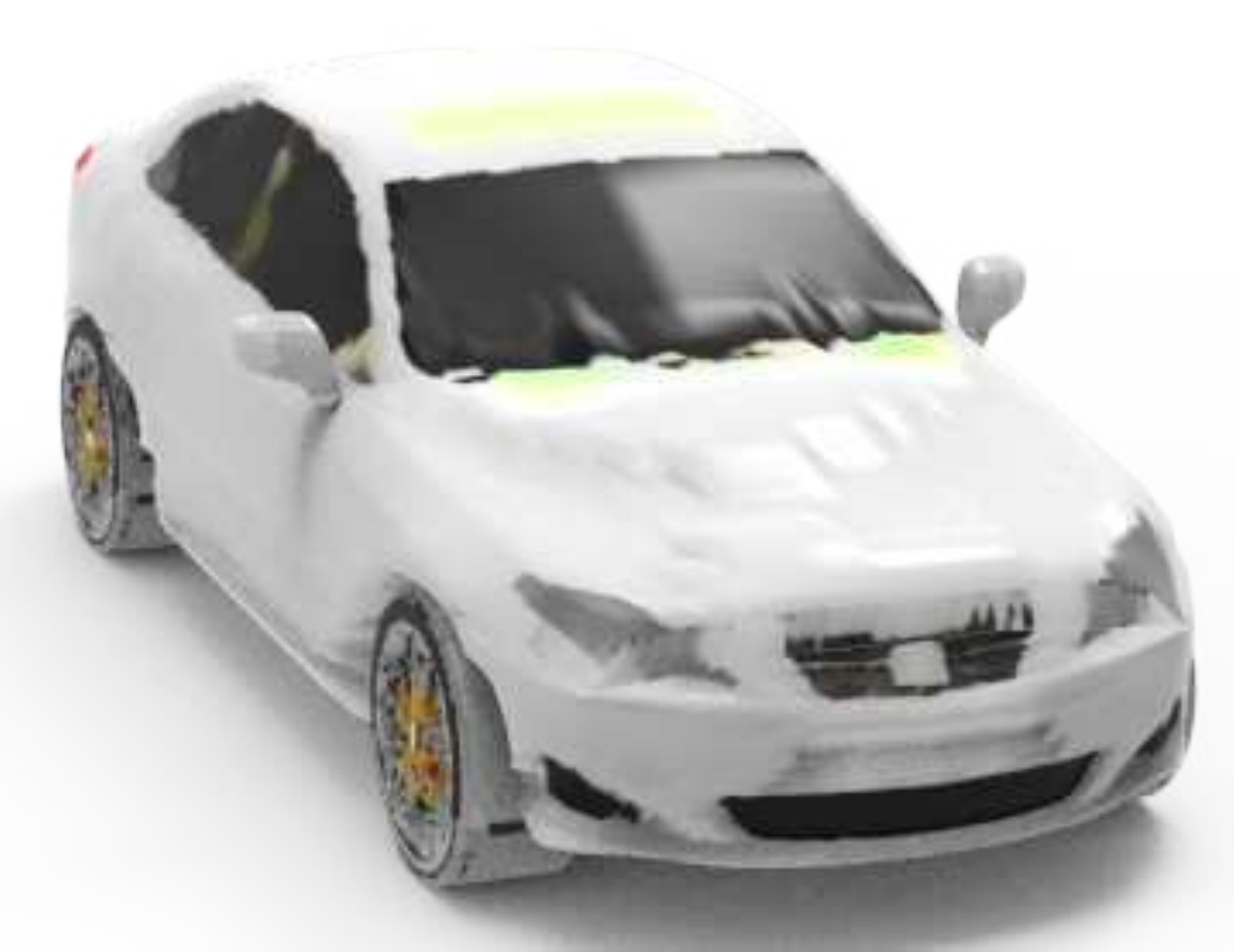}}
        {\includegraphics[width=0.15\linewidth]{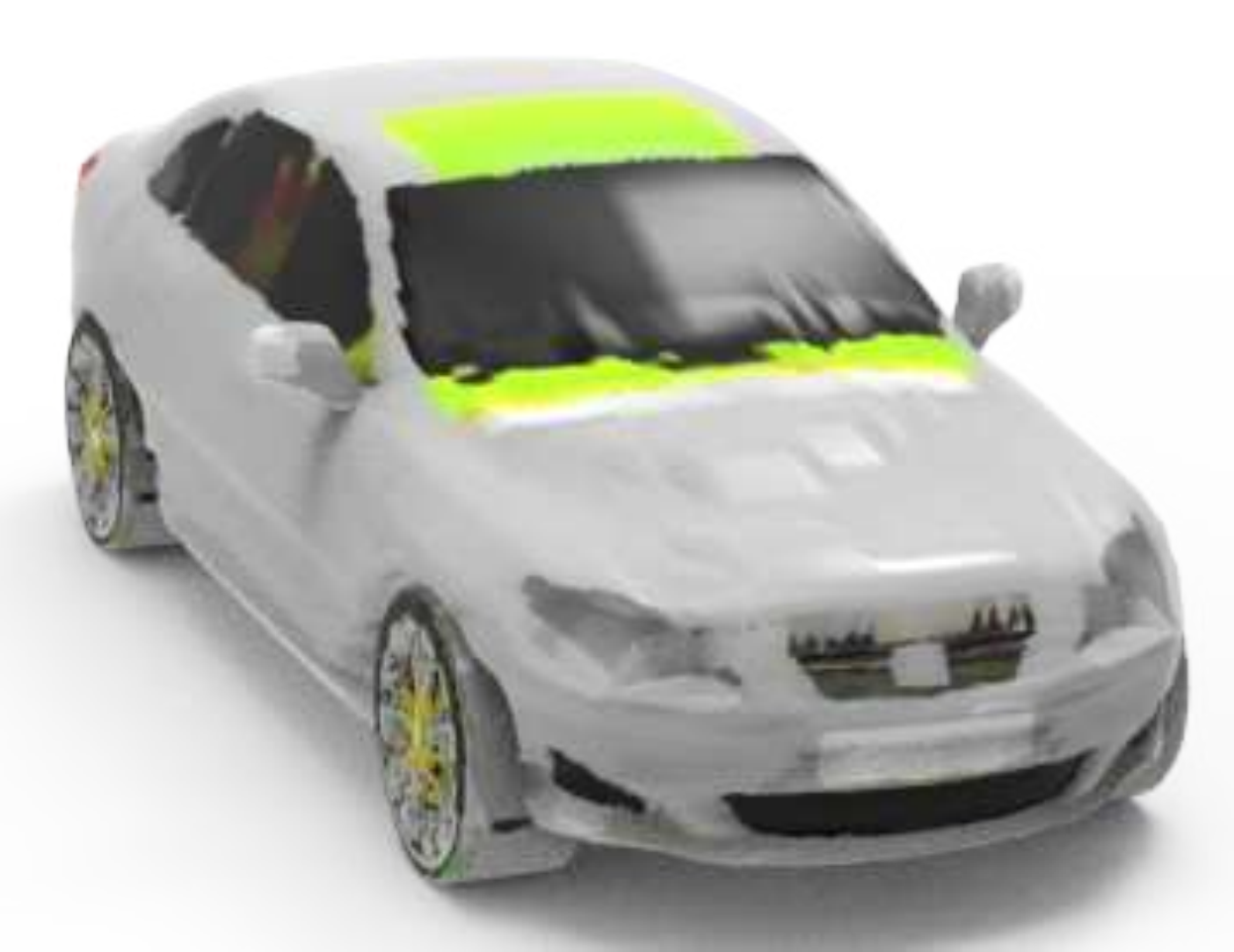}}
        {\includegraphics[width=0.15\linewidth]{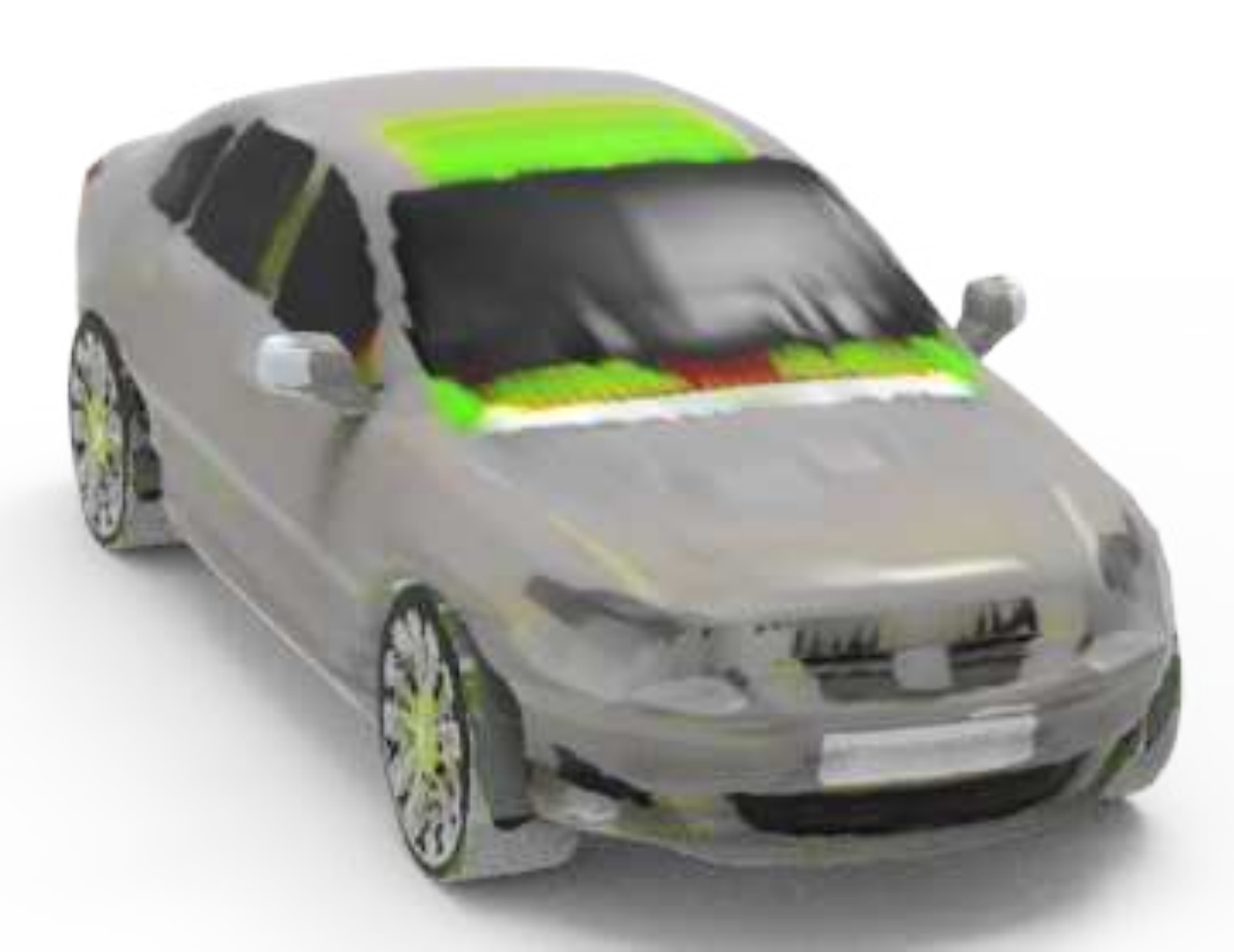}}
        {\includegraphics[width=0.15\linewidth]{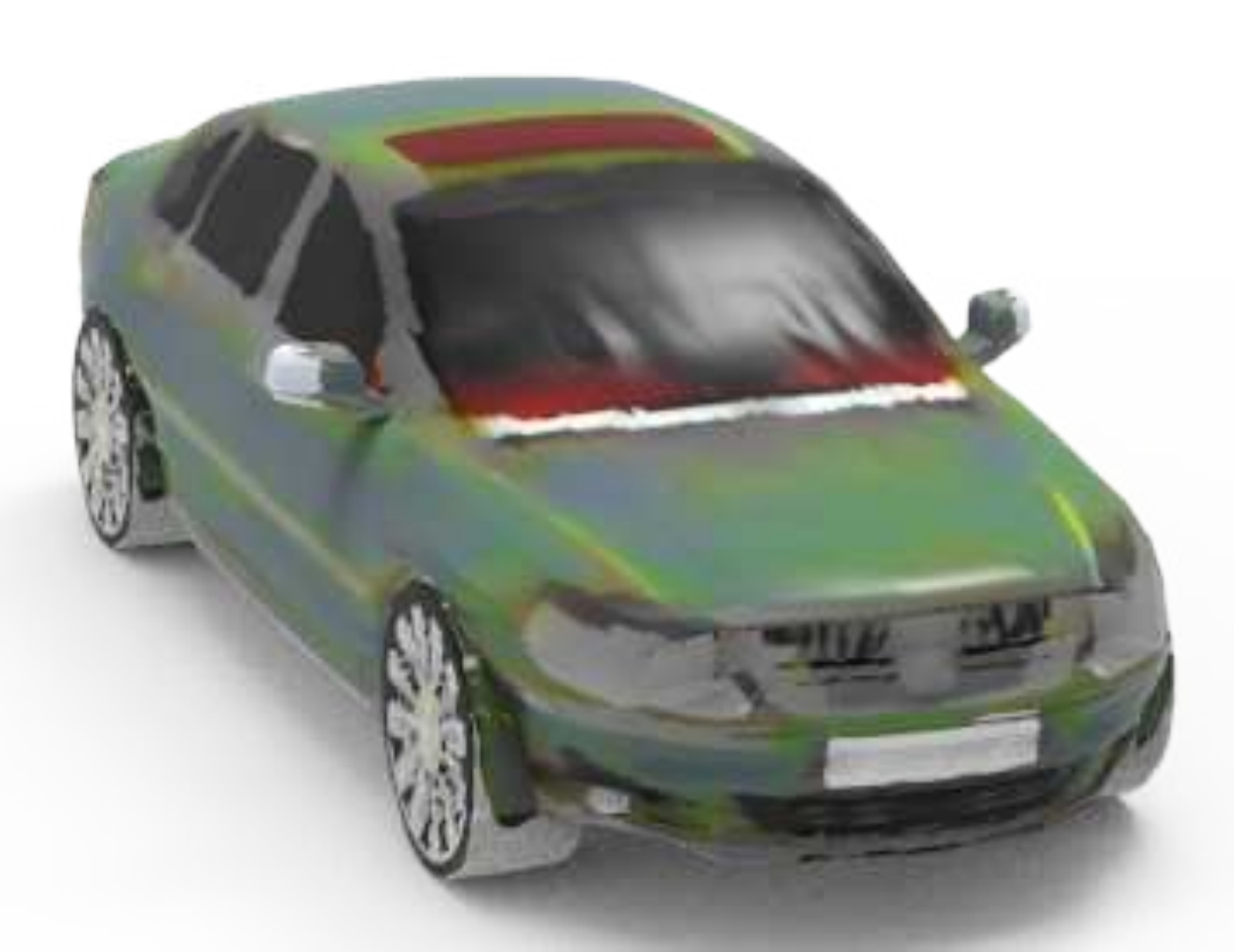}}
        {\includegraphics[width=0.15\linewidth]{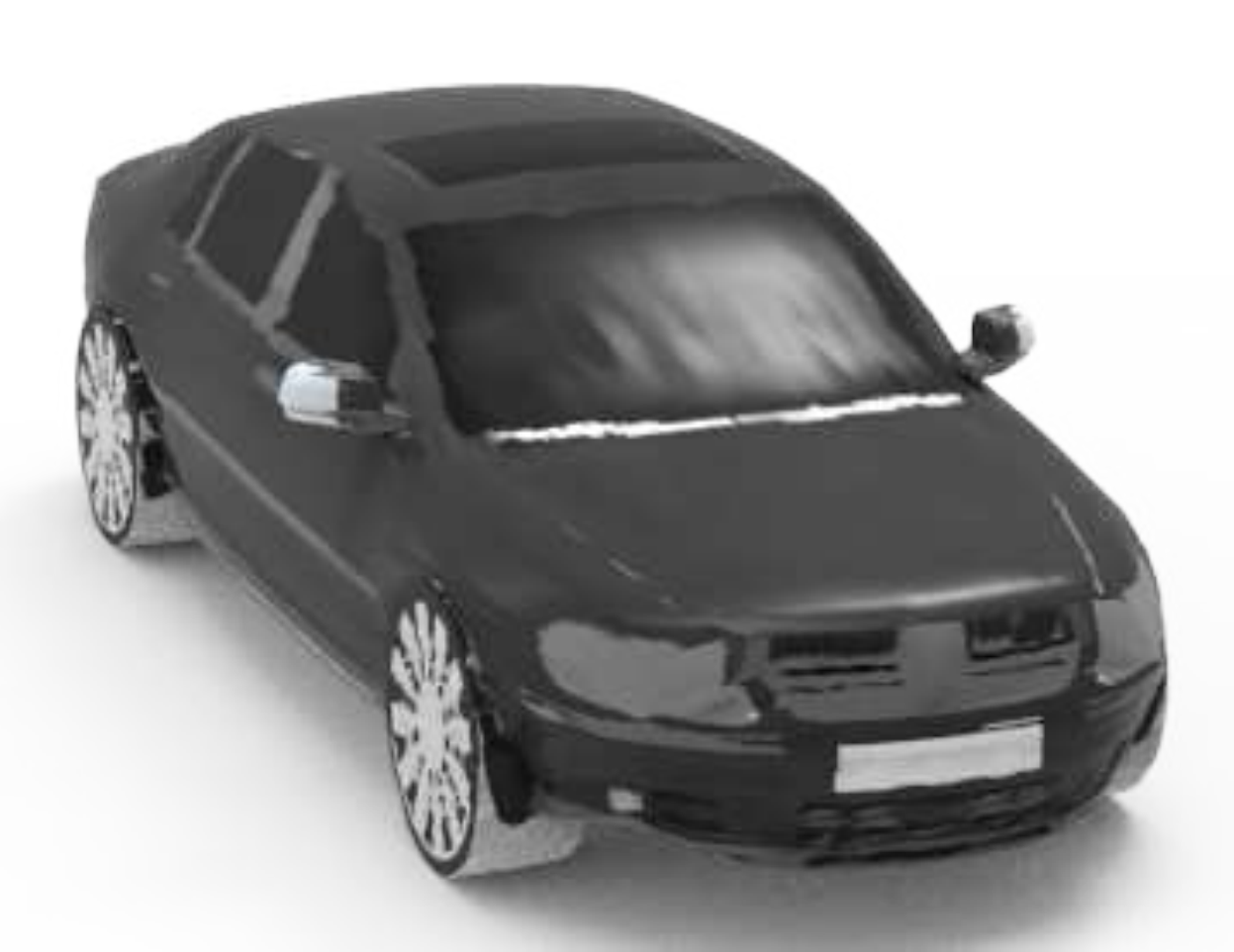}}
        \\
        {\includegraphics[width=0.15\linewidth]{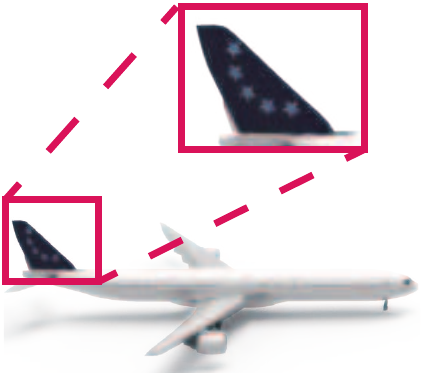}}
        {\includegraphics[width=0.15\linewidth]{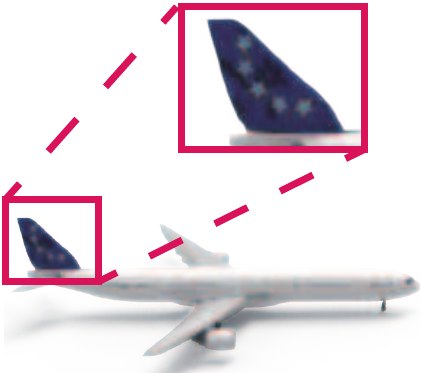}}
        {\includegraphics[width=0.15\linewidth]{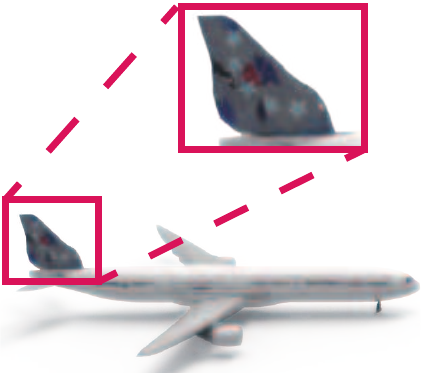}}
        {\includegraphics[width=0.15\linewidth]{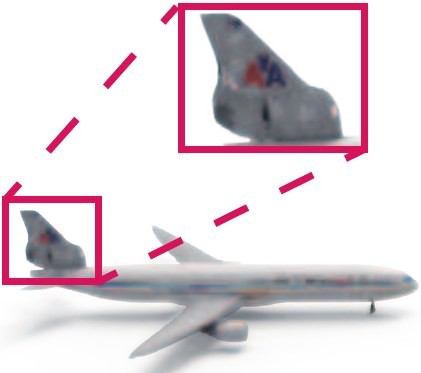}}
        {\includegraphics[width=0.15\linewidth]{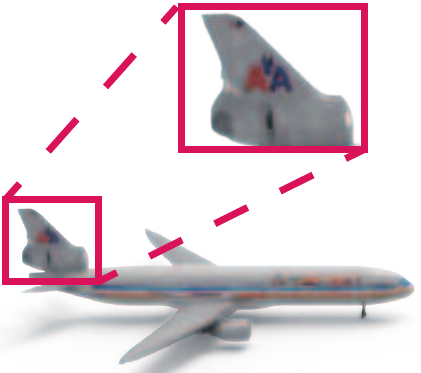}}
        {\includegraphics[width=0.15\linewidth]{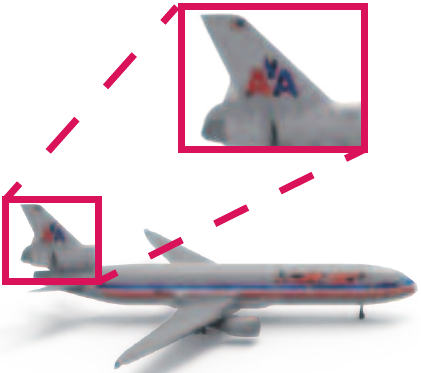}}
        \\
        {\includegraphics[width=0.15\linewidth]{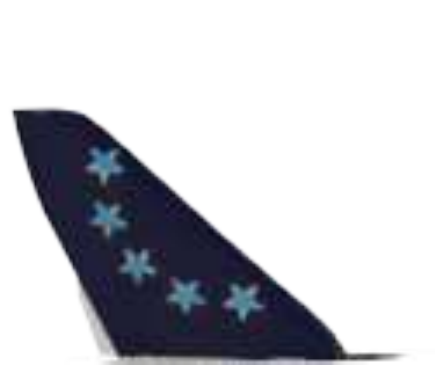}}
        {\includegraphics[width=0.15\linewidth]{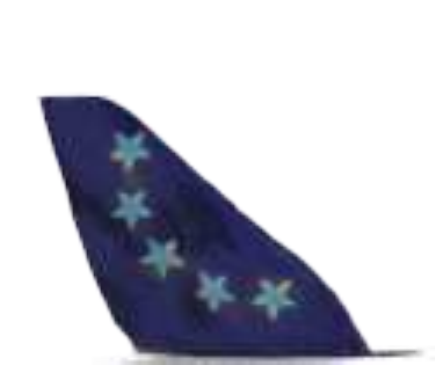}}
        {\includegraphics[width=0.15\linewidth]{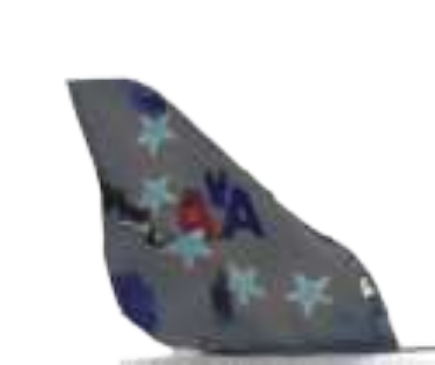}}
        {\includegraphics[width=0.15\linewidth]{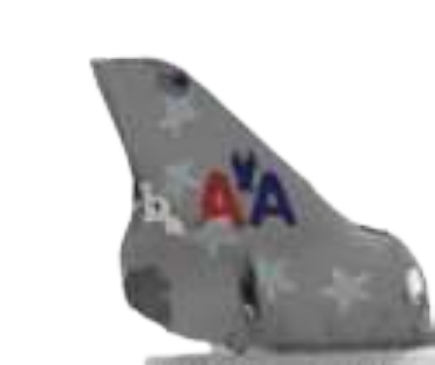}}
        {\includegraphics[width=0.15\linewidth]{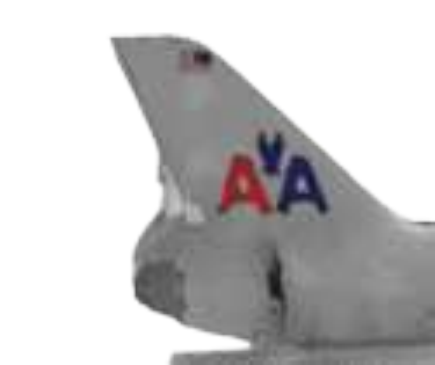}}
        {\includegraphics[width=0.15\linewidth]{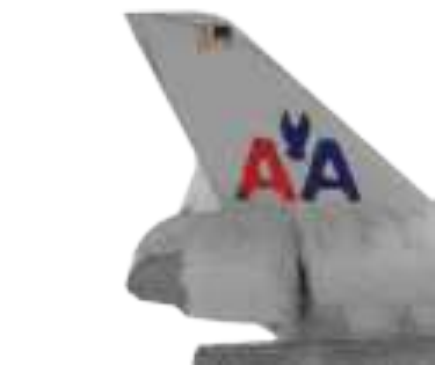}}
	}
    \caption{\rznn{Cross-fading results from latent-space interpolation by TM-NET on a car and a plane example (best viewed by zooming in). The blown-up views from row 2 are further enlarged in row 3 for a better visualization.}}
\label{fig:FailureCaseInterpolation}
\end{figure}

\begin{figure}
    \centering
{
\subfigure[Ground truth.]{\includegraphics[width=0.24\linewidth]{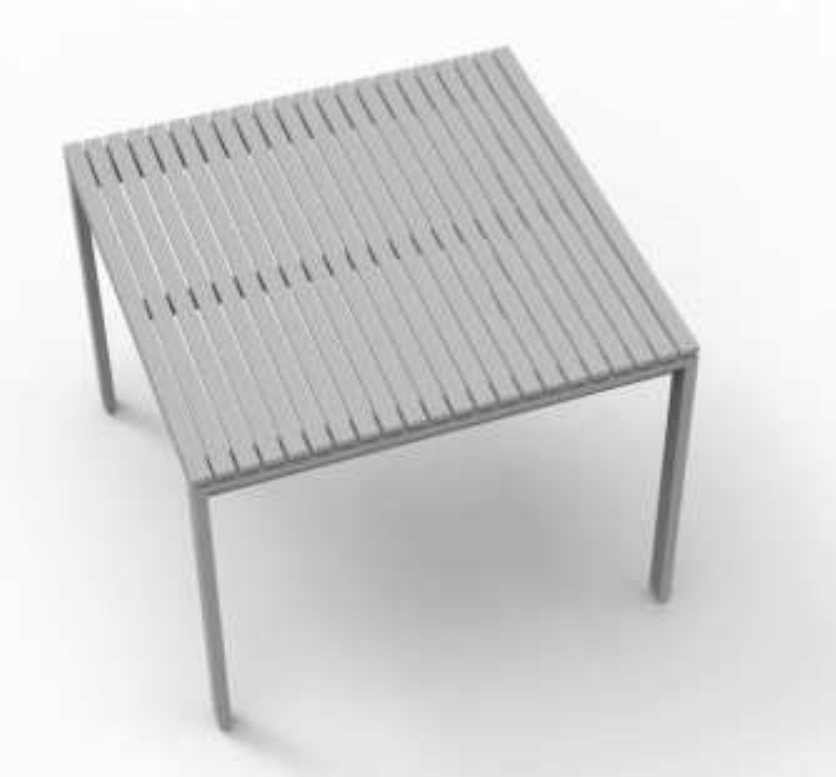}}
\subfigure[Reconstruction.]{\includegraphics[width=0.24\linewidth]{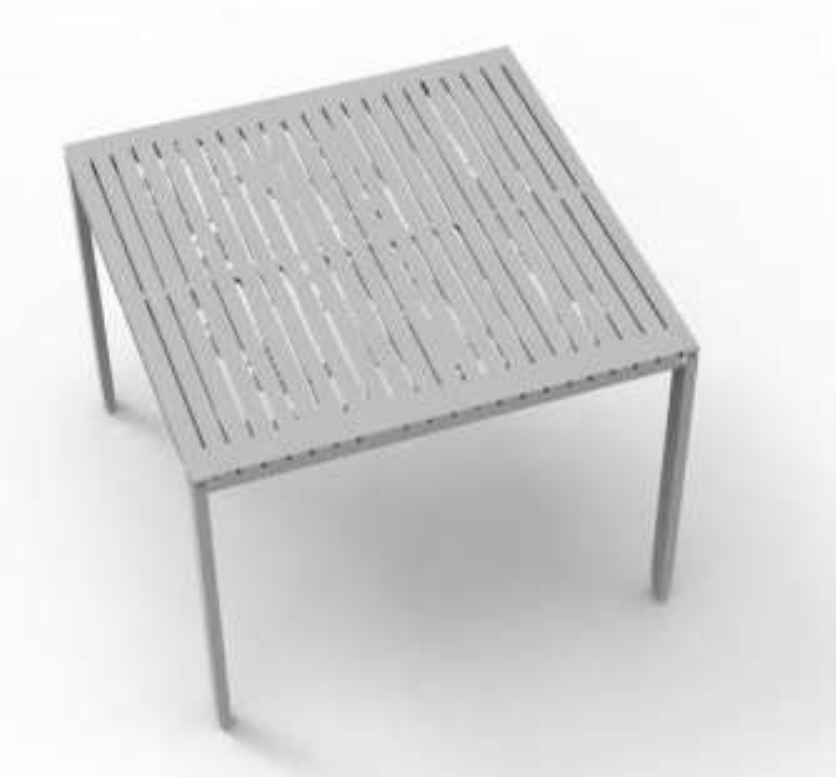}}
\subfigure[Image guidance.]{\includegraphics[width=0.24\linewidth]{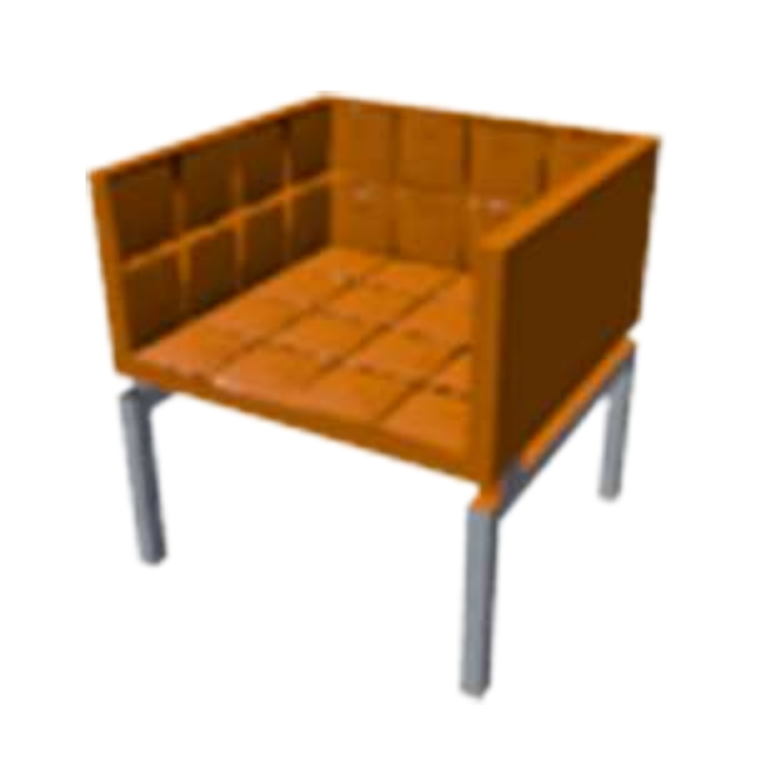}}
\subfigure[Generation.]{\includegraphics[width=0.24\linewidth]{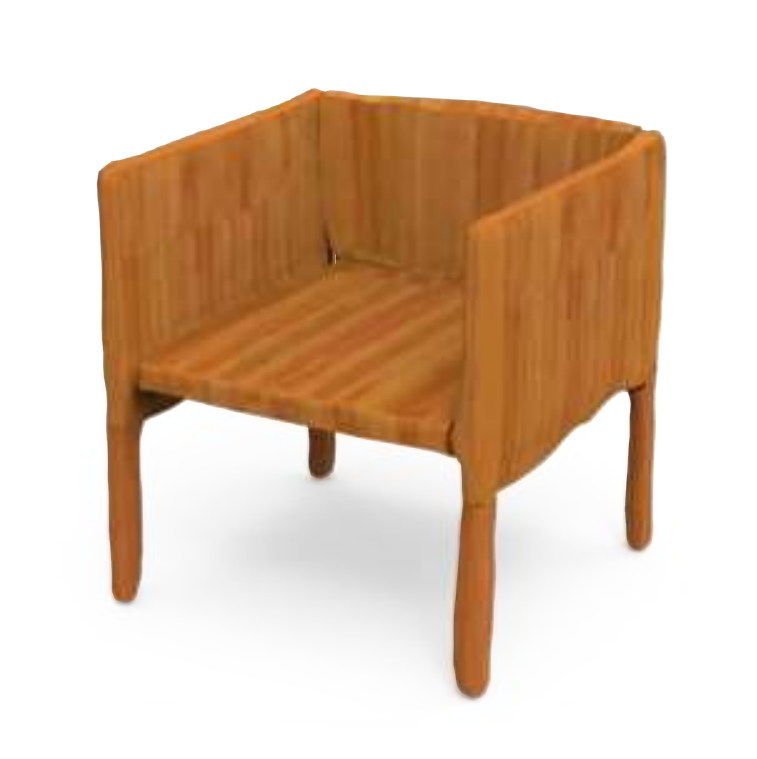}}
}
    \caption{\rzn{Some failure cases. (a-b): TM-NET cannot accurately reconstruct the small slits on top of the ground truth table. (c-d): Image-guided generation of textured shapes by TM-NET may produce unexpected texture. }}
    \label{fig:FailureCase}
\end{figure}

As for other possible directions to explore, we would like to extend our method to other applications,
e.g. single-view 3D reconstruction. Expansions to other types of input, such as line drawings for both
shape and texture, and to other encoding schemes are also interesting to explore. We also plan to
extend our current representation with procedural texture generation, e.g., based on Perlin noise style texture~\cite{henzler2019learning} %
to improve texture generation.
\rznn{Finally, our current shape-conditioned texturing is unable to generate stylish outputs such as the
ones shown in Figure~\ref{fig:incomp_chairs} which exhibit significant deviations between textures in different parts.
Learning {\em texturing styles\/} would be an interesting direction for future work.}

\bibliographystyle{ACM-Reference-Format}
\bibliography{tsdm}


\begin{thebibliography}{56}


\ifx \showCODEN    \undefined \def \showCODEN     #1{\unskip}     \fi
\ifx \showDOI      \undefined \def \showDOI       #1{#1}\fi
\ifx \showISBNx    \undefined \def \showISBNx     #1{\unskip}     \fi
\ifx \showISBNxiii \undefined \def \showISBNxiii  #1{\unskip}     \fi
\ifx \showISSN     \undefined \def \showISSN      #1{\unskip}     \fi
\ifx \showLCCN     \undefined \def \showLCCN      #1{\unskip}     \fi
\ifx \shownote     \undefined \def \shownote      #1{#1}          \fi
\ifx \showarticletitle \undefined \def \showarticletitle #1{#1}   \fi
\ifx \showURL      \undefined \def \showURL       {\relax}        \fi
\providecommand\bibfield[2]{#2}
\providecommand\bibinfo[2]{#2}
\providecommand\natexlab[1]{#1}
\providecommand\showeprint[2][]{arXiv:#2}

\bibitem[\protect\citeauthoryear{Achlioptas, Diamanti, Mitliagkas, and
  Guibas}{Achlioptas et~al\mbox{.}}{2018}]%
        {achlioptas18a}
\bibfield{author}{\bibinfo{person}{Panos Achlioptas}, \bibinfo{person}{Olga
  Diamanti}, \bibinfo{person}{Ioannis Mitliagkas}, {and}
  \bibinfo{person}{Leonidas Guibas}.} \bibinfo{year}{2018}\natexlab{}.
\newblock \showarticletitle{Learning Representations and Generative Models for
  3{D} Point Clouds}. In \bibinfo{booktitle}{\emph{International Conference on
  Machine Learning (ICML)}}, Vol.~\bibinfo{volume}{80}.
  \bibinfo{pages}{40--49}.
\newblock


\bibitem[\protect\citeauthoryear{Berthelot, Raffel, Roy, and
  Goodfellow}{Berthelot et~al\mbox{.}}{2018}]%
        {berthelot2018understanding}
\bibfield{author}{\bibinfo{person}{David Berthelot}, \bibinfo{person}{Colin
  Raffel}, \bibinfo{person}{Aurko Roy}, {and} \bibinfo{person}{Ian
  Goodfellow}.} \bibinfo{year}{2018}\natexlab{}.
\newblock \showarticletitle{Understanding and improving interpolation in
  autoencoders via an adversarial regularizer}.
\newblock \bibinfo{journal}{\emph{arXiv preprint arXiv:1807.07543}}
  (\bibinfo{year}{2018}).
\newblock


\bibitem[\protect\citeauthoryear{Chang, Funkhouser, Guibas, Hanrahan, Huang,
  Li, Savarese, Savva, Song, Su, Xiao, Yi, and Yu}{Chang
  et~al\mbox{.}}{[n.d.]}]%
        {shapenet}
\bibfield{author}{\bibinfo{person}{Angel~X. Chang}, \bibinfo{person}{Thomas
  Funkhouser}, \bibinfo{person}{Leonidas Guibas}, \bibinfo{person}{Pat
  Hanrahan}, \bibinfo{person}{Qixing Huang}, \bibinfo{person}{Zimo Li},
  \bibinfo{person}{Silvio Savarese}, \bibinfo{person}{Manolis Savva},
  \bibinfo{person}{Shuran Song}, \bibinfo{person}{Hao Su},
  \bibinfo{person}{Jianxiong Xiao}, \bibinfo{person}{Li Yi}, {and}
  \bibinfo{person}{Fisher Yu}.} \bibinfo{year}{[n.d.]}\natexlab{}.
\newblock \showarticletitle{{ShapeNet: An Information-Rich 3D Model
  Repository}}.
\newblock \bibinfo{journal}{\emph{arXiv preprint arXiv:1512.03012}}
  (\bibinfo{year}{[n.\,d.]}).
\newblock


\bibitem[\protect\citeauthoryear{Chen, Gao, Ling, Smith, Lehtinen, Jacobson,
  and Fidler}{Chen et~al\mbox{.}}{2019}]%
        {chen2019dibrender}
\bibfield{author}{\bibinfo{person}{Wenzheng Chen}, \bibinfo{person}{Jun Gao},
  \bibinfo{person}{Huan Ling}, \bibinfo{person}{Edward Smith},
  \bibinfo{person}{Jaakko Lehtinen}, \bibinfo{person}{Alec Jacobson}, {and}
  \bibinfo{person}{Sanja Fidler}.} \bibinfo{year}{2019}\natexlab{}.
\newblock \showarticletitle{Learning to Predict 3D Objects with an
  Interpolation-based Differentiable Renderer}. In
  \bibinfo{booktitle}{\emph{NeurIPS}}. \bibinfo{pages}{9605--9616}.
\newblock


\bibitem[\protect\citeauthoryear{Chen, Mishra, Rohaninejad, and Abbeel}{Chen
  et~al\mbox{.}}{2018}]%
        {chen2018pixelsnail}
\bibfield{author}{\bibinfo{person}{Xi Chen}, \bibinfo{person}{Nikhil Mishra},
  \bibinfo{person}{Mostafa Rohaninejad}, {and} \bibinfo{person}{Pieter
  Abbeel}.} \bibinfo{year}{2018}\natexlab{}.
\newblock \showarticletitle{Pixelsnail: An improved autoregressive generative
  model}. In \bibinfo{booktitle}{\emph{International Conference on Machine
  Learning}}. \bibinfo{pages}{864--872}.
\newblock


\bibitem[\protect\citeauthoryear{Chen, Tagliasacchi, and Zhang}{Chen
  et~al\mbox{.}}{2020}]%
        {BSPNet}
\bibfield{author}{\bibinfo{person}{Zhiqin Chen}, \bibinfo{person}{Andrea
  Tagliasacchi}, {and} \bibinfo{person}{Hao Zhang}.}
  \bibinfo{year}{2020}\natexlab{}.
\newblock \showarticletitle{{BSP-NET}: Generating Compact Meshes via Binary
  Space Partitioning}. In \bibinfo{booktitle}{\emph{CVPR}}.
\newblock


\bibitem[\protect\citeauthoryear{Chen and Zhang}{Chen and Zhang}{2019}]%
        {chen2019-IMNET}
\bibfield{author}{\bibinfo{person}{Zhiqin Chen} {and} \bibinfo{person}{Hao
  Zhang}.} \bibinfo{year}{2019}\natexlab{}.
\newblock \showarticletitle{Learning implicit fields for generative shape
  modeling}. In \bibinfo{booktitle}{\emph{CVPR}}. \bibinfo{pages}{5939--5948}.
\newblock


\bibitem[\protect\citeauthoryear{Dosovitskiy and Brox}{Dosovitskiy and
  Brox}{2016}]%
        {VAE-blurry}
\bibfield{author}{\bibinfo{person}{Alexey Dosovitskiy} {and}
  \bibinfo{person}{Thomas Brox}.} \bibinfo{year}{2016}\natexlab{}.
\newblock \showarticletitle{Generating Images with Perceptual Similarity
  Metrics based on Deep Networks}.
\newblock \bibinfo{journal}{\emph{CoRR}}  \bibinfo{volume}{abs/1602.02644}
  (\bibinfo{year}{2016}).
\newblock
\showeprint[arxiv]{1602.02644}
\urldef\tempurl%
\url{http://arxiv.org/abs/1602.02644}
\showURL{%
\tempurl}


\bibitem[\protect\citeauthoryear{Efros and Leung}{Efros and Leung}{1999}]%
        {Efros1999}
\bibfield{author}{\bibinfo{person}{Alexei~A. Efros} {and}
  \bibinfo{person}{Thomas~K. Leung}.} \bibinfo{year}{1999}\natexlab{}.
\newblock \showarticletitle{Texture Synthesis by Non-parametric Sampling}. In
  \bibinfo{booktitle}{\emph{ICCV}}. \bibinfo{pages}{1033--1038}.
\newblock


\bibitem[\protect\citeauthoryear{Gao, Yang, Wu, Yuan, Fu, Lai, and Zhang}{Gao
  et~al\mbox{.}}{2019}]%
        {gaosdmnet2019}
\bibfield{author}{\bibinfo{person}{Lin Gao}, \bibinfo{person}{Jie Yang},
  \bibinfo{person}{Tong Wu}, \bibinfo{person}{Yu-Jie Yuan},
  \bibinfo{person}{Hongbo Fu}, \bibinfo{person}{Yu-Kun Lai}, {and}
  \bibinfo{person}{Hao(Richard) Zhang}.} \bibinfo{year}{2019}\natexlab{}.
\newblock \showarticletitle{{SDM-NET}: Deep Generative Network for Structured
  Deformable Mesh}.
\newblock \bibinfo{journal}{\emph{ACM Trans. on Graphics (SIGGRAPH Asia)}}
  \bibinfo{volume}{38}, \bibinfo{number}{6} (\bibinfo{year}{2019}),
  \bibinfo{pages}{243:1--243:15}.
\newblock


\bibitem[\protect\citeauthoryear{Gatys, Ecker, and Bethge}{Gatys
  et~al\mbox{.}}{2015}]%
        {gatys2015texture}
\bibfield{author}{\bibinfo{person}{Leon Gatys}, \bibinfo{person}{Alexander~S
  Ecker}, {and} \bibinfo{person}{Matthias Bethge}.}
  \bibinfo{year}{2015}\natexlab{}.
\newblock \showarticletitle{Texture synthesis using convolutional neural
  networks}. In \bibinfo{booktitle}{\emph{NeurIPS}}. \bibinfo{pages}{262--270}.
\newblock


\bibitem[\protect\citeauthoryear{Goodfellow, Pouget-Abadie, Mirza, Xu,
  Warde-Farley, Ozair, Courville, and Bengio}{Goodfellow et~al\mbox{.}}{2014}]%
        {NIPS2014_5423}
\bibfield{author}{\bibinfo{person}{Ian Goodfellow}, \bibinfo{person}{Jean
  Pouget-Abadie}, \bibinfo{person}{Mehdi Mirza}, \bibinfo{person}{Bing Xu},
  \bibinfo{person}{David Warde-Farley}, \bibinfo{person}{Sherjil Ozair},
  \bibinfo{person}{Aaron Courville}, {and} \bibinfo{person}{Yoshua Bengio}.}
  \bibinfo{year}{2014}\natexlab{}.
\newblock \showarticletitle{Generative Adversarial Nets}.
\newblock In \bibinfo{booktitle}{\emph{NeurIPS}}. \bibinfo{pages}{2672--2680}.
\newblock


\bibitem[\protect\citeauthoryear{Groueix, Fisher, Kim, Russell, and
  Aubry}{Groueix et~al\mbox{.}}{2018}]%
        {AtlasNet2018}
\bibfield{author}{\bibinfo{person}{Thibault Groueix}, \bibinfo{person}{Matthew
  Fisher}, \bibinfo{person}{Vladimir~G. Kim}, \bibinfo{person}{Bryan Russell},
  {and} \bibinfo{person}{Mathieu Aubry}.} \bibinfo{year}{2018}\natexlab{}.
\newblock \showarticletitle{{AtlasNet: A Papier-M{\^a}ch{\'e} Approach to
  Learning {3D} Surface Generation}}. In \bibinfo{booktitle}{\emph{CVPR}}.
  \bibinfo{pages}{216--224}.
\newblock


\bibitem[\protect\citeauthoryear{Han, Gao, and Yu}{Han et~al\mbox{.}}{2017}]%
        {DeepSketch2Face}
\bibfield{author}{\bibinfo{person}{Xiaoguang Han}, \bibinfo{person}{Chang Gao},
  {and} \bibinfo{person}{Yizhou Yu}.} \bibinfo{year}{2017}\natexlab{}.
\newblock \showarticletitle{{DeepSketch2Face}: a deep learning based sketching
  system for 3D face and caricature modeling}.
\newblock \bibinfo{journal}{\emph{ACM Trans. Graph.}} \bibinfo{volume}{36},
  \bibinfo{number}{4} (\bibinfo{year}{2017}), \bibinfo{pages}{article 126}.
\newblock


\bibitem[\protect\citeauthoryear{Henzler, Mitra, and Ritschel}{Henzler
  et~al\mbox{.}}{2019}]%
        {henzler2019learning}
\bibfield{author}{\bibinfo{person}{Philipp Henzler}, \bibinfo{person}{Niloy~J
  Mitra}, {and} \bibinfo{person}{Tobias Ritschel}.}
  \bibinfo{year}{2019}\natexlab{}.
\newblock \showarticletitle{Learning a Neural 3D Texture Space from 2D
  Exemplars}.
\newblock \bibinfo{journal}{\emph{arXiv:1912.04158}} (\bibinfo{year}{2019}).
\newblock


\bibitem[\protect\citeauthoryear{Heusel, Ramsauer, Unterthiner, Nessler, and
  Hochreiter}{Heusel et~al\mbox{.}}{2017}]%
        {NIPS2017_7240}
\bibfield{author}{\bibinfo{person}{Martin Heusel}, \bibinfo{person}{Hubert
  Ramsauer}, \bibinfo{person}{Thomas Unterthiner}, \bibinfo{person}{Bernhard
  Nessler}, {and} \bibinfo{person}{Sepp Hochreiter}.}
  \bibinfo{year}{2017}\natexlab{}.
\newblock \showarticletitle{GANs Trained by a Two Time-Scale Update Rule
  Converge to a Local Nash Equilibrium}.
\newblock In \bibinfo{booktitle}{\emph{NeurIPS}}. \bibinfo{pages}{6626--6637}.
\newblock


\bibitem[\protect\citeauthoryear{Hu, Lin, Han, and Zwicker}{Hu
  et~al\mbox{.}}{2019}]%
        {tao2019}
\bibfield{author}{\bibinfo{person}{Tao Hu}, \bibinfo{person}{Geng Lin},
  \bibinfo{person}{Zhizhong Han}, {and} \bibinfo{person}{Matthias Zwicker}.}
  \bibinfo{year}{2019}\natexlab{}.
\newblock \showarticletitle{Learning to Generate Dense Point Clouds with
  Textures on Multiple Categories}.
\newblock \bibinfo{journal}{\emph{arXiv:1912.10545}} (\bibinfo{year}{2019}).
\newblock


\bibitem[\protect\citeauthoryear{Kanazawa, Tulsiani, Efros, and Malik}{Kanazawa
  et~al\mbox{.}}{2018}]%
        {cmrKanazawa18}
\bibfield{author}{\bibinfo{person}{Angjoo Kanazawa}, \bibinfo{person}{Shubham
  Tulsiani}, \bibinfo{person}{Alexei~A. Efros}, {and} \bibinfo{person}{Jitendra
  Malik}.} \bibinfo{year}{2018}\natexlab{}.
\newblock \showarticletitle{Learning Category-Specific Mesh Reconstruction from
  Image Collections}. In \bibinfo{booktitle}{\emph{ECCV}}.
  \bibinfo{pages}{371--386}.
\newblock


\bibitem[\protect\citeauthoryear{Kingma and Welling}{Kingma and
  Welling}{2013}]%
        {kingma2013auto}
\bibfield{author}{\bibinfo{person}{Diederik~P Kingma} {and}
  \bibinfo{person}{Max Welling}.} \bibinfo{year}{2013}\natexlab{}.
\newblock \showarticletitle{Auto-encoding variational bayes}.
\newblock \bibinfo{journal}{\emph{arXiv:1312.6114}} (\bibinfo{year}{2013}).
\newblock


\bibitem[\protect\citeauthoryear{Kopf, Chi-Wing~Fu, Deussen, Lischinski, and
  Wong}{Kopf et~al\mbox{.}}{2007}]%
        {Kopf2007}
\bibfield{author}{\bibinfo{person}{Johannes Kopf}, \bibinfo{person}{Daniel
  Cohen-Or Chi-Wing~Fu}, \bibinfo{person}{Oliver Deussen},
  \bibinfo{person}{Dani Lischinski}, {and} \bibinfo{person}{Tien-Tsin Wong}.}
  \bibinfo{year}{2007}\natexlab{}.
\newblock \showarticletitle{Solid texture synthesis from {2D} exemplars}.
\newblock \bibinfo{journal}{\emph{ACM Transactions on Graphics}}
  \bibinfo{volume}{26}, \bibinfo{number}{3} (\bibinfo{year}{2007}),
  \bibinfo{pages}{2}.
\newblock


\bibitem[\protect\citeauthoryear{Li, Xu, Chaudhuri, Yumer, Zhang, and
  Guibas}{Li et~al\mbox{.}}{2017}]%
        {li_sig17}
\bibfield{author}{\bibinfo{person}{Jun Li}, \bibinfo{person}{Kai Xu},
  \bibinfo{person}{Siddhartha Chaudhuri}, \bibinfo{person}{Ersin Yumer},
  \bibinfo{person}{Hao Zhang}, {and} \bibinfo{person}{Leonidas~J. Guibas}.}
  \bibinfo{year}{2017}\natexlab{}.
\newblock \showarticletitle{{GRASS}: Generative Recursive Autoencoders for
  Shape Structures}.
\newblock \bibinfo{journal}{\emph{ACM Trans. Graph.}} \bibinfo{volume}{36},
  \bibinfo{number}{4} (\bibinfo{year}{2017}), \bibinfo{pages}{52:1--52:14}.
\newblock


\bibitem[\protect\citeauthoryear{Li and Zhang}{Li and Zhang}{2021}]%
        {li2021d2im}
\bibfield{author}{\bibinfo{person}{Manyi Li} {and} \bibinfo{person}{Hao
  Zhang}.} \bibinfo{year}{2021}\natexlab{}.
\newblock \showarticletitle{{D2IM-Net}: Learning Detail Disentangled Implicit
  Fields from Single Images}. In \bibinfo{booktitle}{\emph{CVPR}}.
\newblock


\bibitem[\protect\citeauthoryear{Liu, Li, Chen, and Li}{Liu
  et~al\mbox{.}}{2019}]%
        {liu2019softras}
\bibfield{author}{\bibinfo{person}{Shichen Liu}, \bibinfo{person}{Tianye Li},
  \bibinfo{person}{Weikai Chen}, {and} \bibinfo{person}{Hao Li}.}
  \bibinfo{year}{2019}\natexlab{}.
\newblock \showarticletitle{Soft Rasterizer: A Differentiable Renderer for
  Image-based 3D Reasoning}.
\newblock  (\bibinfo{year}{2019}), \bibinfo{pages}{7708--7717}.
\newblock


\bibitem[\protect\citeauthoryear{Lun, Gadelha, Kalogerakis, Maji, and Wang}{Lun
  et~al\mbox{.}}{2017}]%
        {lun2017}
\bibfield{author}{\bibinfo{person}{Zhaoliang Lun}, \bibinfo{person}{Matheus
  Gadelha}, \bibinfo{person}{Evangelos Kalogerakis}, \bibinfo{person}{Subhransu
  Maji}, {and} \bibinfo{person}{Rui Wang}.} \bibinfo{year}{2017}\natexlab{}.
\newblock \showarticletitle{3D Shape Reconstruction from Sketches via
  Multi-view Convolutional Networks}. In \bibinfo{booktitle}{\emph{Proc. of the
  International Conference on 3D Vision (3DV)}}.
\newblock


\bibitem[\protect\citeauthoryear{Martin-Brualla, Pandey, Bouaziz, Brown, and
  Goldman}{Martin-Brualla et~al\mbox{.}}{2020}]%
        {martinbrualla2020gelato}
\bibfield{author}{\bibinfo{person}{Ricardo Martin-Brualla},
  \bibinfo{person}{Rohit Pandey}, \bibinfo{person}{Sofien Bouaziz},
  \bibinfo{person}{Matthew Brown}, {and} \bibinfo{person}{Dan~B Goldman}.}
  \bibinfo{year}{2020}\natexlab{}.
\newblock \showarticletitle{{GeLaTO: Generative Latent Textured Objects}}. In
  \bibinfo{booktitle}{\emph{ECCV}}.
\newblock


\bibitem[\protect\citeauthoryear{Mescheder, Oechsle, Niemeyer, Nowozin, and
  Geiger}{Mescheder et~al\mbox{.}}{2019}]%
        {mescheder2019-Occupancy}
\bibfield{author}{\bibinfo{person}{Lars Mescheder}, \bibinfo{person}{Michael
  Oechsle}, \bibinfo{person}{Michael Niemeyer}, \bibinfo{person}{Sebastian
  Nowozin}, {and} \bibinfo{person}{Andreas Geiger}.}
  \bibinfo{year}{2019}\natexlab{}.
\newblock \showarticletitle{Occupancy networks: Learning {3D} reconstruction in
  function space}. In \bibinfo{booktitle}{\emph{CVPR}}.
  \bibinfo{pages}{4460--4470}.
\newblock


\bibitem[\protect\citeauthoryear{Mo, Guerrero, Yi, Su, Wonka, Mitra, and
  Guibas}{Mo et~al\mbox{.}}{2019a}]%
        {mo2019structurenet}
\bibfield{author}{\bibinfo{person}{Kaichun Mo}, \bibinfo{person}{Paul
  Guerrero}, \bibinfo{person}{Li Yi}, \bibinfo{person}{Hao Su},
  \bibinfo{person}{Peter Wonka}, \bibinfo{person}{Niloy Mitra}, {and}
  \bibinfo{person}{Leonidas Guibas}.} \bibinfo{year}{2019}\natexlab{a}.
\newblock \showarticletitle{{StructureNet}: Hierarchical Graph Networks for
  {3D} Shape Generation}.
\newblock \bibinfo{journal}{\emph{ACM Trans. Graph.}} \bibinfo{volume}{38},
  \bibinfo{number}{6} (\bibinfo{year}{2019}), \bibinfo{pages}{242:1--242:29}.
\newblock


\bibitem[\protect\citeauthoryear{Mo, Zhu, Chang, Yi, Tripathi, Guibas, and
  Su}{Mo et~al\mbox{.}}{2019b}]%
        {mo2018partnet}
\bibfield{author}{\bibinfo{person}{Kaichun Mo}, \bibinfo{person}{Shilin Zhu},
  \bibinfo{person}{Angel~X Chang}, \bibinfo{person}{Li Yi},
  \bibinfo{person}{Subarna Tripathi}, \bibinfo{person}{Leonidas~J Guibas},
  {and} \bibinfo{person}{Hao Su}.} \bibinfo{year}{2019}\natexlab{b}.
\newblock \showarticletitle{PartNet: A Large-scale Benchmark for Fine-grained
  and Hierarchical Part-level 3D Object Understanding}. In
  \bibinfo{booktitle}{\emph{CVPR}}. \bibinfo{pages}{909--918}.
\newblock


\bibitem[\protect\citeauthoryear{Niemeyer, Mescheder, Oechsle, and
  Geiger}{Niemeyer et~al\mbox{.}}{2020}]%
        {DVR}
\bibfield{author}{\bibinfo{person}{Michael Niemeyer}, \bibinfo{person}{Lars
  Mescheder}, \bibinfo{person}{Michael Oechsle}, {and} \bibinfo{person}{Andreas
  Geiger}.} \bibinfo{year}{2020}\natexlab{}.
\newblock \showarticletitle{Differentiable Volumetric Rendering: Learning
  Implicit 3D Representations without 3D Supervision}. In
  \bibinfo{booktitle}{\emph{CVPR}}.
\newblock


\bibitem[\protect\citeauthoryear{Oechsle, Mescheder, Niemeyer, Strauss, and
  Geiger}{Oechsle et~al\mbox{.}}{2019}]%
        {OechsleICCV2019}
\bibfield{author}{\bibinfo{person}{Michael Oechsle}, \bibinfo{person}{Lars
  Mescheder}, \bibinfo{person}{Michael Niemeyer}, \bibinfo{person}{Thilo
  Strauss}, {and} \bibinfo{person}{Andreas Geiger}.}
  \bibinfo{year}{2019}\natexlab{}.
\newblock \showarticletitle{Texture Fields: Learning Texture Representations in
  Function Space}. In \bibinfo{booktitle}{\emph{ICCV}}.
  \bibinfo{pages}{4531--4540}.
\newblock


\bibitem[\protect\citeauthoryear{Park, Florence, Straub, Newcombe, and
  Lovegrove}{Park et~al\mbox{.}}{2019}]%
        {park2019-DeepSDF}
\bibfield{author}{\bibinfo{person}{Jeong~Joon Park}, \bibinfo{person}{Peter
  Florence}, \bibinfo{person}{Julian Straub}, \bibinfo{person}{Richard
  Newcombe}, {and} \bibinfo{person}{Steven Lovegrove}.}
  \bibinfo{year}{2019}\natexlab{}.
\newblock \showarticletitle{{DeepSDF}: Learning Continuous Signed Distance
  Functions for Shape Representation}. In \bibinfo{booktitle}{\emph{CVPR}}.
  \bibinfo{pages}{165--174}.
\newblock


\bibitem[\protect\citeauthoryear{Pavllo, Spinks, Hofmann, Moens, and
  Lucchi}{Pavllo et~al\mbox{.}}{2020}]%
        {pavllo2020convolutional}
\bibfield{author}{\bibinfo{person}{Dario Pavllo}, \bibinfo{person}{Graham
  Spinks}, \bibinfo{person}{Thomas Hofmann}, \bibinfo{person}{Marie-Francine
  Moens}, {and} \bibinfo{person}{Aurelien Lucchi}.}
  \bibinfo{year}{2020}\natexlab{}.
\newblock \showarticletitle{Convolutional Generation of Textured 3D Meshes}.
\newblock \bibinfo{journal}{\emph{arXiv preprint arXiv:2006.07660}}
  (\bibinfo{year}{2020}).
\newblock


\bibitem[\protect\citeauthoryear{Qi, Su, Mo, and Guibas}{Qi
  et~al\mbox{.}}{2017a}]%
        {Qi2017cvpr}
\bibfield{author}{\bibinfo{person}{Charles~Ruizhongtai Qi},
  \bibinfo{person}{Hao Su}, \bibinfo{person}{Kaichun Mo}, {and}
  \bibinfo{person}{Leonidas~J. Guibas}.} \bibinfo{year}{2017}\natexlab{a}.
\newblock \showarticletitle{{PointNet}: Deep Learning on Point Sets for {3D}
  Classification and Segmentation.}. In \bibinfo{booktitle}{\emph{CVPR}}.
  \bibinfo{pages}{77--85}.
\newblock


\bibitem[\protect\citeauthoryear{Qi, Yi, Su, and Guibas}{Qi
  et~al\mbox{.}}{2017b}]%
        {Qi2017nips}
\bibfield{author}{\bibinfo{person}{Charles~Ruizhongtai Qi}, \bibinfo{person}{Li
  Yi}, \bibinfo{person}{Hao Su}, {and} \bibinfo{person}{Leonidas~J. Guibas}.}
  \bibinfo{year}{2017}\natexlab{b}.
\newblock \showarticletitle{{PointNet++}: Deep Hierarchical Feature Learning on
  Point Sets in a Metric Space.}. In \bibinfo{booktitle}{\emph{NeurIPS}}.
  \bibinfo{pages}{5105--5114}.
\newblock


\bibitem[\protect\citeauthoryear{Raj, Ham, Barnes, Kim, Lu, and Hays}{Raj
  et~al\mbox{.}}{2019}]%
        {Raj_2019_CVPR_Workshops}
\bibfield{author}{\bibinfo{person}{Amit Raj}, \bibinfo{person}{Cusuh Ham},
  \bibinfo{person}{Connelly Barnes}, \bibinfo{person}{Vladimir Kim},
  \bibinfo{person}{Jingwan Lu}, {and} \bibinfo{person}{James Hays}.}
  \bibinfo{year}{2019}\natexlab{}.
\newblock \showarticletitle{Learning to Generate Textures on 3D Meshes}. In
  \bibinfo{booktitle}{\emph{CVPR Workshops}}.
\newblock


\bibitem[\protect\citeauthoryear{Razavi, van~den Oord, and Vinyals}{Razavi
  et~al\mbox{.}}{2019a}]%
        {vqvae-2}
\bibfield{author}{\bibinfo{person}{Ali Razavi}, \bibinfo{person}{Aaron van~den
  Oord}, {and} \bibinfo{person}{Oriol Vinyals}.}
  \bibinfo{year}{2019}\natexlab{a}.
\newblock \showarticletitle{Generating diverse high-fidelity images with
  vq-vae-2}. In \bibinfo{booktitle}{\emph{NeurIPS}}.
  \bibinfo{pages}{14866--14876}.
\newblock


\bibitem[\protect\citeauthoryear{Razavi, van~den Oord, and Vinyals}{Razavi
  et~al\mbox{.}}{2019b}]%
        {vqvae}
\bibfield{author}{\bibinfo{person}{Ali Razavi}, \bibinfo{person}{A{\"{a}}ron
  van~den Oord}, {and} \bibinfo{person}{Oriol Vinyals}.}
  \bibinfo{year}{2019}\natexlab{b}.
\newblock \showarticletitle{Generating Diverse High-Resolution Images with
  {VQ-VAE}}. In \bibinfo{booktitle}{\emph{Deep Generative Models for Highly
  Structured Data, {ICLR} 2019 Workshop, New Orleans, Louisiana, United States,
  May 6, 2019}}. \bibinfo{publisher}{OpenReview.net}.
\newblock
\urldef\tempurl%
\url{https://openreview.net/forum?id=ryeBN88Ku4}
\showURL{%
\tempurl}


\bibitem[\protect\citeauthoryear{Richter and Roth}{Richter and Roth}{2018}]%
        {Matryoshka}
\bibfield{author}{\bibinfo{person}{Stephan~R. Richter} {and}
  \bibinfo{person}{Stefan Roth}.} \bibinfo{year}{2018}\natexlab{}.
\newblock \showarticletitle{Matryoshka networks: Predicting 3D geometry via
  nested shape layers}. In \bibinfo{booktitle}{\emph{CVPR}}.
\newblock


\bibitem[\protect\citeauthoryear{Ronneberger, Fischer, and Brox}{Ronneberger
  et~al\mbox{.}}{2015}]%
        {ronneberger2015u}
\bibfield{author}{\bibinfo{person}{Olaf Ronneberger}, \bibinfo{person}{Philipp
  Fischer}, {and} \bibinfo{person}{Thomas Brox}.}
  \bibinfo{year}{2015}\natexlab{}.
\newblock \showarticletitle{U-net: Convolutional networks for biomedical image
  segmentation}. In \bibinfo{booktitle}{\emph{International Conference on
  Medical image computing and computer-assisted intervention}}. Springer,
  \bibinfo{pages}{234--241}.
\newblock


\bibitem[\protect\citeauthoryear{Saito, Wei, Hu, Nagano, and Li}{Saito
  et~al\mbox{.}}{2017}]%
        {saito2017photorealistic}
\bibfield{author}{\bibinfo{person}{Shunsuke Saito}, \bibinfo{person}{Lingyu
  Wei}, \bibinfo{person}{Liwen Hu}, \bibinfo{person}{Koki Nagano}, {and}
  \bibinfo{person}{Hao Li}.} \bibinfo{year}{2017}\natexlab{}.
\newblock \showarticletitle{Photorealistic facial texture inference using deep
  neural networks}. In \bibinfo{booktitle}{\emph{CVPR}},
  Vol.~\bibinfo{volume}{3}. \bibinfo{pages}{5144--5153}.
\newblock


\bibitem[\protect\citeauthoryear{Simonyan and Zisserman}{Simonyan and
  Zisserman}{2014}]%
        {simonyan2014very}
\bibfield{author}{\bibinfo{person}{Karen Simonyan} {and}
  \bibinfo{person}{Andrew Zisserman}.} \bibinfo{year}{2014}\natexlab{}.
\newblock \showarticletitle{Very deep convolutional networks for large-scale
  image recognition}.
\newblock \bibinfo{journal}{\emph{arXiv:1409.1556}} (\bibinfo{year}{2014}).
\newblock


\bibitem[\protect\citeauthoryear{Sitzmann, Zollh{\"o}fer, and
  Wetzstein}{Sitzmann et~al\mbox{.}}{2019}]%
        {sitzmann2019srns}
\bibfield{author}{\bibinfo{person}{Vincent Sitzmann}, \bibinfo{person}{Michael
  Zollh{\"o}fer}, {and} \bibinfo{person}{Gordon Wetzstein}.}
  \bibinfo{year}{2019}\natexlab{}.
\newblock \showarticletitle{Scene Representation Networks: Continuous
  3D-Structure-Aware Neural Scene Representations}. In
  \bibinfo{booktitle}{\emph{NeurIPS}}.
\newblock


\bibitem[\protect\citeauthoryear{Snelgrove}{Snelgrove}{2017}]%
        {Snelgrove:2017:ROD:3145749.3149449}
\bibfield{author}{\bibinfo{person}{Xavier Snelgrove}.}
  \bibinfo{year}{2017}\natexlab{}.
\newblock \showarticletitle{High-Resolution Multi-Scale Neural Texture
  Synthesis}. In \bibinfo{booktitle}{\emph{SIGGRAPH ASIA 2017 Technical
  Briefs}} \emph{(\bibinfo{series}{SA '17})}. \bibinfo{publisher}{ACM}.
\newblock
\showISBNx{978-1-4503-5406-6/17/11}


\bibitem[\protect\citeauthoryear{Sun, Huh, Liao, Zhang, and Lim}{Sun
  et~al\mbox{.}}{2018}]%
        {sun2018multiview}
\bibfield{author}{\bibinfo{person}{Shao-Hua Sun}, \bibinfo{person}{Minyoung
  Huh}, \bibinfo{person}{Yuan-Hong Liao}, \bibinfo{person}{Ning Zhang}, {and}
  \bibinfo{person}{Joseph~J Lim}.} \bibinfo{year}{2018}\natexlab{}.
\newblock \showarticletitle{Multi-view to Novel View: Synthesizing Novel Views
  with Self-Learned Confidence}. In \bibinfo{booktitle}{\emph{ECCV}}.
\newblock


\bibitem[\protect\citeauthoryear{Tan, Gao, Lai, and Xia}{Tan
  et~al\mbox{.}}{2018}]%
        {meshvae2017}
\bibfield{author}{\bibinfo{person}{Qingyang Tan}, \bibinfo{person}{Lin Gao},
  \bibinfo{person}{Yu{-}Kun Lai}, {and} \bibinfo{person}{Shihong Xia}.}
  \bibinfo{year}{2018}\natexlab{}.
\newblock \showarticletitle{Variational Autoencoders for Deforming {3D} Mesh
  Models}. In \bibinfo{booktitle}{\emph{CVPR}}. \bibinfo{pages}{5841--5850}.
\newblock


\bibitem[\protect\citeauthoryear{Tulsiani, Zhou, Efros, and Malik}{Tulsiani
  et~al\mbox{.}}{2017}]%
        {drcTulsiani17}
\bibfield{author}{\bibinfo{person}{Shubham Tulsiani}, \bibinfo{person}{Tinghui
  Zhou}, \bibinfo{person}{Alexei~A. Efros}, {and} \bibinfo{person}{Jitendra
  Malik}.} \bibinfo{year}{2017}\natexlab{}.
\newblock \showarticletitle{Multi-view Supervision for Single-view
  Reconstruction via Differentiable Ray Consistency}. In
  \bibinfo{booktitle}{\emph{CVPR}}. \bibinfo{pages}{2626--2634}.
\newblock


\bibitem[\protect\citeauthoryear{Wang, Zhang, Li, Fu, Liu, and Jiang}{Wang
  et~al\mbox{.}}{2018b}]%
        {wang2018pixel2mesh}
\bibfield{author}{\bibinfo{person}{Nanyang Wang}, \bibinfo{person}{Yinda
  Zhang}, \bibinfo{person}{Zhuwen Li}, \bibinfo{person}{Yanwei Fu},
  \bibinfo{person}{Wei Liu}, {and} \bibinfo{person}{Yu-Gang Jiang}.}
  \bibinfo{year}{2018}\natexlab{b}.
\newblock \showarticletitle{Pixel2mesh: Generating 3d mesh models from single
  rgb images}. In \bibinfo{booktitle}{\emph{ECCV}}. Springer,
  \bibinfo{pages}{52--67}.
\newblock


\bibitem[\protect\citeauthoryear{Wang, Sun, Liu, and Tong}{Wang
  et~al\mbox{.}}{2018a}]%
        {Wang2018ocnn}
\bibfield{author}{\bibinfo{person}{Peng-Shuai Wang}, \bibinfo{person}{Chun-Yu
  Sun}, \bibinfo{person}{Yang Liu}, {and} \bibinfo{person}{Xin Tong}.}
  \bibinfo{year}{2018}\natexlab{a}.
\newblock \showarticletitle{Adaptive {O-CNN}: A Patch-based Deep Representation
  of {3D} Shapes}.
\newblock \bibinfo{journal}{\emph{ACM Trans. Graph.}} \bibinfo{volume}{37},
  \bibinfo{number}{6} (\bibinfo{year}{2018}), \bibinfo{pages}{217:1--217:11}.
\newblock


\bibitem[\protect\citeauthoryear{Wu, Zhang, Xue, Freeman, and Tenenbaum}{Wu
  et~al\mbox{.}}{2016}]%
        {3dgan}
\bibfield{author}{\bibinfo{person}{Jiajun Wu}, \bibinfo{person}{Chengkai
  Zhang}, \bibinfo{person}{Tianfan Xue}, \bibinfo{person}{William~T Freeman},
  {and} \bibinfo{person}{Joshua~B Tenenbaum}.} \bibinfo{year}{2016}\natexlab{}.
\newblock \showarticletitle{Learning a probabilistic latent space of object
  shapes via 3d generative-adversarial modeling}. In
  \bibinfo{booktitle}{\emph{NeurIPS}}. \bibinfo{pages}{82--90}.
\newblock


\bibitem[\protect\citeauthoryear{Wu, Wang, Lin, Lischinski, Cohen-Or, and
  Huang}{Wu et~al\mbox{.}}{2019}]%
        {pageSAGnet19}
\bibfield{author}{\bibinfo{person}{Zhijie Wu}, \bibinfo{person}{Xiang Wang},
  \bibinfo{person}{Di Lin}, \bibinfo{person}{Dani Lischinski},
  \bibinfo{person}{Daniel Cohen-Or}, {and} \bibinfo{person}{Hui Huang}.}
  \bibinfo{year}{2019}\natexlab{}.
\newblock \showarticletitle{{SAGNet}: Structure-aware Generative Network for
  3D-Shape Modeling}.
\newblock \bibinfo{journal}{\emph{ACM Trans. Graph.}} \bibinfo{volume}{38},
  \bibinfo{number}{4} (\bibinfo{year}{2019}), \bibinfo{pages}{91:1--91:14}.
\newblock


\bibitem[\protect\citeauthoryear{Xu, Wang, Ceylan, Mech, and Neumann}{Xu
  et~al\mbox{.}}{2019}]%
        {xu2019disn}
\bibfield{author}{\bibinfo{person}{Qiangeng Xu}, \bibinfo{person}{Weiyue Wang},
  \bibinfo{person}{Duygu Ceylan}, \bibinfo{person}{Radomir Mech}, {and}
  \bibinfo{person}{Ulrich Neumann}.} \bibinfo{year}{2019}\natexlab{}.
\newblock \showarticletitle{Disn: Deep implicit surface network for
  high-quality single-view 3d reconstruction}.
\newblock \bibinfo{journal}{\emph{arXiv preprint arXiv:1905.10711}}
  (\bibinfo{year}{2019}).
\newblock


\bibitem[\protect\citeauthoryear{Zhang, Isola, and Efros}{Zhang
  et~al\mbox{.}}{2016}]%
        {zhang2016colorful}
\bibfield{author}{\bibinfo{person}{Richard Zhang}, \bibinfo{person}{Phillip
  Isola}, {and} \bibinfo{person}{Alexei~A Efros}.}
  \bibinfo{year}{2016}\natexlab{}.
\newblock \showarticletitle{Colorful Image Colorization}. In
  \bibinfo{booktitle}{\emph{ECCV}}.
\newblock


\bibitem[\protect\citeauthoryear{Zhang, Isola, Efros, Shechtman, and
  Wang}{Zhang et~al\mbox{.}}{2018}]%
        {zhang2018perceptual}
\bibfield{author}{\bibinfo{person}{Richard Zhang}, \bibinfo{person}{Phillip
  Isola}, \bibinfo{person}{Alexei~A Efros}, \bibinfo{person}{Eli Shechtman},
  {and} \bibinfo{person}{Oliver Wang}.} \bibinfo{year}{2018}\natexlab{}.
\newblock \showarticletitle{The Unreasonable Effectiveness of Deep Features as
  a Perceptual Metric}. In \bibinfo{booktitle}{\emph{CVPR}}.
  \bibinfo{pages}{586--595}.
\newblock


\bibitem[\protect\citeauthoryear{Zhou, Synder, Guo, and Shum}{Zhou
  et~al\mbox{.}}{2004}]%
        {Zhou2004}
\bibfield{author}{\bibinfo{person}{Kun Zhou}, \bibinfo{person}{John Synder},
  \bibinfo{person}{Baining Guo}, {and} \bibinfo{person}{Heung-Yeung Shum}.}
  \bibinfo{year}{2004}\natexlab{}.
\newblock \showarticletitle{Iso-charts: stretch-driven mesh parameterization
  using spectral analysis}. In \bibinfo{booktitle}{\emph{Symposium on Geometry
  processing}}. \bibinfo{pages}{45--54}.
\newblock


\bibitem[\protect\citeauthoryear{Zhou, Zhu, Bai, Lischinski, Cohen-Or, and
  Huang}{Zhou et~al\mbox{.}}{2018}]%
        {TexSyn18}
\bibfield{author}{\bibinfo{person}{Yang Zhou}, \bibinfo{person}{Zhen Zhu},
  \bibinfo{person}{Xiang Bai}, \bibinfo{person}{Dani Lischinski},
  \bibinfo{person}{Daniel Cohen-Or}, {and} \bibinfo{person}{Hui Huang}.}
  \bibinfo{year}{2018}\natexlab{}.
\newblock \showarticletitle{Non-stationary Texture Synthesis by Adversarial
  Expansion}.
\newblock \bibinfo{journal}{\emph{ACM Trans. Graph.}} \bibinfo{volume}{37},
  \bibinfo{number}{4} (\bibinfo{year}{2018}).
\newblock


\bibitem[\protect\citeauthoryear{Zhu, Zhang, Zhang, Wu, Torralba, Tenenbaum,
  and Freeman}{Zhu et~al\mbox{.}}{2018}]%
        {VON}
\bibfield{author}{\bibinfo{person}{Jun-Yan Zhu}, \bibinfo{person}{Zhoutong
  Zhang}, \bibinfo{person}{Chengkai Zhang}, \bibinfo{person}{Jiajun Wu},
  \bibinfo{person}{Antonio Torralba}, \bibinfo{person}{Joshua~B. Tenenbaum},
  {and} \bibinfo{person}{William~T. Freeman}.} \bibinfo{year}{2018}\natexlab{}.
\newblock \showarticletitle{Visual Object Networks: Image Generation with
  Disentangled 3{D} Representations}. In \bibinfo{booktitle}{\emph{NeurIPS}}.
  \bibinfo{pages}{118--129}.
\newblock


\end{thebibliography}

\end{document}